\documentclass[english,a4paper,11pt]{article}
\pdfoutput=1
\usepackage{jheppub}
\usepackage{xcolor}
\usepackage{float}
\usepackage{amsmath}
\usepackage{amssymb}
\usepackage{graphicx}
\usepackage{mwe}
\usepackage{subcaption}
\usepackage{bbm}
\usepackage{bbold}

\newcommand{\rd}{\mbox{d}}

\newcommand{\bw}{\begin{widetext}}
\newcommand{\ew}{\end{widetext}}

\def\nn{\nonumber\\}

\newcommand{\Ha}{{\cal H}}
\newcommand{\Hh}{{{\cal H}_h}}
\newcommand{\Hl}{{{\cal H}_l}}

\newcommand\be            {\begin{equation}}
\newcommand\ee            {\end{equation}}



\title{\boldmath Approaching the Self-Dual Point of \\
  the Sinh-Gordon model}
\author[a]{Robert Konik,}
\author[b,c]{M\'arton L\'ajer}
\author[d]{and Giuseppe Mussardo}

\affiliation[a]{CMPMS Dept., Bldg. 734, Brookhaven National Laboratory, Upton, NY 11973-5000, USA}
\affiliation[b]{Wigner Research Centre for Physics, Konkoly-Thege Mikl\'os u. 29-33, 1121 Budapest, Hungary}
\affiliation[c]{Institute for Theoretical Physics, Roland E\"otv\"os University, P\'azm\'any s\'et\'any 1/A, 1117 Budapest, Hungary}
\affiliation[d]{SISSA and INFN, Sezione di Trieste, via Bonomea 265, I-34136, Trieste, Italy}
\emailAdd{rmk@bnl.gov}
\emailAdd{lajerm@caesar.elte.hu}
\emailAdd{mussardo@sissa.it}
\abstract{One of the most striking but mysterious properties of the sinh-Gordon model (ShG) is the $b \rightarrow 1/b$ self-duality of its $S$-matrix, of which there is no trace in its Lagrangian formulation. Here $b$ is the coupling appearing in the model's eponymous hyperbolic cosine present in its Lagrangian, $\cosh(b\phi)$.  In this paper we develop truncated spectrum methods (TSMs) for studying the sinh-Gordon model at a finite volume as we vary the coupling constant. We obtain the expected results for $b \ll 1$ and intermediate values of $b$, but as the self-dual point $b=1$ is approached, the basic application of the TSM to the ShG breaks down. We find that the TSM gives results with a strong cutoff $E_c$ dependence, which disappears according only to a very slow power law in $E_c$.  Standard renormalization group strategies -- whether they be numerical or analytic -- also fail to improve upon matters here.  We thus explore three strategies to address the basic limitations of the TSM in the vicinity of $b=1$. In the first, we focus on the small-volume spectrum. We attempt to understand how much of the physics of the ShG is encoded in the zero mode part of its Hamiltonian, in essence how `quantum mechanical' vs `quantum field theoretic' the problem is. In the second, we identify the divergencies present in perturbation theory and perform their resummation using a supra-Borel approximate. In the third approach, we use the exact form factors of the model to treat the ShG at one value of $b$ as a perturbation of a ShG at a different coupling. In the light of this work, we argue that the strong coupling phase $b > 1$ of the Lagrangian formulation of model may be different from what is na\"ively inferred from its $S$-matrix.  In particular, we present an argument that the theory is massless for $b>1$.} 

\begin{document}
\maketitle
\flushbottom

\newpage
\section{Introduction}

The sinh-Gordon model (ShG) is a canonical quantum integrable field theory.  It has a number of different descriptions, but in this work we are going to take as a starting point the Lagrangian formulation of the model given by
\be
{\cal L}_{ShG}\,=\, \frac{1}{16\pi}\left(\partial_\mu \phi\right)^2 - 2\mu  \cosh(b\phi)\,\,\, .
\label{lagrangian}
\ee
Here $\phi(x,t)$ is a real non-compact scalar field, $\mu$ is some dimensionful mass scale and $b$ is a dimensionless coupling constant. Upon quantization, $\mu$ is replaced by a renormalized coupling constant depending on the chosen quantization scheme. The spectrum of the model is exceedingly simple,  consisting of a single massive particle of mass $M_{ShG}$.  This makes the ShG the simplest of {\it interacting} integrable field theories. Much is known about its properties. Its elastic $S$-matrix was 
found in \cite{Arinshtein:1979pb}.  The form factors of local operators were obtained in \cite{Fring:1992pt,Koubek:1993ke}, while the vacuum expectation values of 
exponential operators were found in \cite{Fateev:1997yg}.  The exact relationship between the physical mass $M_{ShG}$ and the renormalized coupling in the perturbed gaussian CFT scheme, here denoted later as $\mu_{ShG}$, as a function of the coupling $b$,
was derived in \cite{Zamolodchikov:1995xk}. Its thermodynamic Bethe ansatz for the ground state and for the excited states was studied in 
\cite{Zamolodchikov:2000kt,Teschner:2007ng}, while the thermal correlation functions of the model were discussed in \cite{Leclair:1999ys,Lukyanov:2000jp,Negro:2013wga}.  A suggestive 
connection between the ShG model and roaming renormalization group trajectories among the minimal models of CFT was studied in \cite{Zamolodchikov:1991pc}, while a direct mapping between the ShG and the Ising model was established in \cite{Ahn:1993dm}. 
Furthermore, beyond simply being a model that is amenable to analytic manipulation, the ShG finds applications in a wide range of areas of physics running from toy models of quantum gravity \cite{Larsen:1996gn}, to cold atomic gases \cite{Kormos:2009yp,Bastianello:2018bbe}, studies of thermalization in classical field theories \cite{DeLuca:2016etx,DelVecchio:2020lxf},
and lattice models with non-compact quantum group symmetries \cite{Bytsko:2006ut}. 
It is also worth stressing that the ShG model is the simplest example of Toda field theories, a large class of models with exponential interactions based on root systems of Lie algebras, see for instance \cite{Mussardo:2020rxh} and references therein. The main difference between the ShG model and the rest of the Toda field theories is that the ShG does not have bound states. 

One of the most striking but mysterious aspects of the ShG model\footnote{The Toda field theories have a similar duality.} is its apparent weak-strong duality:
\be
b \leftrightarrow 1/b \,\,. 
\label{duality}
\ee
In the presence of such a symmetry, the self-dual point $b = 1$ clearly emerges as a special value of the ShG model, for it divides the weak-coupling regime, $ b < 1$, from the strong coupling regime, $b > 1$. It is important to underline that this duality is not at all manifest in the Lagrangian of the theory but is apparent, as discussed later, 
in its S-matrix formulation. It is the primary aim of this paper to develop truncated spectrum methods (TSMs) in order to study the model at finite volume by varying its coupling constant in the vicinity of this self-dual point. For reasons which will become clear later, the obvious regime in which these methods can be implemented is the weak-coupling regime $b <1$, but we can extend them to approach the $b=1$ self-dual point. As we shall see, at $b=1$ the physical mass vanishes and the theory appears to be (at least naively) critical. Ultimately we aim to explore the ShG at $b=1$ and understand what theory is described by the Lagrangian given in eq.\,(\ref{lagrangian}). The theory's duality, as expressed in eq.~(\ref{duality}), is built on results established at $b<1$ which are then subsequently analytically continued to regimes beyond their nominal validity.  It is then an important question to understand whether the Lagrangian corresponding to these analytic continuations is the same as given in eq.~(\ref{lagrangian}) with $b>1$. 

As we explore in this paper, the development of TSM for the ShG nearby the self-dual point  $b=1$  proves to be surprisingly challenging.  
Truncated spectrum methods treat a model by firstly defining it on a finite volume (typically an infinitely long cylinder of width $R$) and then, secondly, introducing a hard UV cutoff, $E_c$, in the number of energy levels which are included in the computation \cite{Yurov:1989yu,Yurov:1991yu,LASSIG1991591,PhysRevLett.98.147205,tsm_review,Coser:2014lla,Bajnok:2015bgw,PhysRevD.91.085011,Rychkov:2015vap,PhysRevD.96.065024,Elias-Miro2017}. Under these two conditions, numerics can be performed (either exact diagonalization or Lanzcos based approaches) and the low lying energy spectrum, together with vacuum expectation values  and matrix elements of several operators, can be computed.  Of course, in this treatment it is crucial to understand the effect of $E_c$ on the computed results. For certain models, even small values of $E_c$ lead to results that are, in effect, independent of the cutoff (i.e. the $c=1/2$ Ising model perturbed by the presence of a magnetic field being a classic example \cite{Yurov:1991yu}). In other cases, as for instance those analysed in Refs.~\cite{LASSIG1991591,LASSIG1991666,konik2011exciton,konik2015predicting,BERIA2013457,KONIK2015547,AzariaPRD16},  the results are instead noticeably affected by $E_c$ and, to ameliorate the effects of the introduction of $E_c$, various renormalization group (RG) strategies have been employed, both analytic and numerical \cite{PhysRevLett.98.147205,feverati2006renormalisation,giokas2011renormalisation,Lencses:2014tba,tsm_review,PhysRevD.91.085011,Rychkov:2015vap, PhysRevD.96.065024,Elias-Miro2017}. 

The premise of all these RG strategies is that cutoff dependent effects are in some sense small. However, in the case of the ShG model, we shall see that such cutoff effects near $b=1$ can be on the contrary extremely large and therefore the traditional RG strategies do not work. In order to deal with this new situation, we propose herein three different approaches to tackle the problem:

\vskip 10pt

\noindent 1. In the first, we explore more carefully the small-volume regime and its `quantum mechanical nature'. In particular, one may expect that the UV behaviour of the spectrum is dominated by the quantum mechanics of the zero mode of the field. By using either this quantum mechanical picture or the TBA equations combined with the mass-coupling relation, it is possible to derive a systematic expansion for certain energy levels (more precisely, their corresponding scaling functions) in terms of $(\ln(\mu_{ShG} R))^{-1}$. The two expansions are however different in subleading orders. As the energies contain an additional $R^{-1}$ factor relative to the scaling function, the difference between TBA and zero mode energy levels eventually diverge for $R\rightarrow 0$ for all $b>0$. We derive an effective potential, partially taking into account the effect of oscillators. At the one hand, we analytically reproduce the exact expansion up to $O(b^{12})$, confirming that the oscillators are able to explain the differences in the log-expansion. On the other hand, we  show that TSM numerics significantly outperforms even the numerical solution of the complete effective potential. We then use this fact to provide a more precise measurement of the IR parameters from TSM, combining UV numerics with the small-volume expansion of TBA.

\vskip 10pt

\noindent 2. In the second strategy, we recast the analytic RG strategy used to remove the effect of the cutoff.  Typically this RG strategy is pursued by initially performing low order perturbation theory in the conformal coupling, here $\mu_{ShG}$. However this fails for the ShG near $b=1$ as the perturbation theory of this model is divergent term by term. These divergences, we show, actually appear for any value of $b$, although for small values of $b$ their appearance is delayed until higher orders.  Facing this, we argue that the diverging perturbative series can be resummed.  However, this is not a Borel resummation per se, as the series is diverging more rapidly than $n!$, but it does nonetheless admit a supra-Borel resummation.  

\vskip 10pt

\noindent 3. In the third and last strategy, we abandon the use of the non-compact boson Hilbert space as a computational basis.  One way to understand the difficulties in using TSM's about $b=1$ is to think of them as arising due to a poor choice of computational basis.  We have already said that the theory becomes critical at $b=1$ (i.e. the mass scale $M_{ShG}$ vanishes at fixed $\mu_{ShG}$). Hence, using a non-interacting field to describe the vicinity of what it could be a non-trivial conformal field theory (presumably strongly interacting) may then simply be inappropriate.  Thus we explore the possibility of using an interacting basis of states as a computational basis.  The natural choice here is to use, as a computational basis, the basis of exact eigenstates of the ShG at one value of $b$ to study the theory at a different (relatively close) value of $b$.

\vskip 10pt

The paper is organized as follows.  In Section \ref{reviewSHG}, we review basic information on the ShG model, pointing out its origin and possible limitations. Although this section reviews previously known results, it is crucial for understanding the rest of the paper. In Section \ref{TSMGENERAL} we discuss truncated spectrum methods, in particular the key role played by the choice of computational basis. In Section \ref{TSMCFT} we then present our particular choice of computational basis.  In Section \ref{TSMresults}, we discuss our numerical results for various quantities including the finite volume spectrum, the $S$-matrix, and the vacuum expectation values of various exponential operators of the model. In Section \ref{TSMresults} we further demonstrate how the standard renormalization group techniques used to improve TSM results fail to do so for the ShG model close to the self-dual point, thus setting up the rationale for the next three sections. In Section \ref{QMreductionssection} we explore in detail the information carried by the zero modes of the theory and the `quantum-mechanical' nature of the ShG model in certain regimes of the coupling and volume. In Section \ref{Supraborelsection} we analyse the nature of perturbation theory which defines the ShG model as a massive deformation of a Gaussian theory and we argue that the perturbative series is badly behaved and is non-Borel resummable. This leads us to consider a supra-Borel resummation in order to give meaning to these divergent sums. In Section \ref{FFTCSASECTION} we come back to the issue of a proper choice of the basis for the TSM and we explore the possibility to study the ShG model at a given coupling $b$ in terms of states and matrix elements of a ShG model defined at a different value of $b$. As we will see, this approach admits of a series of sanity checks. In Section \ref{Conclusionssection} we finally discuss our conclusions and future directions.

\section{Basic Features of the Sinh-Gordon model}\label{reviewSHG}

In this section we briefly review the basic properties of the ShG necessary to understand the TSM results and their interpretation presented in the main body of the paper. 
The scale dimension of the renormalized counterpart of the coupling $\mu$ appearing in the ShG Lagrangian depends on the quantization scheme of the model. 
Hereafter we are going to discuss three such schemes:  (i) a perturbative scheme based on Feynman diagrams; (ii) treating the theory on the same grounds as its analytically continued cousin, the sine-Gordon model, namely as a perturbed Gaussian CFT; and  (iii) as a perturbation of a Liouville quantum field theory.  The model's self-duality is often encoded in the parameter 
\be
Q=b+b^{-1},
\ee
which we record here for the reader to emphasize its importance.

\begin{figure}[b]
\centering
\includegraphics[width=0.55\textwidth]{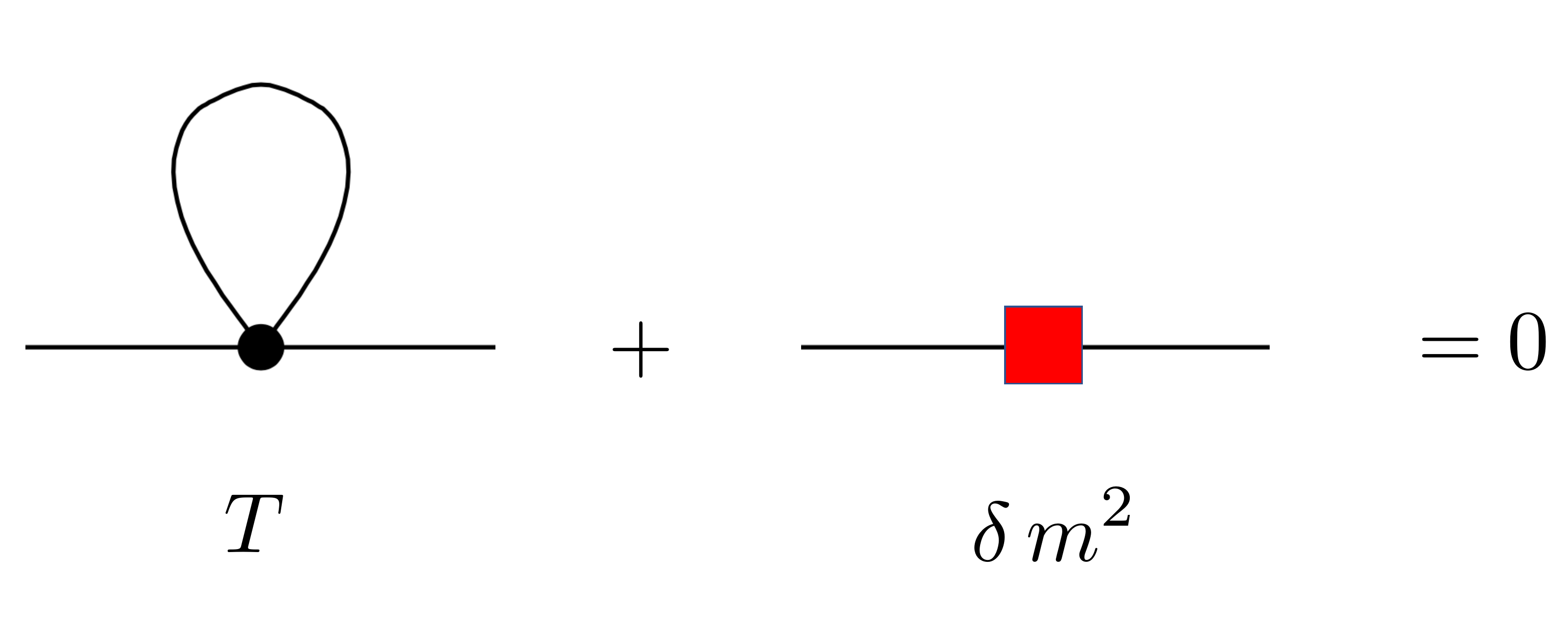}
\caption{Diagrams for the first order tadpole diagram $T$ and the corresponding relative mass counterterm $\delta m^2=-T(m_0)$.}
\label{tadpole}
\end{figure}

\subsection{Feynman Diagrammatic Analysis}
\begin{figure}[t]
\centering
\includegraphics[width=0.65\textwidth]{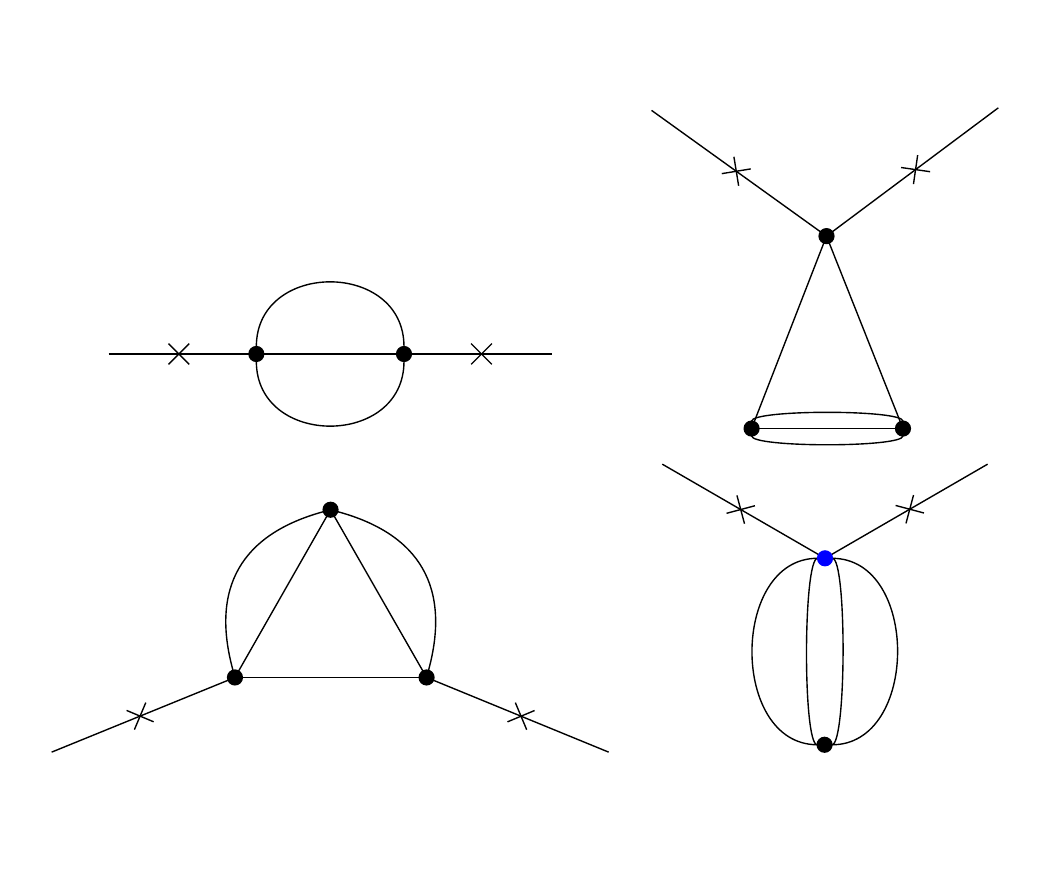}
\caption{Feynman diagrams up to order $b^6$ entering the expansion of $M_{ShG}^2$. The $\phi^6$ vertex is distinguished by a blue dot.}
\label{Feynm}
\end{figure}

In the first scheme, the ShG model is considered by employing perturbation theory in the coupling constant $b$ and evaluating all quantities in terms of Feynman diagrams.  This can be done by introducing a momentum cutoff $\Lambda$ and expanding the potential of the theory in terms of $b$:
\be
2\mu\, \cosh (b \phi) \,=\,2 \mu(\bar\mu,\Lambda,b) \left(1 + \frac{b^2}{2} \phi^2 + \frac{b^4}{4!} \phi^4+ \cdots + \frac{b^{2n}}{(2 n)!} \phi^{2n} + \cdots \right).
\label{expansionb}
\ee
Here $\mu(\bar\mu,\Lambda,b)$ is a bare parameter of dimension mass squared. Using the cutoff $\Lambda$, we have introduced a renormalized coupling $\bar\mu$, which we aim to keep fixed as we tune $\mu$ such that the physical quantities are finite:
\be
\mu(\bar\mu,\Lambda,b)=\bar\mu+O(b^2\log(\Lambda))
\ee
Above the unique ground state of the theory, there is a massive excitation, whose mass at the lowest order in $b$ is $m_0$ given by
\be
m_0^2 \,=\, 16 \pi\, b^2\,\bar\mu \,\,\,. 
\label{lowestmass}
\ee
Of course the actual mass of the particle will get corrections by all the higher order interactions. However, the perturbative series contains divergences. Fortunately, in 1+1d theories with local interactions all divergences come from the tadpole diagrams. These divergences can be cured by introducing a mass counter-term $\delta m^2$ and imposing that, order by order, $\delta m^2$ cancels the infinities coming from the tadpole diagrams.  At the lowest order in $b^2$, for instance, we have the condition expressed in Fig.~\ref{tadpole}, where the tadpole is regularized in terms of the momentum cutoff $\Lambda$ as
\be
T(m_0) \,=\,(8\pi)^2 b^4 \bar\mu\intop_{-\Lambda}^{\Lambda}\frac{dk_1}{2\pi}\intop_{-\infty}^{\infty}\frac{dk_0}{(2 \pi)} \frac{1}{k^2 + m_{0}^2} \,=\,(8\pi)^2 b^4 \bar\mu\frac{1}{2\pi}\log\left(\frac{\Lambda}{m_{0}}+\sqrt{1+\frac{\Lambda^2}{m_{0}^2}}\right).
\ee
The counterterm $\delta m^2=-T(m_{IR})$, involving in general an arbitrary mass scale $m_{IR}$, is absorbed by the bare parameter $\mu$ such that
\be
\mu(\bar\mu,\Lambda,b)=\bar\mu+\frac{\delta m^2}{16\pi b^2}+O(b^4\log^2\Lambda) .\label{mutoorder2} 
\ee
This prescription is equivalent to defining a normal ordering for the Lagrangian (eq.~(\ref{lagrangian})). 
The quantization scheme is fixed by the choice of $m_{IR}$. In particular, setting $m_{IR}=m_0$ leads to the usual scheme of a perturbed massive boson, where normal ordering is with respect to the free mass $m_0$, eliminating altogether the tadpole diagrams at each order.
The exact relation between $\mu$ and $\bar\mu$ in the normal ordering scheme $m_{IR}$ is easily obtained by means of the Baker-Campbell-Hausdorff formula. It reads
\be
\mu(\bar\mu,\Lambda,b)=\bar\mu\left(\frac{\Lambda}{m_{IR}}+\sqrt{1+\frac{\Lambda^2}{m_{IR}^2}}\right)^{-2b^2}.
\ee
In this way,
all $n$-point correlation functions of the theory are finite to all orders in perturbation theory. In particular, one can compute the physical mass $M_{ShG}$ of the theory, as a function of $m_0$ and $b^2$, by looking at the pole of the 2-point correlation function. In the scheme $m_{IR}=m_0\equiv m$, we obtain 
\be M_{ShG}^2\,=\, 
 m^2 \,\left(1 - \frac{b^4}{384 g^2} + \frac{b^6}{g^3} \left(\frac{1}{1536 \pi} + \frac{7 \,\zeta(3)}{3072 \pi^3} - \frac{14 \,\zeta(3)}{3072\pi^3}\right)\right) + {\mathcal O}\left(\frac{b^8}{g^4}\right),
 \label{perturbativeexpansionM}
 \ee
where $g = \frac{1}{8\pi}$
 and each term comes from the Feynman diagrams of Fig.~\ref{Feynm}.
We will point out later that this perturbative analysis is consistent with an exact formula for the mass that we present in Section \ref{subsubShGAnalyticCont}.

We note that at ${\cal O}(b^4)$, the ShG coincides with a $\phi^4$ Landau-Ginzburg model.  Given the repulsive nature of this latter theory, the ShG is expected to have no bound states, its spectrum consisting of multi-particle states of the same particle. As we will see shortly, this conclusion is in agreement with the exact $S$-matrix of the model.

\subsection{Relation with the Sine-Gordon Model}
In the second approach, properties of the ShG model are extracted from a closely related model, the sine-Gordon (SG) model.  The SG model has a Lagrangian given by
\be
{\cal L}_{SG}\,=\, \frac{1}{16\pi}\left(\partial_\mu \phi\right)^2 
+ 2 \mu \, \cos(b\phi)\,\,\, .
\ee
This can be obtained from the ShG Lagrangian (\ref{lagrangian}) by making the substitutions 
\be
\begin{array}{l}
b \rightarrow i b \,\,\,,\\
\mu \rightarrow - \mu \,\,\,.
\end{array}
\label{transformations}
\ee
It is important to stress that, although the two theories are related by this simple transformation, their underlying nature is rather different and there are indeed a series of hidden subtleties behind the innocent looking analytic continuation (\ref{transformations}), some of which are discussed below. 
 
\subsubsection{SG and ShG Models as Deformations of a Gaussian Theory}
Both theories may be regarded as deformations of the Gaussian fixed point action given by the kinetic term in eq.\,(\ref{lagrangian}) 
\be
{\cal A}_0 \,=\,\int \, \frac{1}{16 \pi} (\partial_\mu \phi)^2 \, d^2 x \,\,\,.
\label{gaussianfixedpoint}
\ee  
With respect to this CFT of central charge $c=1$, the chiral conformal dimension of a vertex operator, $V(a) = e^{i a \phi}$, is $\Delta(a) = a^2$. The sine-Gordon model involves the vertex operators $V(\pm b) = e^{\pm i \, b \phi}$ which are compact and bounded, while the sinh-Gordon model employs the vertex operators $V(\mp i b) = e^{\pm b \phi}$ which are instead non-compact and unbounded.  Moreover, while in the sine-Gordon model the conformal dimensions of the vertex operators are positive and given by 
\be
\Delta_{\pm b} \,=\, b^2 \,\,\,,
\label{deltaSG}
\ee 
in the sinh-Gordon model they are instead {\em negative} and given by
\be
\Delta_{\pm i  b} \,=\, - b^2 \,\,\,.
\label{deltaSh}
\ee   
How the sinh-Gordon model turns out to be a unitarity quantum field theory, despite the negative conformal dimension of its basic vertex operators, is one of the 
remarkable aspects of this model.  The way the theory restores its unitarity is through the existence of non-zero vacuum expectation values (VEV), whose exact values are provided in eq.\,(\ref{VEVexp}) below.  With $x_{12} = x_1 - x_2$, consider for instance the operator product expansion (OPE) with respect to the Gaussian fixed point:
\be
\cosh (b \phi(x_1)) \, \cosh (b \phi(x_2)) \,=\,  \frac{1}{|x_{12}|^{4 b^2}} \, \cosh (2 b \phi(x_2)) + | x_{12}|^{4 b^2}\mathbb{1} + \cdots ,
\label{OPE}
\ee
and taking the vacuum expectation value of both terms of this equation, we have 
\be   
\langle \cosh (b \phi(x_1)) \, \cosh (b \phi(x_2)) \rangle \simeq \frac{\langle  \cosh (2 b \phi(0)) \rangle}{|x_{12}|^{4 b^2}} + | x_{12}|^{4 b^2} + \cdots .
\label{effecOPE}
\ee
Hence, if $ \langle  \cosh (2 b \phi(0)) \rangle \neq 0$, we see that the two-point function of the vertex operators $e^{\pm b \phi}$ has effectively the {\em same} leading short-distance singularity as it would have in the case of a {\em positive} $\Delta = b^2$ conformal dimension.  

\subsubsection{Coleman Bound in SG and Its Formal Absence in ShG}

From a renormalization group point of view, the vertex operators which give rise to the sine-Gordon model are relevant operators for $b^2 \leq 1$, where the upper value $b^2=1$ is known in the literature as Coleman's bound \cite{Coleman:1974bu}. The values $0 \leq b^2 \leq 1$
are those for which the SG is ultraviolet stable (i.e. we do not need extra non-trivial counter-terms in its Lagrangian to cure its ultraviolet divergencies). As we already know, the only divergences come from the tadpoles, which can be absorbed by a normal ordering prescription under which the vertex operators get renormalized multiplicatively. Defining $m_{IR}$ as the mass scale by which normal ordering is defined and using $a^{-1}$ as the UV cutoff, the multiplicative renormalization appears as
\be
e^{\pm i b \phi} \,=\, \left[\left(\frac{m_{IR}a}{2}\right)^{2b^2}+O(a^2)\right] \, : \, e^{\pm i b \phi} \, :_{m_{IR}}.
\label{normalorder1}
\ee
When $b^2 > 1$, the vertex operators are irrelevant: hence the SG model becomes essentially a massless theory 
\cite{Amit:1979ab}. 

In the ShG model, the renormalization of the operators (or the coupling) occurs with the {\em inverse factor} of the SG model
 \be
e^{\pm i b \phi} \,=\, \left[\left(\frac{2}{m_{IR}a}\right)^{2b^2}+O(a^2)\right] \, : \, e^{\pm i b \phi} \, :_{m_{IR}}.
\label{normalorder2}
\ee 
Typically in this multiplicative renormalization the power of $a$ that arises is absorbed into the bare coupling $\mu$ so defining a renormalized dimensionful parameter $\mu_{SG/ShG}$:
\begin{equation}\label{renormmu}
\mu_{SG/ShG} = \mu a^{\pm 2b^2},
\end{equation}
where $+$ is for the SG theory and $-$ is for the ShG model.  $\mu_{SG/ShG}$ then has engineering dimension in the two theories of $2\mp 2b^2$.  The scale $m_{IR}$ that appears in this multiplicative renormalization is then typically absorbed into the definition of the normal ordered vertex operator so that the OPE has the conventions expressed in eq.~(\ref{OPE}). Henceforth it is understood as part of the definition of $\mu_{SG/ShG}$ that $m_{IR}$ is chosen this way.  The relation between the free mass $m$ appearing in eq. \eqref{perturbativeexpansionM} and the coupling $\mu_{ShG}$ is \cite{BLSV} 
\be
\mu_{ShG} \,=\, \frac{m^{2 + 2 b^2}}{2^{4 + 2 b^2}\pi b^2} \, e^{2 b^2\gamma_E}\,\,\, ,
\label{lambdamm}
\ee
as derived in Appendix \ref{AppRelatingSchemes}.

Because of the negative conformal dimension of its vertex operators (which makes them relevant operators), at least formally the ShG model does not have a Coleman bound. However,  according to the argument given above, the singularity structure of the OPE for the ShG interaction (eq.~\ref{effecOPE}) is the same as for the SG model.  One thus may suspect that there is in fact a Coleman bound for the ShG model, namely that the theory is properly defined only for $b^2 < 1$, has a singularity at $b^2=1$ and a massless phase for $b^2 > 1$. This is the scenario we will actually present later in the paper.

\subsubsection{The Spectrum of SG model} 

Let us now turn our attention to the spectrum of the SG model.  This quantity is key as the spectrum and S-matrix of the SG model will be connected to that of the ShG model by analytic continuation.  Reproducing this spectrum will be one of the major targets of our TSM studies.

\begin{figure}[t]
\centering
\begin{subfigure}[b]{0.475\textwidth}
\centering
\includegraphics[draft=false,width=\textwidth]{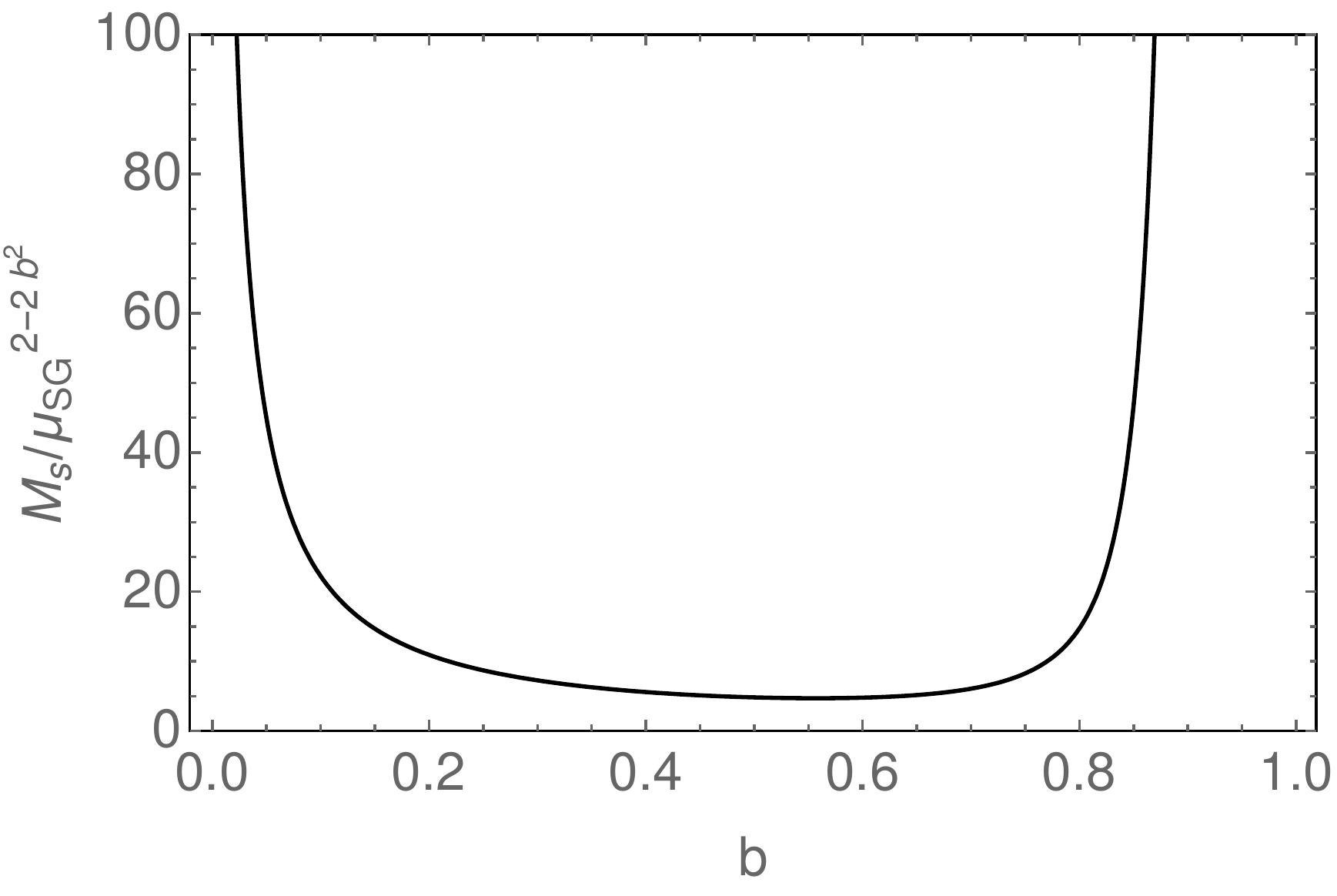}
\caption{Mass of the sine-Gordon soliton as a function of $b$.}
\end{subfigure}
\hfill
\begin{subfigure}[b]{0.475\textwidth}
\centering
\includegraphics[width=\textwidth]{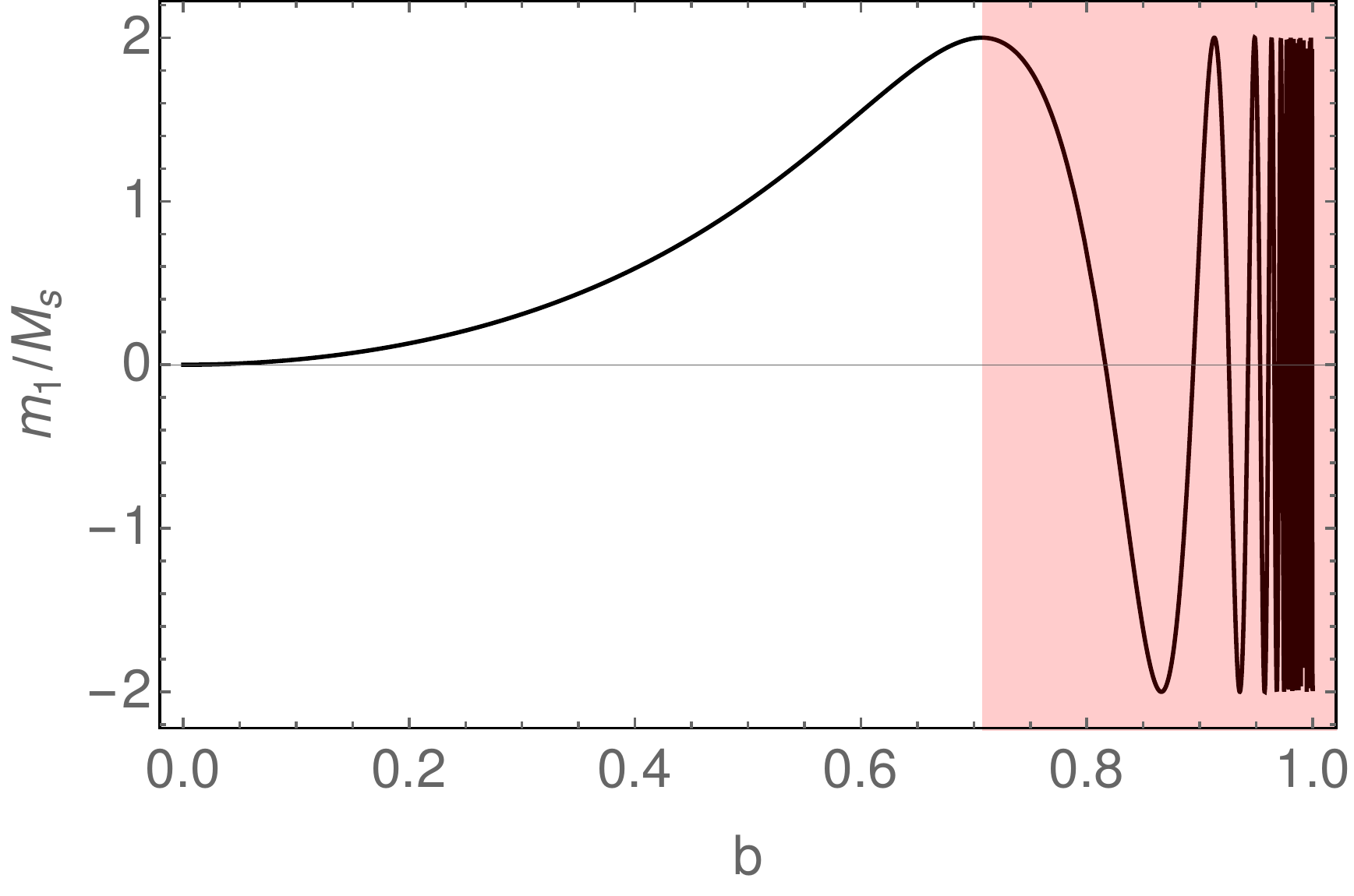}
\caption{Mass of first SG breather vs $b$. The non-physical region is highlighted in red.}
\end{subfigure}
\caption{Masses in the sine-Gordon model.}
\label{SGMASSES}
\end{figure}

We note that this analytic continuation is subtle.   While the sinh-Gordon model has only one vacuum state, the sine-Gordon model has instead an infinite number of vacuum states, $|n \rangle$, which are associated to the minima of the potential, $\phi_n = 2\pi n /b$.   These multiple vacua give rise to solitons and anti-solitons, excitation which interpolate between two neighboring vacua, $| n \rangle $ and $| n \pm 1 \rangle$. For the integrability of the theory, scattering among solitons and anti-solitons is elastic and the relative amplitudes can be computed exactly \cite{Zamolodchikov:1978xm}. Here it is sufficient to remind the reader of the main results
of this analysis. It is convenient to define 
\be
\xi = \frac{b^2}{1-b^2} \,\,\, ,
\label{xi}
\ee
as this parameter controls the spectrum of the SG theory.  The number of neutral soliton-anti-soliton bound states (breathers) is given by 
\be
N \,=\, \left[\frac{1}{\xi}\right],
\ee
where  $\left[x\right]$ denotes the integer part of $x$. Denoting by $M_s$ the mass of the soliton, the breather masses are given by
\be
m_n \,=\,2 M_s \, \sin\left(n \frac{\pi \xi}{2} \right) ,~~~ n = 1, 2, \ldots < \frac{1}{\xi} .\label{massbreather}
\ee
Hence, the first breather exists provided $\xi \leq 1$, namely only in the range $b^2 \leq 1/2$. Ignoring this restriction, we plot $m_1(b)$ vs $b$ for the entire interval $(0,1)$ (see  Fig.~\ref{SGMASSES}.b). Notice that even though the breather does not exist for $b^2> 1/2$, its mass remains positive until $b^2 = 2/3$. After this, its value turns negative and begins to rapidly oscillate, reflecting its possession of an essential singularity at $b^2=1$.

The mass scale $M_s$ can be related to the renormalized coupling of the theory $\mu_{SG}$ (as defined in eq.~(\ref{renormmu})).  In the SG model the ground state energy in finite volume and in the presence of an external field coupled to the topological charge of the model can be computed in two different ways: using the thermodynamic Bethe ansatz (TBA) and using conformal perturbation theory. The former approach employs the physical mass $M_s$ while the latter, the renormalized mass scale $\mu_{SG}$. Comparing the results coming from the two different approaches,  Al. Zamolodchikov \cite{Zamolodchikov:1995xk} was able to obtain an exact formula encoding the $\mu_{SG}-M_s$ relation:
 \be 
 M_s\,=\,\frac{2 \Gamma\left(\frac{\xi}{2}\right)}{\sqrt{\pi} \Gamma\left(\frac{1}{2} + \frac{\xi}{2}\right)} 
 \,\left(\mu_{SG} \, \frac{\pi \Gamma(1-b^2)}{\Gamma(b^2)}\right)^{\frac{1}{2-2 b^2}}
 \,\,\,.
 \label{massformulaSG}
 \ee
We see that this formula is consistent with $\mu_{SG}$ in the SG model having engineering dimension $2-2b^2$.  It is also important to stress that this formula assumes the vertex operators are normalized with the convention of eq.~(\ref{OPE}).

This formula is physical in the interval $0 \leq b^2 \leq 1$.  It has an essential singularity when $b^2 \rightarrow 1$ (i.e. $\xi \rightarrow \infty$)
\be
M_s \,\simeq \, 2\sqrt{2}e^{\frac12-\gamma_E}\mu_{SG}^{\frac{\xi+1}{2}}(\pi\xi)^\frac{\xi}{2}e^{\frac{1}{4\xi}}, ~~~ b^2 \rightarrow 1.
\ee 
It also diverges when $b^2 \rightarrow 0$ as 
\be
M_s \,\simeq\, 4 \sqrt{\frac{\mu_{SG}}{\pi}} \, \frac{1}{b}, ~~~ b^2 \rightarrow 0.
\ee
Its behaviour in the interval $0 \leq b^2 \leq 1$ is shown on  Fig.~\ref{SGMASSES}a.

\begin{figure}[t]
\centering
\begin{subfigure}[b]{0.475\textwidth}
\centering
\includegraphics[draft=false,width=\textwidth]{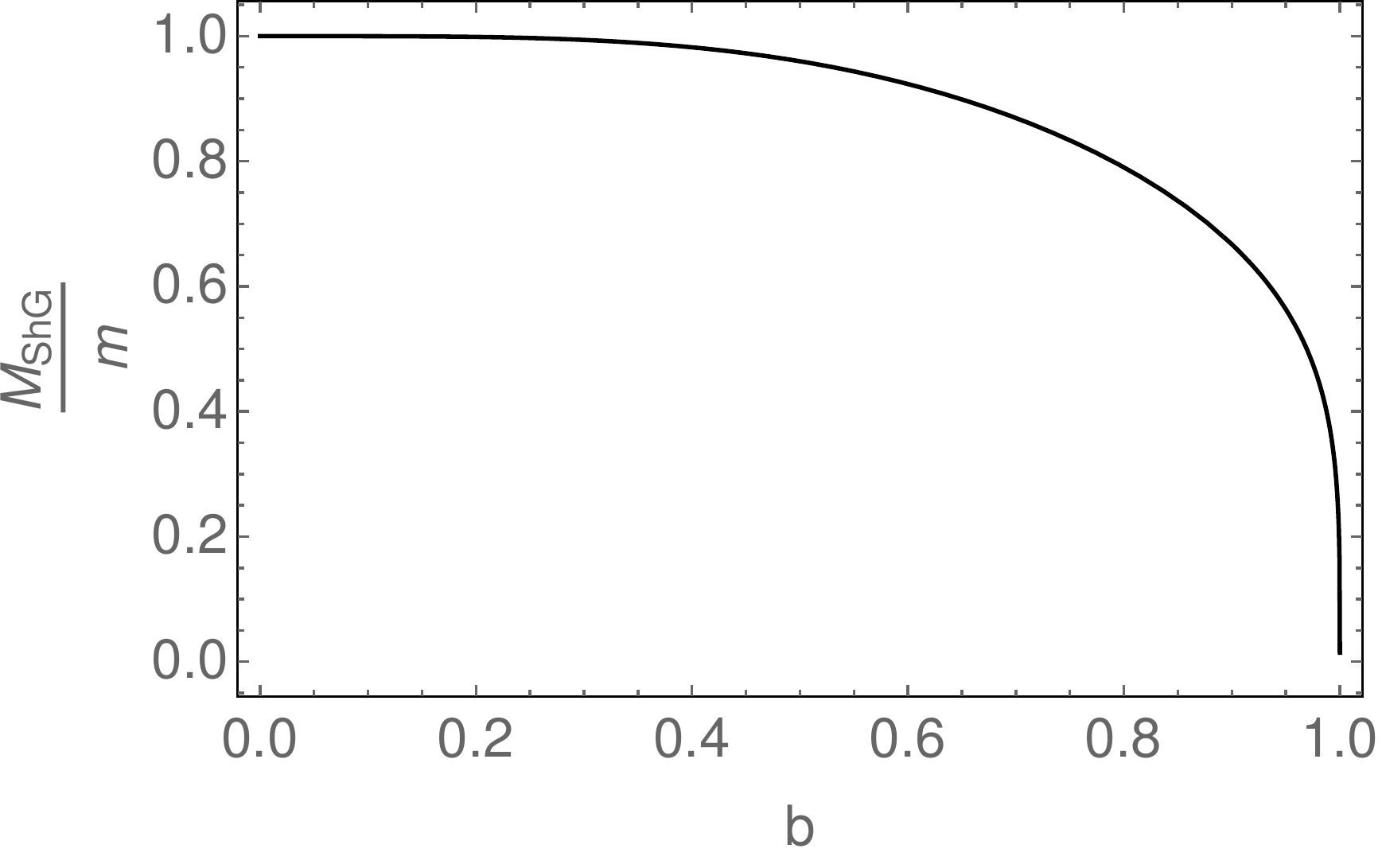}
\caption{Mass of the ShG normalized to $m$, the renormalized mass appearing in perturbation theory.}
\end{subfigure}
\hfill
\begin{subfigure}[b]{0.475\textwidth}
\centering
\includegraphics[width=\textwidth]{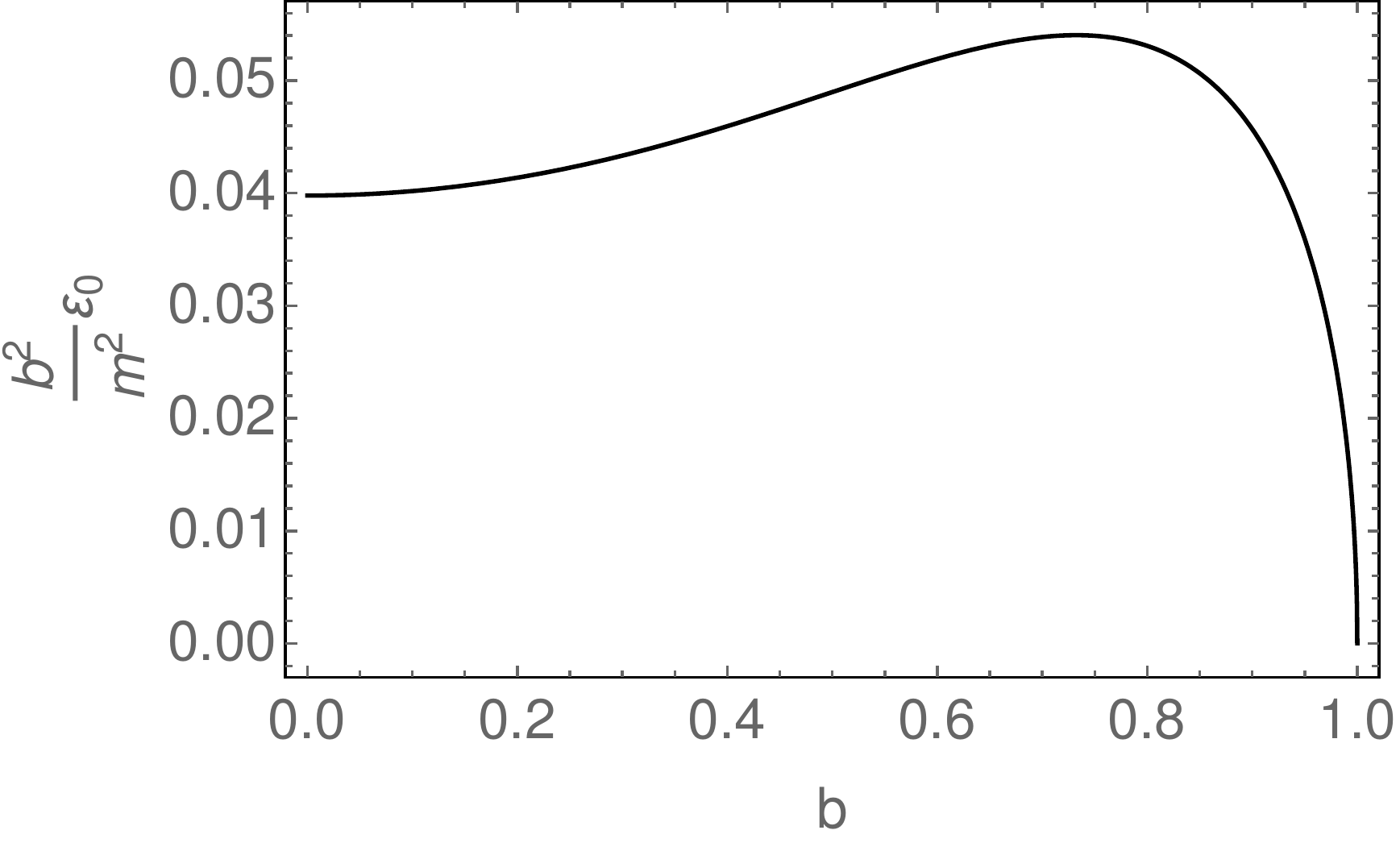}
\caption{Ground state energy density (bulk energy), normalized with respect to $\frac{m^2}{b^2}$.}
\end{subfigure}
\caption{Mass and ground state energy of the ShG model.}
\label{massShG}
\end{figure}

\subsubsection{ShG model as Analytic Continuation of SG model} \label{subsubShGAnalyticCont}
As we have stated, the ShG model can be thought of as the analytic continuation of the SG.  How then to connect the rich spectrum of SG containing topological excitations and their bound states to the much simpler spectrum of ShG consisting of a single parity odd excitation?  The choice typically made is to identify the first breather of the SG model with the massive excitation of ShG. 

To obtain the mass, $M_{ShG}$, of the fundamental excitation in ShG, we then take $M_{ShG}(b) = m_1(ib)$.  Using eqns.~(\ref{massformulaSG}) and (\ref{massbreather}) we arrive at \cite{Fateev:1997yg}
\be 
M_{ShG} \,=\, \frac{4 \sqrt{\pi}}{\Gamma\left(\frac{1}{2+2 b^2}\right) \Gamma\left(1 + \frac{b^2}{2 + 2 b^2}\right)} 
\, \left[- \mu_{ShG}\, \frac{\pi \Gamma(1 + b^2)}{\Gamma(-b^2)}\right]^{\frac{1}{2+2 b^2}}\,\,\,, 
\label{ShGmassformula}
\ee 
where we have replaced $\mu_{SG}$ with $-\mu_{ShG}$ - necessary as only $\mu_{ShG}$ has the correct engineering dimension.
The plot of this quantity can be found on Fig.~\ref{massShG}.  A few remarks are in order:

\begin{enumerate}

\item Keeping $\mu_{ShG}$ fixed, the mass formula (eq.~\ref{ShGmassformula}) is {\em not} invariant under the weak-strong duality $b \rightarrow 1/b$ of the ShG model.  Moreover, its analytic continuation for $b^2 > 1$ gives generally complex values for $M_{ShG}$.

\item For any finite value of $\mu_{ShG}$, the mass $M_{ShG}$ vanishes both at $b^2=0$ and $b^2 =1$. The nature of these zeros is however very different. Indeed, the zero at $b^2=0$ disappears if we rescale $\mu_{ShG} \rightarrow \mu_{ShG}/(8 \pi b^2)$ adopting the more conventional definition of the coupling constant of the model used in the Feynman diagram expansion.  However on the approach to $b^2=1$, we instead find a singular point:
\be
M_{ShG} \simeq (1-b^2)^{1/4}, ~~~ b^2 \rightarrow 1.
\label{mbremasin}
\ee

\item In order to compare with the perturbative expansion (eq.~\ref{perturbativeexpansionM}), it is necessary to connect the renormalization scheme used to define $\mu_{ShG}$ with that used to define $m$ (eq.~(\ref{lagrangian})).  
Substituting eq. \eqref{lambdamm} into (eq.~\ref{ShGmassformula}) and expanding in $b^2$, we have 
\be
M_{ShG}\,=\,m \,\left(1 - \frac{\pi^2}{12}b^4 + \frac{ 2\pi^2-7\zeta(3)}{12}b^6 \right) +
{\mathcal O}\left(b^8\right),
\ee
agreeing with the series expansion of the square root of expression eq.~\ref{perturbativeexpansionM}.
\end{enumerate}

\subsubsection{S-matrix of the ShG model and its duality}

Having argued the fundamental excitation of the ShG model is to be identified with the analytically continued breather of SG, we are now in a position to derive the S-matrix of the ShG's excitation.  This S-matrix is nothing but the analytically continued breather-breather S-matrix, $S_{B_1B_1}(\theta)$, of SG which is given by  
\be
S_{B_1B_1} (\theta) \,=\, \frac{\sinh\theta + i \sin\pi\xi}{\sinh\theta - i \sin\pi\xi}\,\,\,,  
\label{Smatrixbreather}
\ee
where $\theta = \theta_1 - \theta_2$ and $\theta_i$ ($i=1,2$) is the rapidity of each of the breathers involved in the scattering, with energy and momentum given by 
$E_i = m_1 \cosh\theta_i$ and $p_i = m_1 \sinh\theta_i$. This amplitude has a pole at $\theta = i \pi \xi$,
\be
S(\theta) \,\simeq i \, \frac{2 \tan(\pi \xi)}{\theta - i \pi \xi} \,\,\,, 
\label{polesmatrix}
\ee 
and, for $\xi < 1$, its residue is positive, i.e. this pole signals a further bound state, while for $\xi > 1$ the residue changes sign, which can be interpreted as another signal of the absence in the spectrum of the first breather for $\xi > 1$. 

If we now continue this expression analytically by substituting $-b^2$ for $b^2$ in eq.~(\ref{Smatrixbreather}), we obtain for the exact 2-body $S$-matrix of the ShG model 
\be
S(\theta) \,=\, 
\frac{\sinh\theta - i \sin\pi B}{\sinh\theta + i \sin\pi B}\,\,\,,
\label{SmatrixShG}
\ee
where 
\be
B\,=\, \frac{b^2}{1 + b^2}
\label{importantB}
\,\,\,.
\ee
This expression coincides with the $S$-matrix of the ShG model proposed in \cite{Arinshtein:1979pb}.  Although this argument is amazingly simple, the final result is nonetheless surprising because a duality has appeared.  The S-matrix now is invariant under the weak/strong duality $b \leftrightarrow 1/b$ or $B \leftrightarrow 1-B$.  However this duality is nowhere apparent in the Lagrangian (\ref{lagrangian}) of the model.

\subsubsection{Vacuum Expectation Values in the SG and ShG model}
Similar to the mass and S-matrix, the vacuum expectation values (VEVs) of the vertex operators of the ShG model can be obtained as analytic continuations from the corresponding expressions for the SG model, the eponymous FLZZ formula \cite{Lukyanov:1996jj,Fateev:1997yg}:

\begin{eqnarray}\label{VEVexp}
G(\alpha) \,=\, \langle e^{\alpha \phi} \rangle & \,\equiv\, & M_{ShG}^{-2 \alpha^2} \, {\mathcal G}(\alpha) \,=\, 
M_{ShG}^{-2 \alpha^2} \left[\frac{\Gamma\left(\frac{1}{2+2 b^2}\right) \Gamma\left(1 + \frac{b^2}{2 + 2 b^2}\right)} 
{4 \sqrt{\pi}}\right]^{- 2 \alpha^2} \cr\cr
&& \hskip -.5in \times\exp\left\{
\int_{0}^{\infty} \frac{dt}{t} \left[
- \frac{\sinh^2(2 \alpha  b t)}{2 \sinh(b^2 t) \sinh t \, \cosh((1+b^2)t)} + 
2 \alpha^2\, e^{-2 t}\right]\right\},
\end{eqnarray}
with $G(\alpha) = G(-\alpha)$. Notice that the integral above converges for  
\be
|\alpha| \,< \,\frac{1}{2} \, Q, 
\,\,\,
\ee
a bound conceived for physical operators by N. Seiberg in his study of the allied Liouville problem \cite{Seiberg:1990eb}.
For values of $\alpha$ beyond the Seiberg bound, one can exploit an analytic continuation of ${\mathcal G}(\alpha)$.  Obtaining this continuation is facilitated by the expression \cite{Lashkevich:2011ne}:
\begin{eqnarray} 
{\mathcal G}(\alpha) &\, = \, & e^{-2 \gamma_E \alpha^2} \, \cos\frac{\pi \alpha}{Q} \label{analyticVEV}\\
& \times &
 \prod_{k=1}^\infty 
 \frac{ e^{\frac{2 \alpha^2}{k}} \, \Gamma^2\left(\frac{1}{2} + \frac{k B}{2}\right) 
 \,  \Gamma^2\left(\frac{1}{2} + \frac{k (1- B)}{2}\right)}{
  \Gamma\left(\frac{1}{2} -\frac{\alpha}{Q} + \frac{k B}{2}\right)\,
   \Gamma\left(\frac{1}{2} + \frac{\alpha}{Q} + \frac{k B}{2}\right)\, 
    \Gamma\left(\frac{1}{2} -\frac{\alpha}{Q} + \frac{k (1 - B)}{2}\right)
    \,\Gamma\left(\frac{1}{2} +\frac{\alpha}{Q} + \frac{k (1 - B)}{2}\right) 
    }.\nonumber \label{eq:VEVproduct}
    \end{eqnarray}
From it one can see that the VEV, as a function of $\alpha$, does not have poles but only zeros. Besides the zero at $\alpha = Q/2$, there is an infinite set of generically simple zeros located at:
 \begin{eqnarray}
 && \alpha \,=\, \pm \alpha_{n,m} \,=\, \pm \frac{Q}{2} \mp \left(  \frac{m}{2} \,b^{-1}  + \frac{n}{2} \, b\right) ,\\
&&{\rm with} \, \,\, m \geq n \geq 2 \hspace{3mm} n \in 2 \,\mathbb{Z}
 \hspace{8mm} 
 {\rm or} \,\, \,\,\,\,\,\,\,\,n > m \geq 2 \hspace{3mm} m \in 2 \,\mathbb{Z} .\nonumber 
 \end{eqnarray}
Formula \ref{eq:VEVproduct} is positive for $-\frac{Q}{2}<\alpha<\frac{Q}{2}$, but it changes sign at its zeroes. The vertex operators, being the exponentials of Hermitian operators, are positive (semi-)definite. This means that, outside the above domain, the analytic continuation cannot directly correspond to the expectation value. We will therefore consider these values "unphysical".
The function ${\mathcal G}(\alpha)$ itself is self-dual, i.e. invariant under $ b \rightarrow 1/b$ - for the proof one may benefit from the identity
\[
\int_0^{\infty} \frac{dt}{t} \left[e^{-2 t} - e^{-2 t /b}\right] \,=\, - \log \,b \,\,\,, 
\]
but the VEV is not itself self-dual because of the presence of $M_{ShG}^{-2 \alpha^2}$. From the dependence on $\alpha$ of this term, we can infer that the scaling dimension of the vertex operator 
$V(\alpha)$ is $\Delta(\alpha) = -\alpha^2$, a value which coincides with its conformal dimension with respect to the Gaussian fixed point. Using the VEV (\ref{VEVexp}) we can compute the expectation value of the trace of the stress-energy tensor, an operator that, on general terms, is defined as 
\be
\Theta(x) \,=\, 2 \pi \beta(\mu_{ShG}) \, {\mathcal O} \,\,\,,
\label{tracestressenergytensor}
\ee
where $\beta(\mu_{ShG})$ is the $\beta$-function of the coupling $\mu_{ShG}$ which perturbs a critical point and ${\mathcal O}$ its conjugate field. For the case at hand, we have 
\be
\Theta(x) \,=\, 8 \pi (1 +  b^2) \,\mu_{ShG} \,\cosh b \phi(x) .
\ee 
Using the mass formula (eq.~\ref{ShGmassformula}) and the simplified expression of the VEV at $\alpha = b$ 
\be
\langle e^{\pm b \phi} \rangle \,=\, - M_{ShG}^{-2 b^2} \,\frac{\pi}{16 (1+b^2)}\, \frac{1}{\sin\pi B} \,\frac{\Gamma(1+b^2)}{\Gamma(-b^2)} 
\left[\frac{\Gamma\left(\frac{1}{2+2 b^2}\right) \Gamma\left(1 + \frac{b^2}{2 + 2 b^2}\right)} 
{4 \sqrt{\pi}}\right]^{- (2+2 b^2)},
\ee 
we end up with
\be
\langle \Theta \rangle \,=\,\frac{\pi M_{ShG}^2}{2 \sin\pi B}\,\,\,. 
\label{vevtrace}
\ee
 
\subsubsection{Questions Arising from the Analytic Continuation $b\leftrightarrow ib$}

In this section we have presented a number of results for the ShG model (its spectrum, its S-matrix, and the VEVs of its exponential operators) that are arrived at by analytically continuing results from SG.  The question of the validity of these analytic continuations has to be raised.  While the S-matrix of the ShG model is physically sensible for all $b$, the expressions for the mass and VEVs are not.  Given that the fundamental excitation of the ShG is identified with the breather of SG and the SG breather ceases to exist for $b<1/\sqrt{2}$, what exactly can be said for $b>1/\sqrt{2}$ is not entirely clear.  And certainly the mass formula for $M_{ShG}$ for $b>1$ breaks down entirely giving complex-valued results.

That we are able to match perturbative computations of the mass formula with the exact expression is thus important.  This gives us some confidence that the results for $M_{ShG}$ are valid for $b<1$.  This confidence will be increased in the following section where we discuss the ShG model as a perturbation of a Liouville theory.  However the validity and interpretation of formulae at $b>1$ including the S-matrix arising from the analytic continuation remains, in our opinion, an open question.

\subsection{ShG and Liouville Models}
We have now considered the ShG model from the perspective of perturbation theory and an analytically continued SG model.  We now present a third way to look at the ShG model: as a deformation of a Liouville field theory \cite{Zamolodchikov:1995aa,Fateev:1997yg,Mussardo:1993ut}.  This third way will be essential for us in what follows and results presented here will be used in our discussion of quantum mechanical reductions of the ShG model in Section \ref{QMreductionssection}.

The Liouville conformal field theory is defined by the action \cite{Zamolodchikov:1995aa,Fateev:1997yg}
\be
S_{Liouville} \,=\, \int d^2x \bigg(\frac{1}{16 \pi} (\partial_{\mu} \phi)^2 +\mu_L\, :e^{b \phi}:\bigg) + Q\,\phi_\infty
\,\,\,,
\label{liouvilleS}
\ee
Here the operator $e^{b\phi}$ has conformal dimension 1 and $\mu_L$ is dimensionless.  This vertex operator has this dimension because we have coupled the field to a background charge of strength $Q= b + b^{-1}$.  In general, in the presence of the charge at infinity, the conformal dimension of the vertex operator $e^{\alpha \phi}$ becomes 
\be
\Delta(\alpha) \,=\, \alpha (Q  - \alpha) .
\label{newliouville dimension}
\ee 
In order to ensure the theory is IR finite, the theory can be placed on a Riemann sphere (of area $A$) with a metric 
\begin{equation}
g_{\mu\nu}(x) = \rho(x)\delta_{\mu\nu}; ~~~ \rho(x) \,=\, (1 + \pi |x |^2/A)^{-2} .
\label{metric}
\end{equation}
The ShG model then can be obtained as a perturbation of the Liouville action:
\begin{equation}\label{perturbedLT}
S_{ShG} = S_{Liouville} + \mu_L d^{-4-4b^2}\int d^2x \, \rho(x)^{2+2b^2} :e^{-b\phi}: .
\end{equation}
Note that we have made a scale $d$ that comes from normal ordering the vertex operator explicit instead of absorbing it into $\mu_L$. We use $d$ here to distinguish this scale from ones (i.e. $a$) previously introduced in normal ordering vertex operators in different (Gaussian) schemes. $\mu_L$ here is the same dimensionless constant that appears in the original Liouville action, eq.~(\ref{liouvilleS}).

A key property of the Liouville field theory is that the exponential operators are pairwise identified as \cite{Zamolodchikov:1995aa} 
\be 
e^{\alpha \phi (x)} \,=\, R(\alpha) \, e^{(Q - \alpha) \phi (x)} \,\,\,,
\label{identification}
\ee
where $R(\alpha)$ is related to the Liouville reflection amplitude $S_L(P)$ as 
\be
R\left(\frac{Q}{2} + i P \right) \,=\,S_L(P) \,=
- \left(\frac{\pi \mu_L \Gamma(b^2)}{\Gamma(1-b^2)}\right)^{-2 i P/b} \,
\frac{\Gamma(1 + 2 i P/b) \, \Gamma(1 + 2 i P b)}{\Gamma(1 - 2 i P/b) \, \Gamma(1 - 2 i P b)} \,\,\,. 
\label{LiouvilleSmatrix}
\ee
This identification of the operators implies that their VEV must satisfy the reflection relations
\be
\begin{array}{l}
G_L(\alpha) \,\hspace{3mm}=\, R(\alpha) \, G_L( Q - \alpha ),\\
G_L(-\alpha) \,=\, R(\alpha) \, G_L(-Q  + \alpha).
\end{array}
\label{reflectionampl}
\ee
Here the L-subscripts indicate that we are taking these VEVs as defined in the perturbed Liouville formulation of ShG and are not assuming these relations are the same as in eq.~(\ref{analyticVEV}).
A solution of these equations can be obtained by an infinite iteration
\be
G_L(\alpha) \,\propto \,\prod_{n=0}^{\infty} R(\alpha - n Q) .
\ee
With the further assumption of minimality, we can find a result equivalent to combining (\ref{ShGmassformula}) and (\ref{VEVexp}).  This provides below an alternate way of understanding these formulae without resorting to analytic continuation from results derived for SG.

In order to present this argument, we need to trace carefully the dimensions of the quantities involved.   This is an issue because in the Liouville approach the exponential operator, $e^{a\phi}$, has dimension (\ref{newliouville dimension}) whereas the dimension of this operator in the perturbed Gaussian formulation of the ShG model is instead $-\alpha^2$. To understand this dimensional transmutation, we follow along with Ref.\,\cite{Fateev:1997yg} and interpret properly the results. 

To begin, we use the action (eq.~(\ref{perturbedLT}) to write the following perturbative expansion of the VEV of $e^{a\phi}$:
\begin{equation}
G_L (\alpha) = 
Z^{-1} \, 
\sum_{n=2}^\infty \frac{(-\mu_L)^n}{n!} \, 
\int d^2y_1 \ldots d^2 y_n \, 
\langle e^{\alpha \phi}(x) \, 
e^{-b \phi}(y_1) \ldots e^{-b \phi}(y_n) \rangle_{{\rm L}}\,\,\,.\label{eq:GlLiouv}
\end{equation}
Since in Liouville field theory the coupling $\mu_L$ can be absorbed into the field $\phi$ via a redefinition of this field by an additive shift, it is easy to obtain the explicit $\mu_L$ dependence of all correlators appearing in eq.~(\ref{eq:GlLiouv}). With the IR regulator in place, this expression can be written as the following series:
 \be
G_L(\alpha) = 
\mu_L^{-\alpha/b} \, A^{\alpha (\alpha - Q)} \, \sum_{n=2}^\infty 
\left[\mu_L^2 \left(\frac{A}{d^2}\right)^{2(1+b^2)}\right]^n \tilde G_{Ln}(\alpha) \,\,, \label{eq:GlGltilde}
\ee
 where $\tilde G_{Ln}(\alpha)$ is independent of $\mu_L$.
The prefactor $\mu_L^{-\alpha/b}$ also ensures the satisfiability of the reflection relations (\ref{reflectionampl}), given the dependence on $\mu_L$ of the refection amplitude $R(\alpha)$, see  eq.\,(\ref{LiouvilleSmatrix}). We see that our IR regulator appears in a dimensionless combination with the scale $d$ in this perturbative expansion.
If we assume that this expression has a sensible large $A$ limit, the above series must behave asymptotically as
\be
\tilde G_L(\alpha,t) \equiv \sum_{n=2}^\infty 
t^n \, \tilde G_{Ln}(\alpha)  \rightarrow \tilde G_L(\alpha) \, t^{-\frac{\alpha (\alpha -Q)}{2 (1+b^2)}}, ~~~ t= \mu_L^2 \left(\frac{A}{d^2}\right)^{2(1+b^2)}.
\ee
Thus in the large $A$ limit, we obtain
\be
G_L(\alpha) = d^{2\alpha (\alpha -Q)}  \,\mu_L^{-2 \alpha^2/(2 + 2 b^2)}  \tilde G_L (\alpha).
\ee 
We thus see the VEV has dimension $-2\alpha(\alpha-Q)$ as set by the UV cutoff $d$ - the dimension set by the Liouville CFT.  At the same time its dependence upon $\mu_L$ is exactly what would be expected from thinking of the ShG model as a perturbation of a free non-compact boson.

We can now complete the argument showing how eq.~(\ref{ShGmassformula}) and eq.~(\ref{VEVexp}) can be determined, at least in combination.  Using eq.~(\ref{reflectionampl}), we obtain \cite{Fateev:1997yg} for the function $\tilde G_L(\alpha)$:
\begin{eqnarray}
\tilde G_L(\alpha) &=& \bigg(\frac{\pi\Gamma(1+b^2)}{\Gamma(-b^2)}\bigg)^{-\alpha^2/(1+b^2)}\cr\cr
&& \times \exp\left\{
\int_{0}^{\infty} \frac{dt}{t} \left[
- \frac{\sinh^2(2 \alpha  b t)}{2 \sinh(b^2 t) \sinh t \, \cosh((1+b^2)t)} + 
2 \alpha^2\, e^{-2 t}\right]\right\}\,\,\,,
\end{eqnarray}
If we compare $\tilde G_L(\alpha)$ with the expression for $G(\alpha)$ presented in eq.~(\ref{VEVexp}) and eq.~(\ref{ShGmassformula}), we see that we are consistent, i.e.
\begin{equation}
\frac{G(\alpha)}{\mu_{SG}^{-2\alpha^2}} = \tilde G_L(\alpha)\,\,\,. 
\end{equation}
If we identify the couplings in the two formulations via 
$$
\mu_{ShG} = \mu_L d^{-2-2b^2}\,\,\,,
$$
we can identify the expressions for the VEVs in the two formulations via
\begin{equation}
G_L(\alpha) = d^{-2\alpha Q}G(\alpha)\,\,\,.
\end{equation}
The appearance of the factor $d$ in this expression is a reflection of the different normal ordering schemes in the Gaussian vs Liouville pictures.

\subsection{Generalized TBA equations for Ground and Excited State Energies} \label{SubsecTBA}

One of the tools that we will use extensively in characterizing our TSM data is the thermodynamic Bethe ansatz (TBA).
The exact ground state energy, $E_0(R)$, at a finite volume $R$ from the TBA by using the `finite volume-finite temperature' equivalence of the partition function $Z$: 
\be
E_{0}\left(R\right)\,=\, R\,\mathcal{E}_{0}-M_{ShG}\int_{-\infty}^{\infty}\frac{du}{2\pi}\,\cosh u\,\log(1+e^{-\epsilon(u)}) \,\,\,,\label{tba_gs}
\ee
where $\mathcal{E}_{0}=-\lim_{R\rightarrow\infty}\frac{1}{R}\ln Z\left(R\right)$ is the `vacuum' energy density equal to the VEV of eq.~(\ref{VEVexp}):
\be
\mathcal{E}_{0}\,=\,\frac{M_{ShG}^{2}}{8\sin\pi B} \,\,\,.
\label{vacumenergy}
\ee
The pseudo-energy $\epsilon\left(\theta\right)$ in eq.~(\ref{tba_gs}) is defined as the solution of the nonlinear integral equation
\be
\epsilon(\theta)\,=\, M_{ShG} R\cosh\theta - \int_{-\infty}^{\infty}\frac{dv}{2\pi}\phi(\theta-v)\log(1+e^{-\epsilon(v)}) \,\,\,,
\ee
where the kernel is related to the $S$-matrix as
\be
\phi\left(\theta\right)\,=\,-i\frac{d}{d\theta}\log S(\theta) \,\,\,.
\ee
For excited states, the exact finite volume energies can be obtained either by careful analytic continuations of the ground state TBA (following
\cite{Dorey:1996re}), or by examining the continuum limit of an integrable lattice regularization \cite{Teschner:2007ng}. A finite volume $n$-particle
state can thus be described by the \emph{multiparticle} pseudo energy $\epsilon(\theta\vert\{\vartheta_{j}\}_{j=1}^{n})\equiv\epsilon(\theta\vert\{\vartheta\})$
and a set of quantization numbers $\left\{ I_{j}\right\} _{j=1}^{n}$ satisfying the non-linear integral equation 
\begin{equation}
\epsilon(\theta\vert\{\vartheta\})=M_{ShG}R\cosh\theta+\sum_{j}\log S(\theta-\vartheta_{j}-\frac{i\pi}{2})-\int_{-\infty}^{\infty}\frac{dv}{2\pi}\phi(\theta-v)\log(1+e^{-\epsilon(v\vert\{\vartheta\})}),\label{eq:shGTBA}
\end{equation}
together with the additional quantization conditions 
\begin{equation}
Q_{j}(\{\vartheta\})=2\pi I_{j}, ~~~ Q_{j}(\{\vartheta\})=-i\epsilon(\vartheta_{j}+\frac{i\pi}{2}\vert\{\vartheta\})-\pi, ~~~ j=1,\dots,n.\label{eq:shGQQ}
\end{equation}
Given quantization numbers $I_{j}\in\mathbb{Z}$, the rapidities $\{\vartheta\}$ and the pseudo energy $\epsilon(\theta\vert\{\vartheta\})$ can be
determined self-consistently by efficient numerical methods. Note that the above phase convention is `bosonic' in the sense that it is permitted to have $I_j=I_k,\:j\neq k$ (see also \cite{Bajnok_2019}). Nevertheless, the resulting rapidities are always different, reflecting the fermionic nature of the particles. The above ingredients provide the finite volume energy of the multiparticle state
as
\begin{equation}
E_{\{I_{j}\}}(R)\,=\, R\,\mathcal{E}_{0} + M_{ShG}\sum_{j}\cosh\vartheta_{j} - M_{ShG}\int_{-\infty}^{\infty}\frac{du}{2\pi}\,\cosh u\,\log(1+e^{-\epsilon(u\vert\{\vartheta\})})\label{eq:shGE}\,\,\,.
\end{equation}
Notice that, neglecting all terms exponentially suppressed in $M_{ShG}R$ in the finite volume expression of the energies, these are nothing else but the well-known Bethe ansatz equations of the ShG model. Moreover, regarding the mass $M_{ShG}$ just as a parameter of these equations (i.e. as a quantity independent on $b$), these equations are invariant under the duality $b \leftrightarrow 1/b$ present in the $S$-matrix. 

\subsection{Summary}

Let us summarise the main points of this section: 
\begin{itemize}
\item
As a Lagrangian theory, the ShG model has three equivalent descriptions: (i) one arising from perturbation theory in $b$; (ii) one as a deformation of a Gaussian $c=1$ CFT; and (iii) finally one as a deformation of a Liouville theory. 
\item In the Gaussian picture, the ShG model can be thought of as an analytic continuation of the SG model.  In this way, results can be derived for the mass spectrum, the S-matrix, and the VEVs of the vertex operators.
\item The S-matrix so obtained suggests the theory has a weak-strong duality: $b \leftrightarrow 1/b$.  However the mass formulas (and so the VEVs) are not invariant under this duality.
\item The validity of this analytic continuation for values of $b>1$ is unclear and there are doubts about its validity even for $b>1/\sqrt{2}$.
\item There is however reason to believe the expressions for the mass and the VEV up to $b=1$ because of the availability of an alternate derivation of the VEVs using the Liouville picture as well as consistency with perturbation theory.
\item For the ShG model, we have a formalism that allows us to compute in a numerically exact fashion the excited state energies at any volume $R$.  These equations are invariant under duality if we consider $M_{ShG}$ to be merely a parameter
\end{itemize}

In the next two sections we will present the results coming from the truncated spectrum method and testing the various expressions for the mass $M_{ShG}$, the S-matrix, $S(\theta)$, and the VEVs $\langle e^{\alpha\phi}\rangle$ presented here.  This analysis will give us some insight, even if not definitive, on the nature of the theory for $b>1$.

\section{Truncated Spectrum Methods}\label{TSMGENERAL}

The purpose of this section is to give the reader an overview of truncated spectrum methods (TSMs) and their application to the sinh-Gordon model.  TSMs were introduced by Yurov and Zamolodchikov \cite{Yurov:1989yu} to study the low-energy spectrum of 2D perturbed conformal field theories. However the method is able to study the spectrum and matrix elements of any theory whose Hamiltonian can be conveniently written as a sum of two terms 
\begin{equation}\label{eIIIi}
H \,=\, H_0 + \int^R_0 dx\, V_{\rm pert}(x)\,\,,
\end{equation}
where $H_0$ is a base theory of which we assume to have a complete control of its energy eigenvalues and eigenstates $|E_n\rangle_0$. In particular, we assume that we are able to 
write down the matrix elements of the second term in the full Hamiltonian, $V_{\rm pert}(x)$, in the basis $\{|E_n\rangle_0\}$ of eigenvectors of $H_0$. From a computational point of view, 
the actual implementation of the method requires both a denumerable set of energy states and finite-dimensional subspaces of the Hilbert space of the model: the former condition is typically achieved by putting the model onto a cylinder of finite circumference $R$ in the spatial direction; the latter condition is satisfied by restricting the set of eigenstates of $H_0$ to those whose energies fall below a cutoff  $E_c$.  Once a finite basis is obtained in this way, the truncated Hamiltonian is constructed.  This operator possesses the same matrix elements as the original Hamiltonian in the truncated subspace, but acts trivially in the orthogonal subspace. Having this in hand, one then solves, numerically, the eigenproblem of the truncated Hamiltonian. Assuming for the moment that the dependence of the data on the cutoff $E_c$ is smooth, once the truncated Hamiltonian is diagonalized for different volumes, the infinite volume quantities can be obtained by extrapolation via L\"uscher's  principles \cite{Luscher:1985dn,Luscher:1986pf,LASSIG1991666,Klassen:1990ub}.

TSMs were first applied to the scaling Lee-Yang model \cite{Yurov:1989yu} and the Ising model \cite{Yurov:1991yu}. In both cases the numerical results reported therein were strongly convergent in $E_c$. In the study of the perturbed tri-critical Ising theory, though, it was argued that the convergence in $E_c$ of the TSM results depends on the scaling dimension of the perturbing operator \cite{LASSIG1991591,LASSIG1991666}.  Various renormalization group approaches have been advocated to treat cases where convergence in $E_c$ is suboptimal.  These strategies are both numerical \cite{PhysRevLett.98.147205} and analytical \cite{feverati2006renormalisation,giokas2011renormalisation, Lencses:2014tba, Hogervorst:2015}
(For a comprehensive review of such strategies see \cite{tsm_review}.)  We will demonstrate later that these strategies require modification (at the very least) for the case of the ShG. 

The performance of TSMs is dependent on the choice of the computational basis used to perform the calculations (or, in other words, how we split the Hamiltonian $H$ into $H_0$ and $\int dx V_{\rm pert}$).  As with any variational method, we want to use a computational basis that captures at the start features of the physics of the model at hand.  In this paper we study the ShG model by means of two different choices of $H_0$ or computational bases:

\vskip 10pt 

\noindent 1. In the first, discussed in the Section \ref{TSMCFT}, we consider the ShG model as a deformation of a Gaussian CFT and the corresponding non compact bosonic field expanded in terms of an infinite number of oscillators and a single zero mode. When $H_0$ is a compact CFT on a cylinder and $V$ is a relevant operator, one typically has control on the magnitude of the interaction between different energy scales in the theory. The ShG model is, however, different, and the low and high energy scales in the problem become strongly coupled on the approach towards $b=1$.

\vskip 10pt 

\noindent 2. This leads us to our second basis choice, outlined in Section \ref{FFTCSASECTION}, where we use the basis of the ShG model itself as the computational basis.  In particular, we use the basis of the ShG model at one value of $b$ to compute the properties of the model at a different value of $b$.  In this scheme, $V_{\rm pert}$ is the difference between two hyperbolic cosines.  This approach does immediately raise questions of circularity.  We are, after all, using conjectured information
about the model at a point $b_{0}$ as input, to obtain results at point $b_{1}\neq b_{0}$.  We will address this question in Section \ref{FFTCSASECTION}, 
and attempt to ameliorate this concern.

\section{TSM for the ShG using a Non-Compact Massless Bosonic Basis}\label{TSMCFT}

In our first attempt to study the ShG model using TSMs, we employ a computational basis based on a non-compact massless basis.  In this section we review the details surrounding this choice of basis.

\subsection{Non-Compact Massless Boson}

In describing this basis the starting point is the mode expansion of the massless non-compact bosonic field on an infinite cylinder of radius $R$:
\begin{equation}\label{eIIIii}
\varphi\left(x,t\right)=\varphi_{0}+8\pi\frac{\Pi_{0}}{R}t+\sum_{n\neq0}\sqrt{\frac{2}{\left|n\right|}}\left(a_{n}e^{i\left(k_{n}x-\left|k_{n}\right|t\right)}+a_{n}^{\dagger}e^{-i\left(k_{n}x-\left|k_{n}\right|t\right)}\right);\quad k_{n}=\frac{2\pi n}{R},
\end{equation} where the oscillators are subject to the usual Fock commutator relations, 
\begin{eqnarray}
\left[a_{n},a_{m}^{\dagger}\right] &=& \delta_{nm}\,\,\,,
\end{eqnarray}
while the zero mode defines an effective 1D quantum mechanical system with the canonical commutator $\left[\varphi_{0},\Pi_{0}\right]=i$.

The computational basis of states follows from the specification of $H_0$ in eq.~(\ref{eIIIi}).  Here we will divide $H_0$ into a zero mode $H_{ZM}$ and non-zero mode 
$H_{NZM}$ part:
\begin{eqnarray}\label{eIIIiii}
H_0 &=& H_{ZM} + H_{NZM}\,\,;\cr\cr
H_{ZM} &=&\frac{4\pi}{R}\Pi_{0}^{2}+\mu_{ShG} R\left(\frac{R}{2\pi}\right)^{2b^{2}}\left[:e^{b\varphi_{0}}: + :e^{-b\varphi_{0}}:\right]\,\,;\cr\cr 
H_{NZM}&=&\frac{2\pi}{R}\left(L_{0}+\bar{L}_{0}-\frac{1}{12}\right)\,\, ,
\end{eqnarray}
where the Virasoro generators $L_{0}$ and $\bar{L}_{0}$ appearing in $H_{NZM}$ are related to the Fock mode operators as 
\[L_{0}\,=\,\sum_{n>0}na_{n}^{\dagger}a_{n}
\hspace{5mm}
,
\hspace{5mm}
\bar{L}_{0}=\sum_{n<0}\left|n\right|a_{n}^{\dagger}a_{n}\,\,\,.
\]  
Notice that $H_{ZM}$, unlike $H_{NZM}$, is an interacting
Hamiltonian. With this writing of $H_0$, our computational eigen-basis
has a tensor product structure composed of a zero mode and an
oscillator sector:
\begin{equation}\label{eIIIiv}
\mathcal{H} = \mathcal{H}_{ZM}\otimes\mathcal{H}_{osc}\equiv \mathcal{H}_{ZM}\otimes\mathcal{H}_{R}\otimes\mathcal{H}_{L}\,\,\, .
\end{equation}
Here $\mathcal{H}_{osc}$ is decomposed into a chiral subspace,
$\mathcal{H}_{R}$, is spanned by right-moving particles 
$$
\mathcal{H}_{R}=\left\{ a_{n_{1}}^{\dagger}\dots a_{n_{k}}^{\dagger}\left|0\right\rangle, n_i>0 \right\} \,\,\,,
$$
and an anti-chiral subspace, $\mathcal{H}_{L}$, is spanned by left-moving particles 
$$
\mathcal{H}_{L}=\left\{ a_{-n_{1}}^{\dagger}\dots a_{-n_{k}}^{\dagger}\left|0\right\rangle, n_i>0  \right\}\,\,\,.
$$  

\subsection{Zero modes}

Unlike the non-interacting $H_{NZM}$, we have chosen a form for $H_{ZM}$ that is non-trivial.  We do so following \cite{Rychkov:2015vap,Bajnok:2015bgw, BLSV} so that $\mathcal{H}_{ZM}$ consists of a countable (i.e. discrete) basis of states.  We will denote this basis of states as follows
\begin{equation}\label{eIIIv}
\mathcal{H}_{ZM} = \{|m\rangle\,\,,\,\, H_{ZM}|m\rangle = E_{ZM,m}|m\rangle \}.
\end{equation}
Unlike the states in $\mathcal{H}_{NZM}$, in our implementation of the TSM the eigenstates $|m\rangle$ will be found numerically. To do so, we need to choose a computational basis to represent $H_{ZM}$ in eq.~(\ref{eIIIiii}) and, for this aim, we choose the position basis in the zero mode coordinate
\begin{equation}\label{eIIIvi}
{\cal H}_{ZM,computational} \,=\, \{|\phi_0\rangle, \phi_0 = -L + na, n=0,\cdots,2L/a\}\,\,\,,
\end{equation}
where $2L$ is the length of the truncated zero mode space (rather than having a non-compact zero mode, we assume it lies between $-L$ and $L$) and $a$ is our spatial  discretization parameter. In performing our computations here, we have always taken $L$ large enough and $a$ small enough so that the eigenvalues and eigenstates (or at least their matrix elements) of $H_{ZM}$ have converged completely.

\subsection{Truncated Hilbert Spaces}

Having determined ${\cal H}_{ZM}$, we are now in a position to define the truncated basis, ${\cal H}_T$, for the problem as a whole.  We truncate each part of the Hilbert space separately.  In particular we write
\begin{eqnarray}\label{eIIIvii}
\mathcal{H}_T &=& \mathcal{H}_{ZM,T}\otimes\mathcal{H}_{R,T}\otimes\mathcal{H}_{L,T};\cr\cr
\mathcal{H}_{R,T} &=& \left\{ a_{n_{1}}^{\dagger}\dots a_{n_{k}}^{\dagger}\left|0\right\rangle, n_i>0,\sum^k_{i=1} n_i \leq N_c \right\} ;\cr\cr
\mathcal{H}_{L,T} &=& \left\{ a_{-m_{1}}^{\dagger}\dots a_{-m_{l}}^{\dagger}\left|0\right\rangle, m_i>0,\sum^l_{i=1} m_i \leq N_c \right\} ;\cr\cr
\mathcal{H}_{ZM,T} &=& \left\{|m\rangle, m=1,\cdots,N_{ZM}, E_{ZM,1}\leq\cdots\leq E_{ZM,N_{ZM}}\right\}.
\end{eqnarray}
The cutoff is then implemented in terms of two separate parameters, $N_{c}$, the level of which we cut off the chiral oscillator mode part of the Hilbert space, and $N_{ZM}$, the number of zero mode eigenstates of smallest energy (w.r.t. to $H_{ZM}$) that we keep.  We typically work not in this full space, but its zero-momentum counterpart composed of tensored states from $\mathcal{H}_{R,T}$ and $\mathcal{H}_{L,T}$ that satisfy
$$
\sum^k_{i=1}n_i = \sum^l_{i=1}m_i.
$$
The last step of forming the Hamiltonian matrix consists of specifying the interaction part of the Hamiltonian and its corresponding matrix elements.  In the zero-momentum subspace, $P=0$, we can write 
\begin{equation}\label{eIIIviii}
\int^R_0 dx \,V_{pert}(x) = \delta_{P,0} \mu_{ShG} \bigg(\frac{R}{2\pi}\bigg)^{2b^2} R\left[e^{b\varphi_{0}}\left(:e^{b\tilde{\varphi}\left(0\right)}:-1\right)+e^{-b\varphi_{0}}\left(:e^{-b\tilde{\varphi}\left(0\right)}:-1
\right)\right],
\end{equation}
where $\delta_{P,0}$ reflects the projection onto the zero momentum subspace and we have separated out the zero mode from the field:
\begin{eqnarray}\label{eIIIix}
\varphi (x,\tau) &=& \varphi_{0}(\tau)+\tilde{\varphi}\left(x,\tau\right) \equiv \varphi_0(\tau) + \varphi_R(x,\tau) + \varphi_L(x,\tau)  + \varphi^\dagger_R(x,\tau) + \varphi^\dagger_L(x,\tau);\cr\cr
\varphi_0(\tau) &=& \phi_0 -i\frac{8\pi\tau}{R}\Pi_0 .
\end{eqnarray}
In the above, normal ordering is defined as
\begin{equation}\label{eIIIx}
:e^{b{\varphi}\left(x,\tau\right)}:\, \equiv\, e^{b\phi_0(\tau)}e^{b\varphi_{R}^\dagger(x,\tau)}e^{b\varphi_{R}(x,\tau)}e^{b\varphi_{L}^\dagger(x,\tau)}e^{b\varphi_{L}(x,\tau)}\,\,\,,
\end{equation}
where
\begin{align}\label{eIIIxi}
\varphi_{R}\left(x,\tau\right) & =\sum_{n>0}\sqrt{\frac{2}{\left|n\right|}}a_{n}e^{ik_{n}x-k_n\tau},\quad\varphi_{L}\left(x,\tau\right)=\sum_{n<0}\sqrt{\frac{2}{\left|n\right|}}a_{n}e^{ik_{n}x-|k_n|\tau}\,\,\,.
\end{align}
The matrix elements of the chiral parts of $:e^{b\tilde{\varphi}\left(x\right)}:$ 
admit a closed analytic expression
\begin{eqnarray}\label{eIIIxii}
\left\langle n_{1},\dots,n_{k}\left|e^{b\varphi_{R}^\dagger}e^{b\varphi_{R}}\right|m_{1},\dots,m_{k}\right\rangle &=&\cr\cr
&& \hskip -2in =\,\prod_{q=1}^{k}\frac{1}{\sqrt{n_{q}!m_{q}!}}\left\{ \sum_{n_{1q}=0}^{\min\left(n_{q},m_{q}\right)}n_{1q}!\binom{n_{q}}{n_{1q}}\binom{m_{q}}{n_{1q}}\left(b\sqrt{\frac{2}{q}}\right)^{n_{q}+m_{q}-2n_{1q}}\right\},
\end{eqnarray}
where the chiral state vector $\left|n_{1},\dots,n_{k}\right\rangle $ is a normalized state having $n_{q}$ right-moving
particles with momentum $\frac{2\pi q}{R}$ for each $q\in\left\{ 1,\dots,k\right\} $.  Using this expression together with the knowledge of the (numerical) zero mode matrix elements
$$
\langle m | e^{b\varphi_0} | n\rangle\,\,\,,
$$
we can construct the full matrix elements of $H_{int}$.

\subsection{Methods of diagonalization}

Once we have collected all the matrix elements of both $H_0$ and $H_{int}$, for any truncated space we have a finite dimensional matrix to diagonalise.  To find its eigenstates and eigenvalues  we can proceed in two ways: 
\begin{itemize}
\item We can use exact diagonalization perhaps augmented with a numerical renormalization group.  This latterprocedure will be discussed further in Section \ref{RGIMPR}.
\item We can also use iterative methods that, thanks to their reverse communication protocols, do not require us to store the full Hamiltonian in memory. The only cost that we need to pay is that we are restricted here to computing the low-lying eigenvalues.  However for our purposes here this is not a limitation. Using the Jacobi-Davidson method and exploiting the tensor product structure (i.e. eq.~(\ref{eIIIiv})) of the Hilbert space, we can treat matrices arising from truncation parameters of up to $(N_c=20,N_{ZM}=24)$, corresponding to a truncated Hilbert space of size approximately $2.4\times 10^7$.  We elaborate on the usage of the tensor product structure in Appendix \ref{sec:Improved-exact-TCSA}. We specifically use the JDQMR\textunderscore{ETOL} algorithm\footnote{An earlier version of the method (without reverse communications protocols) was presented to check exponential finite volume corrections of matrix elements in sinh-Gordon theory
\cite{BLSV}, anticipating the present paper. There the computations were restricted to the small-coupling regime.}
 provided in the package PRIMME \cite{PRIMME,svds_software}.
\end{itemize}

\section{TSM Results for ShG Model} \label{TSMresults}

\begin{figure}
\begin{subfigure}[b]{0.475\textwidth}
\centering
\includegraphics[draft=false,width=\textwidth]{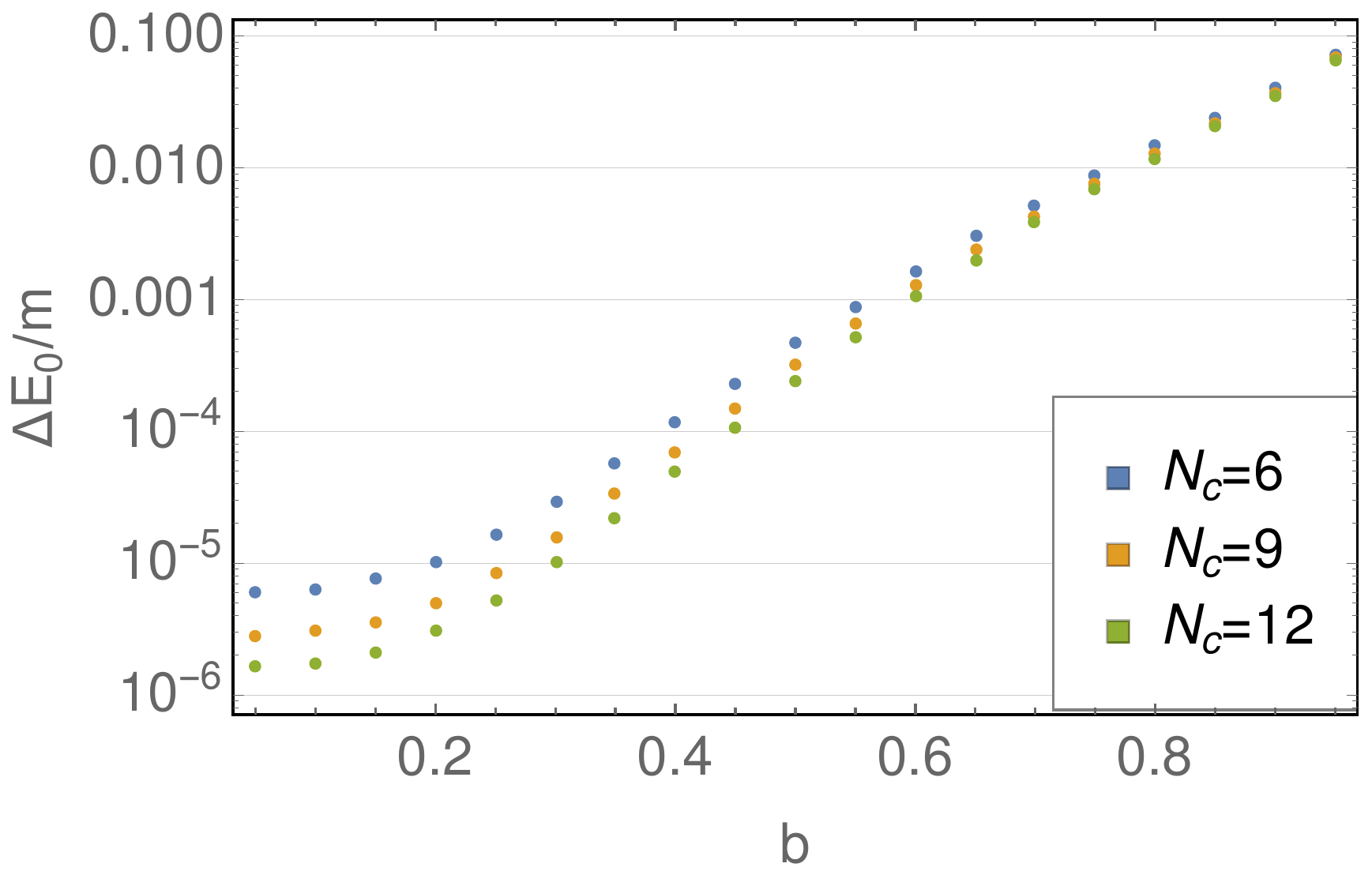}
\caption{$E_{gs}$ at different $b$'s.}
\end{subfigure}
\hfill
\begin{subfigure}[b]{0.475\textwidth}
\centering
\includegraphics[draft=false,width=\textwidth]{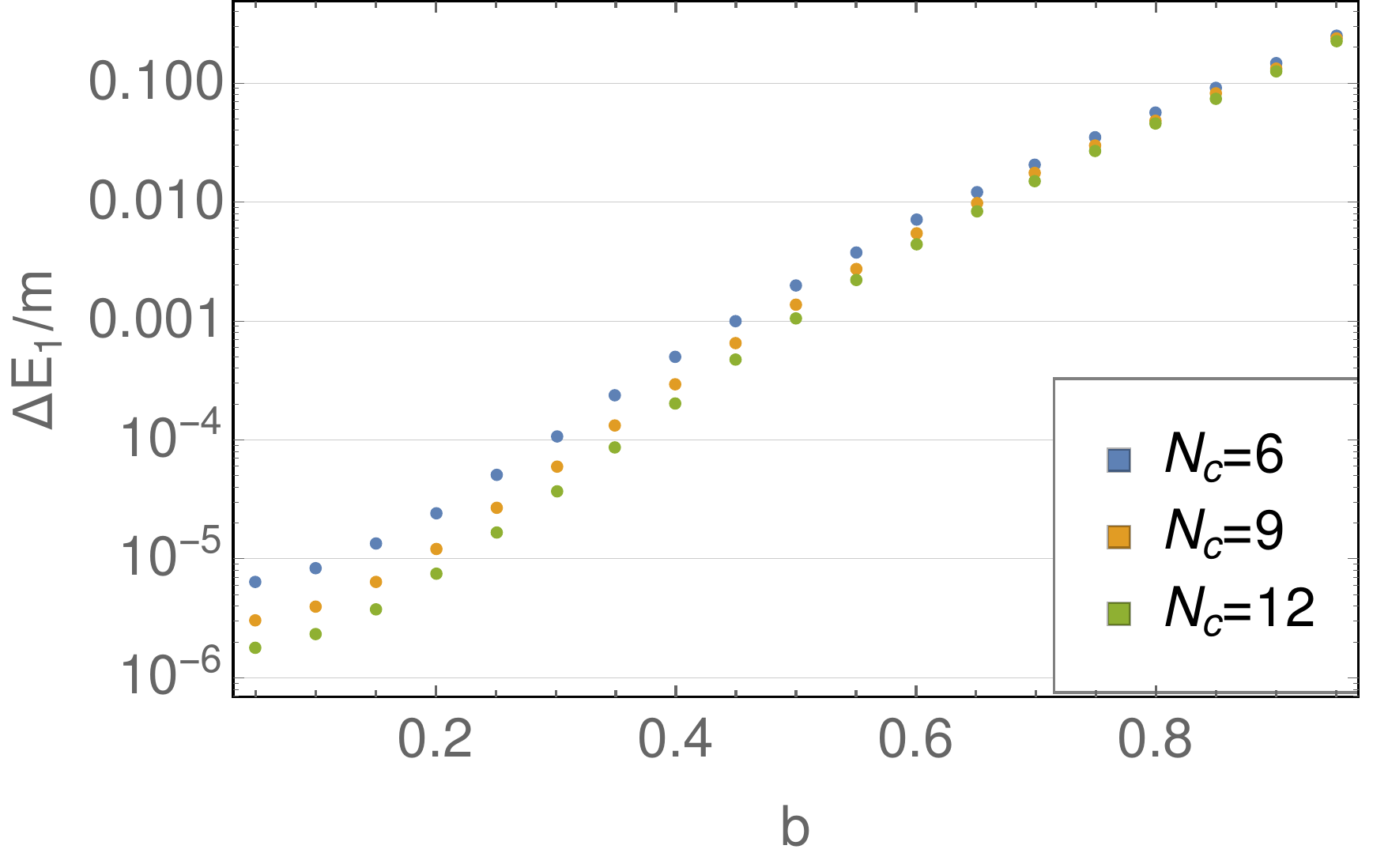}
\caption{$E_{1,exc}$ at different $b$'s.}
\end{subfigure}
\caption{
Here we present results on the behavior of the ground state energy (a) and first excited state (b) as a function of $b$ at fixed $M_{ShG}R=1$.  We present results for 3 different values of the cutoff ($N_c=6,9,12,N_{ZM}$=24). The energies are normalized with respect to the free mass $m$. We have plotted the results as differences between the numerical values and the expected exact values from TBA.}
\label{FigRawCut1}
\end{figure}

In this section we present our TSM results based on the non-compact bosonic computational basis for various quantities in the ShG model.  In the first part, we show how the TSM results for the ShG are robust at small b $(b\ll 1)$, but begin to have strong cutoff effects in $N_c$ (and, to a lesser extent, $N_{zm}$), deviating noticeably from exact results predicted by the thermodynamic Bethe ansatz (TBA), as $b$ exceeds $1/2$.  We then discuss our first strategy in dealing with these cutoffs: a power law extrapolation in $N_c$ (and $N_{zm}$).  We show that this procedure in fact produces robust results - albeit still imperfect as $b\rightarrow 1$ is approached.  We then turn to other quantities that we are able to measure using TSM methods, such as the VEVs of the exponential operators and the S-matrix.

In the second part of this section we consider standard renormalization group strategies for alleviating the effects of the cutoff. We show that while these strategies work at small values of $b$, they lead to sub-optimal (or even unphysical) results at larger values of $b$ which are closer to the self-dual point.  However this failure provides the motivating drive to consider other strategies for treating the sensitivity of TSM results to the cutoff that form the next three sections that follow this one. It will also set the scene for understanding why the power law extrapolation used in the first part of this section is robust.

\begin{figure}[t]
\begin{subfigure}[b]{0.475\textwidth}
\centering
\includegraphics[draft=false,width=\textwidth]{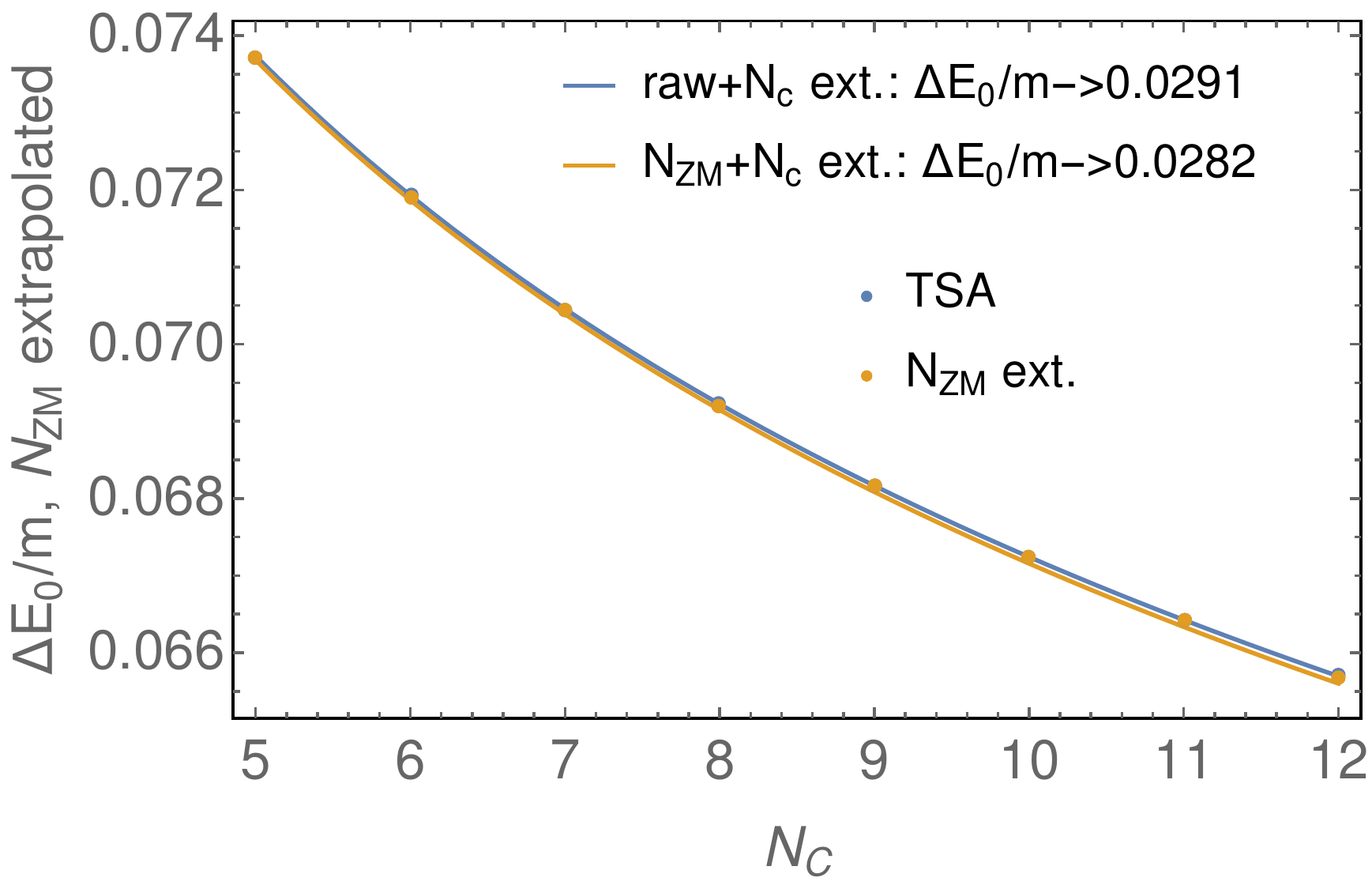}
\caption{ Raw TSM ($N_{zm}=42$) and $N_{zm}$-extrapolated values of $E_{gs}$ as a function of $N_c$.}
\end{subfigure}
\hfill
\begin{subfigure}[b]{0.475\textwidth}
\centering
\includegraphics[draft=false,width=\textwidth]{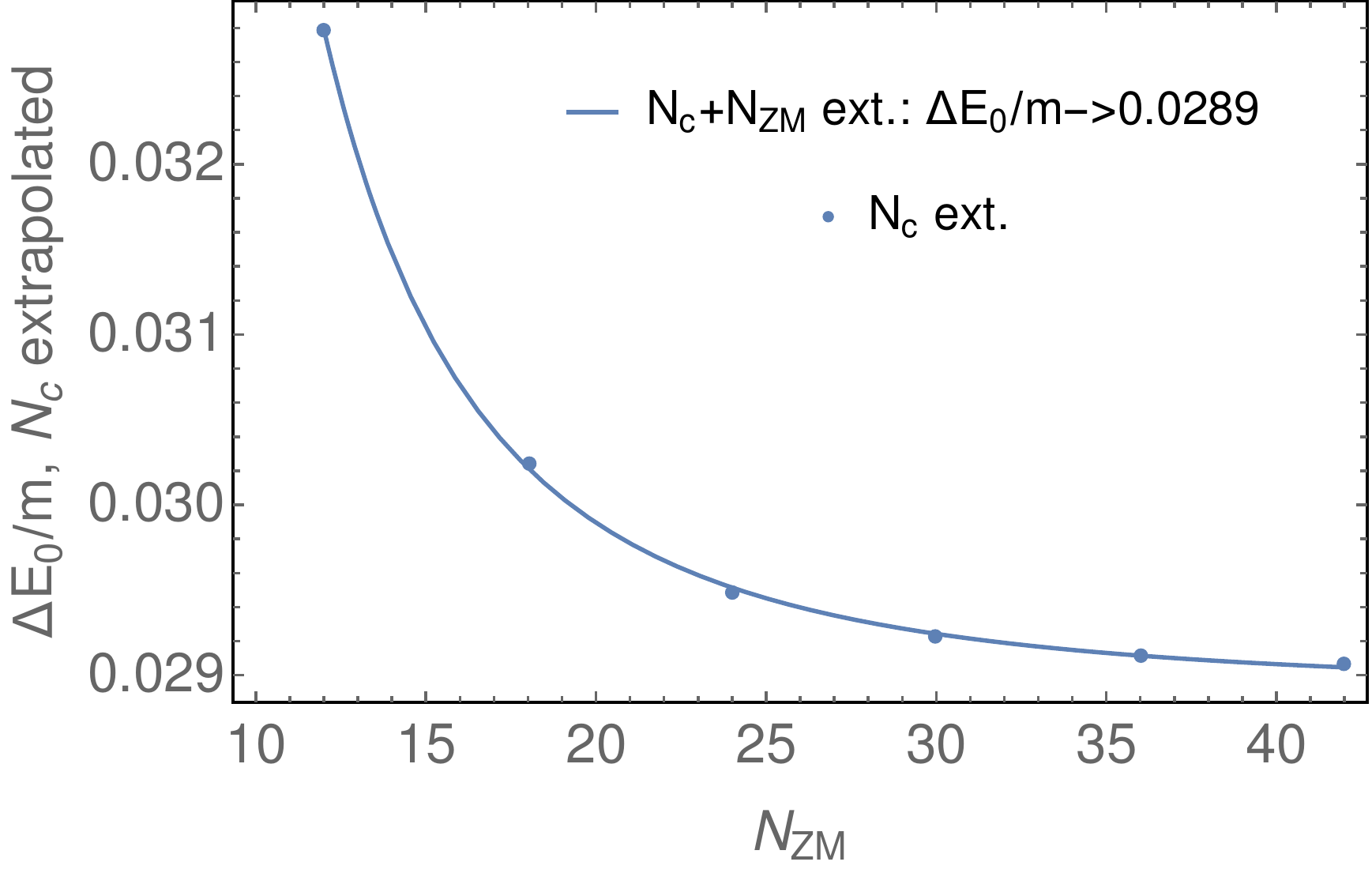}
\caption{Values of $E_{gs}$ after extrapolation in $N_c$, as a function of $N_{zm}$.}
\end{subfigure}
\caption{
Here we present the two extrapolation schemes for the cutoffs $N_{c}$ and $N_{ZM}$.  We perform the extrapolation for data of the ground state energy at $b=0.95$, $M_{ShG}R=1$). In a) we first extrapolate in $N_{ZM}$, plotting the results at each $N_c$, then performing a further extrapolation in $N_c$.  The results of this extrapolation is shown in the legend.  In b) we reverse the order of extrapolations.  Note that the data has a much stronger dependence on $N_c$ than on $N_{ZM}$. }\label{FigCutExtrap}
\end{figure}

\subsection{Results}

 We present our numerical results with the aim to answer the following specific questions:
\begin{enumerate}
\item What is the performance of the TSM applied to the ShG model? What level of precision can be achieved below the self-dual point (compared directly to finite volume theoretical quantities), and how does it depend on the coupling $b$ and the volume $R$?
\item Is there a simple extrapolation that robustly improves the accuracy of the numerics? How much does it improve?
\item Not assuming any special properties of the ShG model (in other words, relying only on standard TSM analysis), to what extent can the conjectured infinite volume parameters (mass, vacuum energy density, S-matrix) be reproduced?
\item How effectively does TSM reproduce the one-point functions of vertex operators? In particular, what happens when we probe them outside the region of validity of the FLZZ formula, eq.~(\ref{VEVexp})?
\end{enumerate}
The following results are organized according to the four points listed above.

\begin{figure}[t]
\centering
\begin{subfigure}[b]{0.475\textwidth}
\centering
\includegraphics[draft=false,width=\textwidth]{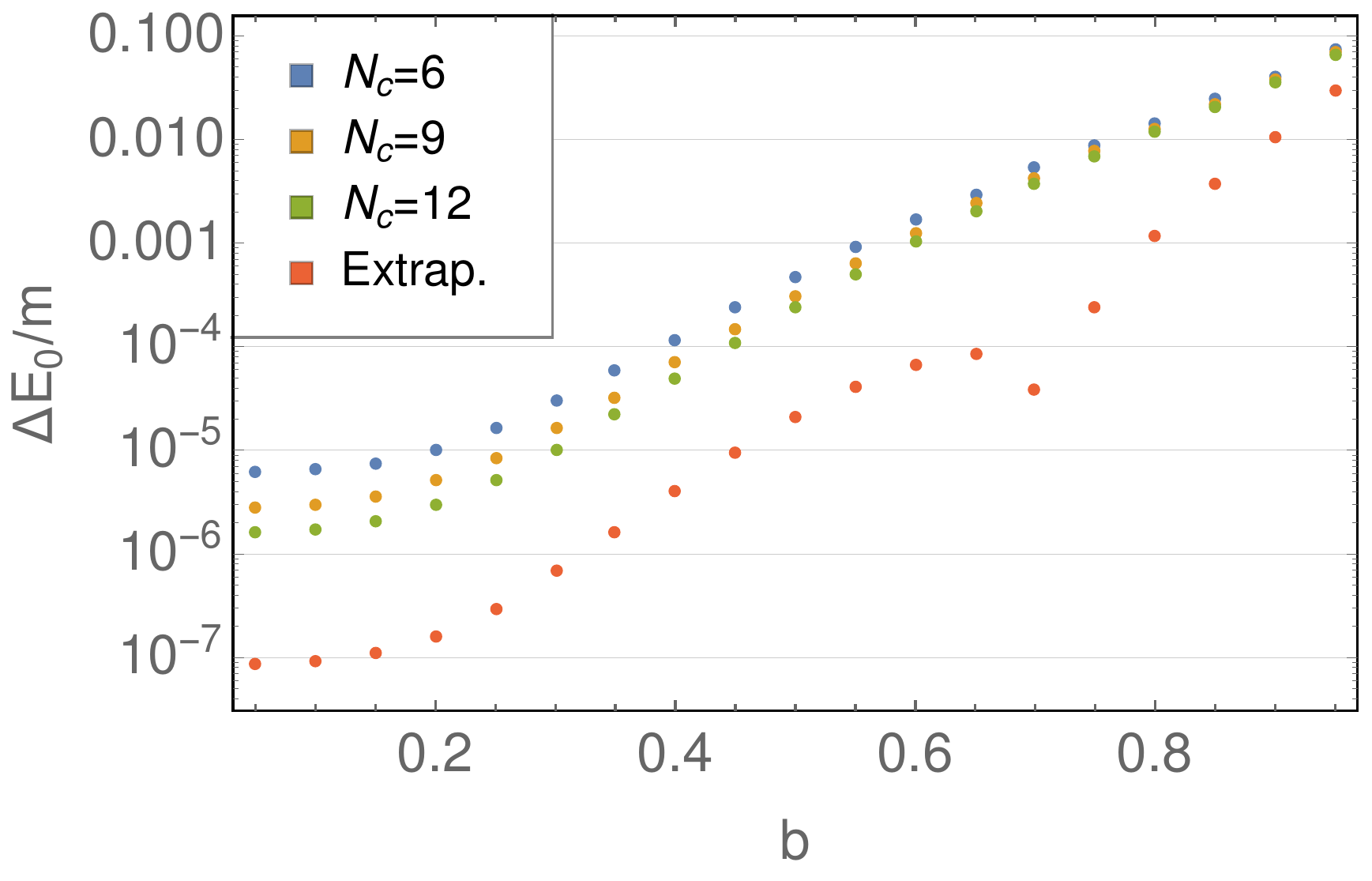}
\caption{Raw TSM data together with extrapolation in $N_{c}$ for $E_{gs}$ at different $b$'s.}
\end{subfigure}
\hfill
\begin{subfigure}[b]{0.475\textwidth}
\centering
\includegraphics[width=\textwidth]{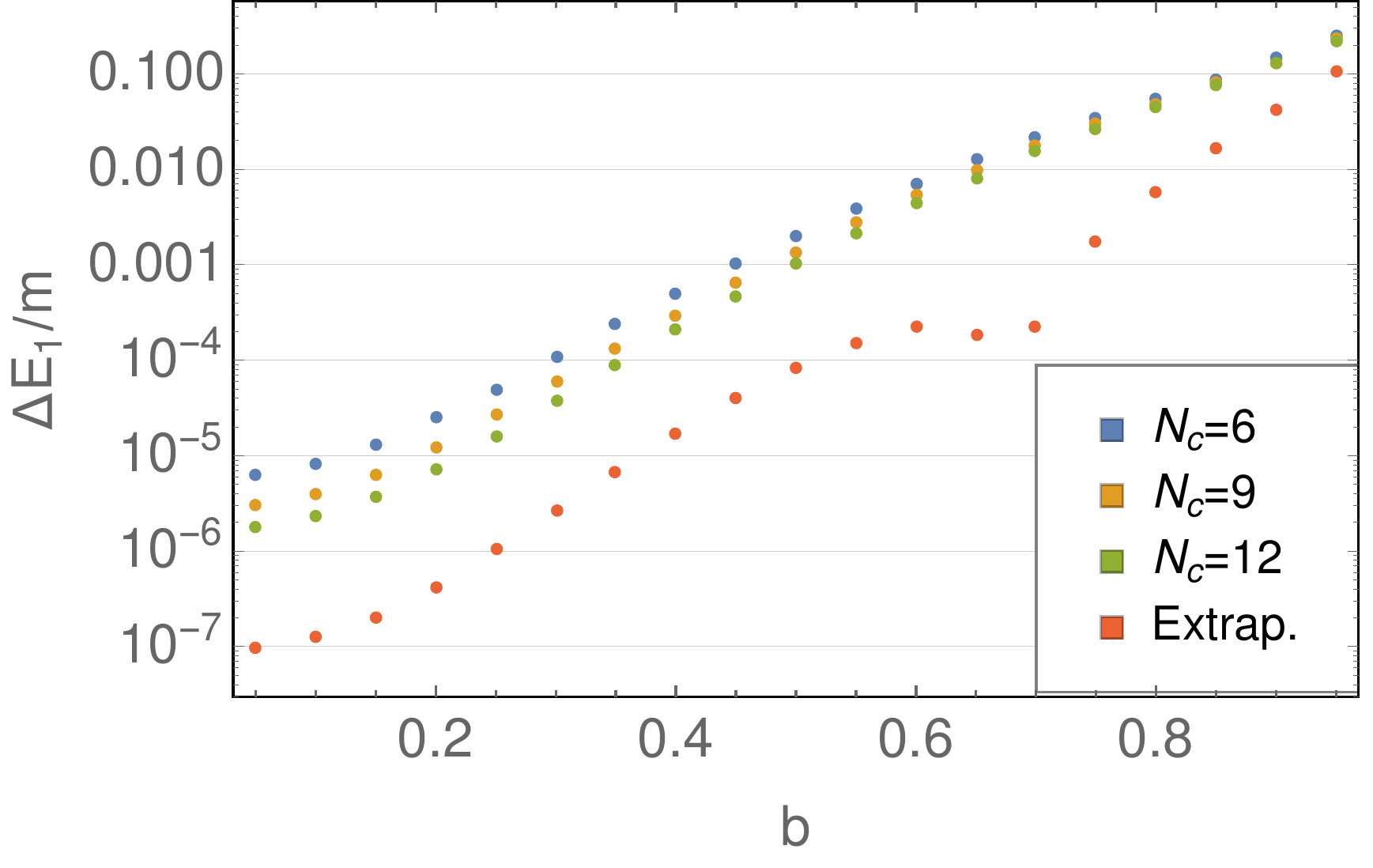}
\caption{Raw TSM data together with extrapolation in $N_{c}$ for $E_{exc}$ at different $b$'s.}
\end{subfigure}
\caption{The behaviour of the $N_c$-extrapolated TSM data as a function of b. }
\label{FigExtrapRes}
\end{figure}

\subsubsection{Finite volume spectrum}

\begin{figure}[b]
\centering
\includegraphics[width=0.45\textwidth]{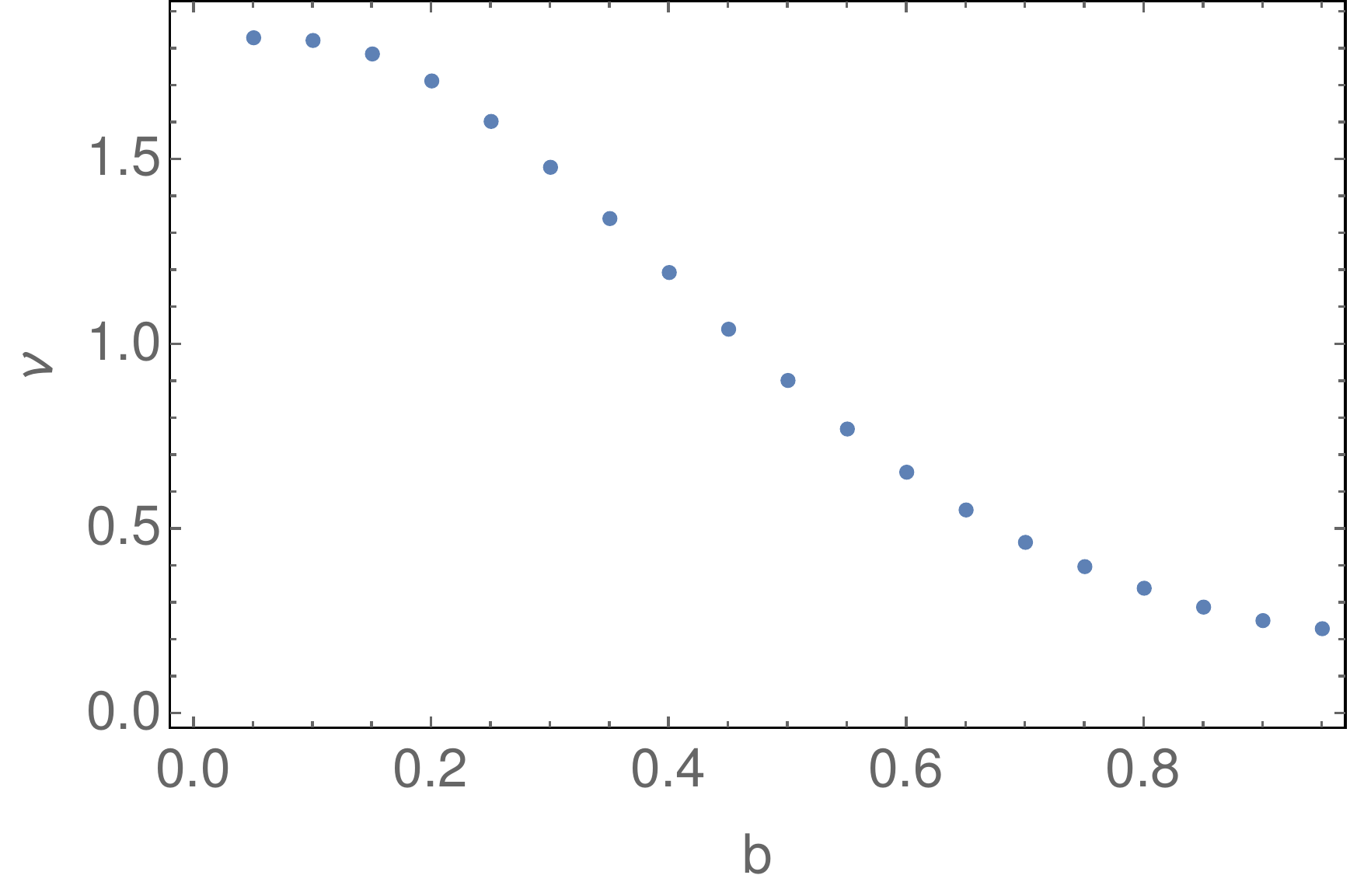}
\caption{Power law exponents as a function of $b$ for the extrapolation in $N_{c}$ given in eq.~(\ref{extrap}).}
\label{FigExtrapPowlaw}
\end{figure}

Let us begin with the finite volume spectrum. In Fig.~\ref{FigRawCut1} we present results for the ground state energy and the first excited state energy for different values of $b$ at fixed volume $M_{ShG}R=1$, where $M_{ShG}$ is the physical mass. The computations were done using different chiral cut-offs $N_c$, at a fixed zero mode cutoff $N_{ZM}$. On the other hand, we have also calculated the corresponding quantities by numerically integrating the (excited state) TBA equations eq.~(\ref{eq:shGE}). We consider the difference between TSM and TBA (taking into account the vacuum energy density \eqref{vacumenergy}) to be the error of the former. The energy levels are normalized with respect to the free mass $m$ defined in eq. (\ref{lambdamm}) and we plot the differences between the TBA computations and the TSM data on a log scale. The largest cutoff, $N_c=12$, $N_{ZM}=24$, at which data are presented has been obtained using a truncated Hilbert space of size $3\times 10^5$.

It is apparent that the errors are slightly different for the ground state and the excited state, but the overall pattern is very similar: for small $b$, even a raw cutoff can produce precise results, and a reasonable increase in the cutoff $N_c$ actually has a strong positive effect on the precision. On the other hand, the error increases exponentially in increasing the coupling constant $b$ and, at the same time, the precision becomes less sensitive to the cutoff.  In the immediate vicinity of the self-dual point, the error essentially becomes $O(1)$, indicating that the naive TSM is limited to a region below the self-dual point.  
As a first step to improve the results, we propose an (at this point `empirical') extrapolation scheme, which involves fitting the numerical results with power laws in $N_c$ and $N_{ZM}$.

In Fig.~\ref{FigCutExtrap}, we present two implementations of this power law fitting at $b=0.95$, $M_{ShG}R=1$. In the first, presented in Fig.~\ref{FigCutExtrap}(a), we first extrapolate in zero mode number $N_{ZM}$ and then perform a further extrapolation in $N_c$.  Fig.~\ref{FigCutExtrap}(a) then shows the result of this first extrapolation in $N_{ZM}$ at each $N_c$.  We see that the results of this extrapolation differ little from the raw data determined at $N_{ZM}=42$.   Having done this, we then perform a separate extrapolation with respect to the chiral cutoff $N_c$.  The result, reported in the legend of Fig.~\ref{FigCutExtrap}(a) is about a 3\% error in units of the free mass $m$.  We also perform the same extrapolation in $N_c$ for the raw data obtained at $N_{ZM}=42$.  The result is essentially identical.  In Fig.~\ref{FigCutExtrap}(b), we consider the second implementation.  This is obtained by performing the two extrapolations in the opposite order but for the same set of input data.  Shown in Fig.~\ref{FigCutExtrap}(b) is the result of the first extrapolation in $N_c$ at fixed $N_{ZM}$.  The second extrapolation in $N_{ZM}$ lead to results essentially the same as the results reported with the first scheme.

We now consider TSM data over a range of values of $b$.  Having seen that the data is essentially converged at $N_{ZM}$ sufficiently large, we work at fixed $N_{ZM}$ and only consider extrapolations in $N_c$.   In particular we use the fitting function:
\begin{equation}
E(N_c) \,=\, E_{\mathrm{extrap}}+d\cdot{N_c^{-\nu}} \,\,\, . 
\label{extrap}
\end{equation}
In Fig.~\ref{FigExtrapRes} we report our results for the ground state and first excited state energies at $M_{ShG}R=1$.  We show the raw TSM data at different $N_c$ together with the extrapolated values (red dots).
In most cases, the extrapolation improves the numerical data by at least an order of magnitude. We note that the cusp-like feature appearing in the extrapolated data is due to a sign change of the extrapolated error, of which we take the absolute value to produce the log-scale plot. The precise position of the cusp is also volume-dependent.
In Fig.~\ref{FigExtrapPowlaw} we present the fitting exponent $\nu(b)$ (see eq.~(\ref{extrap})) coming from these extrapolations.  We see that at small $b$ the exponent is large indicating that the data is rapidly converging in $N_c$ while at values of $b$ approaching the self-dual point, the exponent becomes much smaller.  We will provide a partial explanation for the behavior of the power law in Section \ref{Supraborelsection}.

\begin{figure}
\begin{subfigure}[b]{0.475\textwidth}
\centering
\includegraphics[draft=false,width=\textwidth]{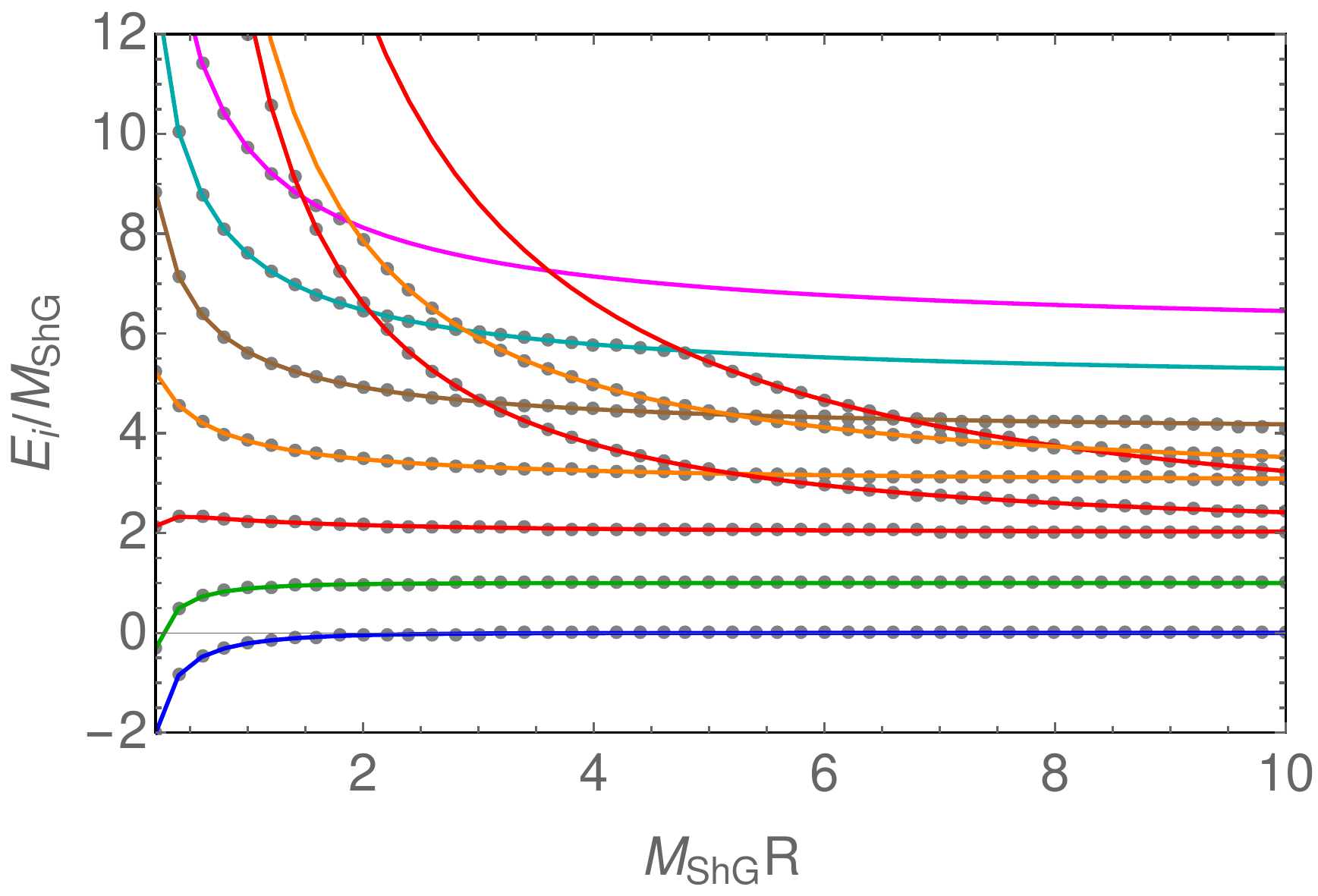}
\caption{Energy levels at $b=0.4$.}
\end{subfigure}
\hfill
\begin{subfigure}[b]{0.475\textwidth}
\centering
\includegraphics[draft=false,width=\textwidth]{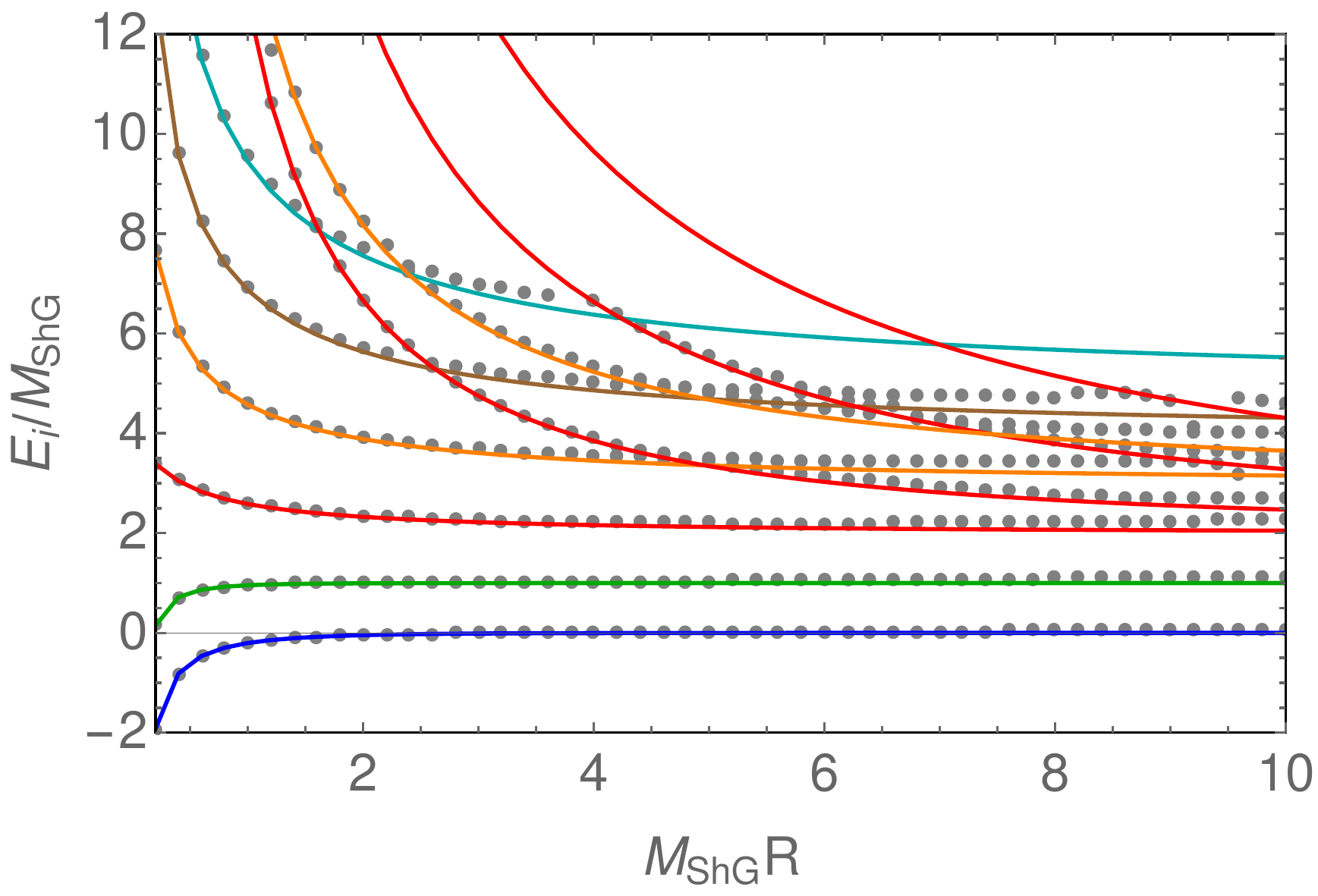}
\caption{Energy levels at $b=0.8$.}
\end{subfigure}
\caption{Here we present TSM data (dots) for the first $8$ energy levels after $N_c$ extrapolation
as functions of the dimensionless volume $M_{ShG}R$ in the zero momentum sector.
The vacuum energy $\mathcal{E}_0$ is subtracted and energies are normalized with respect to the mass $M_{ShG}$.  The color coding of the TBA (solid) curves is as follows: the ground state is depicted in blue, the one-particle state in green, two-particle states in red, three-particle states in orange, four-particle states in brown, five-particle states in teal, and six-particle states in pink. 
} \label{FigFiniteVSpec}
\end{figure}

Having presented results as a function of $b$, we now consider the spectrum as a function of $R$. 
In Fig.~\ref{FigFiniteVSpec} we present data for the low-lying finite volume spectrum (after extrapolation in $N_c$) after subtraction of the exact vacuum energy density \eqref{vacumenergy} for two different couplings, $b=0.4$ and $b=0.8$. The TSM data is plotted against the numerical solution of the exact TBA equations (shown in the plots with continuous curves). It is apparent that TSM follows very closely the theoretical excited-TBA data for $b=0.4$, while small discrepancies become visible at $b=0.8$, especially at larger volumes. Contrary to the previous plots, here we opted for normalizing the energies with respect to the physical mass $M_{ShG}$, owing to the emphasis of finite volume corrections presented in these plots.

We close the first part of this subsection with a contour plot which shows the order of magnitude of errors as a function of both $b$ and the dimensionless volume $M_{ShG}R$ at the same time. The error can be smaller than $10^{-8}$ in the small-$R$ region of the perturbative sector, and remains below $10^{-4}$ over a wide range of couplings and volumes. On the other hand, the error increases exponentially as either the volume or the coupling is increased. We note that the apparent `islands' on the top and the `valley' around $M_{ShG}R=9$ are due to the same sign-changing phenomenon that causes the cusps in Fig.~\ref{FigExtrapRes}.

\begin{figure}[t]
\centering
\includegraphics[width=0.55\textwidth]{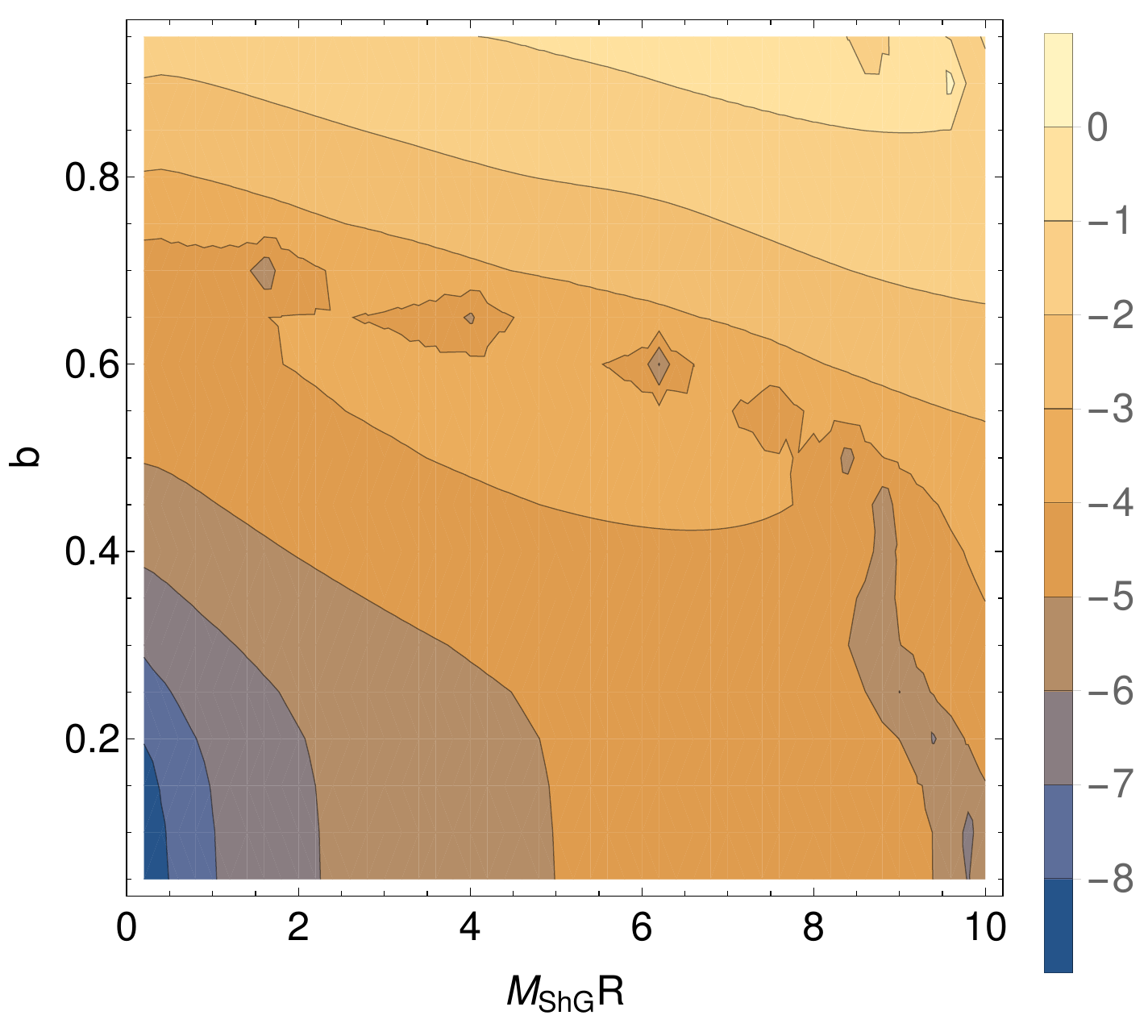}
\caption{Base 10 logarithm of the normalized differences between the TSM ground state energies and their TBA counterparts for different couplings and volumes. }
\label{contour}
\end{figure}

\subsubsection{Determination of Mass, Bulk energy and $S$-matrix}

In this subsection, we present and discuss the TSM numerical results for the particle mass, $M_{ShG}$, the bulk energy density, and the $S$-matrix of the ShG model. 

\vspace{3mm}
\noindent {\bf Physical Mass:} 
We have measured the physical mass $M_{ShG}$ of the ShG model through taking the difference of the two lowest energy levels.  Ideally, this difference converges to the physical mass in the $R\rightarrow\infty$ limit. In practice, for large volumes, truncation effects produce an overestimate for the mass. On the other hand, small-volume effects also produce an overestimate. As a consequence, we determine the mass as the minimum of the volume-dependent energy difference.
\begin{figure}
\centering
\begin{subfigure}[b]{0.475\textwidth}
\centering
\includegraphics[draft=false,width=\textwidth]{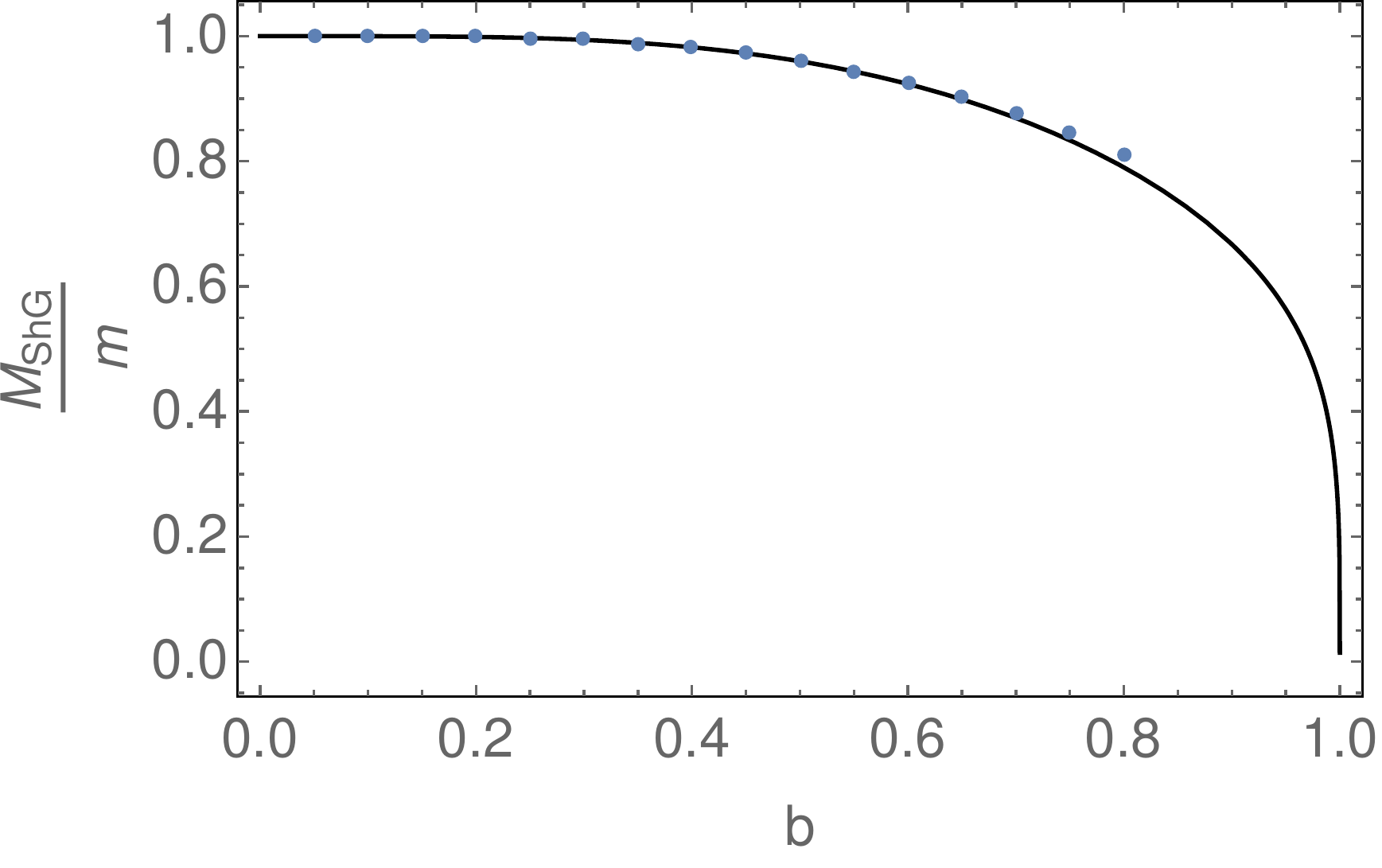}
\caption{Measured versus theoretical masses, normalized with respect to $m$.}
\end{subfigure}
\hfill
\begin{subfigure}[b]{0.475\textwidth}
\centering
\includegraphics[width=\textwidth]{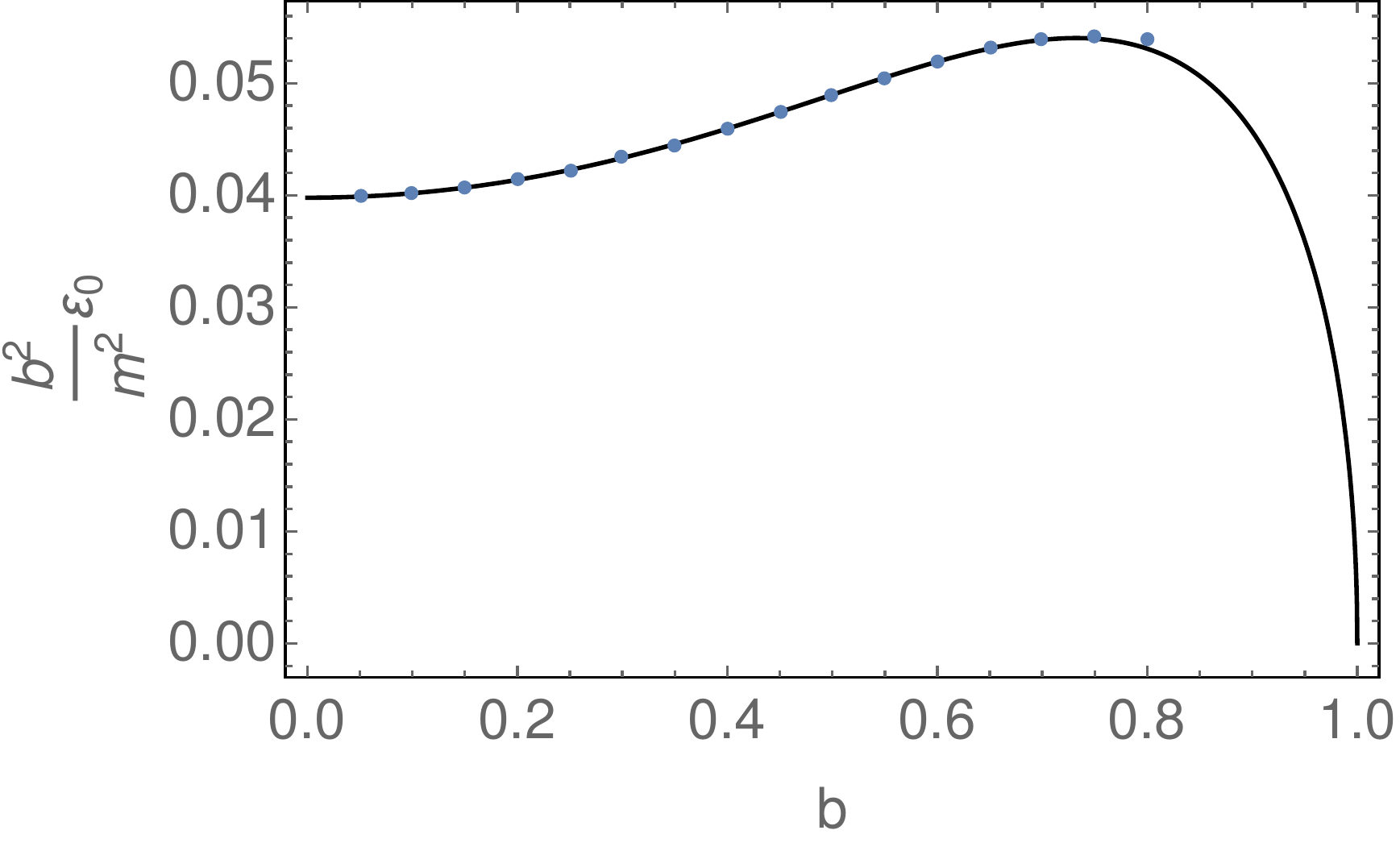}
\caption{Measured versus theoretical vacuum energies, normalized with respect to $\frac{m^2}{b^2}$.}
\end{subfigure}
\begin{subfigure}[b]{0.475\textwidth}
\centering
\includegraphics[draft=false,width=\textwidth]{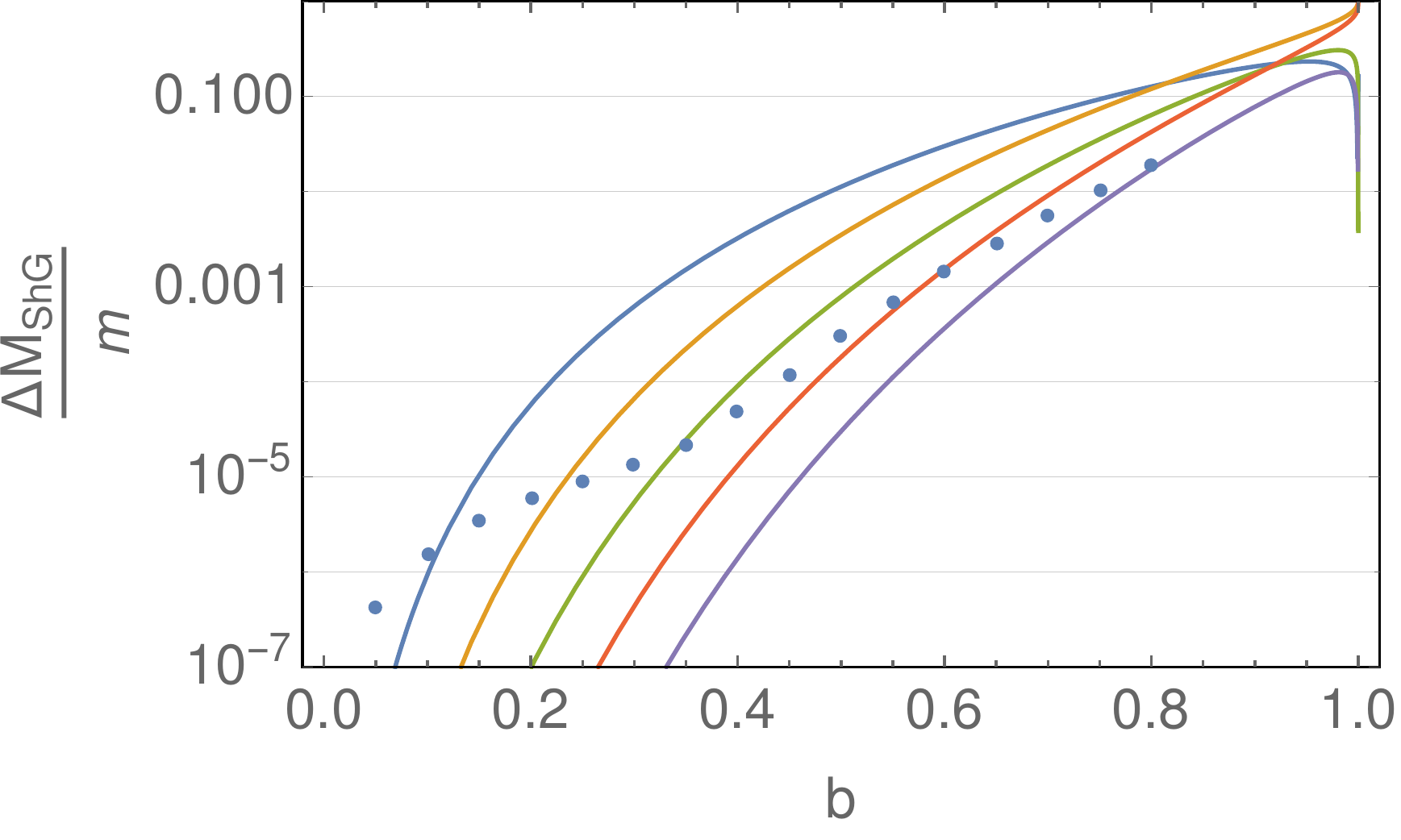}
\caption{Error in TSM determination of normalized masses.}
\end{subfigure}
\hfill
\begin{subfigure}[b]{0.475\textwidth}
\centering
\includegraphics[width=\textwidth]{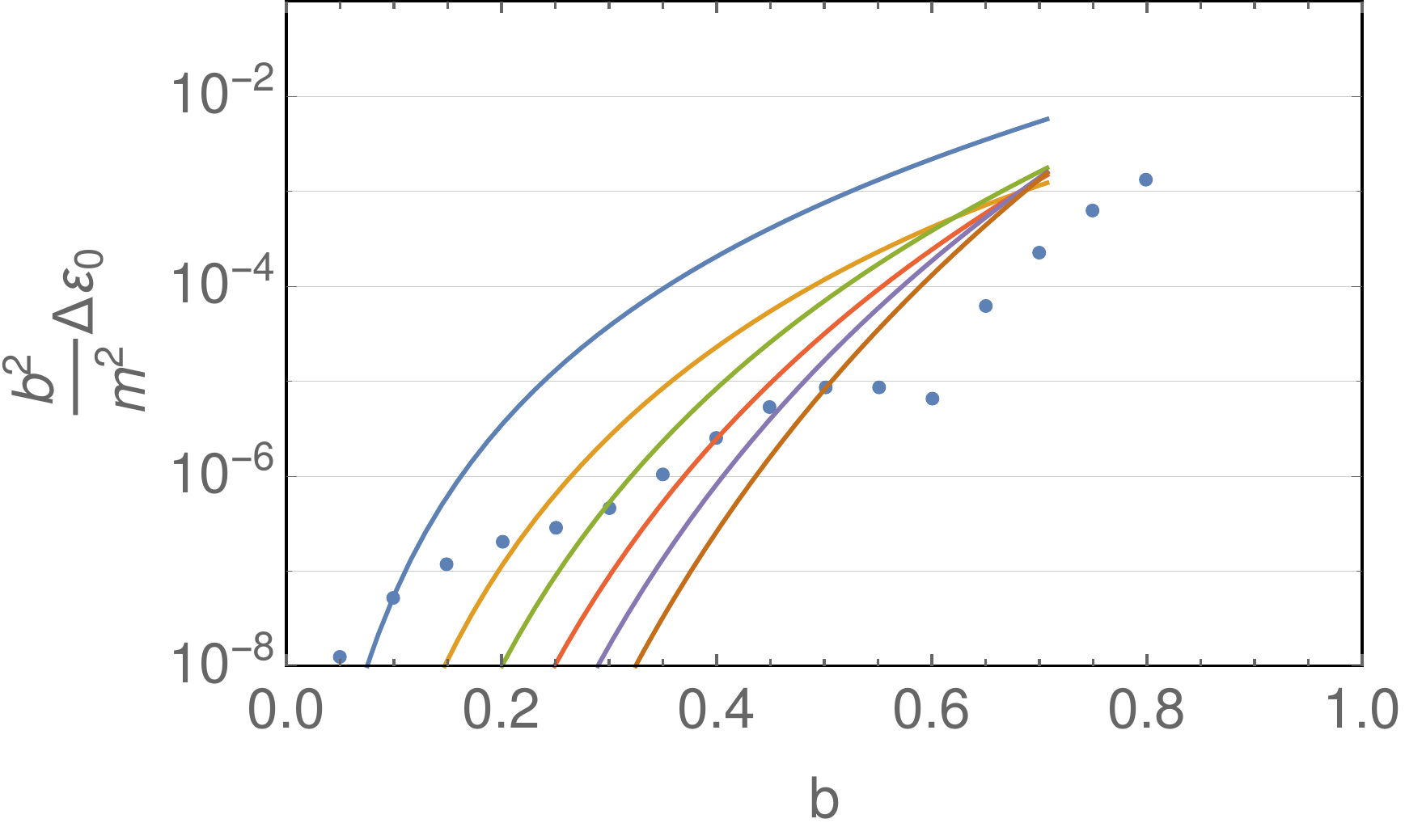}
\caption{Error in TSM determination of normalized vacuum energies.}
\end{subfigure}
\caption{Measured mass (a) and vacuum energy (b) from TSM (blue dots) plotted against the solid theoretical curve.  In panels (c) and (d), the error in the TSM results (i.e. the normalized difference between the TSM results and the theoretical values) are presented.   We also present the differences between a Feynman perturbative expansion in $b$ with the exact formula.  Shown are differences for orders $b^4$ (blue), $b^6$ (orange), $b^8$ (green), $b^{10}$ (red), $b^{12}$ (purple), and $b^{14}$ (brown).
 }\label{massandVE}
\end{figure}

The results are shown as the function of $b$ in Fig.~\ref{massandVE}(a) and (c).   In (a) we plot the TSM data (extrapolated) against the theoretical curve expected from eq.~(\ref{ShGmassformula}).  In (c) we plot, on a logarithm scale, the absolute value of the differences between the measured mass and the same theoretical value.  For comparison here, we show the differences of the perturbative expansions of the exact mass formula, truncated at various orders with their exact counterpart. This gives an idea of the relative precision of TSM with respect to a perturbative expansion. We see that for intermediate couplings, the TSM outperforms a $5$-loop (up to and including $O(b^{10})$) perturbative expansion of the mass, coming close to $6$-loop accuracy.   Approaching the self-dual point, the region of viable TSM data shrinks to a region in $M_{ShG}R$ where the exponential corrections become relevant and therefore the standard TSM methods are not available. (Of course, in the sinh-Gordon model everything is supposedly known about these exponential corrections, but for now we intentionally opt for neglecting any {\it a priori} knowledge on the integrability of the model.)

\vspace{3mm}
\noindent
{\bf Vacuum energy density:}
Let us now turn our attention to the vacuum or bulk energy density, $\mathcal{E}_0$. The measurement of this quantity proceeds by measuring the slope of the ground state energy $E_0(R)\approx\mathcal{E}_0 R$. For small $R$, this function enjoys a conformal $R^{-1}$ dependence up to logarithms, and is monotonically increasing. For intermediate volumes, it is essentially linear. The bulk energy needs to be measured in this linear region since for larger volumes, truncation errors are expected to dominate. Therefore the best first approximation to $\mathcal{E}_0$ is the minimum of the numerical derivative of $E_0(R)$.
In a general field theory, the leading exponential (L{\"u}scher) correction to the ground state energy is of the form
\begin{equation}
E_0(R)\,=\, R\mathcal{E}_0-M\intop_{-\infty}^{\infty}\frac{du}{2\pi}\cosh{u}\:e^{-MR\cosh{u}}\,\,\,.
\end{equation}
Substituting the mass measured previously and subtracting this correction from the numerical ground state energy improves the precision.
The results as a function of $b$ are shown in Fig.~\ref{massandVE}(b) and (d) in the same fashion as presented in (a) and (c) of this same figure for the physical mass.  Like with Fig.~\ref{massandVE}(c), we show the differences between a finite order Feynman perturbative computation and the exact value for orders $b^4$ through $b^{14}$.
Note that the convergence radius of the series for $\mathcal{E}_0$ is only $b_{\mathrm{max}}=1/\sqrt{2}$ as opposed to $1$ for the mass, $M_{ShG}$.\footnote{This
can be seen from the analytic structure of eq. \ref{vacumenergy}.}  We thus only show the perturbative curves up to this point in $b$. Conventional TSM is able to measure the bulk energy with an error of $10^{-3}$ even in this strongly coupled region.

\begin{figure}
\centering
\begin{subfigure}[b]{0.475\textwidth}
\centering
\includegraphics[draft=false,width=\textwidth]{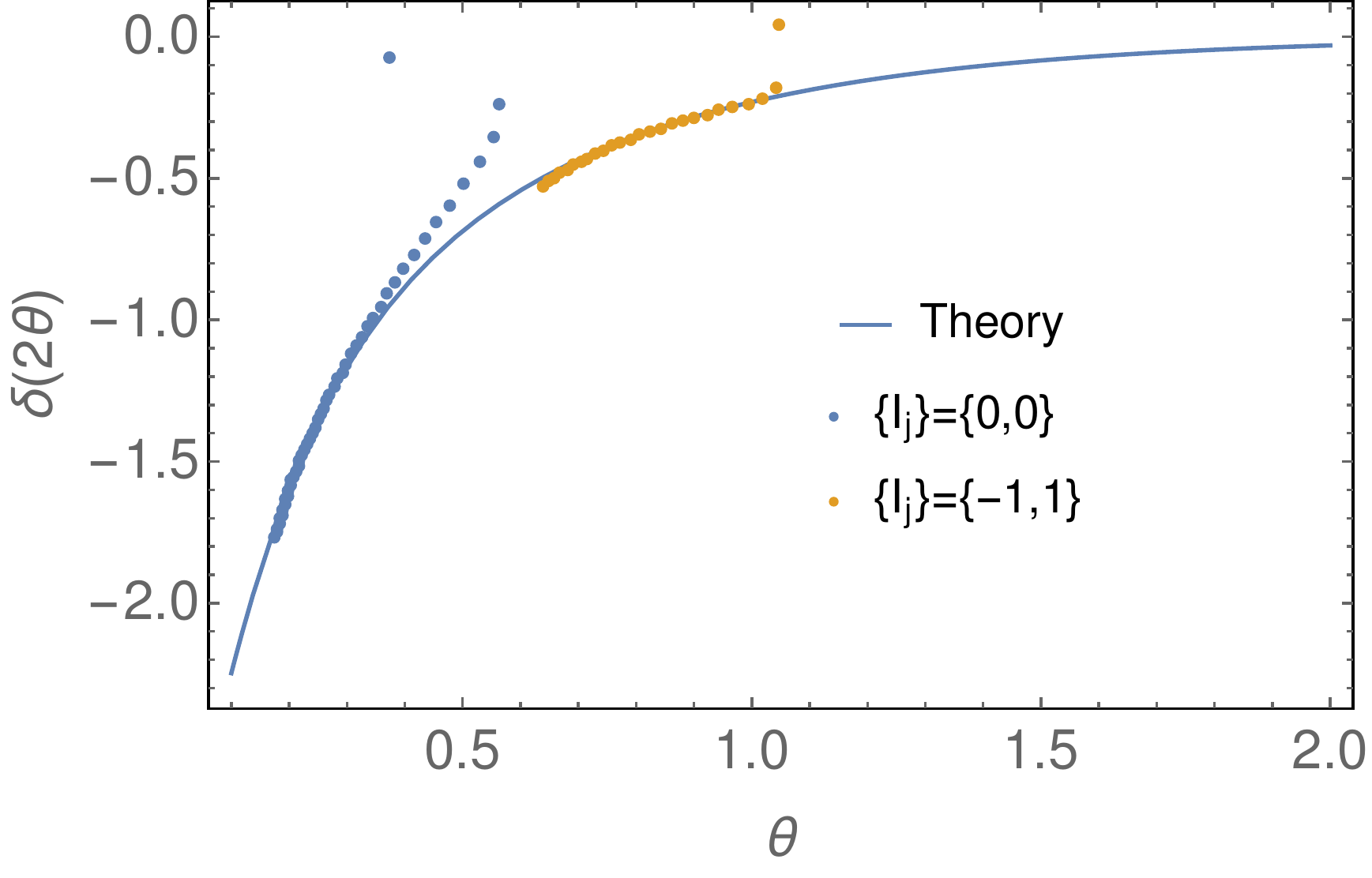}
\caption{Measured versus theoretical phase shifts obtained from the
$n=0$ and $n=1$ two-particle states at $b=0.4$.}
\end{subfigure}
\hfill
\begin{subfigure}[b]{0.475\textwidth}
\centering
\includegraphics[draft=false,width=\textwidth]{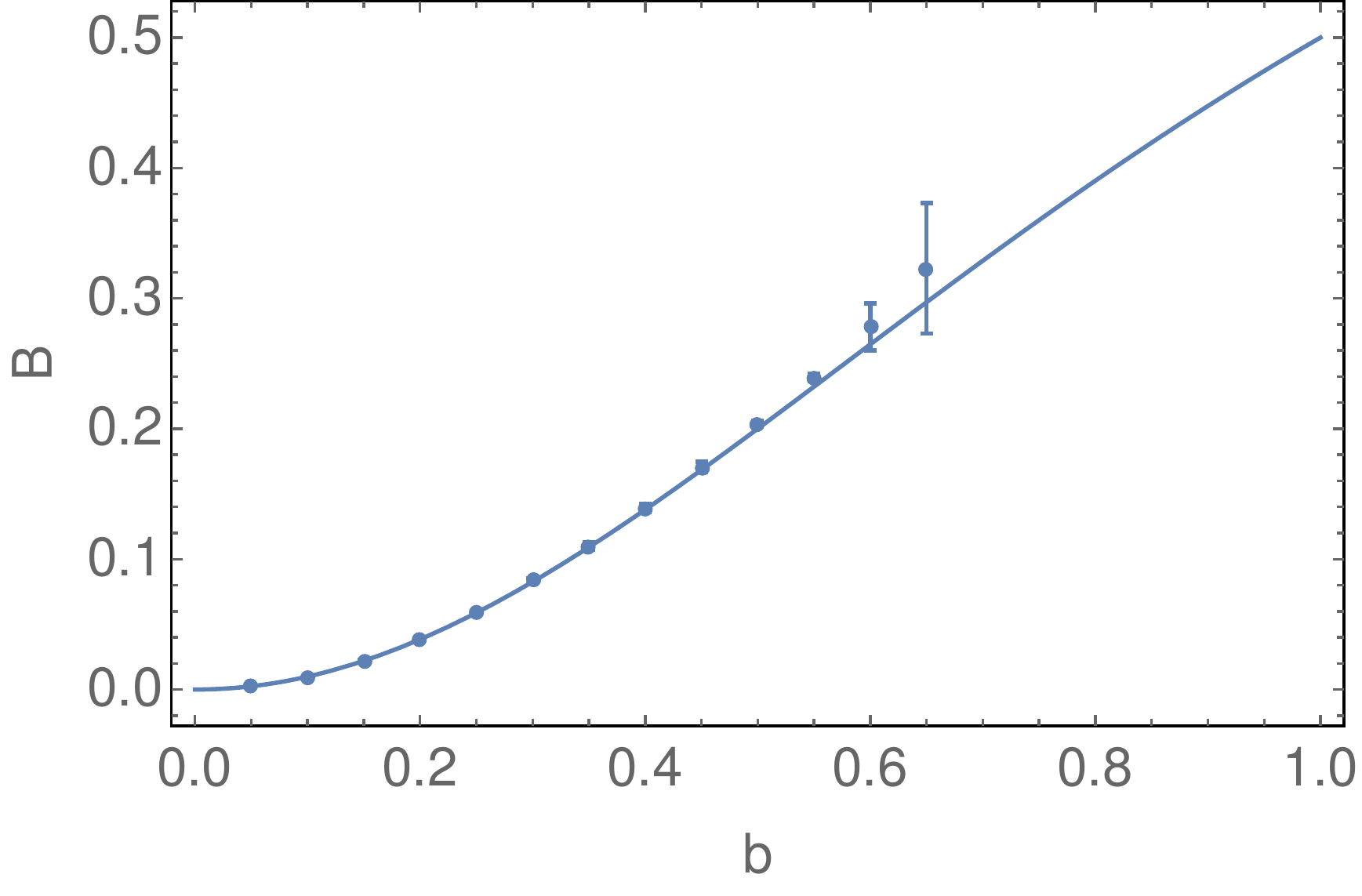}
\caption{Measurement of the parameter $B$ of the $S$-matrix as a function of $b$.\\ \:}
\end{subfigure}
\caption{Measurement of the S-matrix. }\label{SmatFig}
\end{figure}

\vspace{3mm}
\noindent
{\bf $S$-matrix:} Finally, let us consider the measurement of the S-matrix from TSM data. For asymptotically large volumes, two-particle states with zero overall momentum (each particle having rapidity $\pm\theta$) are quantized by the requirement that the multi-particle wave-function be one-valued on the cylinder
\begin{equation}
e^{i M_{ShG} R\sinh{\theta}}\, S(2\theta)\,=\,1\,\,\,,
\end{equation}
which, after taking the logarithm, provides the Bethe-Yang quantization condition
\begin{equation}
\delta(2\theta) + M_{ShG} R \, \sinh(\theta)\,=\, 2\pi n,\quad n\geq 0,\label{BetheYang}
\end{equation}
where we have introduced the phase shift $S(\theta) = e^{i\delta(\theta)}$. Eq.\,(\ref{BetheYang}) is a quantization condition which determines the rapidity $\theta$. In fact, it is the large-volume limit of the TBA quantization condition \eqref{eq:shGQQ} for the state $\{I_1=-n,I_2=n\}$. Once this quantity is known, we have access to the energy of the two-particle state since, up to exponential corrections, the energy is a sum of one-particle terms
\begin{equation}
E_{\mathrm{large R}}\,=\,R\mathcal{E} + M_{ShG} \, \sum_{j=1}^{k}\cosh(\theta_j)\,\,\,.
\end{equation}
We focus on the lowest energy two-particle states in the zero-momentum sector by taking $n=0,1$ in (\ref{BetheYang}). 
Numerically, for large enough volumes, this corresponds to the fourth lowest energy level. In this domain, we express the rapidity $\theta$ in terms of the energy difference between the two-particle state and the vacuum:
\begin{equation}
\theta\,=\, \mathrm{Arcosh}\left(\frac{E-E_0}{2M_{ShG}}\right)\,\,\,.
\end{equation}
Thus we can directly measure the phase shift appearing in eq.~(\ref{BetheYang}). The result extracted from the $n=0$ and $n=1$ two-particle states is shown on Fig.~\ref{SmatFig}(a).

In the following, we have assumed that the S-matrix indeed consists of a single CDD-like factor, namely 
\begin{equation}
S(\theta) \,=\,\frac{\sinh\theta - i \, \sin\pi B}{\sinh\theta + i \, \sin\pi B}\,\,\,,
\label{CDDD}
\end{equation}
but we have treated the quantity $B$ in the $S$-matrix amplitude as a parameter to be fitted to the numerical data.  To perform this fitting, we have only utilized our $n=1$ TSM data in region of $R$ where no level crossings occur. The numerical phase shift obtained in this way is more robust than that coming from the $n=0$ state.  We estimate the phase shift by two methods. In the first method, we increase the parameter $B$ until one of the numerically determined phases coincide with the theoretical curve.  The value of $B$ at this point is the estimate.  In the second method, we instead look at the two largest rapidities available from the $n=1$ data.  We then find the value of $B$ for which eq.~(\ref{CDDD}) best approximates the values of $\delta(\theta)$ at these rapidities.
The difference of the results of the two methods is considered to be the error of the measurement.

We will return to the measurement of the above quantities by an alternative method in Section \ref{SubsecUVmeasurement}.

\subsubsection{One-Point Functions}

TSMs can be used to measure one-point functions (either on the vacuum or on exited states) by sandwiching Schrödinger-picture operators between the numerically obtained eigenstates. Since the method directly uses the eigenvectors, the resulting precision is inevitably more limited as compared to the energy spectrum. Nevertheless, it is possible to compare the numerical estimate of the VEVs of the exponential operators to their theoretical FLZZ formula eq.~(\ref{VEVexp}) in a wide region. 

\begin{figure}
\centering
\begin{subfigure}[b]{0.475\textwidth}
\centering
\includegraphics[draft=false,width=\textwidth]{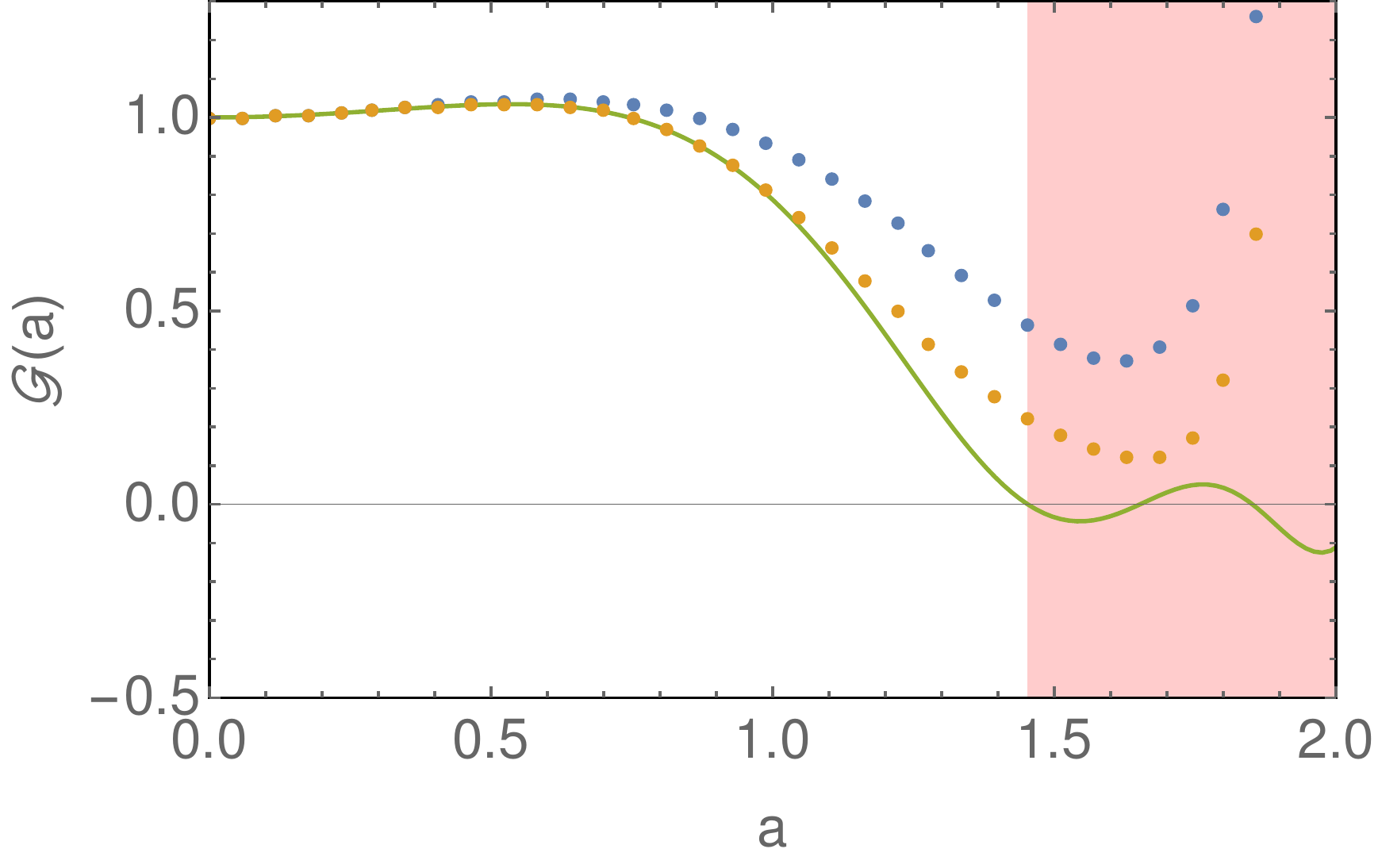}
\caption{Expectation values for $b=\frac{2}{\sqrt{8\pi}}\approx 0.4$.}
\end{subfigure}
\hfill
\begin{subfigure}[b]{0.475\textwidth}
\centering
\includegraphics[width=\textwidth]{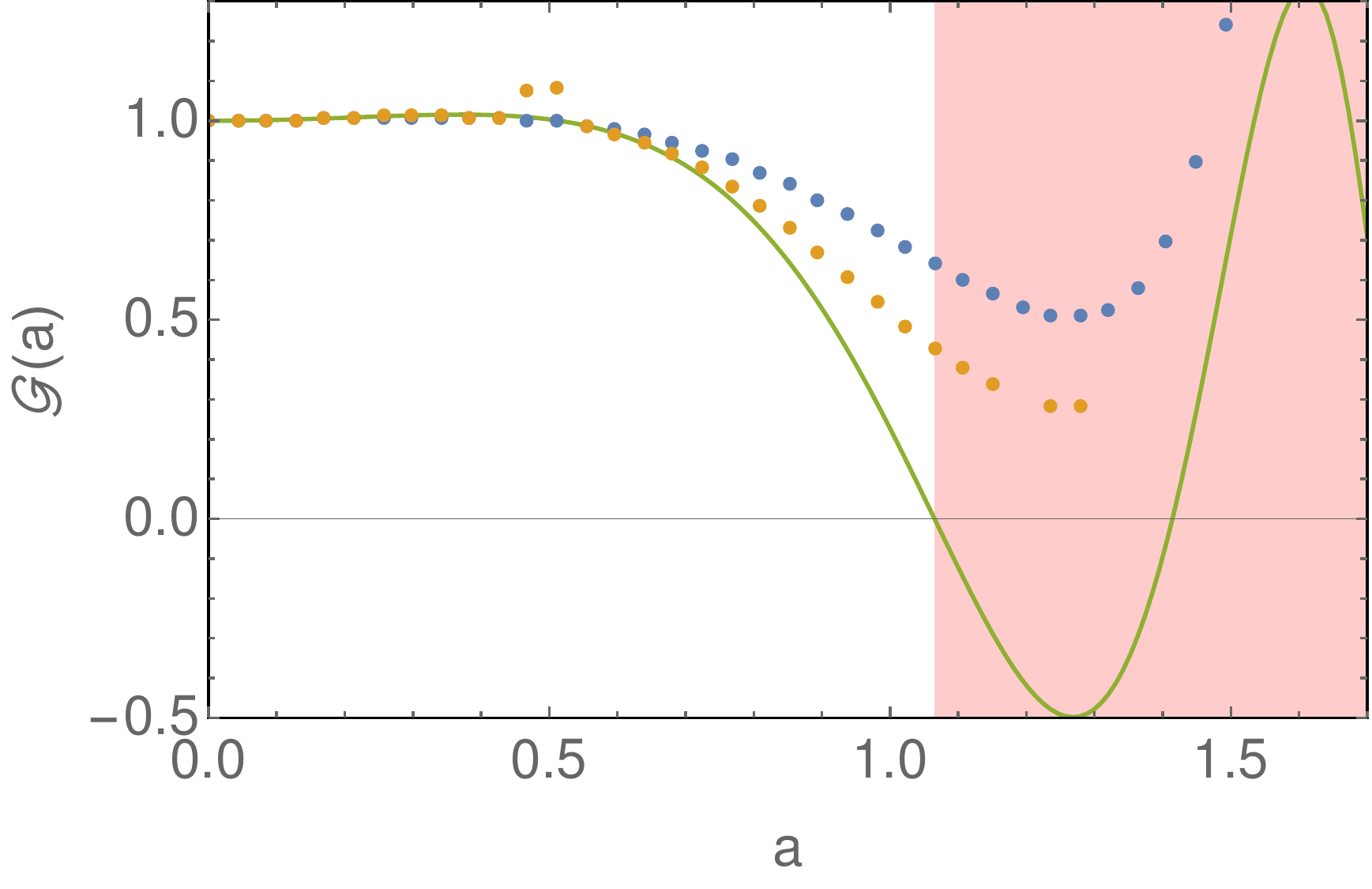}
\caption{Expectation values for $b=\frac{4}{\sqrt{8\pi}}\approx 0.8$.}
\end{subfigure}
\caption{Vacuum expectation values for $\mu_{ShG}=0.1, R=6$.
The solid green curve is predicted by the FLZZ formula. Raw TSM results are shown with blue dots, while their power-extrapolated counterparts are shown orange. The $a > Q/2$ (unphysical) domain is highlighted with red. } \label{VEVfig1}
\end{figure}

\begin{figure}
\centering
\begin{subfigure}[b]{0.475\textwidth}
\centering
\includegraphics[draft=false,width=\textwidth]{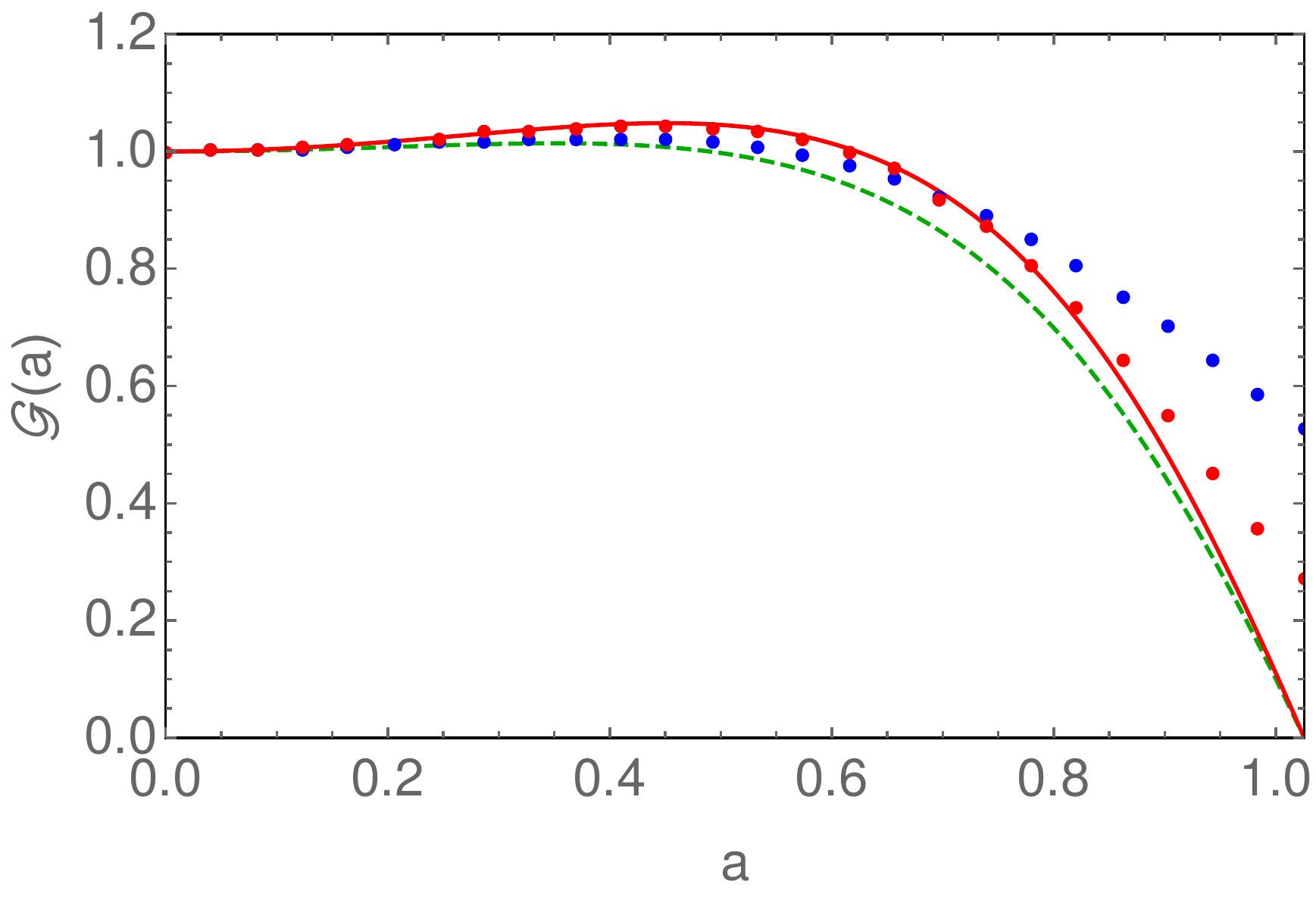}
\caption{Expectation values for $b=\frac{4}{\sqrt{8\pi}}\approx 0.8$.}
\end{subfigure}
\hfill
\begin{subfigure}[b]{0.475\textwidth}
\centering
\includegraphics[width=\textwidth]{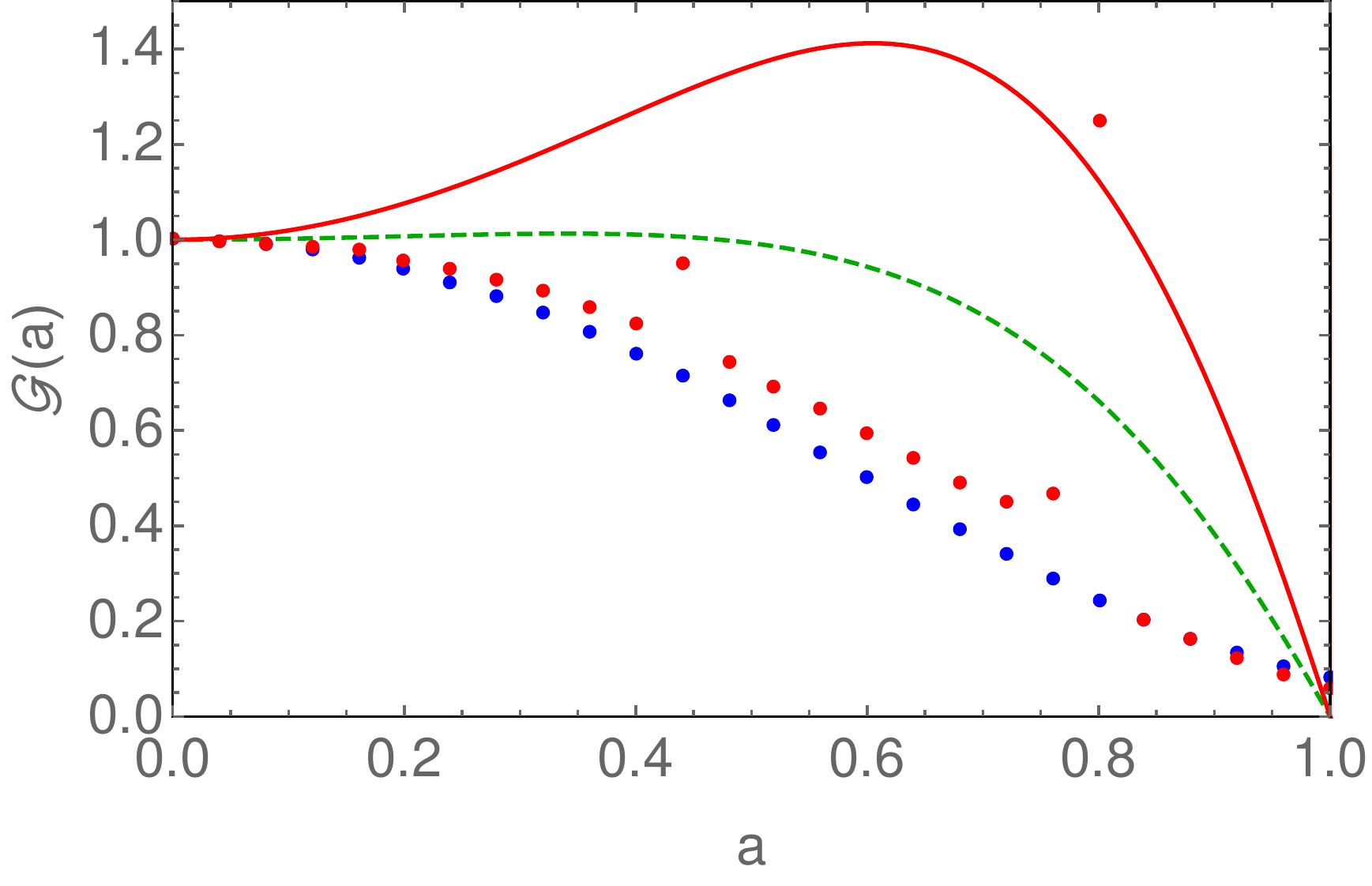}
\caption{Expectation values for $b=\frac{5}{\sqrt{8\pi}}\approx 0.99$.}
\end{subfigure}
\caption{Vacuum expectation values for $\mu_{ShG}=0.1, R=2$.
Raw TSM results are shown with blue dots, while their power-extrapolated counterparts are depicted with red dots. The $R\rightarrow\infty$ theoretical value of the VEV is plotted as a dashed green line.
The finite volume corrected version involving up to 2nd order terms in the
Leclair-Mussardo series is shown as a red solid curve. 
 }\label{VEVfigsmallR}
\end{figure}

Beyond the usual $b$ and $R$-dependence, the convergence of the VEV of the exponential operators, $\langle e^{a\varphi}\rangle$, heavily depends on the exponent $a$ of the operator. We have again tried to control the cutoff dependence of the one-point functions by power-law fits on the chiral cutoff $N_c$. The functional form eq.~(\ref{eq:plawextrap}) is helpful as long as $b$ is small enough. Generally the cutoff extrapolation becomes less stable as the coupling or the volume is increased. For larger couplings, we have found it advantageous to use a sum of two power laws as a fitting function:
\begin{equation}
E(N_c) = E_{\mathrm{extrap}}+d_1\cdot{N_c^{-\nu_1}}+d_2\cdot{N_c^{-\nu_2}}\label{eq:plawextrap}.
\end{equation}
We show the dimensionless quantity $\mathcal{G}(a)$ as a function of $a$ in Fig.~\ref{VEVfig1}, for two different couplings and at $R=6$.  For small to moderate $a$, the extrapolated numerics agrees with the FLZZ formula to at least to $1$\%.  Regardless of the coupling, the error (and the cutoff dependence) always becomes significant before reaching the first pair of zeros (located at the Seiberg bounds $\pm Q/2$) of the analytic VEV formula.
For larger couplings, it is possible to achieve higher precision by performing the computations at a small volume. However, in this case one needs to take into account the finite volume corrections of the VEV, which are available through the LeClair-Mussardo formula \cite{Leclair:1999ys} in the form of an infinite series:
\begin{equation}
\mathcal{G}(a,R)=\mathcal{G}(a)\left( 1+\sum_{n=1}^{\infty}\frac {1}{n! (2\pi)^n}\intop\prod_{i=1}^n\frac{d\theta_i}{1+e^{\epsilon(\theta_i)}}C_n^a(\left\{\theta_i\right\})\right),\label{eq:LecMus}
\end{equation}
where $C_n^a$ is the \emph{connected} evaluation of the $n$-particle diagonal form factor for the operator $e^{a\varphi}$:
\begin{equation}
\lim_{\forall\epsilon_j\rightarrow 0}F_{2n}^a(\theta_n+i\pi+\epsilon_n,\dots,\theta_1+i\pi+\epsilon_1,\theta_1,\dots,\theta_n)=C_n^a(\left\{\theta_i\right\})+O(\epsilon^{-1}).
\end{equation}
In Fig.~\ref{VEVfigsmallR}, we show the numerical results for $R=2$. Exploiting finite volume corrections extends the availability of TSM estimates of $\mathcal{G}(a)$ up to $b\approx0.8$, as long as $a\ll Q/2$.

\subsection{Renormalization Group Improvements}\label{RGIMPR}

In the previous section, we presented a series of results for various quantities in the ShG model as measured with TSM.  We showed in general that as one moves towards the self-dual point, the quality of the results found using TSM deteriorates in comparison to the available exact predictions.  This deterioration is largely due to the presence of finite cutoff effects.  There are a standard set of renormalization group-like  techniques that are employed in ameliorating the effects of a finite cutoff.  We show in this section that these strategies are suboptimal for the sinh-Gordon model close to the self-dual point.  However this failure is instructional and point the way to a better understanding of some of the peculiarities of the model and new strategies to tackle them. The strategies take two forms: analytical and numerical.  We discuss the analytic form first.

\subsubsection{Analytic Renormalization Group} \label{subsubAnalyticRG}

In presenting how one can analytically take into account the effects of states above the cutoff, we follow the discussion in Ref.~\cite{Hogervorst:2015}. The first step is to divide the Hilbert space, $\Ha$ into two parts: $\Ha$ = $\Hl$ $\otimes$ $\Hh$.  Here $\Hl$, the low energy Hilbert space, consists of all states of the form
\begin{eqnarray}\label{eIVsec2_1}
\Hl &=& \left\{  a_{n_{1}}^{\dagger}\dots a_{n_{k}}^{\dagger}\left|0\right\rangle \otimes a_{-m_{1}}^{\dagger}\dots a_{-m_{l}}^{\dagger}\left|0\right\rangle \otimes |s\rangle;\right.\cr\cr
&& \hskip 1in \left.\sum^k_{i=1} n_k \leq N_c,~~ \sum^l_{m=1} n_l \leq N_c; ~~s=1,\cdots,N_{ZM}\right\},
\end{eqnarray}
while $\Hh$, the high energy part of the Hilbert space consists of states where 
\begin{eqnarray}\label{eIVsec2_2}
\Hh &=& \left\{  a_{n_{1}}^{\dagger}\dots a_{n_{k}}^{\dagger}\left|0\right\rangle \otimes a_{-m_{1}}^{\dagger}\dots a_{-m_{l}}^{\dagger}\left|0\right\rangle \otimes |s\rangle\right.; \cr\cr
&& \hskip 1in \left. \sum^k_{i=1} n_k > N_c {~~\rm or ~~}  \sum^l_{m=1} n_l > N_c; ~~s=1,\cdots,N_{ZM}\right\}.
\end{eqnarray}
Here note we have expressed $\Hl$ and $\Hh$ in a way reflective of the tensor nature of the computational Hilbert space (at least as conceived as that of a free massless non-compact boson).
We are also working at a fixed number of zero mode states, $N_{ZM}$, assuming in effect, that this number of zero mode states leads to completely convergent results (an assumption borne out by our numerics reported in the previous section).  In particular the high and low energy parts of the Hilbert space have the same zero mode content.

We can thus write our Hamiltonian in the following manner:
\begin{equation}\label{eIVsec2_3}
H = \begin{bmatrix}
H_{ll} & H_{lh} \\
H_{hl} & H_{hh} 
\end{bmatrix},
\end{equation}
where $H_{ij}$ ($i,j=h,l$) corresponds to the Hamiltonian matrix restricted to the two subdivisions of the Hilbert space. If we have an eigenstate 
\be
\begin{bmatrix}
c_l \\
c_h 
\end{bmatrix},
\ee
with energy $E$, we can write the Schr\"odinger equation as
\begin{eqnarray}\label{eIVsec2_4}
H_{ll}c_l + H_{lh}c_h &=& Ec_l,\cr\cr
H_{hl}c_l + H_{hh}c_h &=& Ec_h .
\end{eqnarray}
By eliminating $c_h$ from the above set of equations, we have 
\begin{equation}\label{eIVsec2_5}
\bigg(H_{ll} + H_{lh}\frac{1}{E-H_{hh}}H_{hl}\bigg)c_l = (H_{ll} + \delta H )c_l = Ec_l.
\end{equation}
In doing so, we have reformulated the eigenvalue problem in terms of coefficients of states that live in the low energy Hilbert space alone. 

Now we are studying a Hamiltonian of the form $H = H_0 + \mu_{ShG} V$ with $V$ given by
\begin{equation}\label{Vpert}
V \,=\, 2 \left(\frac{R}{2\pi}\right)^{2b^2} \int^R_0 dx \left[:\cosh b\phi(x): - \cosh(b\phi_0)\right] .
\end{equation}
We can then expand $\delta H$ in powers of $\mu_{ShG}$, giving
\begin{eqnarray}\label{eIVsec2_6}
\delta H &=& -\mu_{ShG}^2 V_{lh} \frac{1}{H_0-E} V_{hl} \cr\cr
&& + \mu_{ShG}^3 V_{lh}\frac{1}{H_0-E}V_{hh}\frac{1}{H_0-E}V_{hl} + O(\mu_{ShG}^4).~~
\end{eqnarray}
Introducing the (imaginary) time dependence of operators in the interaction picture,
\be
{\cal O}(\tau) = e^{H_0\tau}{\cal O}(0)e^{-H_0\tau},
\ee
we can rewrite eq.~(\ref{eIVsec2_6}) as 
\begin{eqnarray}\label{eIVsec2_7}
\delta H &=& -\mu_{ShG}^2 \sum_{c\in \Hh} \int^\infty_0 \rd\tau\, e^{(E-H_0)\tau} V(\tau)|c\rangle\langle c| V(0) + O(\mu_{ShG}^3) \cr\cr
&\equiv &
\delta H_2 +  O(\mu_{ShG}^3).
\end{eqnarray}
From here on we are going to focus upon the most singular (in $N_c$) contribution to $\delta H_2$ and so drop from $V$ in eq.~(\ref{Vpert}) the term proportional to $\cosh(b\phi_0)$.  Of course if we were interested in using $\delta H_2$ in a quantitative fashion, we would need to include this term.

We can readily analyze $\delta H_2$ through the use of OPEs.  OPEs allow us to take into account the insertion of the partial resolution of identity in eq.~(\ref{eIVsec2_7})
that involves only the states from the high energy part of Hilbert space, ${\cal H}_h$.   Following the procedure outlined in \cite{tsm_review},
the matrix elements of $\delta H_2$ satisfy
\begin{eqnarray}\label{eIVsec2_8}
(\delta H_2)_{ab} & \approx & -4  \delta_{P_a,P_b}\mu_{ShG}^2 R^2 \bigg(\frac{R}{2\pi}\bigg)^{4b^2} \sum_\varphi C_{\varphi} \sum_{n>N_c}S^2\bigg(n,-2b^2-\Delta_\varphi\bigg)\cr\cr
&& \hskip .75in \bigg(\frac{1}{E_a-E+\frac{2\pi}{R}(2n-2b^2)}\bigg) \langle a|\varphi(0,0)|b\rangle,\nn 
\end{eqnarray}
where we have defined
\begin{equation}\label{eIVsec2_9}
S(n,a) \equiv \frac{1}{n!}\frac{\Gamma(a+n)}{\Gamma(a)}.
\end{equation}
In eq.~(\ref{eIVsec2_8}), the states $a,b$ are drawn from $\Hl$, $\delta_{P_a,P_b}$ enforces momenta conservation, the sum $\sum_{\varphi}$ runs over all fields, $\phi$ (of chiral dimension $\Delta_\phi$),  that appear in the OPE of $\cosh(b\phi)$ with itself, and $C_{\varphi}$ are the corresponding structure constants.  Here the relevant OPEs are given by
\begin{eqnarray}\label{ope}
:\cosh (b\phi(x_1,\tau))::\cosh (b\phi(x_2,\tau_2)): &=& \frac{1}{2}|z_1-z_2|^{4b^2}|z_1|^{-2b^2}|z_2|^{-2b^2}\mathbb{1}\cr\cr
&+& \frac{1}{4}|z_1-z_2|^{-4b^2}|z_1|^{-2b^2}|z_2|^{6b^2} :e^{2b\phi(x_2,0)}:\cr\cr
&+& \frac{1}{4}|z_1-z_2|^{-4b^2}|z_1|^{-2b^2}|z_2|^{6b^2} :e^{-2b\phi(x_2,0)}:\cr
&&
\end{eqnarray}
where 
\be
z_{1,2} = e^{-\frac{2\pi}{R}(\tau_{1,2} + ix_{1,2})}, \quad \bar z_i = z_i^*.
\ee
These OPEs are obtained by combining the OPEs of the oscillator part and the zero mode part (see eq.~(\ref{eIIIix})) of the field, i.e.
\begin{eqnarray}
:e^{b\tilde\phi(x_1,\tau_1)}::e^{\sigma b\tilde \phi(x_2,\tau_2)}: &=& \frac{|z_2|^{4\sigma b^2}}{|z_1-z_2|^{4\sigma b^2}} e^{b(1+\sigma)\tilde \phi(x_2,\tau_2)} + \cdots \cr\cr
e^{b\phi_0(\tau_1)}e^{\sigma b\phi_0(\tau_2)} &=& |z_2|^{2b^2}|z_1|^{-2b^2} e^{b(1+\sigma)\phi_0(\tau_2)} + \cdots ,
\end{eqnarray}
where $\sigma=\pm 1$.  As discussed in Ref.~\cite{tsm_review}, we obtain the sum $\sum_{n>N_c}$ appearing in eq.~(\ref{eIVsec2_8}) by expanding the term $|z_1-z_2|^{-4\sigma b^2}$ that appears in the OPE of the oscillator part of the fields into a Taylor series in $|z_1/z_2|$.  We then only keep the terms in the series at order $N_c+1$ and above.

We now focus on the part of $\delta H_2$ involving the operator $\cosh(2b\phi)$:
\begin{equation}
(\delta H_2)_{ab} = 4\pi^2\mu_{ShG}^2 \bigg(\frac{R}{2\pi}\bigg)^{3+4b^2}\frac{N_c^{4b^2-2}}{2-4b^2} (:\cosh(2b\phi):)_{ab}.\label{dH2expr}
\end{equation}
We can see that this term diverges as $b^2\rightarrow 1/2$ and that furthermore for $b^2>1/2$, the correction tends to $\infty$ as the chiral cutoff, $N_c$, tends to $\infty$.  This means any strategy to compute corrections to TSM results {\it perturbatively} due to states coming from above the cutoff fails for values of $b$ close to the self-dual point.

This result is actually worse than the second order result implies.  At the third order, we can again use OPEs and find that the most singular third order contribution to $\delta H_3$ goes as
\begin{equation}
(\delta H_3)_{ab} \sim \delta_{P_a,P_b}\mu_{ShG}^3 R^{5+6b^2} N_c^{12b^2-4} (:\cosh(3b\phi):)_{ab}.
\end{equation}
Here we see the third order term has a pathological dependence on $N_c$ when $b^2>1/3$, even further away from the self-dual point.  We can continue this to n-th order, finding
\begin{equation}
(\delta H_n)_{ab} \sim \delta_{P_a,P_b}\mu_{ShG}^n R^{2n-1+2b^2n} N_c^{2(n^2-n)b^2-2n+2} (:\cosh(nb\phi):)_{ab}.
\end{equation}
Here we see that the situation becomes worse and worse as we go to higher and higher perturbative order: at n-th order, the correction diverges as $N_c\rightarrow \infty$ for $b^2>1/n$.

From this analysis we can see that the perturbative series developed here is essentially a small-volume expansion in the parameter $R^{2+2b^2}$. This implies that the ground state energy does not have a proper expansion in powers of $R^{2+2b^{2}}$ around the CFT limit $R\rightarrow 0$.  We also want to remark that the pathologies identified here for the ShG do not apply to its analytically continued cousin, the sine-Gordon model.  In the sine-Gordon, this perturbative analysis will give rise to divergences for $b^2>1/2$.  However these divergences occur in the identity channel in terms of the OPE of eq.~(\ref{ope}).  This means the most singular part of $\delta H_n$ is proportional to $\mathbb{1}$ and so leads only to corrections to the energies that are state independent, i.e. energies measured relative to the ground state energy are unaffected.  Alternatively one can add a single counterterm to the sine-Gordon Hamiltonian to remove this divergent behavior.

\subsubsection{Numerical Renormalization Group}

\begin{figure}
\centering
\includegraphics[width=\textwidth]{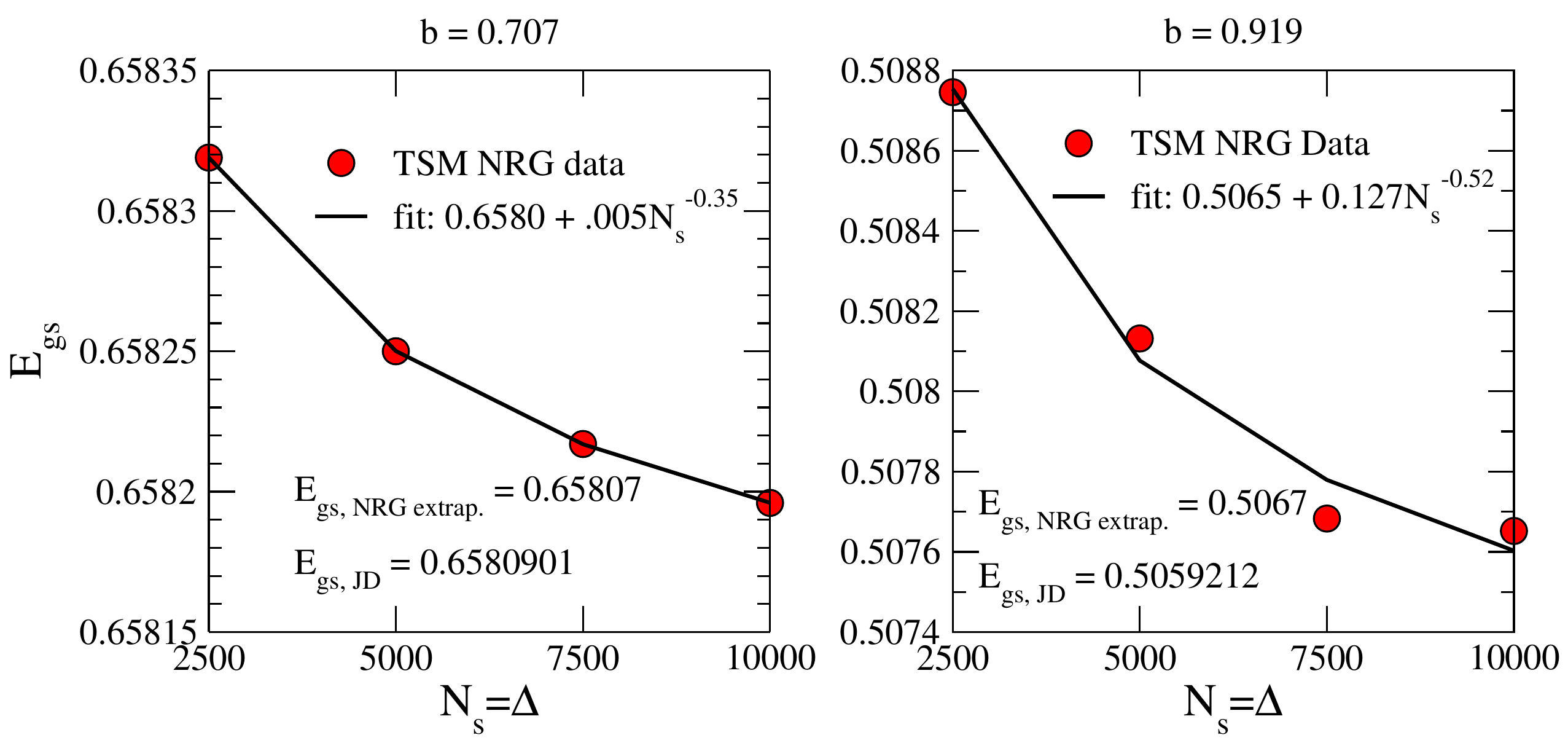}
\caption{The ground state energy computed at $N_c=14$ and $N_{ZM}=24$ as a function of NRG block sizes $N_s=\Delta$.  Left panel: Computation at parameters $b=0.707, R=3.0, \mu_{ShG}=0.1$.  Extrapolating the NRG data to where it would correspond to a JD computation we obtain $E_{\rm gs,~NRG~extrap.} = 0.65807$ in comparison to the exact JD value, $E_{\rm gs,~JD} = 0.6580901$. Right: Computation at $b=0.919, R=3.0, \mu_{ShG}=0.1$.  Here we find $E_{\rm gs,~NRG~extrap.} = 0.5067$ and $E_{\rm gs,~JD} = 0.5059212$.}\label{nrg}
\end{figure}

In Section \ref{subsubAnalyticRG} we demonstrated that a perturbative analytic renormalization group is not a tool that can be used to take into account the states above the truncation $N_c$.  In this section we show that the non-perturbative numerical renormalization group \cite{PhysRevLett.98.147205}, while not beset by pathological divergences, also is challenged for values of $b$ close the self-dual point.  The basic idea of the numerical renormalization group (NRG) for the TSM is to adapt the Wilsonian renormalization group invented to attack the Kondo problem to the case at hand.  Normally in applying TSM, one introduces a cutoff, here $N_c$, and either does a single exact diagonalization or uses Jacobi-Davidson (JD) methods to obtain the energies.  With the NRG, one trades a large single diagonalization for a sequence of smaller exact diagonalizations.  This sequence is determined by two parameters $N_s$ and $\Delta$.  The size of matrices that one diagonalizes in the sequence is $(N_s+\Delta) \times (N_s+\Delta)$.  These parameters should be thought of as variational in nature.  In general the larger these parameters are, the closer one gets to reproducing the exact diagonalization result. 
We will not describe this procedure in further detail here but refer the reader to Refs. \cite{PhysRevLett.98.147205,tsm_review}.

In Fig.~\ref{nrg} we present results for the computation of the ground state energy at two different $b$'s.  We do so at cutoffs of $N_c=14$ and $N_{ZM}=24$.  The Hilbert space size at such cutoffs is 492888.  In Fig.~\ref{nrg} we show the results of the NRG computation for different values of $N_s=\Delta$ ranging from 2500 to 10000 (i.e. we are diagonalizing sequences of matrices with size from $5000\times 5000$ to $20000 \times 20000$).  We see that at even the largest value of $N_s=\Delta=10000$ considered, the results are not converged.  We thus fit a power law to the evolution of $E_{gs}$ as a function of $N_s=\Delta$ and extrapolate the power law to where it would correspond to solving the problem exactly (i.e. finding the low lying eigenenergies of a $492888\times 492888$ matrix, corresponding to evaluating the power law fit function at $N_s=492888/2$).  The result is reported in Fig.~\ref{nrg}.  We see that for $b=1/\sqrt{2}$, the agreement between the NRG at the largest value of $N_s$ considered, $N_s=\Delta=10000$, and the exact JD value is good to 3 significant digits.   Upon NRG extrapolation, this improves to 4 significant figures.  For $b=0.919$, much closer to the self dual point, agreement before extrapolation is only at 2 significant digits and remains at 2 significant digits after (even if the extrapolation does improve the NRG result). 

The performance of the NRG close to the self-dual point is considerably worse than for a model like the sine-Gordon model where we see agreement at the 5 significant digit level for NRG block sizes far smaller than those considered here ($N_s=1500,\Delta=500$) and without extrapolation (for example compare the results here with Table IV of Section VI of Ref.~\cite{tsm_review}).  This we believe is a manifestation of the slow convergence as a function of $N_c$ in the ShG model that we have observed elsewhere in this paper.  While obtaining 4 significant digits is usually sufficient (say at $b=1/\sqrt{2}$), we are of course interested in further extrapolating our results in $N_c$.  These $N_c$-extrapolations turn out to be sensitive to the errors on the order of $10^{-3}$.  This makes the use of NRG-based data, at least for values of $b$ close to $1$, problematic.

\section{Quantum Mechanical Reductions of the Sinh-Gordon Model} \label{QMreductionssection}
In Section \ref{TSMresults} we argued that a straightforward implementation of analytical RG improvements is hindered because the small volume expansion of energy levels is not perturbative
with respect to the parameter $\mu_{ShG} R^{2+2b^2}$. In this section we take a closer look at the small-$R$ UV spectrum. Since the energy of oscillators behave as $R^{-1}$ in the $R\rightarrow0$ limit,
one might expect that the UV behaviour of the spectrum is dominated by the quantum mechanics of the zero mode of the field. Using this quantum mechanical picture, a systematic expansion for certain energy levels (more precisely, their corresponding scaling functions) can be developed in terms of $\frac{1}{\ln(\mu_{ShG} R)}$. Alternatively, one can expand the TBA equations, yielding a similar expansion, but involving IR parameters (the physical mass, $M_{ShG}$, and the S-matrix parameter $B$). Using the mass-coupling relation and
expressing $B$ in terms of the coupling $b$, we get another expansion in $\frac{1}{\ln(\mu_{ShG} R)}$, which is however different from the zero mode expansion in subleading orders. As the energies contain an additional $2\pi R^{-1}$ factor relative to the scaling function, the difference between TBA and zero mode energy levels eventually diverge for $R\rightarrow 0$ for all $b>0$. 

In Subsection \ref{subseceffpot} we derive an effective potential, partially taking into account the effect of oscillators. On one side, we analytically reproduce the exact expansion up to $O(b^{12})$, confirming that the oscillators are able to explain the differences in the log-expansion. On the other side, we  show that the TSM numerics significantly outperforms even the numerical solution of the complete effective potential. We then use this fact to provide an alternative measurement of the IR parameters from TSM, combining UV numerics with the small-volume expansion of the TBA.

\subsection{Semiclassical Reflection Amplitude} \label{SemiclassSubsec}

\begin{figure}[t]
\centering
\includegraphics[width=0.5\textwidth]{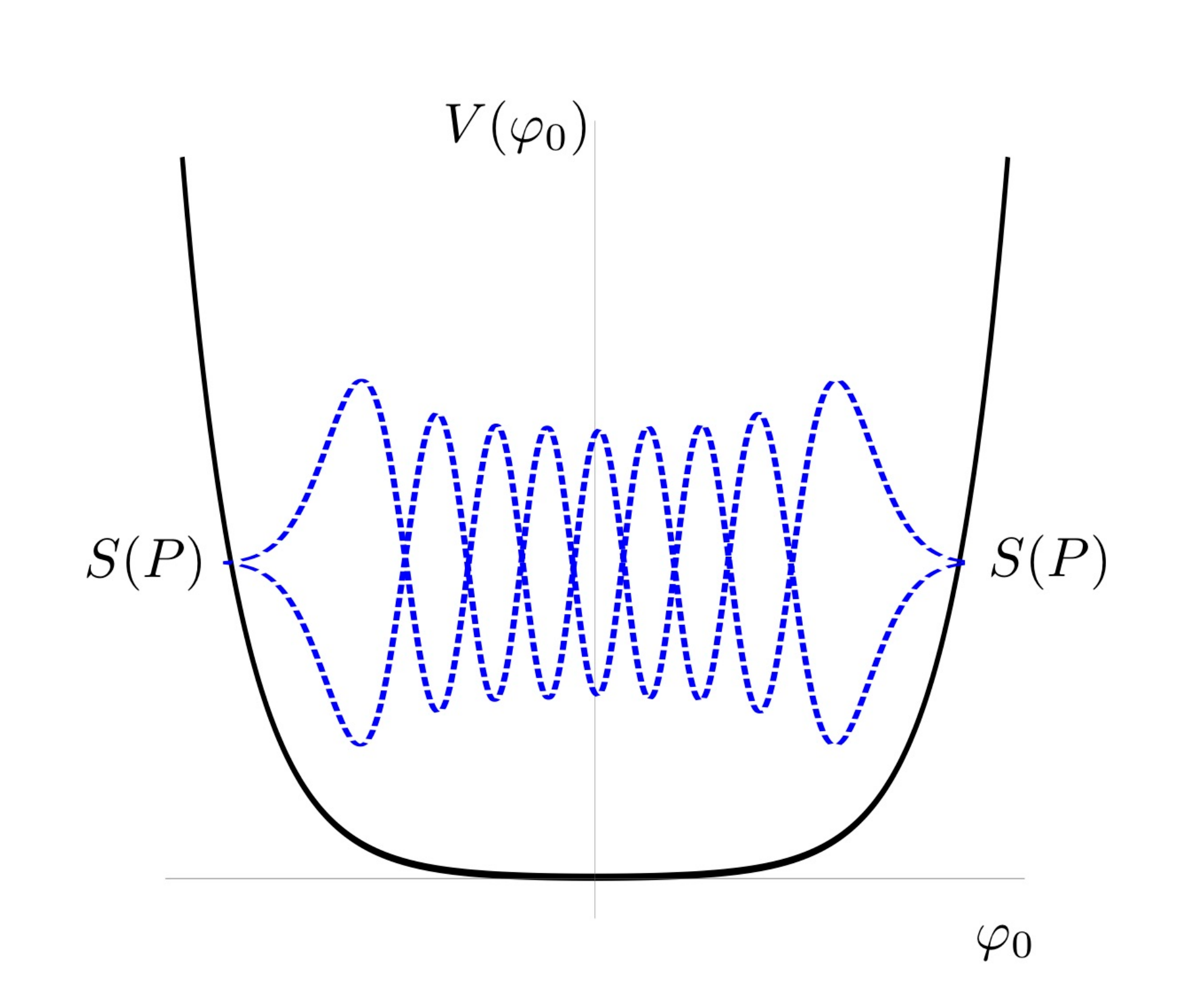}
\caption{Standing waves of the zero mode ruled by the quantization condition eq.~(\ref{S0eq}).} 
\label{wavestanding}
\end{figure}

In the semiclassical limit $b\rightarrow0$ the small volume behavior of energy levels is dominated by the contribution of the zero mode \cite{Zamolodchikov:1995aa}. For $R\rightarrow0$ the potential walls that the zero mode sees (i.e. the points where the $\mu_{ShG}R^{2+2b^2}\cosh(b\phi_0)$ potential exceeds 1) are far from one another. Then it is sensible to consider first the quantum mechanical problem of a particle reflecting from a single wall:
\begin{equation}
H_{\mathrm{exp}}=\frac{2\pi}{R}\left(2\Pi_{0}^{2}+Me^{b\varphi_{0}}\right);\quad M=2\pi\mu_{ShG}\biggl(\frac{R}{2\pi}\biggr)^{2+2b^{2}}.
\end{equation}
Introducing the coordinate representation $\varphi_{0}\equiv x$, $\Pi_{0}=-i\partial_{x}$, it is possible to solve the Schr\"odinger
equation
\begin{equation}
H_{\mathrm{exp}}\psi\,=\,E\psi\label{eq:TISE}.
\end{equation}
Its general solution is given by modified Bessel functions. Requiring that the wave function vanishes at $x\rightarrow\infty$
and evaluating the $x\rightarrow-\infty$ asymptotics, we can write the relative phases of the left-moving and right-moving wave as
\begin{equation}
\psi\left(x\right)\,\simeq \, e^{iPx}+e^{-iPx}S_{sc}\left(P\right);\quad P\,=\, \sqrt{\frac{RE}{4\pi}}\,\,\,,\label{Pdef}
\end{equation}
where the semi-classical reflection amplitude is defined as
\begin{equation}
S_{sc}\left(P\right)\,=\,\left(-\biggl(\frac{R}{2\pi}\biggr)^{-4iPQ}\left(\frac{\pi\mu_{ShG}}{b^{2}}\right)^{-\frac{2iP}{b}}\frac{\Gamma\left(1+2iP/b\right)}{\Gamma\left(1-2iP/b\right)}\right);\quad Q=b+\frac{1}{b}.
\end{equation}
This expression is the semi-classical $b\rightarrow 0$ limit of the Liouville reflection amplitude (\ref{LiouvilleSmatrix}). 
As the other exponential term is turned on, we can get an approximate quantization condition for the energy levels of the full potential through the quantization condition of the wave number $P$ 
according to the reflection equation (see Fig.~\ref{wavestanding}):
\begin{equation}
S_{sc}\left(P\right)^{2}\,=\, 1.\label{S0eq}
\end{equation}
Denoting $S_{sc}\left(P\right)\,=\,-e^{i\delta\left(P\right)}$ and taking the logarithm, the quantization condition (\ref{S0eq}) reads 
\begin{equation}
\delta\left(P\right) \,=\, n \,\pi,\quad n\geq1 \label{SCquantcond}
\end{equation}
and the branch cuts of $\delta(P)$ are to be chosen such that it is continuous for real $P$ and $\delta(0)=0$. The equation for the ground state wave number $P_0$ is then given by  
\begin{equation}
\delta\left(P_{0}\right)\,=\, \pi\,\,\, ,
\end{equation}
Making a formal expansion of $\delta\left(P\right)$
\begin{equation}
\delta\left(P\right)=\delta_{1}P+\delta_{3}P^{3}+\delta_{5}P^{5}+\dots
\end{equation}
we can expand the ground state momentum as a function of $z=\delta_{1}^{-1}$:
\begin{equation}
P_{0}=\pi z-\pi^{3}\delta_{3}z^{4}-\pi^{5}\delta_{5}z^{6}+\dots ,
\end{equation}
where the parameter $z$ reads explicitly
\begin{equation}
z\,=\, -\frac{1}{\frac{4}{b}\gamma_{E}+\frac{2}{b}\ln\left(\frac{\pi\mu_{ShG}}{b^{2}}\left(\frac{R}{2\pi}\right)^{2+2b^{2}}\right)}\,\,\,.
\end{equation}
Hence,  the (semi-classical) ground state energy admits the expansion
\begin{eqnarray}
E_{0} &=& \frac{2\pi}{R}b^{2}\pi^{2}\left(\frac{u^{2}}{2}-\kappa_S u^{3}+\frac{3}{2}\kappa_S^{2}u^{4}-2\left(\kappa_S^{3}-\frac{\pi^{2}}{3}\zeta\left(3\right)\right)u^{5}+\dots\right) \label{E0semiclas};\cr\cr
\kappa_S &=& \left(2\gamma_{E}+\ln\frac{\pi}{b^{2}}\right);\cr\cr
u &=& \left[\ln\left(\mu_{ShG}\left(\frac{R}{2\pi}\right)^{2+2b^{2}}\right)\right]^{-1}.
\end{eqnarray}
It is important to notice that the terms of this series contain $\log\left(\mu_{ShG}\right)^{-1}$ factors. Therefore, it is not so surprising if a power expansion in $\mu_{ShG}$ around $\mu_{ShG}=0$ 
turns out to be pathological.

\subsection{Quantization Condition from the UV limit of TBA}
A similar quantization condition exists for the exact energy levels and can be obtained from the small-volume expansion of the TBA system
\cite{Teschner:2007ng}. In this section we denote the coupling appearing in TBA by $\bar{b}$, to emphasize that this parameter is directly
related to the S-matrix parameter $B=\frac{\bar{b}^{2}}{1+\bar{b}^{2}}$, and not (immediately) to the parameter $b$ appearing in the Hamiltonian.

In the small-$R$ limit, the TBA equations decouple into a right- and a left-moving part. To obtain these equations, one first performs
a shift in the rapidity variables $\left\{ \theta,\vartheta\right\} \rightarrow\left\{ \theta\pm\left|\ln MR\right|,\vartheta\pm\left|\ln MR\right|\right\} $,
leading to a pair of volume-independent equations up to $O\left(MR\right)$ corrections. One then neglects the $O\left(MR\right)$ corrections
and reverses the previous rapidity shift. Let us introduce the notation
$Y_{\pm}\left(\theta\mid\left\{ \vartheta\right\} \right)=e^{-\epsilon_{\pm}\left(\theta\mid\left\{ \vartheta\right\} \right)}$,
where the $\pm$ denotes the right- and left-moving solutions. In
the following it is advantageous to construct the so-called $\mathcal{Q}$-functions,
defined through the pair of functional relations
\begin{align}
\mathcal{Q}\left(\theta+\frac{i\pi a}{2}\right)\mathcal{Q}\left(\theta-\frac{i\pi a}{2}\right) & =Y\left(\theta\mid\left\{ \vartheta\right\} \right)\label{eq:Qfunc1} \,\,\,,\\
\mathcal{Q}\left(\theta+\frac{i\pi}{2}\right)\mathcal{Q}\left(\theta-\frac{i\pi}{2}\right) & =1+Y\left(\theta\mid\left\{ \vartheta\right\} \right)\label{eq:Qfunc2}\,\,\,,
\end{align}
with $a=1-2B$, where we have suppressed the $\vartheta$-dependence of the $\mathcal{Q}$'s. These functions can be obtained by taking the
logarithm of eqs.~\ref{eq:Qfunc1}-\ref{eq:Qfunc2} and (carefully) performing a Fourier transform. In the UV limit, corresponding to
the decoupling of the TBA equations, we get a pair of functions $\mathcal{Q}_{\pm}$ of the form
\begin{equation}
\ln\mathcal{Q}_{\pm}\left(\theta\right)=-\frac{MR}{4\sin\pi B}e^{\pm\theta}+\intop\frac{d\theta^{\prime}}{2\pi}\frac{\ln\left(1+Y^{\pm}\left(\theta^{\prime}\mid\left\{ \vartheta\right\} \right)\right)}{\cosh\left(\theta-\theta^{\prime}\right)}+\left(\text{source terms}\right).\label{eq:Qfuncintrep}
\end{equation}
Note that the source terms and quantization conditions of the TBA system can be understood as a prescription for the zeros of the function
$1+Y$. Correspondingly, $\mathcal{Q}_{\pm}$ needs to have an analogous set of zeroes to be compatible with eq.~(\ref{eq:Qfunc2}). It was shown
in \cite{bytsko2013} that the $\mathcal{Q}$-functions obtained from the decoupled TBA equations possess the asymptotic form
\begin{equation}
\mathcal{Q}_{\pm}\left(\theta\right)\underset{\theta\rightarrow\mp\infty}{\sim}\frac{\cos\left[2PQ\left(\bar{b}\right)\theta\pm\Theta\left(P\right)\right]}{\sqrt{\sinh\left(2\pi\bar{b}P\right)\sinh\left(2\pi\bar{b}^{-1}P\right)}}\label{eq:Qasymptotic}\,\,\,,
\end{equation}
where $P$ is a real parameter and $\Theta\left(P\right)$ is an antisymmetric phase (to be obtained below). For small volumes, the asymptotic eq.~(\ref{eq:Qasymptotic})
is expected to dominate the $\theta$-dependence over a wide region. The parameter $P$ is quantized by the requirement that $\mathcal{Q}_{+}$
and $\mathcal{Q}_{-}$ corresponds to the \emph{same} $Y$-function
$Y_{+}=Y_{-}$, which is nothing else but the UV limit of the $Y\left(\theta\mid\left\{ \vartheta\right\} \right)=e^{-\epsilon\left(\theta\mid\left\{ \vartheta\right\} \right)}$.

Let us focus on the zero mode sector defined by restricting the Bethe quantum numbers to be $I_{j}=0,\:\forall j$. Taking into account that $\pm\mathcal{Q}$ leads to the same $Y$, we arrive at the condition
\begin{equation}
2\,\Theta\left(P\right)=n \pi,\quad n\in\mathbb{Z}\label{eq:quantcond}\,\,\,.
\end{equation}
Due to the antisymmetry of $\Theta$, it is sufficient to restrict to the cases $n\geq0$ (subsequently we will see that the ground state corresponds to $n=1$).  In the zero mode sector, energy levels behave in the UV as
\begin{equation}
E_{n}=\frac{2\pi}{R}\left(2P_{n}^{2}-\frac{1}{12}\right)\,\,\,,
\end{equation}
as can be seen by direct integration (for details, see Appendix C of Ref.~\cite{Teschner:2007ng}).

It is hard to extract the phase $\Theta\left(P\right)$ directly from eq.~(\ref{eq:Qfuncintrep}). However, as shown in \cite{Fateev_2006}, there exists an ingenious trick to
obtain its explicit expression. To this aim, consider the second-order ODE
\begin{equation}
-\psi^{\prime\prime}\left(x\right)+\kappa^{2}\left(e^{2x}+e^{-\frac{2x}{\bar{b}^{2}}}\right)\psi\left(x\right)\,=\,p^{2}\psi\left(x\right)\,\,\,,
\end{equation}
and the pair of solutions defined by their asymptotical behaviour
\begin{align}
\psi_{-}\left(x\right) & \underset{x\rightarrow-\infty}{\sim}\frac{1}{\sqrt{2\kappa}}\exp\left(\frac{x}{2\bar{b}^{2}}-\bar{b}^{2}\kappa e^{-\frac{x}{b^{2}}}\right) \,\,\,,\\
\psi_{+}\left(x\right) & \underset{x\rightarrow+\infty}{\sim}\frac{1}{\sqrt{2\kappa}}\exp\left(-\frac{x}{2}-\kappa e^{x}\right)\,\,\,.
\end{align}
Hence, the Wronskian, $W\left(p\right)$, constructed in terms of these solutions
\begin{equation}
W\left(p\right)\,=\,\psi_{+}\frac{d}{dx}\psi_{-}-\psi_{-}\frac{d}{dx}\psi_{+},
\end{equation}
satisfies the same set of functional equations as the $\mathcal{Q}$-system, provided that the parameters $\kappa$ and $p$ are tuned appropriately.
Let us focus on the right-moving part. $W\left(p\right)$ can be evaluated in a small-$\kappa$ expansion (using the reflection quantization) and
a large-$\kappa$ expansion (by means of the WKB approximation). It is convenient to parametrize $\kappa$ as $\kappa=c\,e^{\theta}$. Then,
comparing the form of the pseudo-energy obtained from the small-$\kappa$ expansion ($\theta\rightarrow-\infty$) to the asymptotic formula
eq.~(\ref{eq:Qasymptotic}), we can fix
\begin{equation}
p\,=\,\frac{2P}{\bar{b}}\,\,\,,
\end{equation}
while, from the large-$\kappa$ expansion, we obtain
\begin{equation}
c\,=\,MR\,\frac{\sqrt{\pi}}{4\sin\frac{\pi B}{2}}\frac{\Gamma\left(\frac{3-B}{2}\right)}{\Gamma\left(\frac{2-B}{2}\right)}\,\,\,.
\end{equation}
Finally, the phase $\Theta\left(P\right)$ is obtained by comparing the small-$\kappa$ expansion to eq.~(\ref{eq:Qasymptotic}) and is given by
\begin{equation}
e^{2i\Theta\left(P\right)}\,=\, - \bar{b}^{\frac{8iP}{\bar{b}}}\rho^{-4iPQ\left(\bar{b}\right)}\,\frac{\Gamma\left(1+2iP\bar{b}\right)\Gamma\left(1+2iP\bar{b}^{-1}\right)}{\Gamma\left(1+2iP\bar{b}\right)\Gamma\left(1+2iP\bar{b}^{-1}\right)}\,\,\,,
\end{equation}
where
\begin{equation}
\rho\,=\,\frac{R}{2\pi}\frac{M}{4\sqrt{\pi}}\Gamma\left(\frac{1-B}{2}\right)\Gamma\left(\frac{2+B}{2}\right)\,\,\,.
\end{equation}

Notice that if we take advantage of the mass-coupling relation and equate $\bar{b}\equiv b$, as it was observed earlier \cite{Zamolodchikov:1995aa,AHN1999505}, the quantization condition eq.~(\ref{eq:quantcond}) can be expressed as
\begin{equation}
S_{L}^{2}\left(P\right)\,=\, 1\label{eq:LiouvQuantCond}\,\,\, ,
\end{equation}
where $S_L(P)$ is the Liouville reflection amplitude given in eq.~\,(\ref{LiouvilleSmatrix}). This is analogous to the semiclassical formula eq.~(\ref{S0eq}) but with the miraculously
appearing Liouville reflection amplitude which replaces the quantum mechanical amplitude $S_{sc}(P)$ introduced in Subsection \ref{SemiclassSubsec}. At the same time,
$S_{L}\left(P\right)$ reduces to the semiclassical $S_{sc}\left(P\right)$ in the $b\rightarrow0$ limit. Comparing to the quantization condition \eqref{SCquantcond} and assuming continuity of energy eigenvalues as functions of $b$, it is then natural to exclude $n=0$ from the quantization condition eq.~(\ref{eq:quantcond}). Repeating the UV expansion of the
energy levels using the condition eq.~(\ref{eq:LiouvQuantCond}), we get
a modified expansion for the ground state energy
\begin{eqnarray}
E_{0} &=& \frac{2\pi}{R}b^{2}\pi^{2}\left(\frac{u^{2}}{2}-\kappa_{L}u^{3}+\frac{3}{2}\kappa_{L}^{2}u^{4}-2\left(\kappa_{L}^{3}-\frac{\pi^{2}}{3}\left(1+b^{6}\right)\zeta\left(3\right)\right)u^{5}+\dots\right)\label{eq:EnfromLiouv};\cr\cr
\kappa_{L} &=& \left(2\left(1+b^{2}\right)\gamma_{E}+\ln\frac{\pi\Gamma\left(b^{2}\right)}{\Gamma\left(1-b^{2}\right)}\right);\cr\cr
u &=& \left[\ln\left(\mu_{ShG}\left(\frac{R}{2\pi}\right)^{2+2b^{2}}\right)\right]^{-1}.
\end{eqnarray}
Even though the leading term of this expansion coincides with the semiclassical expansion eq.~(\ref{E0semiclas}), the $R^{-1}$ factor in the front ensures
that a difference in any term of the $u$-expansion leads to a singular discrepancy in the $R\rightarrow0$ limit. This small-volume discrepancy 
is not trivially accounted for perturbatively. The $\mu_{ShG}$-expansion
introduces terms that vanish in the $R\rightarrow0$ limit. On the
other hand, in the massive scheme of eq.~(\ref{eq:Massivehamiltonian}), where an expansion in $b$ is natural,
the coefficient of the perturbing operator diverges in the $R\rightarrow0$
limit.

In the next subsection, we overcome these obstacles by deriving an effective potential which partially takes into account the corrections appearing in eq.~(\ref{eq:EnfromLiouv}).

\subsection{An Effective Quantum Mechanical Potential} \label{subseceffpot}
In the previous subsections, we compared the small-volume expansion of the ground-state energy obtained from the zero mode quantum mechanics to the UV expansion of TBA, and found that they differ by $R^{-1}(\ln{R})^{-k}$ type terms. It is then an important question whether the oscillator states neglected in the zero mode calculation can account for these \emph{inverse logarithmic} differences, or is this a sign that we are missing additional terms from the Lagrangian. In any case it is not straightforward to reproduce these terms perturbatively.
In the following we derive an effective quantum mechanical potential from the Lagrangian, which provides a more precise description of the UV spectrum of the zero mode subspace, by partially taking into account the effect of oscillators. This is done by means of a Bogoliubov transformation applied to the Hamiltonian, involving the oscillators only (keeping the zero mode intact).
Following the notations of Appendix \ref{AppRelatingSchemes}, we can start by adding and subtracting an auxiliary quadratic term to the Hamiltonian eq.~(\ref{eIIIi}),
\begin{align}
H^{\left(\mathrm{ShG}\right)} & =H_{\mathrm{cyl}}^{\left(0\right)}+\frac{m_{\mathrm{eff}}^{2}}{16\pi}\intop_{0}^{R}dx:\tilde{\varphi}^{2}\left(x\right):\nonumber \\
 & +2\mu_{ShG}\left(\frac{R}{2\pi}\right)^{2b^{2}}\intop_{0}^{R}dx:\cosh\left(b\varphi\left(x,0\right)\right):-\frac{m_{\mathrm{eff}}^{2}}{16\pi}\intop_{0}^{R}dx:\tilde{\varphi}^{2}\left(x\right):,\label{eq:HshgEffPotStart}
\end{align}
but considering the zero mode $\varphi_{0}$ as a free parameter with $m_{\mathrm{eff}}\equiv m_{\mathrm{eff}}\left(\varphi_{0}\right)$ a function that 
depends on it. Upon transforming to the oscillator eigenbasis of the Hamiltonian given by the first two terms of eq.~(\ref{eq:HshgEffPotStart}),
the normalization of the $\cosh$ term changes
\begin{align*}
H^{\left(\mathrm{ShG}\right)} & =\left(\frac{4\pi}{R}\Pi_{0}^{2}-\frac{\pi}{6R}\right)+\sum_{n\neq0}\omega\left(m_{\rm eff}\right)a_{n}^{\dagger}a_{n}\\
 & +2\mu_{ShG} R\delta_{P}\left(\frac{R}{2\pi}\right)^{2b^{2}}e^{\frac{2\pi}{R}b^{2}S_{1}\left(m_{\mathrm{eff}},R\right)}:\cosh\left(b\varphi\left(x,0\right)\right):_{m_{\mathrm{eff}}}\\
 & -\frac{m_{\mathrm{eff}}^{2}}{16\pi}\intop_{0}^{R}dx:\tilde{\varphi}^{2}\left(x\right):_{m_\mathrm{eff}}dx+\tilde{S}_{2}\left(m_{\mathrm{eff}},R\right)
\end{align*}
where $S_{1}\left(m,R\right)$ is defined in Subsection \ref{subsec:The-sumS1} and $\tilde{S}_{2}\left(m,R\right)$ is defined in eq.~(\ref{eq:S2tdef}).  Here $::_{m_{\rm eff}}$ implies that we are normal ordering w.r.t. to massive (of mass $m_{\rm eff}$) oscillator modes).
The value of $m_{\mathrm{eff}}$ is then given by the requirement that the explicit quadratic term precisely cancels that of the cosh
function. This leads to the transcendental equation 
\begin{equation}
m_{\mathrm{eff}}^{2}\left(\varphi_{0}\right)\,=\, 16\pi\mu_{ShG}b^{2}\left(\frac{R}{2\pi}\right)^{2b^{2}}e^{\frac{2\pi}{R}b^{2}S_{1}\left(m_{\mathrm{eff}},R\right)}\,\cosh b\varphi_{0},\quad\forall\varphi_{0}.\label{eq:meffeq}
\end{equation}
The effective Hamiltonian $H_{ZM}^{\mathrm{eff}}$ is then obtained by dropping all higher order oscillator terms of the cosh interaction. This leads to
\begin{align}
H_{ZM}^{\mathrm{eff}} & =\left(\frac{4\pi}{R}\Pi_{0}^{2}-\frac{\pi}{6R}\right)+2\mu_{ShG} R\left(\frac{R}{2\pi}\right)^{2b^{2}}e^{\frac{2\pi}{R}b^{2}S_{1}\left(m_{\mathrm{eff}}\left(\varphi_{0}\right),R\right)}\cosh\left(b\varphi_{0}\right)+\nonumber \\
 & + \tilde{S}_{2}\left(m_{\mathrm{eff}}\left(\varphi_{0}\right),R\right).\label{eq:Heffpotfinal}
\end{align}
consisting of the contribution of a Bogoliubov ground state energy plus a correction due to the change of normalization
of the cosh term.

Let us now focus on the small volume limit of the spectrum of $H_{ZM}^{\rm eff}$. The effective mass of this Hamiltonian is then approximated by
\begin{equation}
m_{\mathrm{eff}}^{2}\left(\varphi_{0}\right)=\frac{1}{R^{2}}\left(\mu_{ShG} R^{2+2b^{2}}\frac{16\pi b^{2}}{\left(2\pi\right)^{2b^{2}}}\cosh b\varphi_{0}+O\left(\mu_{ShG}^{2}R^{4+4b^{2}}\right)\right),\label{eq:meffexpand}
\end{equation}
while the sums $S_{1}$ and $\tilde{S}_{2}$ admit the small-volume behavior
\begin{eqnarray}
S_{1}\left(m,R\right) &= &-\frac{1}{m}\left[\frac{\zeta\left(3\right)}{8\pi^{3}}\left(mR\right)^{3}+O\left(\left(mR\right)^{5}\right)\right]\,\,\, \label{eq:S1S2smallV},\\
\tilde{S}_{2}\left(m,R\right)& = & m\left[\frac{\zeta\left(3\right)}{64\pi^{3}}\left(mR\right)^{3}+O\left(\left(mR\right)^{5}\right)\right]\nonumber\,\,\,.
\end{eqnarray}
Inserting the expansions eq.~(\ref{eq:meffexpand}) and eq.~(\ref{eq:S1S2smallV}) into the effective Hamiltonian eq.~(\ref{eq:Heffpotfinal}) and using
the coordinate representation, we obtain a Schr\"odinger equation with the asymptotic form
\begin{equation}
-y^{\prime\prime}\left(x\right)+re^{\pm x}y\left(x\right)-\epsilon^{2}e^{\pm2x}y\left(x\right)=c^{2}y\left(x\right),\quad x\rightarrow\pm\infty,
\label{schr}
\end{equation}
where $y$ is a zero mode wave-function. In writing (\ref{schr}) we have introduced the notations
\begin{eqnarray}
x &=& b\varphi_{0}+\ln a,\quad a=\mu_{ShG} R\left(\frac{R}{2\pi}\right)^{2b^{2}} \,\,,\quad  r \,=\,\frac{R}{4\pi b^{2}},\cr\cr
\epsilon^{2} &=& \left(\frac{R}{2\pi}\right)^{2}b^{2}\zeta\left(3\right) \,\,, \quad c^{2}  \,=\, \frac{P^{2}}{b^{2}}.
\end{eqnarray}
The potential $V(x\rightarrow \pm \infty) = re^{\pm x} - \epsilon^{2}e^{\pm 2 x}$ that we obtain in this way does not possess normalizable eigenfunctions.
This is an artefact of expanding the effective mass according to eq.~(\ref{eq:meffexpand}). However, for small $R$, the potential builds up a flat plateau around
$x=0$, which is bounded by a large peak on either side. WKB analysis predicts that tunneling through the peaks can be neglected as long as $P\ll\frac{1}{4\sqrt{\zeta(3)}b^2}$. In this
domain, the amplitude of reflection from the peaks can be approximated
as
\begin{equation}
S_{2}\left(P\right)\,=\,-\left(\frac{R}{2\pi}\right)^{-4iPQ}\left(4\pi i b\,\mu_{ShG}\sqrt{\zeta\left(3\right)}\right)^{-\frac{2iP}{b}}\frac{\Gamma\left(1+\frac{2iP}{b}\right)}{\Gamma\left(1-\frac{2iP}{b}\right)}\frac{\Gamma\left(\frac{1}{2}-\frac{iP}{b}-\frac{i}{4b^{3}\zeta\left(3\right)}\right)}{\Gamma\left(\frac{1}{2}+\frac{iP}{b}-\frac{i}{4b^{3}\zeta\left(3\right)}\right)}\label{eq:Seffpot}\,\,\,.
\end{equation}
For $b\ll1$, the absolute value of $S_{2}\left(P\right)$ is essentially $1$ over a wide interval of $P$. To now determine the energy levels, we can once more follow the tactic outlined in Subsection \ref{SemiclassSubsec}, writing a quantization condition for the phase shift $\delta_{\mathrm{eff}}\equiv-i\ln{\left(-S_{2}(P)\right)}=n\pi$. Expanding this in powers of $P$ and using \eqref{Pdef}, we find the small-volume expansion:
\begin{equation}
E_{0} \,=\, \frac{2\pi}{R}b^{2}\pi^{2}\left(\frac{u^{2}}{2}-\kappa_{2}u^{3}+\frac{3}{2}\kappa_{2}^{2}u^{4}-2\left(\kappa_{2}^{3}-\frac{\pi^{2}}{3}\left(1+b^{6}\right)\zeta\left(3\right)+4\pi^{2}\zeta\left(3\right)^{2}b^{12}\right)u^{5}+\dots\right),\label{eq:EgsfromEpot}
\end{equation}
where 
\[
\kappa_2  =\left(2\gamma_{E}+\ln\frac{\pi}{b^{2}}-\frac{2}{3}b^{6}\zeta\left(3\right)\right)\,\,\,, \quad 
u  =\left[\ln\left(\mu_{ShG}\left(\frac{R}{2\pi}\right)^{2+2b^{2}}\right)\right]^{-1}\,\,\,.
\]
Notice that while the difference between the zero mode expansion eq.~(\ref{E0semiclas}) and the small-volume expansion eq.~(\ref{eq:EnfromLiouv}) from the exact
UV quantization condition is order $O\left(b^{8}\right)$, the difference between eq.~(\ref{eq:EgsfromEpot}) and eq.~(\ref{eq:EnfromLiouv}) is instead only order 
$O\left(b^{12}\right)$.

\begin{figure}
\centering
\begin{subfigure}[b]{0.45\textwidth}
\centering
\includegraphics[draft=false,width=\textwidth]{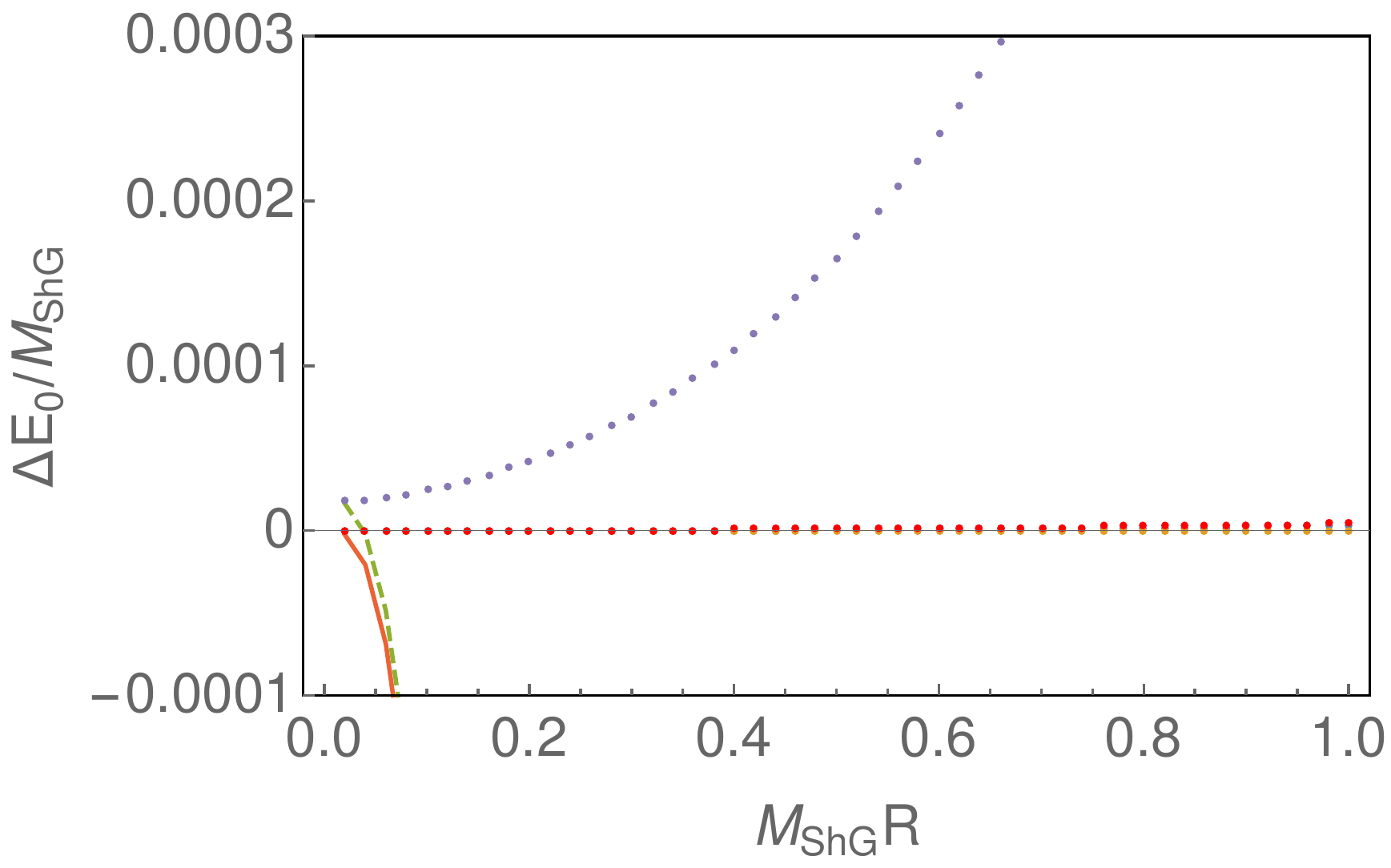}
\caption{$b=0.2$}
\end{subfigure}
\begin{subfigure}[b]{0.45\textwidth}
\centering
\includegraphics[draft=false,width=\textwidth]{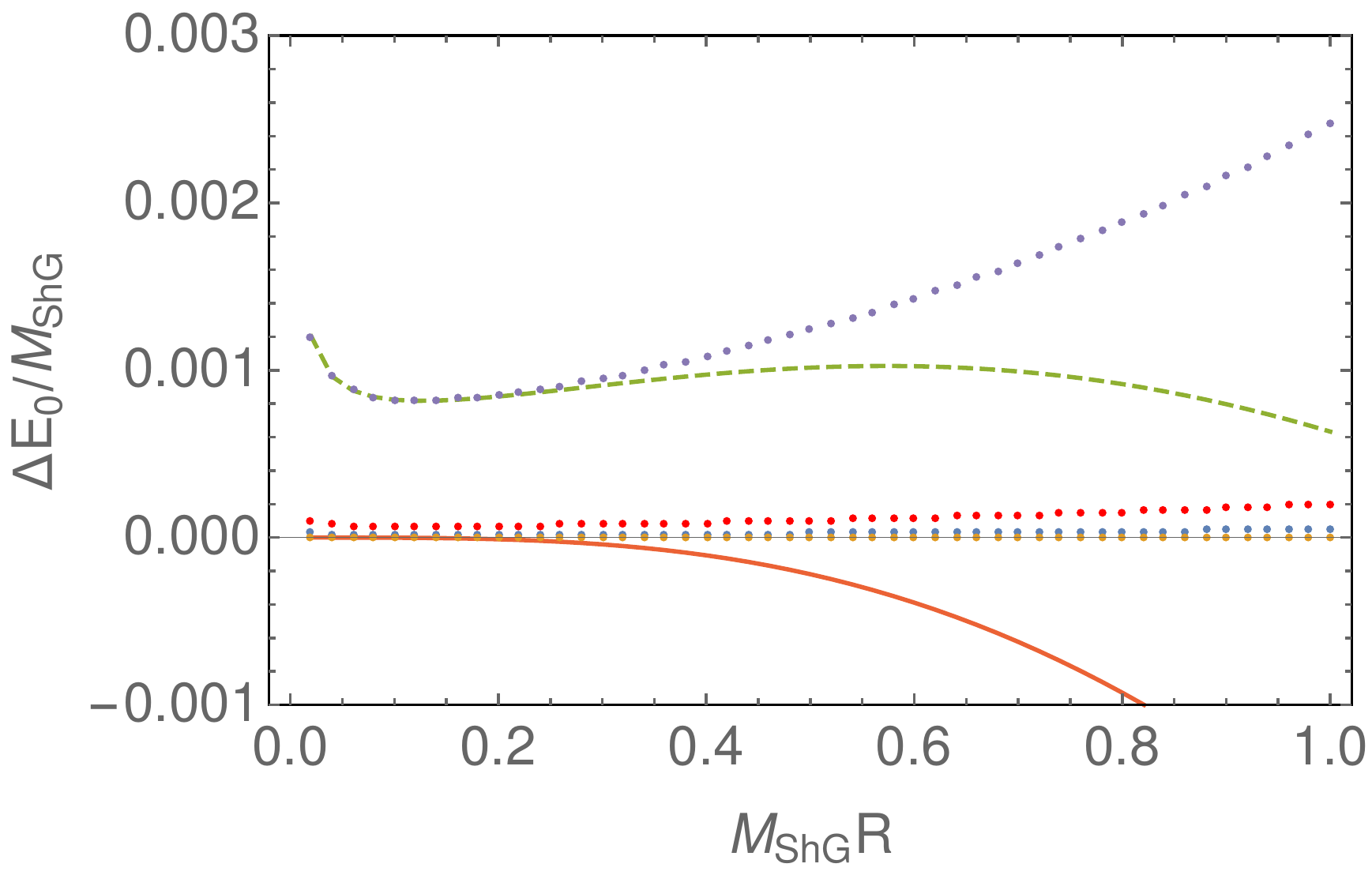}
\caption{$b=0.4$}
\end{subfigure}
\begin{subfigure}[b]{0.45\textwidth}
\centering
\includegraphics[draft=false,width=\textwidth]{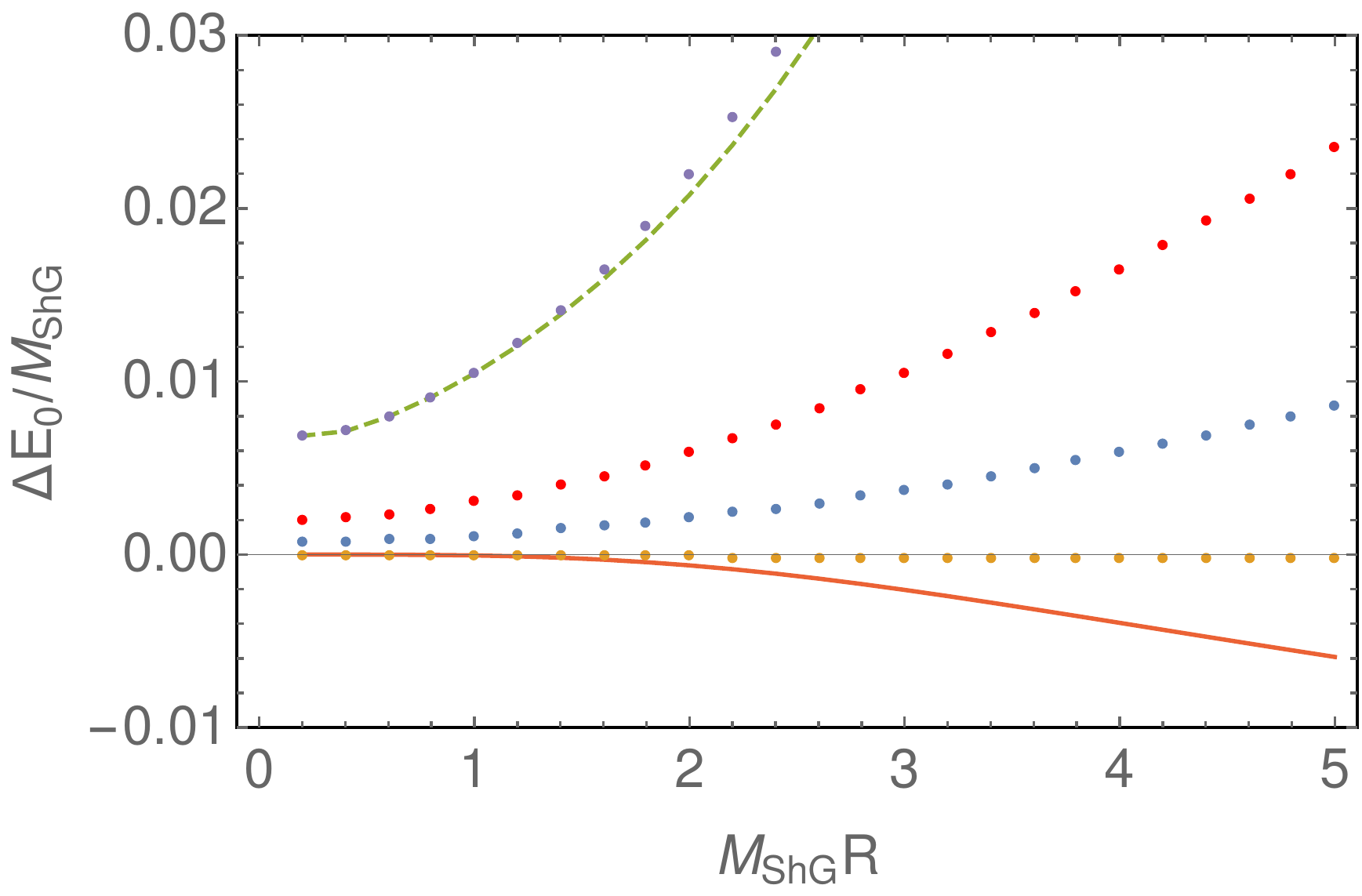}
\caption{$b=0.6$}
\end{subfigure}
\begin{subfigure}[b]{0.45\textwidth}
\centering
\includegraphics[draft=false,width=\textwidth]{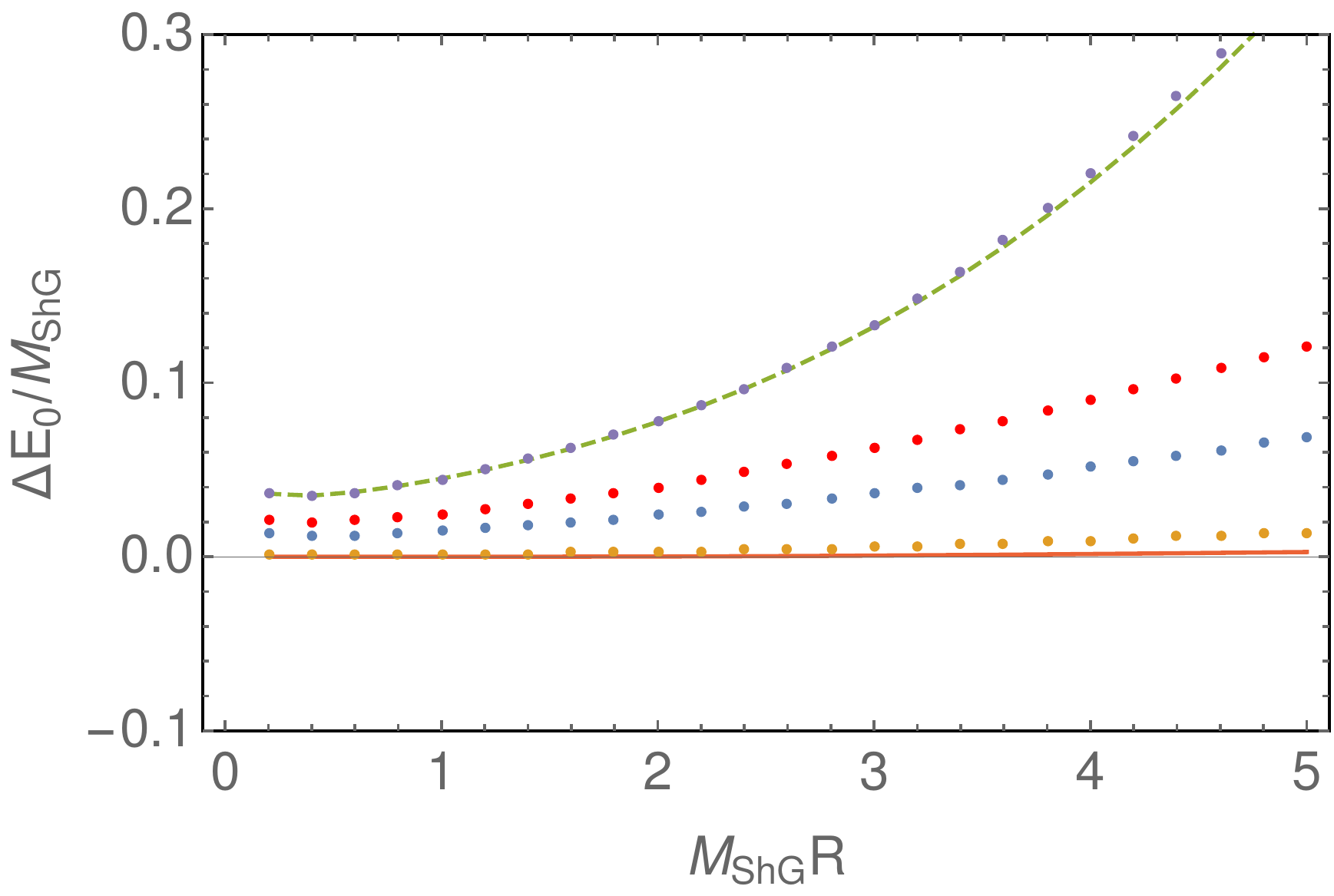}
\caption{$b=0.8$}
\end{subfigure}
\caption{Comparison of the six different approaches available to estimate the UV behavior of the
ground state energy at four different values of $b$.  All data is presented with the exact TBA values subtracted.  Blue dots: Raw ($N_c=12$) TSM data; orange dots: extrapolated TSM data; purple dots: 
diagonalization of $H_{ZM}$; dashed green curve: semi-classical reflection quantization; continuous red curve: Liouville reflection quantization; and red dots: diagonalization of effective zero mode Hamiltonian, $H_{ZM}^{\rm eff}$. }\label{FigComparisonOfUV}
\end{figure}

\subsection{Numerical Performance in the UV}

We have now at hand six different ways to estimate the spectrum of the ShG model in the UV small-$R$ limit: i) raw TSM data; ii) extrapolated TSM data; iii) diagonalization of the zero mode Hamiltonian, $H_{ZM}$ (i.e. eq.~(\ref{eIIIiii})); iv) reflection quantization using the semi-classical reflection amplitude; v) reflection quantization using the full Liouville reflection amplitude; and finally vi) diagonalization of the effective zero mode Hamiltonian, $H_{ZM}^{\rm eff}$, derived in Section \ref{subseceffpot}.  In this subsection we make an effort to compare the numerical accuracy of these different approaches.  We present our results in Fig.~\ref{FigComparisonOfUV} for four different values of $b$.
Here the ground state energy corresponding to the Hamiltonian in eq.~(\ref{eq:Heffpotfinal}) was obtained numerically and compared to the raw
zero mode potential using a real-space basis of size $16000$ -- see eq.~(\ref{eIIIvi}).
Let us summarise what we learn from these plots.

\begin{itemize}
\item The UV limit of TBA is indeed different from that calculated from the zero mode Hamiltonian $H_{ZM}$ alone, apart in the $b\rightarrow0$ limit.
\item In precisely the $b\rightarrow 0$ limit, the validity of the  reflection quantization method shrinks to extremely small volumes. This is intuitive at least in the
semiclassical case as the potential increases relatively mildly around the minimum, so neglecting the overlap of the two exponentials in the middle is
not well-founded.
\item For small couplings, the effective potential eq.~(\ref{eq:Heffpotfinal}) efficiently accounts for the discrepancy between the eigenvalues of $H_{ZM}$ and the exact TBA energies.
\item As the coupling $b$ is increased, higher order oscillator terms in  $\mu_{ShG}$ not included in eq.~(\ref{eq:Heffpotfinal}) become significant for arbitrarily small volumes and our expression
for the effective potential leads to inaccurate results.
\item In this same regime of large $b$ (but always $b < 1$), the accuracy of the ground state energy as obtained from reflection quantization extends to ever larger volumes well into the IR regime of the model.
\item However, TSMs (especially after power-law extrapolation) are able to reproduce the TBA result surprisingly well, even for larger $b$. This provides an important confirmation of the efficiency of TSMs as applied to the sinh-Gordon model.
\end{itemize}
\begin{figure}[b]
\begin{subfigure}[b]{0.475\textwidth}
\centering
\includegraphics[draft=false,width=\textwidth]{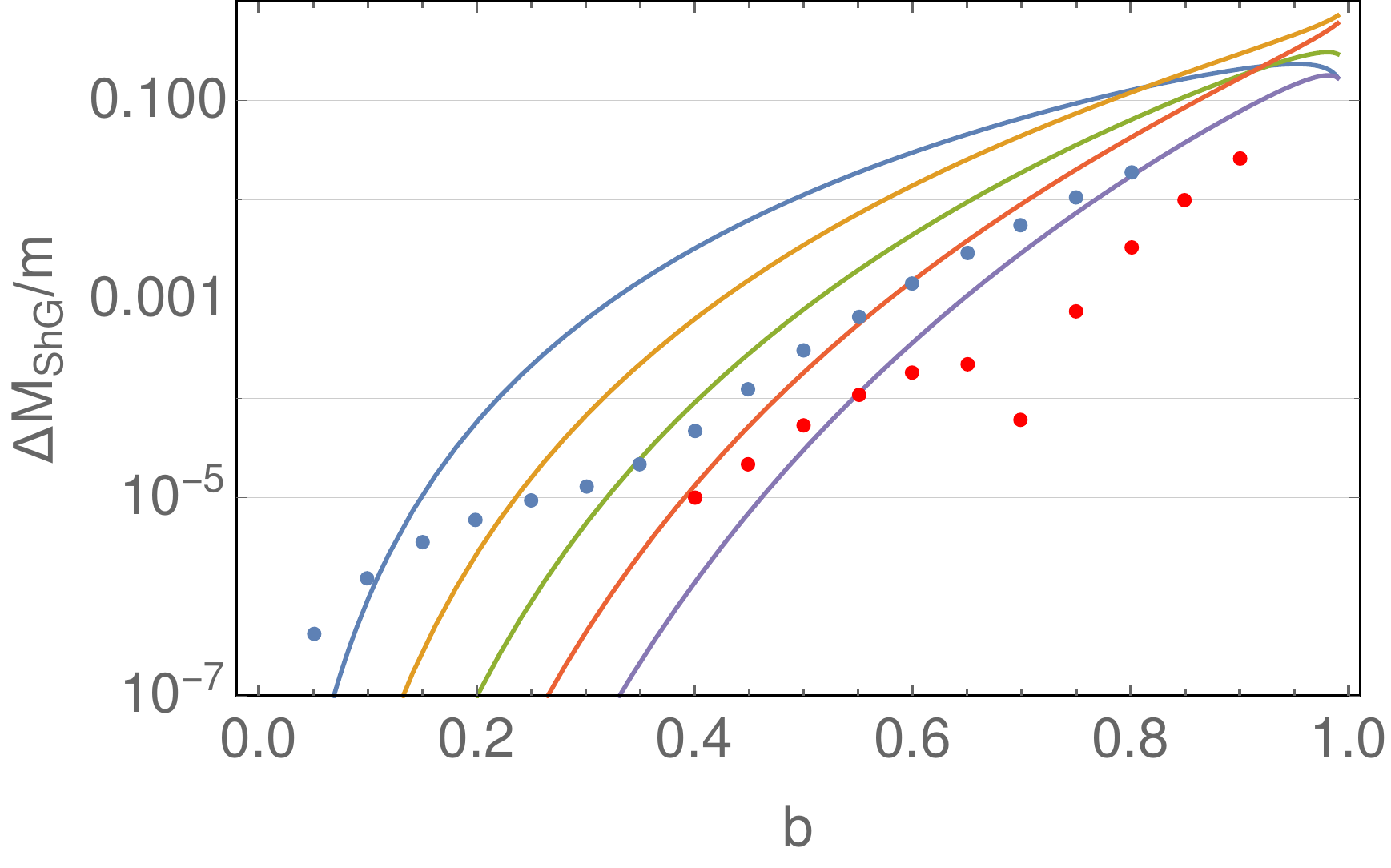}
\caption{Physical mass, $M_{ShG}$.}
\end{subfigure}
\hfill
\begin{subfigure}[b]{0.475\textwidth}
\centering
\includegraphics[draft=false,width=\textwidth]{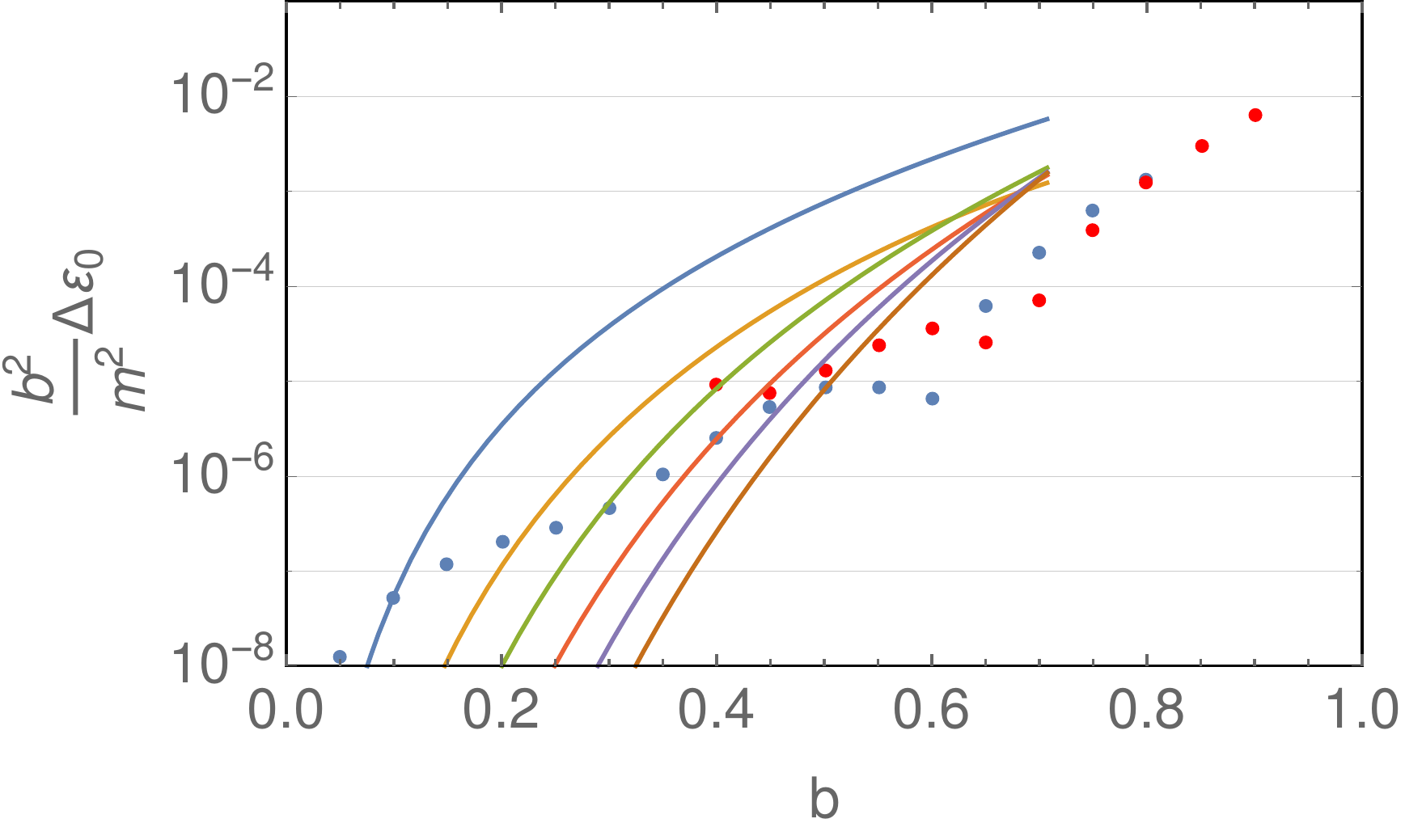}
\caption{Bulk energy, $\mathcal{E}_0$.}
\end{subfigure}
\caption{a) Presentation of $M_{ShG}$ as computed in Section \ref{SubsecUVmeasurement} (red dots) vs. its determination from extrapolated TSM data (blue dots).  b) The same but for the bulk ground state energy density, $\mathcal{E}_0$.  Solid curves shown are the same as in Fig.~11c-d.  Results are presented as differences with the numerically exact TBA values.
}
\label{MassE0fromUV}
\end{figure}
\subsection{Extracting $M_{ShG}$, $\mathcal{E}_0$ and $B$ from UV spectrum} \label{SubsecUVmeasurement}
\begin{figure}[b]
\begin{subfigure}[b]{0.475\textwidth}
\centering
\includegraphics[draft=false,width=\textwidth]{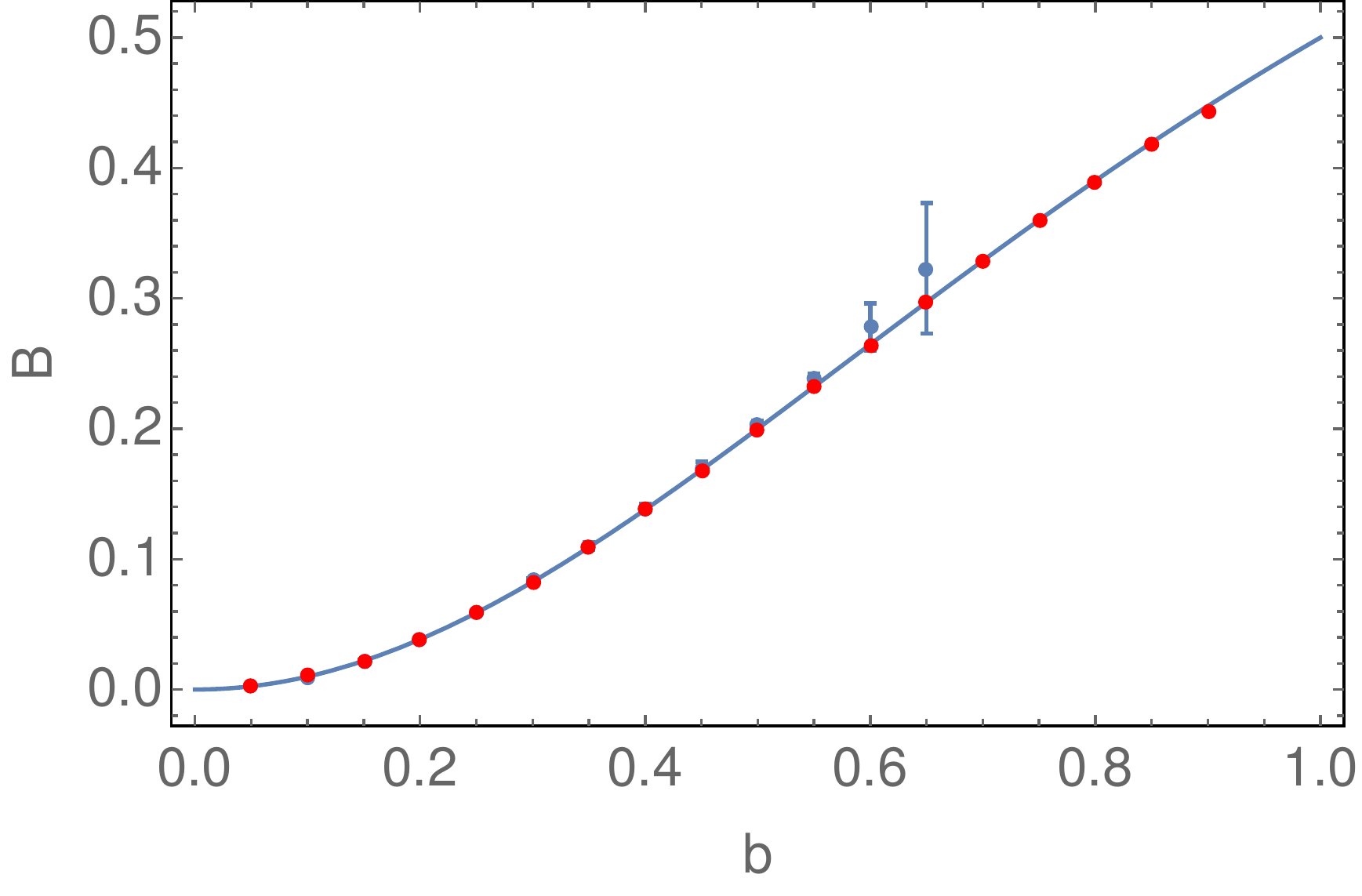}
\caption{S-matrix parameter $B$}
\end{subfigure}
\hfill
\begin{subfigure}[b]{0.475\textwidth}
\centering
\includegraphics[draft=false,width=\textwidth]{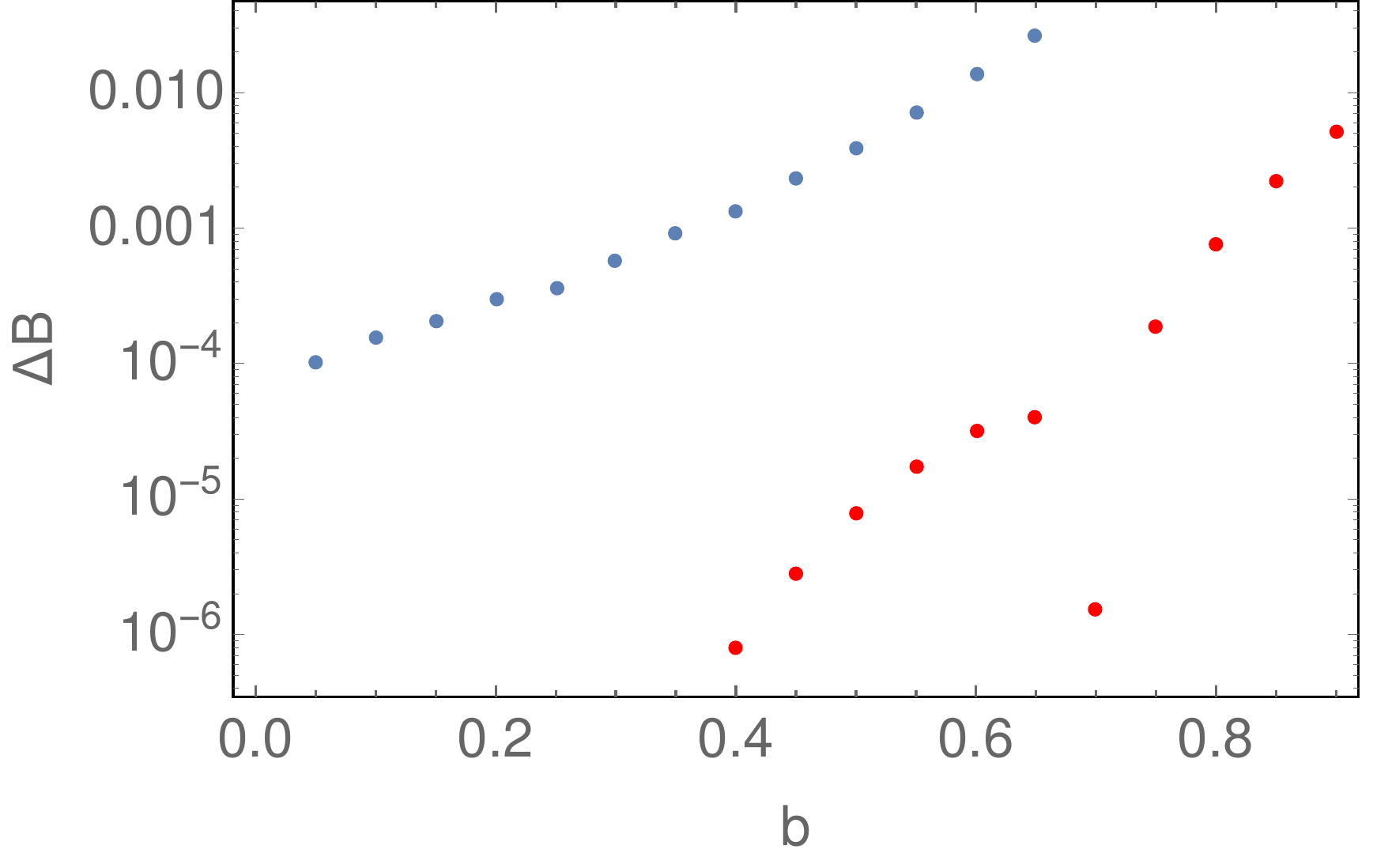}
\caption{Difference $\Delta B$}
\end{subfigure}
\caption{Determination of S-matrix parameter $B$. a) Absolute values of B. Blue dots (TSM - see Fig.~12b), red dots (method described in Section \ref{SubsecUVmeasurement}), solid curve (exact).  b) Differences of determined values of B with exact value.}
\label{BparamfromUV}
\end{figure}

We have shown that the TSM is able to reproduce the UV limit of TBA equations remarkably well.   It is thus reasonable to take advantage of this fact and to try to extract infinite volume
parameters combining TSM numerics with the quantization condition eq.~(\ref{eq:quantcond}).  As the exact quantization condition is expressed in terms of the IR parameters, $M_{ShG}$, $\mathcal{E}_0$, and $B$,
it is possible to fit these parameters using TSM data.  To do so we use the lowest two energy levels (the vacuum and the one-particle state) and
two values of the volume ($mR=0.1$ and $mR=0.2$), and then we minimize the function
\begin{equation}
\sum_{i,j=1}^{2} \left(E_{i}^{\mathrm{TBA}}(R_j)-E_{i}^{\mathrm{TSM}}(R_j)\right)^2,
\end{equation}
as function of $M_{ShG}$, $\mathcal{E}_0$ and $B$.
In Fig.~\ref{MassE0fromUV} we compare the mass and bulk energy obtained in this way to the standard TSM methods discussed in Section \ref{TSMresults}. It is worth stressing that this method produces an estimate for the mass that is an order of magnitude better than the extrapolated TSM data reported in Fig.~11 (see panel (a) of Fig.~\ref{MassE0fromUV}).  The determination of $\mathcal{E}_0$ in provides a precision comparable to the previous analysis.
The real power of the UV method is revealed when we consider the measurement of the S-matrix parameter $B$. Here the improvement of the error generally exceeds two orders of magnitude (see Fig.~\ref{BparamfromUV}). 

\subsection{What Have We Learned?}
In this section we have put the UV behavior of sinh-Gordon theory under scrutiny. After pointing out the incompleteness of the zero mode in describing the small-volume limit of the ground state energy for any finite $b$, we have derived an effective quantum mechanical potential. This effective potential accounts for the above discrepancies up to an $O(b^{12})$ error, by partially taking into account the oscillators neglected from the zero mode. 

Comparing the zero mode, the effective zero mode, and the TBA-Liouville reflection quantization results to TSM and exact TBA numerics, we realized that the effective potential significantly outperforms the zero mode. On the other hand, numerical TSM (especially after extrapolation) significantly outperforms the effective potential, providing an even more effective incorporation of the oscillators.

At the same time we noticed that the validity of the reflection quantization approximation extends to larger volumes as the coupling $b$ is increased. This lead to a more precise, UV-based measurement of the mass and the S-matrix parameter $B$, and even provided a consistency check for the energy density $\mathcal{E}_0$. 

We emphasize that both the S-matrix parameter $B$ and (up to the mass scale) the Liouville reflection amplitude are self-dual quantities, which are successfully reproduced directly from the Lagrangian, which is not in any sense manifestly self-dual. In this sense, the `self-duality' of the Lagrangian ShG model (i.e. the dependence of the model on coupling constant as encoded in the expression as $b^2/(1+b^2)$)  is confirmed up to the precision of TSM. This result, together with the partial analytical reproduction of the reflection amplitude, indicate that the discrepancy in the ground state energy as derived in the UV expansion coming from the semi-classical and Liouville quantizations is solely due to the effect of oscillators. Importantly, no extra terms (like $\cosh(b^{-1}\phi)$) in the Lagrangian are necessary to account for it.

However, note that all the above measurements were done below the self-dual point. As the self-dual point is traversed, the problems arising from the mass-coupling relation are inherited by the Liouville quantization condition as well. In the next section, we develop a supra-Borel resummation technique to motivate the power-law fit and to reach a better understanding of the cutoff dependence in applying TSMs to the ShG model.  In Section \ref{FFTCSASECTION}, we actually provide an argument that if the Lagrangian with a finite positive $\mu_{ShG}$ parameter defines a meaningful theory at all for $b>1$, this theory should actually be massless.

\section{Supra-Borel Resummation} \label{Supraborelsection}

In Section \ref{subsubAnalyticRG} we demonstrated pathologies in the corrections to the TSM results coming from taking into account perturbatively states above the cutoff.  In particular we demonstrated that at n-th order in perturbation theory in $\mu_{ShG}$, the states above the cutoff, $N_c$, contributed a factor behaving as
\begin{equation}
\delta E_{n} \sim N_c^{2(n^2-n)b^2-2n+2} \mu_{ShG}^n.
\end{equation}
That perturbation theory in $\mu_{ShG}$ is pathological should come as no surprise.  Because the $\mu_{ShG}\cosh(b\phi)$ potential is (strongly) unbounded from below if $\mu_{ShG}<0$, the radius of convergence of any perturbation theory in $\mu_{ShG}$ is 0.  Thus we expect any series to be asymptotic.  What is unexpected is that this series is not Borel resummable in the standard sense, namely the coefficients at order $n$ are diverging more quickly than $n!$.  We show in this section that nonetheless it is possible to resum this series in a meaningful way.  From this resummation we will obtain some understanding of the slow convergence of TSM results in $N_c$ as well as a partial justification of the power law scaling that we observed in Section \ref{TSMresults}.

\subsection{Minimal Resummation}\label{min_resum}

To start we are going to assume that a perturbative expansion in $\mu_{ShG}$ for the ground state energy has the following form:
\begin{eqnarray}\label{Eseries}
E(z=\tilde\mu_{ShG} N_{c}^{-2(1+b^2)}) = -\frac{N_c^2}{R}\sum_{n=1}^\infty a_n z^n e^{\frac{n^2}{16\gamma}};~~~
\gamma \equiv \frac{1}{16(2b^2\log(N_c) + \log(c))},
\end{eqnarray}
where $\tilde\mu_{ShG} = \mu_{ShG} \big(\frac{2\pi}{R}\big)^{-2-2b^2}$ is the dimensionless ShG coupling and $a_n,c$ are dimensionless constants independent of $R,N_c$.  This form for the energy, under certain approximations, is derived in Appendix \ref{sec:pertAppendix}.

Despite this series not being Borel resummable, it admits a supra-Borel resummability, something G. Hardy in his classical treatise on divergent series \cite{hardy1949divergent}, termed resummation via moment constant methods.  In this procedure, one supposes that one has the asymptotic series $S(z) = \sum_k a_k z^k$.  In order to resum it, one chooses a function $\rho$ and then defines its moments on $\mathbb{R}_+$ as
\begin{equation}
r_k = \int^\infty_0 t^k\rho (t) dt.
\end{equation}
The series, $S(z)$, is then said to be $r-\rho$ resummable if 
$$
B(t) = \sum_k \frac{a_k}{r_k} t^k,
$$
converges in some neighbourhood of $t=0$ and $B(t)$ has an analytic continuation to a neighbourhood of the positive real axis.
If the integral,
$$
g(z) = \int^\infty_0 B(zt) \rho(t)dt,
$$
is convergent for a $z\neq 0$, then the function $g(z)$ exists, is analytic in some domain $-\infty< Re(\log(z)) < c_0$, and has the asymptotic Taylor series $S(z)$.  $S(z)$ can then be identified as the resummation of $S(z)$ in this same domain.

For the case at hand we choose the function $\rho(t)$ as
\begin{equation}
\rho(t) = \frac{1}{t} e^{-4\gamma\log^2(t)},
\end{equation}
with moments
\begin{equation}
\mu_k = \sqrt{\frac{\pi}{4\gamma}} e^{\frac{k^2}{16\gamma}}.
\end{equation}
As discussed in Ref.~\cite{Grecchi1984} this choice arises generically in problems with exponential potentials. Ref.~\cite{Grecchi1984} discusses the technical conditions needed for resummation for this choice of $\rho(t)$.  Assuming for the moment that they are met, we can resum our expression for the ground state energy and rewrite it in the form:
\begin{eqnarray}\label{resum}
E &=& \frac{a_1\tilde\mu}{R}-\frac{N_c^2}{R\sqrt{4\gamma}}\int^\infty_{-\infty} dx e^{-x^2}B(\tilde\mu N_c^{-(2+2b^2)}e^{-\frac{x}{2\sqrt{\gamma}}});\cr\cr
B(t) &=& \bigg(\frac{4\gamma}{\pi}\bigg)^{1/2}\sum^\infty_{n=2}a_n t^n .
\end{eqnarray}
In Appendix \ref{sec:pertAppendix}, we show that in fact $a_n = (-\alpha)^n$ with $\alpha>0$.  Assuming this, we can then write $B(t)$ as
\begin{equation}\label{btransform}
B(t) = \bigg(\frac{4\gamma}{\pi}\bigg)^{1/2}\frac{\alpha^2t^2}{1+\alpha t},
\end{equation}
and so is analytic along all of the positive real axis.

Now that we have resummed our asymptotic series, let us investigate its properties as a function of $N_c$.  It is not a priori obvious that this resummation will necessarily lead to a sensible result, i.e. something that converges as $N_c\rightarrow \infty$.  But nonetheless it does.  To investigate the asymptotics, we 
write the integral expression for $E$ in a suggestive form:
\begin{eqnarray}
E &=&  \frac{a_1\tilde\mu_{ShG}}{R}-\frac{\tilde\mu_{ShG}\alpha c}{R\sqrt{\pi}} \int^\infty_{-\infty}dx \frac{e^{-x^2}}{1+\alpha^{-1}e^{\beta(x-\mu_{B})}};\cr\cr
\beta &\equiv& \frac{1}{2\sqrt{\gamma}};\cr\cr
\mu_B &\equiv& \frac{1}{4\sqrt\gamma} + 2\sqrt\gamma\log (\tilde\mu_{ShG} N^{-2-2b^2}_c) .
\end{eqnarray}
We can now do a Sommerfeld-type expansion and obtain asymptotics in large $N_c$ of the form:
\begin{eqnarray}\label{asymp}
E &=&  \frac{a_1\tilde\mu_{ShG}}{R}-\frac{\alpha\tilde\mu_{ShG} c}{R\sqrt{\pi}}\bigg[\int^{\mu_B}_{-\infty}dx e^{-x^2} + \frac{1}{\beta}\int^\infty_0 dy e^{-\mu_B^2}\big(\frac{1}{1+\alpha^{-1}e^y}-\frac{1}{1+\alpha e^y}\big) + {\cal O}(\frac{e^{-\mu_B^2}}{\beta^2})\bigg]\cr\cr
&=& \left\{\begin{array}{lr}
\frac{a_1\tilde\mu_{ShG}}{R} -  \frac{\alpha\tilde\mu_{ShG} c}{R\sqrt{\pi}}e^{-\mu_B^2} \left( \frac{1}{2|\mu_B |} +\frac{1}{\beta}\log{\alpha}\right) + C \frac{e^{-\mu_B^2}}{R\beta^2} , & \,\,\,\,\,\,\,\,b^2<1;\\
\frac{a_1\tilde\mu_{ShG}}{R} - \frac{\alpha\tilde\mu_{ShG} c}{R\sqrt{\pi}}e^{-\mu_B^2} \left(\sqrt{\pi} - \frac{1}{2|\mu_B |} +\frac{1}{\beta}\log{\alpha}\right) + C' \frac{e^{-\mu_B^2}}{R\beta^2}, & \,\,\,\,\,\,\,\,b^2>1,
\end{array}\right.
\end{eqnarray}
where $C,C'$ are constants.  We note that in the large-$N_c$ limit, the Borel `chemical potential', $\mu_B$, can be written as
\begin{equation}\label{powerlaw}
\mu_B \approx\frac{\sqrt{\log N_c}(b^2-1)}{\sqrt{2}b},
\end{equation}
and so $|\mu_B|$ in this limit is invariant under $b \rightarrow 1/b$.  We also note that as $N_c\rightarrow\infty$, the resummed part (i.e. involving terms $n\geq 2$) of $E$ vanishes if $b<1$, but tends to a finite constant if $b>1$.  

\begin{figure}[b]
\centering
\includegraphics[width=0.45\textwidth]{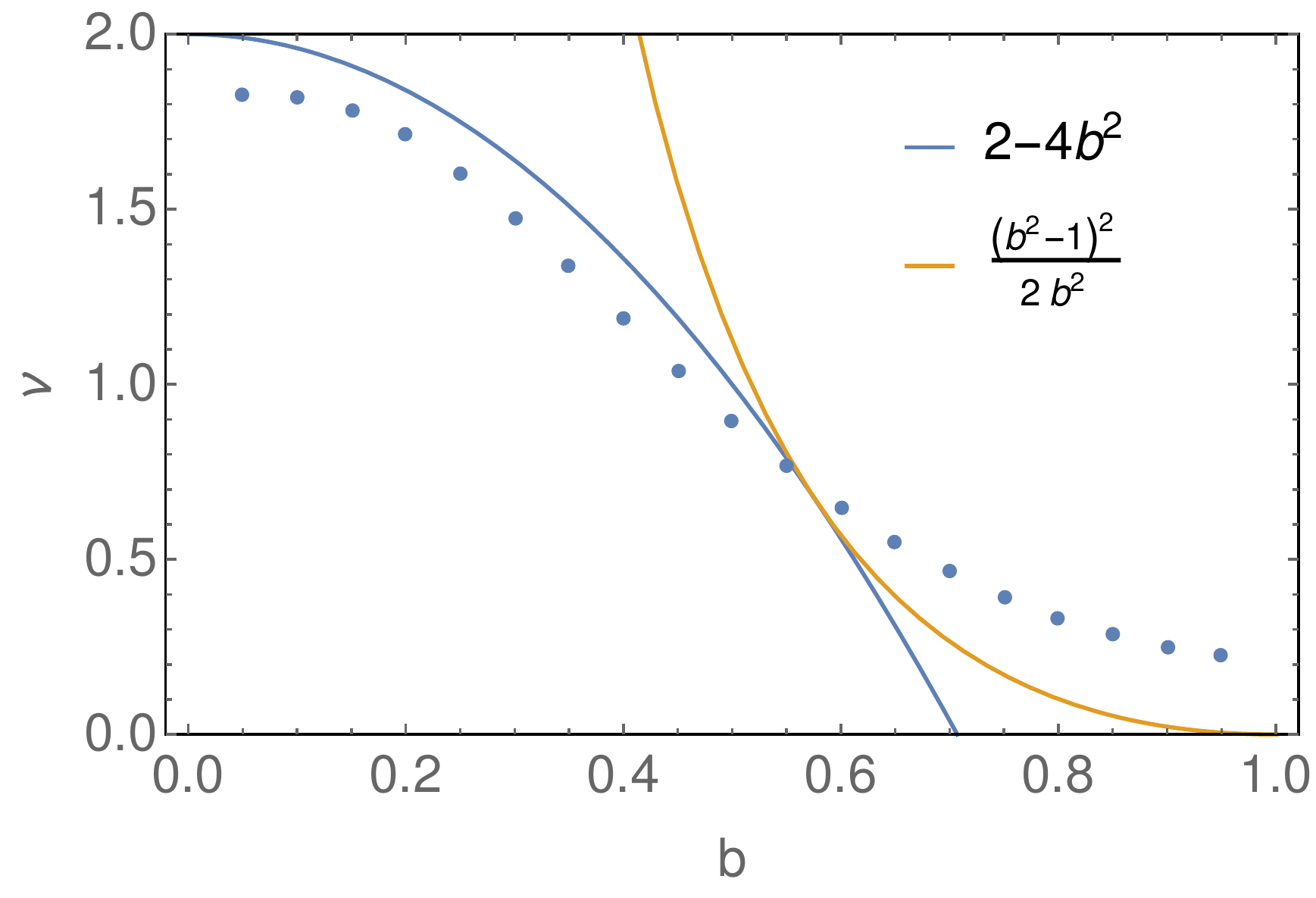}
\caption{Power law exponents as a function of $b$ as determined from TSM extrapolation (blue dots) vs those predicted by eq.~(\ref{exp}) (solid orange curve) and the exponent of the leading na\"ive RG correction eq.~(\ref{dH2expr}).}
\label{Exponents}
\end{figure}

We thus expect that for large $N_c$ that 
\begin{equation}
E \sim N_c^{-\nu}, ~~ \nu = \frac{(b^2-1)^2}{2b^2}. \label{exp}
\end{equation}  
We compare in Fig.~\ref{Exponents} the exponents predicted by this result to those determined from extrapolating TSM data (see Fig.~8 in Section \ref{TSMresults}).
We see that as the self-dual point is approached, the exponent predicted by the asymptotics describes approximately the trend observed from power law fits of the data.  It however fails to describe the fitted exponents at small $b$.  Instead these are described well by the exponent coming from the first perturbative correction to the TSM data (see eq.~(\ref{dH2expr})).

\subsection{General Supra-Borel Approximates}

In the previous section, we considered the resummation of a perturbative expansion of the ground state energy.  This expansion, discussed in Appendix \ref{sec:pertAppendix}, and given in eq.~(\ref{Eseries}), makes certain approximations to the evaluation of the constants $c$ and $a_n$.  How important are these approximations to the final result?  We show here that in fact the final answer is sensitive to the exact form of these constants.

We begin by developing a simple constraint on the resummation arising from the behaviour of the energy as a function of $N_c$.
Let us call the energy of a state without truncation $E_\infty$ and the energy of the state at truncation $N_c$, $E_{N_c}$.  Then we can write
\begin{equation}
E_\infty = E_{N_c} + \delta E(\infty) - \delta E(N_c),
\end{equation}
i.e. if $\delta E(N_c)$ is considered to the correction to the energy due to including {\it all} states up to truncation $N_c$, and $E_{N_c}$ is the energy we obtain from the TSM, then we subtract off $\delta E(N_c)$ so that we do not double count the contribution of states up to truncation $N_c$.
This means
that the fitting form for the TSM energies is
\begin{equation}
E_{N_c} = E_\infty + \delta E(N_c) - \delta E(\infty).
\end{equation}
In general we expect $E_{N_c}$ approaches $E_\infty$ from above, a consequence of the variational principle.  So this means $\delta E(N_c)-\delta E(\infty)$ must be positive.  For $b>1$ we see from asymptotics, i.e. eq.~(\ref{asymp}), that this is indeed the case.  But for $b<1$ it is not, and we thus conclude that the simple form computed for $E$ in eqs.~\ref{Eseries} cannot strictly be correct.  For the purpose of resummation, we require more precise forms for $a_n$'s and the constant $c$.

In general it is difficult to compute to high precision the $a_n$'s and so derive an exact form to the supra-Borel transform that we presented in eq.~(\ref{btransform}).  One way to resolve this problem is to suppose the Borel transform has a more general form than in eq.~(\ref{btransform}).  The simplest possibility is a Pad\'e form:
\begin{equation}
B_M(t) = \frac{\sum_{i=1}^M r_i t^{i+1}}{1+\sum^M_{i=1}s_i t^i}, ~~~M>1.
\end{equation}
We will only look at forms of $B(t)$ whose large t asymptotics are $B(t\rightarrow\infty) \propto t$ as this form leads to a naive cancellation of the powers of $N_c$ in eq.~(\ref{resum}).
With this more general form, we can arrange as we will see in the next section, through an appropriate choice of couplings, that
\begin{equation}
\delta E(N_c) - \delta E(\infty) > 0,
\end{equation}
despite the overall minus sign.  To ensure this, we allow the $r_i$'s to be of either sign and so that we have no poles on the positive real axis, we assume the $s_i$'s to be positive.  So, for example, with $M=2$, we take $r_1>0$, $r_2<0$, and $s_{1,2}>0$.  In principle these free parameters can be fixed by carefully computing the $a_n$'s to finite order.

\begin{figure}
\centering
\begin{subfigure}[b]{0.45\textwidth}
\centering
\includegraphics[width=\textwidth]{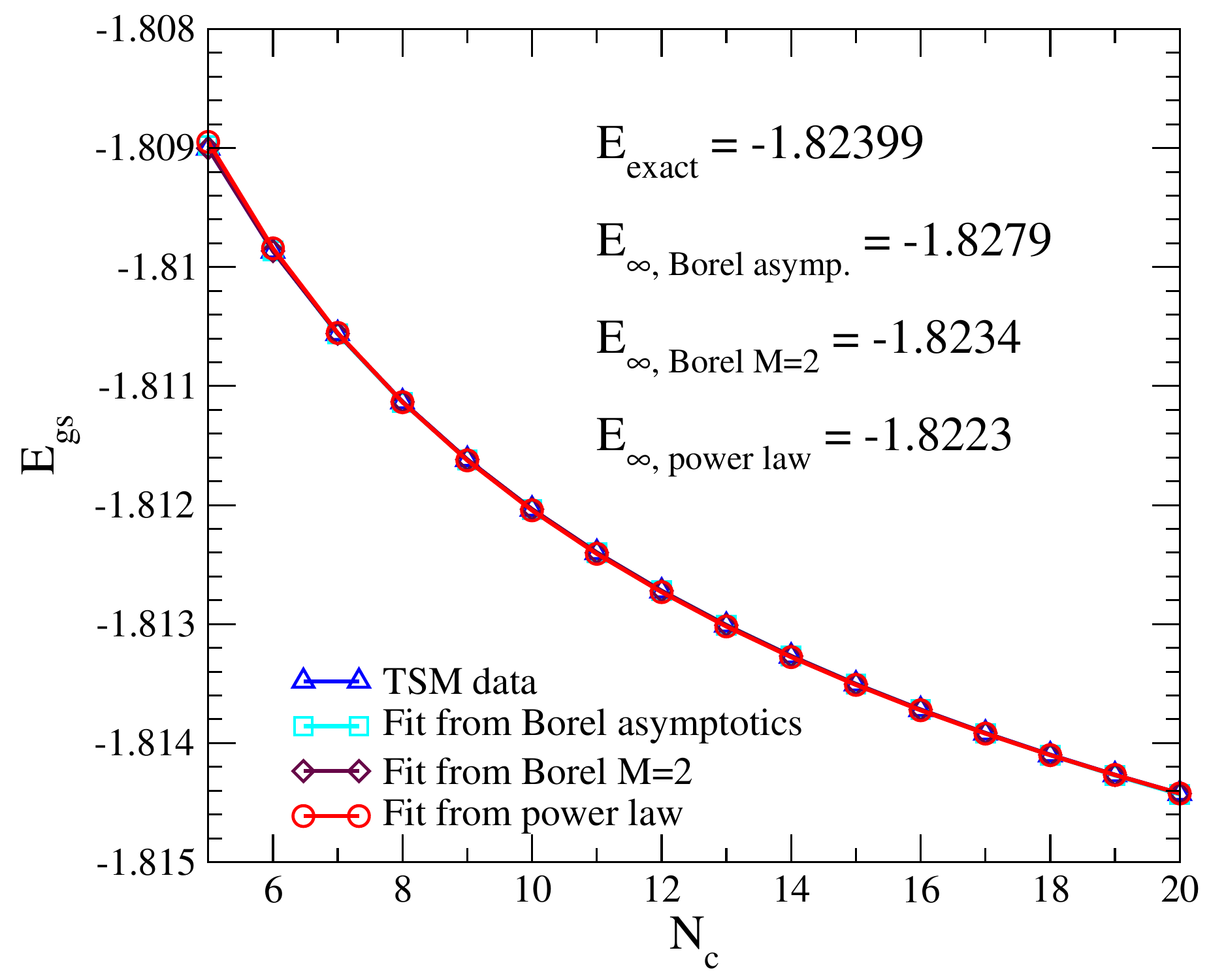}
\caption{$b=0.778,R=0.2$: Linear scale in $N_c$.}
\end{subfigure}
\begin{subfigure}[b]{0.455\textwidth}
\centering
\includegraphics[width=\textwidth]{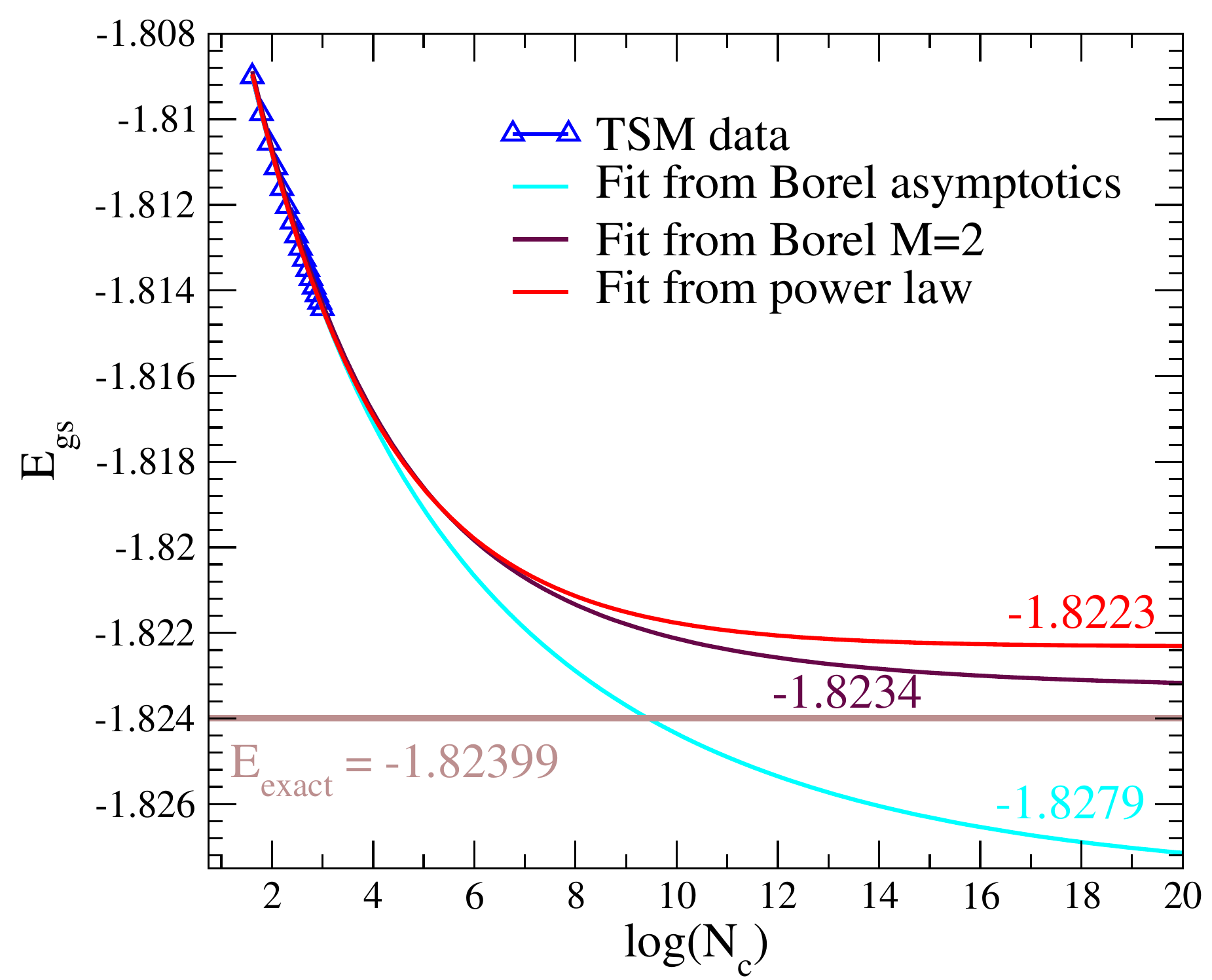}
\caption{$b=0.778,R=0.2$: Log scale in $N_c$.} 
\end{subfigure}
\begin{subfigure}[b]{0.45\textwidth}
\centering
\includegraphics[width=\textwidth]{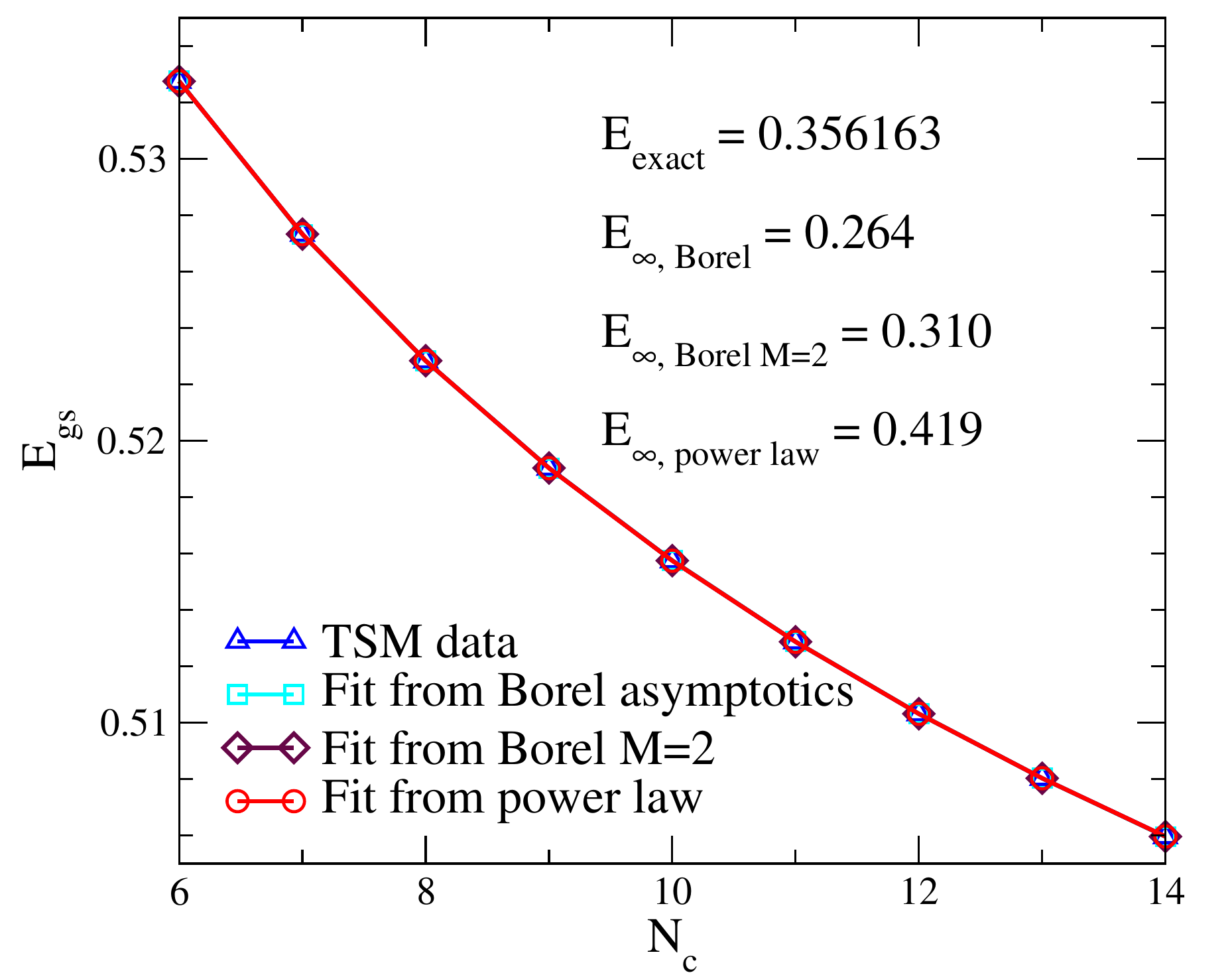}
\caption{$b=0.919,R=3$: Linear scale in $N_c$.} 
\end{subfigure}
\begin{subfigure}[b]{0.45\textwidth}
\centering
\includegraphics[width=\textwidth]{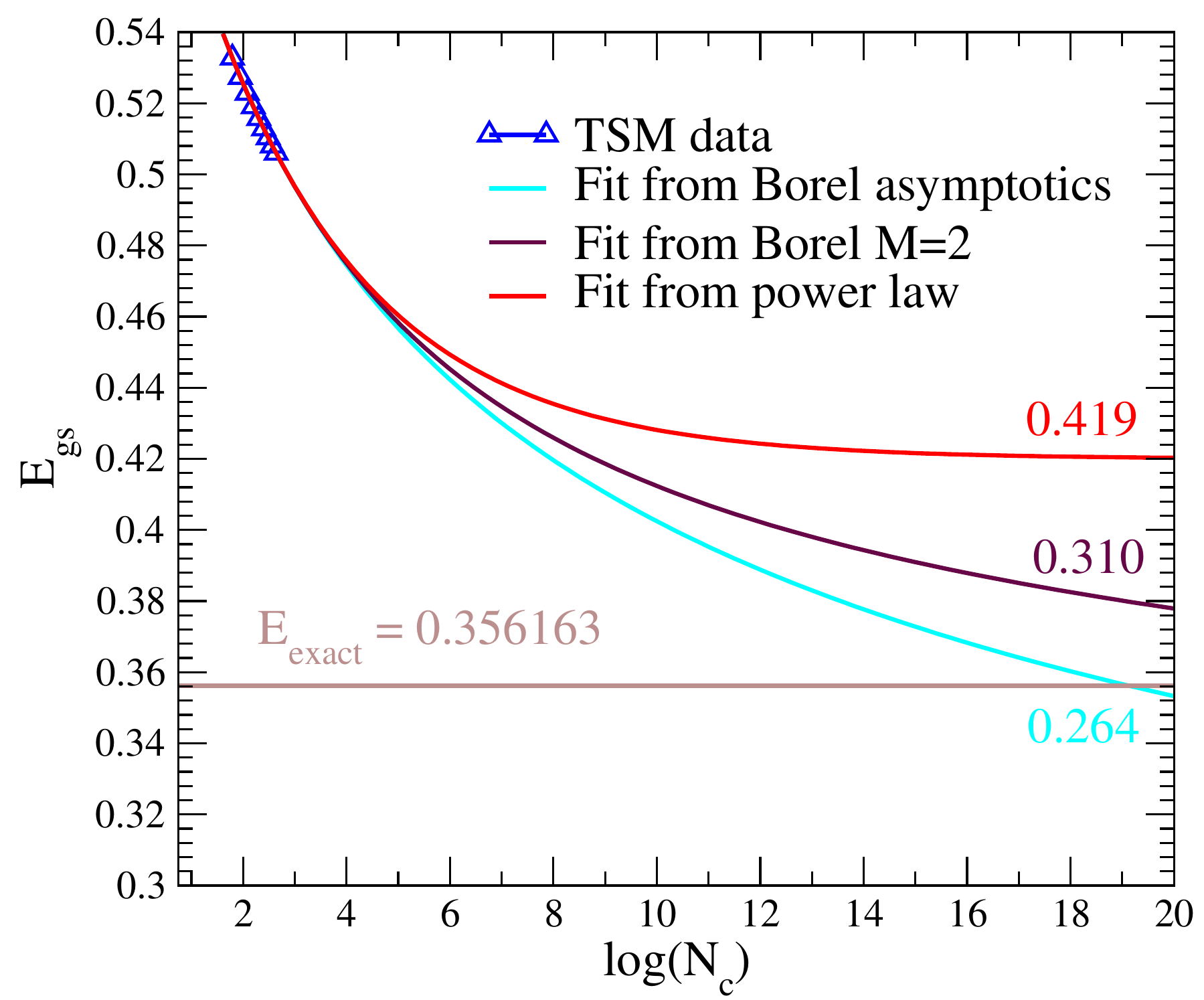}
\caption{$b=0.919,R=3$: Log scale in $N_c$.} 
\end{subfigure}
\caption{Top panels: Borel and power-law fits for $R=0.2,b=0.778$.  The exponent $\nu$ given in eq.~(\ref{pl}) of the power law fit is $0.38$.  This can be compared with exponent predicted by the Borel asymptotics, $\nu=0.13$.  Bottom panels: Borel and power law fits for $R=3,b=0.919$.   The exponent (eq.~(\ref{pl})) of the power law fit is $\nu=0.215$ while that expected from the Borel asymptotics is $\nu\approx 0.014$. }
\label{borelfits}
\end{figure}

\subsection{Fits and Discussion of Limitations}

In this section we look at select examples where we fit as a function of $N_c$ our TSM data to the functional forms suggested by the Borel resummations.  We compare these fits to a simpler power law fit.  We look at two fits suggested by the Borel resummations.  In the first we look at a fit that incorporates only the asymptotics suggested by the minimal resummation discussed in Section \ref{min_resum}:
\begin{equation}\label{basymp}
E(N_c) = E_{\infty} + A \frac{e^{-\mu_{ShG}^2(N_c,\tilde\mu_{ShG},b,c) }} {|\mu_{ShG}(N_c,\tilde\mu_{ShG},b,c)|}.
\end{equation}
In the second we look at a fit to a form involving an $M=2$ supra-Borel transform
\begin{eqnarray}\label{m2borel}
E(N) &=& E_{\infty} + N_c^2\int^\infty_{-\infty}e^{-x^2}B_2(\tilde\mu_{ShG} N_{c}^{-(2+2b^2)}e^{-\frac{x}{2\sqrt{\gamma}}});\cr\cr
B_2(t) &=& \frac{r_1t^2+r_2t^3}{1+s_1t+s_2t^2}.
\end{eqnarray}
Finally we consider a power law fit:
\begin{equation}\label{pl}
E(N_c) = E_{\infty} + AN_c^{-\nu}.
\end{equation}
The latter two forms are generically applicable a priori.  However the form in eq.~(\ref{basymp}) is not.  The Borel resummation `chemical potential', $\mu_B$, has a zero, $N_{c0}$ as a function of $N_c$.  Thus if we are to use this form to fit our numerical TSM data we need to be sure our data runs over values of $N_c > N_{c0}$.  $N_{c0}$ is given by 
\begin{equation}
N_{c0}= (c^2\tilde\mu_{ShG})^{1/(2-2b^2)}.
\end{equation}
Because $\tilde\mu_{ShG} = R^{2+2b^2}\mu_{ShG}$, we see that at large $R$ the regime where the Borel asymptotic form is expected to apply recedes to ever larger $N_c$.  We note in general that if we are in a regime where the Borel asymptotics applies then the exponent from power law fit should approximate that predicted by eq.~(\ref{powerlaw}).  We will see in the examples below for larger values of $b$ that this does not happen.

\begin{table}[h]
\caption{Fits including an uncertainty analysis.  All data is for $\mu_{ShG}=0.1$.}
\centering
\begin{tabular}{| c | c | c | c | c | c |}
\hline\hline
b & R & Borel Asymptotic & Borel $M=2$ & Power Law & Exact Value\\
\hline
0.778 & 0.2 & $-1.8277 \pm 0.0001$ & $-1.8233\pm 0.0004$ & $-1.825 \pm 0.002$ & $-1.82399$ \\
0.919 & 0.3 & $-1.1490 \pm 0.0006$ & $-1.113 \pm 0.007$ & $-1.1204\pm 0.0008$ & $-1.14084$ \\
0.919 & 3.0 & $0.269\pm0.004$ & $0.34 \pm 0.03$ & $0.419\pm 0.001$ & $0.35613$ \\
\hline
\end{tabular}\label{fitdata}
\end{table}

In Fig.~\ref{borelfits} we show the results of these fits for two different values of $b$: $b=0.778$ and $b=0.919$.  On the left hand side of the figure (panels a and c), we show the fits over the existing range of the data (for $b=0.778$, $5\leq N_c \leq 20$, and for $b=0.919$, $6\leq N_c \leq 14$) on a linear scale.  We see that in all three cases (Borel asymptotic, eq.~(\ref{basymp}), $M=2$ Borel, eq.~(\ref{m2borel}), and power law, eq.~(\ref{pl}), the fits appear identical.  What is different however is the value of $E_\infty$ inferred from the fits.  These differ markedly between the fits.  We see in comparison to the exact value (as determined from the TBA), the power law extrapolation, $E_{\infty,\rm{power~law}}$, is greater than the exact value (i.e. there is an under extrapolation) and the Borel asymptotic fit leads to a value $E_{\infty,{\rm Borel~asymp.}}$ that is less than the exact value (i.e. there is an over extrapolation).  The $M=2$ Borel fit for these examples seems to perform best (see below, however), leading to an asymptotic value, $E_{\infty,{\rm Borel~M=2}}$, that is closest to the exact value.  One conclusion that we draw from this is that the values of $N_c$ at which our data is computed are decidedly not in the large-$N_c$ asymptotic regime.  Our data does obey a power law form (as our fits indicate), but this power law is different from the one derived from the Borel asymptotics (i.e. eq.~(\ref{powerlaw})).  In particular the power law being obeyed by the data at $N_c$ in the range $(10,20)$ is greater than the one indicated by eq.~(\ref{powerlaw}).
This is consistent with our finding that the power law fit under extrapolates and leads to a value of $E_\infty$ that is too large while the Borel asymptotics do the opposite.
That the $M=2$ Borel form does better is thus not surprising.  However this should not be taken (at all) as the final word of which fitting form to use.  Certainly we have made no real attempt to explore the space of supra-Borel transforms, $B(t)$, and leave this to later work.

To give the reader a better picture of the extrapolations that are being performed by the fits, we have also plotted the fits on a semi-log scale (the right hand side of Fig.~\ref{borelfits}).
Here we explicitly show the extrapolations to large $N_c$, $N_c\sim e^{20}$.  We see that the different fits only begin to diverge from one another for $N_c$ far in excess of our ability to perform numerical TSM computations.  We also see the necessity for the extrapolations.  Even the most rapidly converging of the fits, the power law fits, only see convergence for $N_c \gg e^{10}$ at the values of $b$ that we are considering in Fig.~\ref{borelfits}.  We do note that the convergence at $b=0.778$ is much more rapid than at $b=0.919$, something we already expected from the power law inferred from the Borel asymptotics eq.~(\ref{powerlaw}).

In the fits described in Fig.~\ref{borelfits}, we made no attempt to assign an uncertainty to our fits.  In Table~\ref{fitdata} we attempt to do this.  For each of the three different forms, we perform four fits. Each of these four fits is over different subsets of our numerical data.  The different subsets are obtained by dropping data points for the lowest values of $N_c$ at which we have numerical data.  For the $b=0.778$ fit, the subsets are formed from data at values of $N_c$ corresponding to the sets $\{5,\cdots,20\}, \{6,\cdots,20\},\{7,\cdots,20\}$, and $\{8,\cdots,20\}$, while for the $b=0.919$ fits the $N_c$-subsets are given by $\{6,\cdots,14\}, \{7,\cdots,14\},\{8,\cdots,14\}$, and $\{9,\cdots,14\}$.  Having done these four fits, we take the average and standard deviation of the obtained values of $E_\infty$ and report these in Table~\ref{fitdata}.  In this table we provide an analysis for one more pair of $(b,R)$, $(0.919,3.0)$, beyond those considered in Fig.~\ref{borelfits}.

\subsection{What Have We Learned?}

We have shown in this section that the divergences observed in the perturbative corrections to the TSM can be made sense of.  In particular, we have shown that these corrections can be both computed at all orders in the coupling $\mu_{ShG}$ and resummed with a resulting finite value via a supra-Borel procedure.  We note that the ShG model is not the only theory where resummation is needed.  The ability to do so, for example, might be useful for studies of the $\Phi^4$ (and other polynomial) theories.  To date at least partial resummations in the context of TSM for the $1+1d$ $\Phi^4$ theory have been attempted in Refs.~\cite{PhysRevD.96.065024,Elias-Miro2017}.

We have shown the functional form of the resummed corrections gives us important insights into the large $N_c$ asymptotics of the TSM.  It provides a partial explanation at values of $b$ close to the self-dual point for the power law fit that we have observed (see Fig. 20).  It also provides a guide to how to extrapolate TSM results to $N_c=\infty$.   In the section we considered two fitting functions suggested by the supra-Borel resummation to use with our TSM data.  In the first, we employed only the asymptotic form (valid for large $N_c$) suggested by the resummation.  In the second we used a Pad\'e form for the Borel approximant.  Unsurprisingly the Pad\'e form outperformed the asymptotic form.  The asymptotic form is likely only really appropriate for values of $N_c$ beyond that we have performed TSM computations.

In presenting this resummation, we have really only scratched the surface.  We are fairly certain that there are more appropriate Borel approximates than used here that are better tailored to the properties of the ShG model.  We view this an important open problem presented by this work.

\section{Form Factor Truncated Spectrum Method}\label{FFTCSASECTION}

In the previous sections we have focused on numerical TSM data arrived at from a computational basis based on a non-compact free massless boson and its zero mode.
We have discussed in detail the strong dependence of the data from the cutoff, $E_c$, in particular approaching the self-dual point $b=1$.  In light of these results, it is natural to
ask the following question:

\begin{center}
{\em Is there a starting point $H_0$ for the TSM much `closer' to the model of interest?}
\end{center}

\noindent In this context, `closer' means having a set of eigenstates such that TSMs is better suited to
approximate the spectrum of the perturbed theory.  For example, instead of separating the zero mode, we might have started with
the Hamiltonian \eqref{eq:Massivehamiltonian} and so used a computation basis built on a standard massive Fock basis.
Having considered this, a more radical possibility suggests itself.  This, however, comes at the cost of abandoning the comfort provided by the \emph{free} nature of $H_0$.

\subsection{Using the ShG Basis as a Computational Basis}
The main idea is as follows.  Our target is a ShG model at coupling $b_{1}$ and we choose as our starting point a ShG model at a \emph{different} coupling $b_{0}$, setting $H_0\,=\,H_{b_{0}}^{\left(\text{shG}\right)}$.   While such a starting point does suffer from the critique that we are using as input information that is itself non-trivial and is meant to be, in part, verified by our TSM studies, we believe that there are good reasons to explore this path:
\begin{itemize}
\item As we will see soon,  this method provides a set of a highly nontrivial self-consistency checks.
\item It provides useful insights regarding the validity of the exact VEV formulae in certain domains.
\item It is a non-traditional starting point for the application of the TSM to interacting massive field theories\footnote{Certainly the TSM has a long history of employing an $H_0$ that is interacting.  However $H_0$ in such cases is an interacting massless CFT \cite{Yurov:1989yu,LASSIG1991591,LASSIG1991666,Klassen:1990ub,KLASSEN1992511,BERIA2013457,KONIK2015547,DOREY1998641,Kormos_2009}. } and insight gained here will be useful for the study of deformations of other massive integrable field theories.
\end{itemize}
In the approach taking in here, we write the Hamiltonian as:
\begin{equation}
H\,=\, H_{b_{0}}^{\left(\text{shG}\right)} + H_1 \,\,\,,
\label{newwww}
\end{equation}
where the solvable piece is given by
\begin{equation}
H_{b_{0}}^{\left(\text{shG}\right)} \,=\,
H_{0}+2\mu_{0}\int^R_0 dx\cosh(b_0\phi) ,
\end{equation}
where $H_{0}$ is the Hamiltonian of the non-compact $c=1$ boson, while the perturbation is given by
\begin{equation}
H_1 = \int^R_0 dx \big(2\mu_1\cosh(b_1\phi(x) )- 2\mu_0\cosh(b_0\phi(x))\big).
\label{eq:FFham}
\end{equation}

In order to proceed with the diagonalization of the Hamiltonian (\ref{newwww}) on a cylinder of finite circumference $R$ we need two highly nontrivial
sets of data:
\begin{itemize}
\item Finite volume energy levels of $H_{b_{0}}^{\left(\text{shG}\right)}$;
\item Finite volume matrix elements of $H_{1}$ between eigenstates of $H_{b_{0}}^{\left(\text{shG}\right)}$.
\end{itemize}
Fortunately, both pieces of data are available in the ShG model (as well as a number of other integrable models), at least up to exponentially small corrections in the volume. Indeed, according to L\"uscher's principle, all information about finite volume quantities are encoded in infinite volume data, like the physical mass $M$ and the $S$-matrix. Indeed, the exact energies of all finite volume eigenstates can be computed efficiently from the excited-state thermodynamical Bethe ansatz (see Section \ref{SubsecTBA}).  Formulae for the matrix elements, at least up to exponential small corrections \cite{BBLW,BLSV}, are also available. In the special case of diagonal matrix elements, one can use a generalization of the Leclair-Mussardo formula \cite{Pozsgay_2015} to systematically compute exponential corrections.
In the following, we are going work mostly in the regime $5\leq M_{ShG} R \leq15$, where exponential corrections should be small and can be neglected.

\subsubsection{Finite Volume Energies}
We have already discussed in Section \ref{SubsecTBA} the excited state TBA that provides the finite volume energies of $H_0$.   However we revisit this discussion to establish notation that is needed to discuss the finite volume matrix elements.
The finite energy eigenstates can be described by a finite number $n$ of particles with momenta $p_{j}=M\sinh\theta_{j}$ quantized due to finite volume $R$. We can
assign an integer quantum number $I_{j}$ to each momenta. The Bethe-Yang equations essentially impose the one-valuedness of the quantum mechanical
multi-particle wave-function in presence of the purely elastic two-particle scattering $S\left(\theta\right)=e^{i\delta\left(\theta\right)}$
\begin{eqnarray}
Q_{j}\left(\left\{ \theta\right\} \right) &\equiv& p_{j}L+\sum_{k\neq j}^{n}\delta\left(\theta_{j}-\theta_{k}\right)=2\pi I_{j}+O\left(e^{-ML}\right),\quad j\in\left\{ 1,\dots,n\right\} ;\cr\cr\label{eq:BYE}
\delta (\theta) &=& 2\tan^{-1}\bigg(\frac{\sinh(\theta)}{\sin(\pi B)}\bigg).
\end{eqnarray}
A finite volume state, $|\{I_{i}\} _{i=1}^{k}\rangle$, is given by the set $I_{j}$, $j\in\left\{ 1,\dots,n\right\} $. The system eq.~\ref{eq:BYE} is solved for the set of rapidities $\theta_{j}$ giving the
total energy and momentum of the state as:
\begin{equation}
E=\sum_{j=1}^{n}M\cosh\theta_{j},\quad P=\sum_{j=1}^{n}M\sinh\theta_{j}
\end{equation}

\subsubsection{Finite Volume Matrix Elements}

What we have not considered to date are matrix elements in finite volume.  Here the Pozsgay-Tak\'acs formulae \cite{Pozsgay:2007kn,Pozsgay:2007gx} come to our aid. For completely non-diagonal form factors, where all rapidities
are pairwise different between the bra and the ket states, the formula expresses the finite volume form factors
\begin{align}
\langle \{ I_{i}\} _{i=1}^{k}|\mathcal{O}(x,0)|\{ \tilde{I}_{j}\} _{j=1}^{l}\rangle _{L} & =\frac{e^{iM(\sum_{j=1}^{l}\sinh\vartheta_{j}-\sum_{i=1}^{k}\sinh\theta_{i})x}\langle \{ \theta_{i}\} _{i=1}^{k}|\mathcal{O}(0,0)|\{ \vartheta_{j}\} _{j=1}^{l}\rangle }{\sqrt{\rho_{k}(\{ \theta_{i}\} )\rho_{l}(\{ \vartheta_{j}\} )}\mathcal{N}_{k}^{*}(\{ \theta_{i}\} )\mathcal{N}_{l}(\{ \vartheta_{j}\} )}\label{eq:PozsTak}\\
 & +O(e^{-ML})\nonumber
\end{align}
in terms of infinite volume form factors: $\langle \{ \theta_{i}\} _{i=1}^{k}|\mathcal{O}(0,0)|\{ \vartheta_{j}\} _{j=1}^{l}\rangle$ \cite{Fring:1992pt,Koubek:1993ke} 
 (see Appendix \ref{sec:FFappendix}).
Here ${\cal N}_k$ is defined as
\begin{equation}
\mathcal{N}_{k}(\{ \theta_{i}\} )=\sqrt{\prod_{r=1}^{k}\prod_{s=r+1}^{k}S(\theta_{r}-\theta_{s})},\label{eq:Ncoef}
\end{equation}
while $\rho_{k}$ is given in terms of a Gaudin determinant:
\begin{equation}
\rho_{k}(\{ \theta_{i}\} _{i=1})=\det R_{pq},\quad R_{pq}=\frac{\partial Q_{q}}{\partial\theta_{p}}.
\end{equation}
The normalization coefficients of \eqref{eq:Ncoef} make all matrix elements (\ref{eq:PozsTak}) real (apart from the $e^{iMx}$ factors).

We still need to treat the case of when rapidities in matrix elements happen to coincide.  In such a case, the infinite volume form factors have singularities that require regulation.
This requires then significantly more complicated formulae than eq.~\ref{eq:PozsTak}. 
Coinciding rapidities happen in two different circumstances. In the first the bra and ket states are the same state  - so called diagonal form
factor). 
The second possibility is that the two different states have an odd number of particles $2n+1$ and $2m+1$, respectively, with quantization
numbers 
$$
\{ I_{j}\} _{j=1}^{n}\cup\{ 0\} \cup\{ -I_{j}\} _{j=1}^{n}, ~{\rm and}~\{ \tilde{I}_{j}\} _{j=1}^{m}\cup\{ 0\} \cup\{ -\tilde{I}_{j}\} _{j=1}^{m}.
$$
In the latter case, a particle of exactly zero momentum is present
in both states. There is, however, a simple trick \cite{BajnokWu} to circumvent the
complications in numerical computations by exploiting the factorization
property of exponential operators. Introducing an auxiliary particle with rapidity $\theta_\Lambda\rightarrow\infty$, 
\be
\frac{\langle \theta_{\Lambda},\{ \theta_{i}\} _{i=1}^{k}|\mathcal{O}|\{ \vartheta_{j}\} _{j=1}^{l}\rangle }{\langle \mathcal{O}\rangle } \rightarrow \frac{\langle \theta_{\Lambda}|\mathcal{O}|\text{vac}\rangle \langle \{ \theta_{i}\} _{i=1}^{k}|\mathcal{O}|\{ \vartheta_{j}\} _{j=1}^{l}\rangle }{\langle \mathcal{O}\rangle ^{2}}
\ee
and 
\be
\rho_{k+1}(\theta_{\Lambda},\{ \theta_{i}\} ) \rightarrow \rho_{1}(\theta_{\Lambda})\rho_{k}(\{ \theta_{i}\} ).
\ee
This permits us to isolate the matrix element of interest even if it contains coinciding rapidities. With all $\{ I_{i}\} $
and $\{ \tilde{I}_{j}\} $ finite,
\be
\lim_{\Lambda\rightarrow\infty}\langle \{ I_{i}\} _{i=1}^{k}\cup\{ \Lambda\} |\mathcal{O}(0,0)|\{ \tilde{I}_{j}\} _{j=1}^{l}\rangle _{L}=\frac{\langle\Lambda|\mathcal{O}|\mathrm{vac}\rangle_L}{\langle\mathcal{O}\rangle_L}\langle \{ I_{i}\} _{i=1}^{k}|\mathcal{O}(0,0)|\{ \tilde{I}_{j}\} _{j=1}^{l}\rangle _{L}\label{diagFFlimit}
\ee
where $\Lambda$ is the momentum quantization number of the auxiliary particle. Choosing instead a large but finite $\Lambda$, the equality is not
exact, but the error can be made arbitrarily small by increasing $\Lambda$.
In practice we need to solve the set of BYE \eqref{eq:BYE} twice:
once for $k+1$ particles involving the extra quantization number
$\Lambda$, and once for the quantization number set $\{ \tilde{I}_{j}\} _{j=1}^{l}$.
Thus we obtain rapidities of the form $\theta_{i}=\vartheta_{i}+\epsilon_{i}$,
where $\epsilon_{i}$ are small corrections depending on all the $I_{i}$
and $\Lambda$. These are then put back into the left hand side of eq.~\ref{diagFFlimit}, yielding the sought for matrix element.

\subsection{Sanity checks}
The setup of the TSM in the basis of the ShG model itself allows us to make some non-trivial consistency checks. The first such check is offered by an infinitesimal
change in the coupling constant $b_{1}=b_{0}+\delta b$.

\subsubsection{Comparison with Form Factor Perturbation Theory} 
Denote $\mu_{0} =\mu(b)$  and $\mu_{1}=\mu(b+\delta b)$.
Let us calculate the first order correction of the mass of the particle in form factor perturbation theory \cite{Delfino:1996xp}. It is convenient to keep the system in a finite
(very large) volume $R$.  Here the volume factors from spatial integration and the Hilbert state norms cancel.  The first
perturbative correction to the mass is then:
\begin{equation}
\delta M_{ShG} =\langle M_{ShG}|H_{1}|M_{ShG}\rangle -\langle 0|H_{1}|0\rangle \label{eq:dm}\,\,\,,
\end{equation}
where $|M_{ShG}\rangle $ denotes the one-particle state of zero momentum and $|0\rangle $ is the vacuum. The diagonal
form factor contains a disconnected term which be cancelled by a similar term arising from the VEV of eq.~\ref{eq:dm}:
\begin{eqnarray}
\langle M_{ShG} |H_{1} | M_{ShG} \rangle &= & \langle 0| H_{1}|0 \rangle \cr\cr
&& \hskip -1in + \frac{2}{M_{ShG}}(-\mu(b)\langle e^{b\varphi}\rangle F_{2}^{\exp(b\varphi)}(i\pi,0)+\mu(b+\delta b)\langle e^{(b+\delta b)\varphi}\rangle F_{2}^{\exp((b+\delta b)\varphi)}(i\pi,0)).\nonumber
\end{eqnarray}
Inserting the explicit form of the two-particle form factor calculated using eq.~\ref{F_n}, we arrive at the relation:
\begin{eqnarray}
M_{ShG}(b)[M_{ShG}(b+\delta b)-M_{ShG}(b)] & = & 8(-\mu(b)\langle e^{b\varphi}\rangle \sin(\pi B)\\
&+& \mu(b+\delta b)\langle e^{(b+\delta b)\varphi}\rangle \frac{\sin^{2}((1+\frac{\delta b}{b})\pi B)}{\sin(\pi B)})
\nonumber \,\,\,,
\end{eqnarray}
where the vacuum expectation values are given by the FLZZ formula \eqref{VEVexp}.
Therefore, for an infinitesimal $\delta b$, we can write down a differential equation for the mass shift equation
\begin{align}
\frac{\partial[M_{ShG}(b)^{2}]}{\partial b} &= 16\mu(b)\langle e^{b\varphi}\rangle (\frac{\delta\langle e^{(b+\delta b)\varphi}\rangle }{\langle e^{b\varphi}\rangle }\sin(\pi B)+\frac{2\pi B}{b}\cos(\pi B)+\frac{\partial(\log\mu)}{\partial b}\sin(\pi B))\label{eq:diffeq}
\end{align}
where
\begin{eqnarray}
\frac{\delta\langle e^{(b+\delta b)\varphi}\rangle }{\langle e^{b\varphi}\rangle } &=& -2b(1-B)\ln(\frac{-\mu(b)\pi\Gamma(1+b^{2})}{\Gamma(-b^{2})})\label{eq:dVEV}\cr\cr
 && +\intop_{0}^{\infty}dt\bigg[-b\frac{\sinh(4b^{2}t)}{\sinh(b^{2}t)\sinh t\cosh((1+b^{2})t)}+\frac{4b}{t}e^{-2t}\bigg].
\end{eqnarray}
The physical mass $M_{ShG}$, given by the mass-formula (\ref{ShGmassformula}), numerically agrees with the result of integrating equation \eqref{eq:diffeq}. In the
special case $\lim_{b\rightarrow0}\mu\propto b^{-2}$, we also have to explicitly specify $M_{ShG}(b=0)$. The integral appearing
in the RHS of \eqref{eq:dVEV} can be evaluated explicitly in terms
of a digamma function $\Psi(z)$:
\begin{eqnarray}
\intop_{0}^{\infty}dt[-b\frac{\sinh(4b^{2}t)}{\sinh(b^{2}t)\sinh t\cosh((1+b^{2})t)}+\frac{4b}{t}e^{-2t}] &=& -2b\bigg\{\frac{1}{B}-\Psi(1-b^{2})-\Psi(1+b^{2})\cr\cr
&& \hskip -2in -\frac{1-B}{2}\big(\Psi(\frac{1}{2}-\frac{B}{2})-\Psi(\frac{B}{2})+\Psi(\frac{B-1}{2})-\Psi(1-\frac{B}{2})\big) \bigg\},
\end{eqnarray}
A somewhat tedious but straightforward calculation shows that substituting
eq.~\ref{eq:dVEV} into the RHS of eq.~\ref{eq:diffeq} indeed agrees
with the derivative of the exact mass formula (eq.~\ref{ShGmassformula}), once squared.

\subsubsection{Massive Boson Limit} 
In the second check, we fix $\mu(b_{i})=\frac{m^{2}}{2b_{i}^{2}}$ and
take the limit $b_{0}\rightarrow0$.  Here we want to show that we then obtain the Hamiltonian in the massive boson basis. In particular in this limit
all matrix elements of the perturbation vertex operators need to approach the limit:
\begin{eqnarray} \langle \{ I_{i}\} _{i=1}^{k}|:e^{a\varphi}(0,0):|\{ \tilde{I}_{j}\} _{j=1}^{l}\rangle _{L} \rightarrow &&\cr\cr
&& \hskip -1.5in \prod_{k}\frac{1}{\sqrt{n_{k}!m_{k}!}}\bigg\{ \sum_{n_{1k}=0}^{\min(n_{k},m_{k})}n_{1k}!\binom{n_{k}}{n_{1k}}\binom{m_{k}}{n_{1k}}(\frac{a}{\sqrt{2L\omega_{k}}})^{n_{k}+m_{k}-2n_{1k}}\bigg\} \label{eq:massivelimit}
\end{eqnarray}
with $a=\pm b_{1}$ and the quantities on the LHS are understood with
respect to theory $b_{0}$. In formula eq.~\ref{eq:massivelimit} we
denoted the number of particles with quantum number $k$ as $n_{k}$
in the bra vector and $m_{k}$ in the ket vector.  Furthermore,
matrix elements of vertex operators with $a=\pm b_{0}$ need to approach
those of the operator $-\frac{m^{2}}{2}:\varphi^{2}:$.

These relations are highly nontrivial, especially for diagonal matrix elements, but we
have confirmed numerically their validity on a large number of matrix elements.
 Note that the $b_{0}\rightarrow0$ limit of the form factors is numerically unstable and so the precision often
needs to be increased from the usual machine (double) precision.

\subsection{Implementation and results} \label{subsecFFimpres}

\begin{figure}[t]
\begin{subfigure}[b]{0.475\textwidth}
\centering
\includegraphics[draft=false,width=\textwidth]{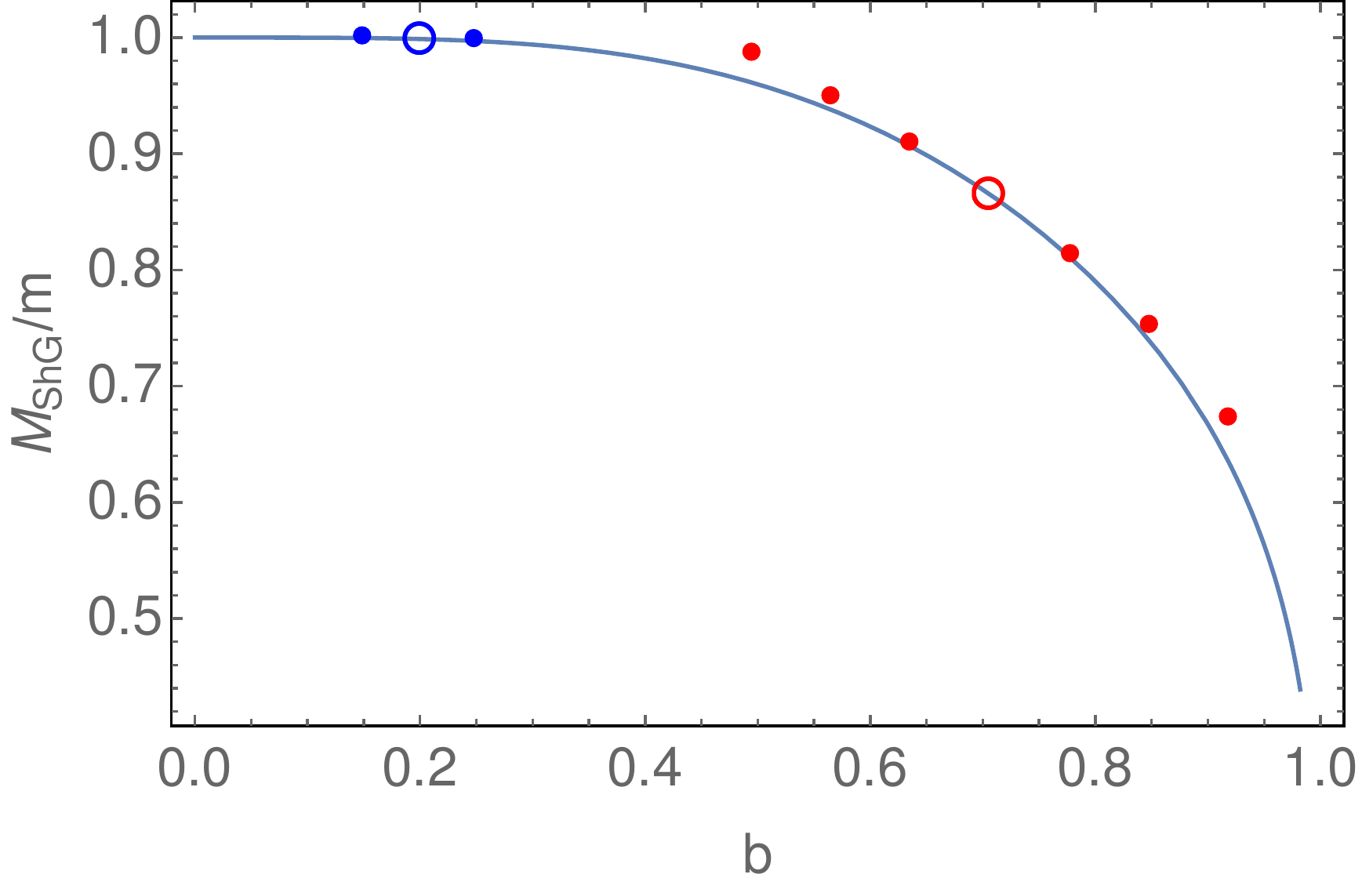}
\caption{Blue: $b_0=0.2$, Red: $b_0=\frac{1}{\sqrt{2}}$}
\end{subfigure}
\hfill
\begin{subfigure}[b]{0.475\textwidth}
\centering
\includegraphics[draft=false,width=\textwidth]{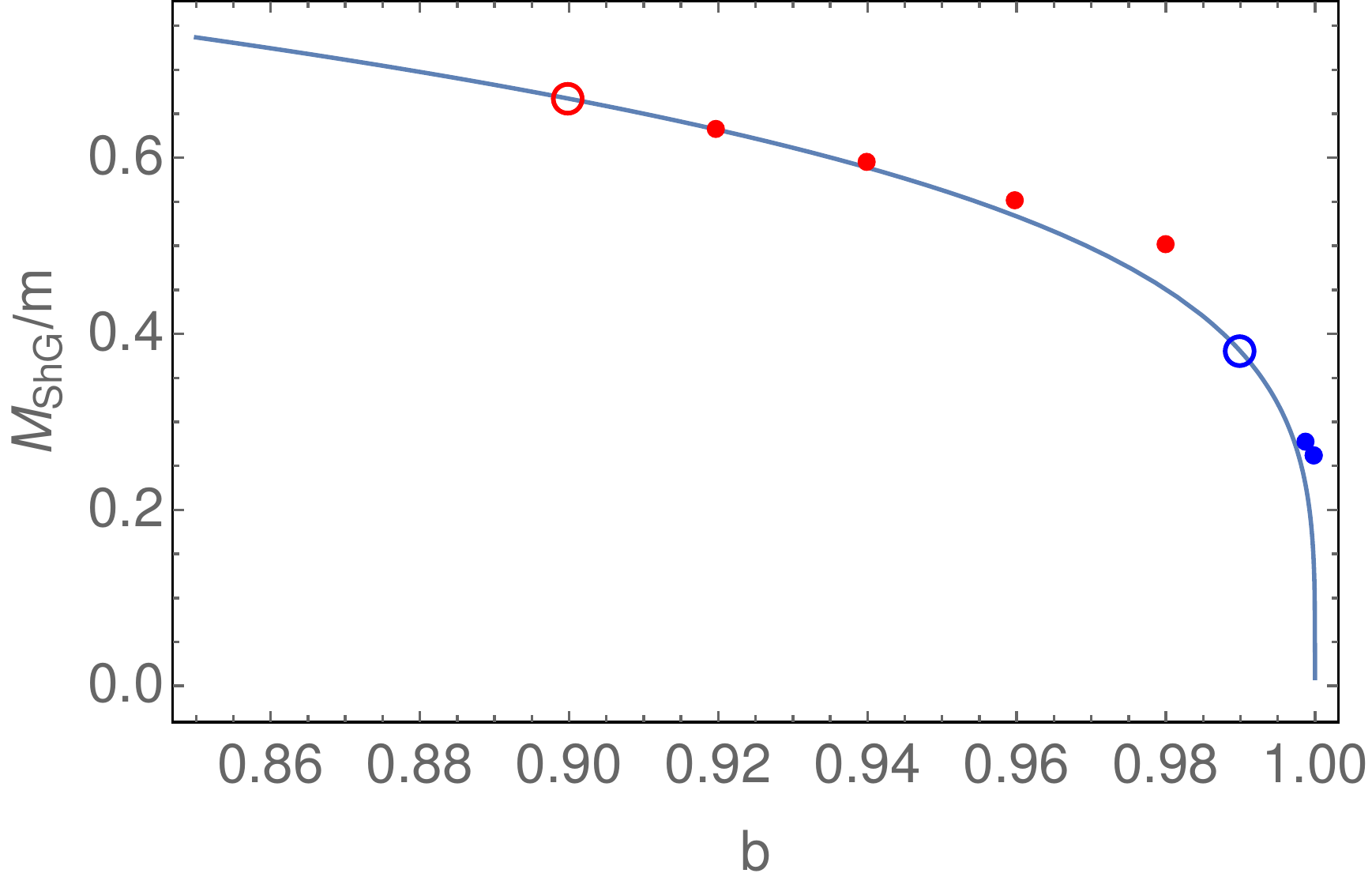}
\caption{$b_0=0.9$ (red) and $b_0=0.99$ (blue)}
\end{subfigure}
\caption{The mass-coupling relation as measured using $H_0=H_{b_0}^{ShG}$ at four different values of $b_0$ (two in the left panel and two in the right).  The mass of the unperturbed theory at $b_0$ is plotted as a large
unfilled circle while the masses derived at neighbouring values of $b_1$ are shown as dots of the same color.
}
\label{FigFFTCSA}
\end{figure}
Due to the high numerical precision needed, especially in taking diagonal
limits, we opted for an implementation in Mathematica. The truncated
basis was selected with two cutoffs, a momentum cutoff $k_{\text{max}}$
and a limit on the number of particles $N_{\text{max}}.$
Fig.~\ref{FigFFTCSA} summarizes the numerical results obtained starting
from theories with four different values of $b_0$.

To facilitate comparison of different bases, we introduce the following double cutoff. We limit the maximum number of particles $N_\mathrm{max}$ as well as the sum of Bethe-Yang quantization numbers of each sign: $\sum_{j: I_j>0} I_j <= k_\mathrm{max}$, .
Let us remark that enlarging the basis size through the increase of $k_\mathrm{max}$ only is computationally easier than increasing $N_\mathrm{max}$. This is because the latter results in having to evaluate much more complicated expressions for the matrix elements. However, from numerical tests we deduce that the majority of cutoff dependence is found when changing $N_\mathrm{max}$ and $k_\mathrm{max}$ simultaneously. Therefore in the following we fix $N_\mathrm{max}=k_\mathrm{max}$.

In Fig. \ref{FigFFTCSA} we present sample results for the mass gap, starting from four different basis theories (denoted by unfilled circles). Computations were done with a raw cutoff $N_\mathrm{max} = 6$. This resulted in Hilbert spaces of dimension $203$ in the even particle number sector and $124$ in the odd. Numerics were performed at $M_\mathrm{ShG} R = 6$ with $M_\mathrm{ShG}$ being the physical mass of
the unperturbed (basis) theory.

\begin{figure}[b]
\begin{subfigure}[b]{0.475\textwidth}
\centering
\includegraphics[draft=false,width=\textwidth]{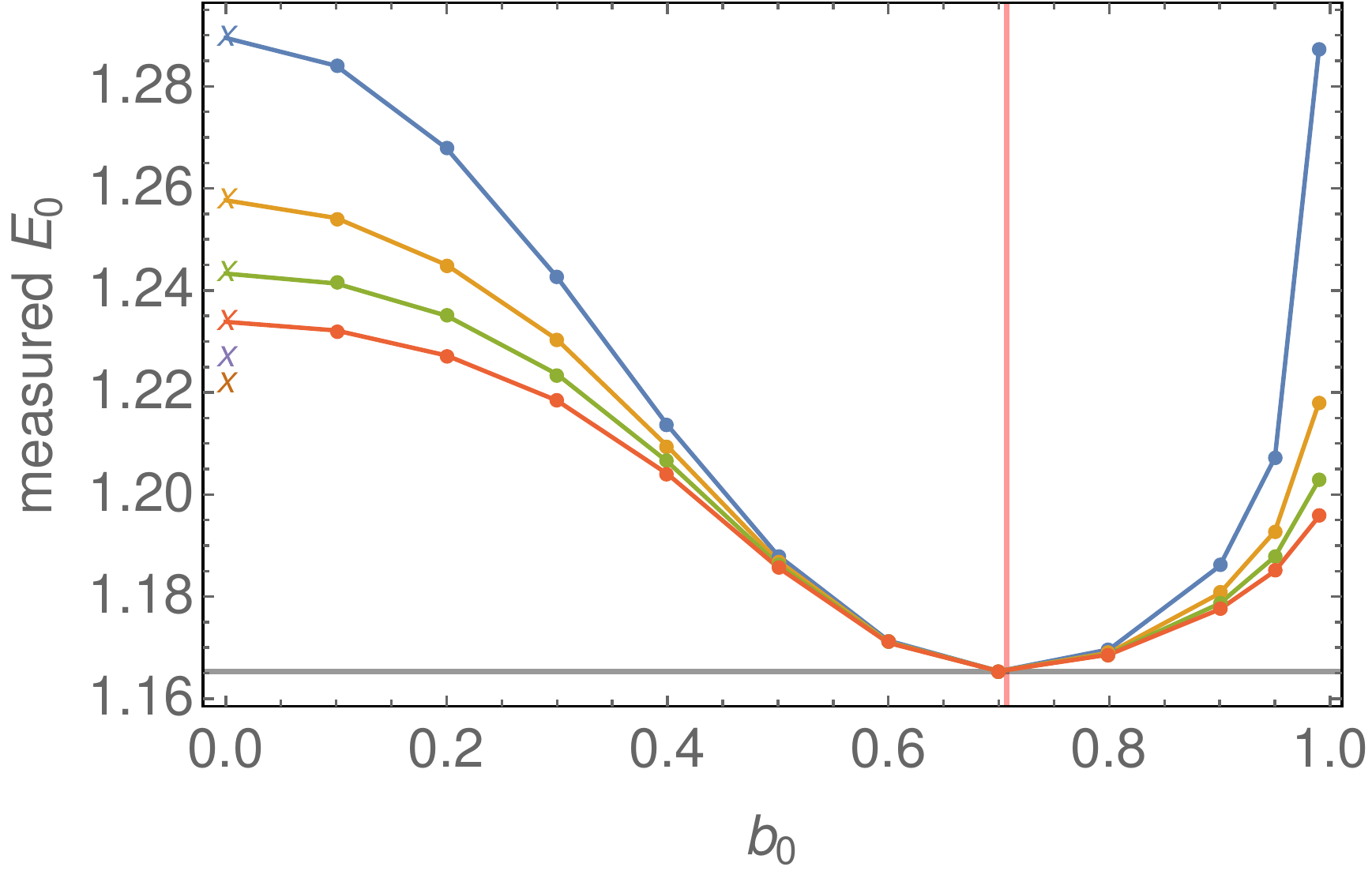}
\caption{$b_1=\frac{1}{\sqrt{2}}$. $b_0=b$ is shown with a red line. Theoretical value is shown with a horizontal line. }
\end{subfigure}
\hfill
\begin{subfigure}[b]{0.475\textwidth}
\centering
\includegraphics[draft=false,width=\textwidth]{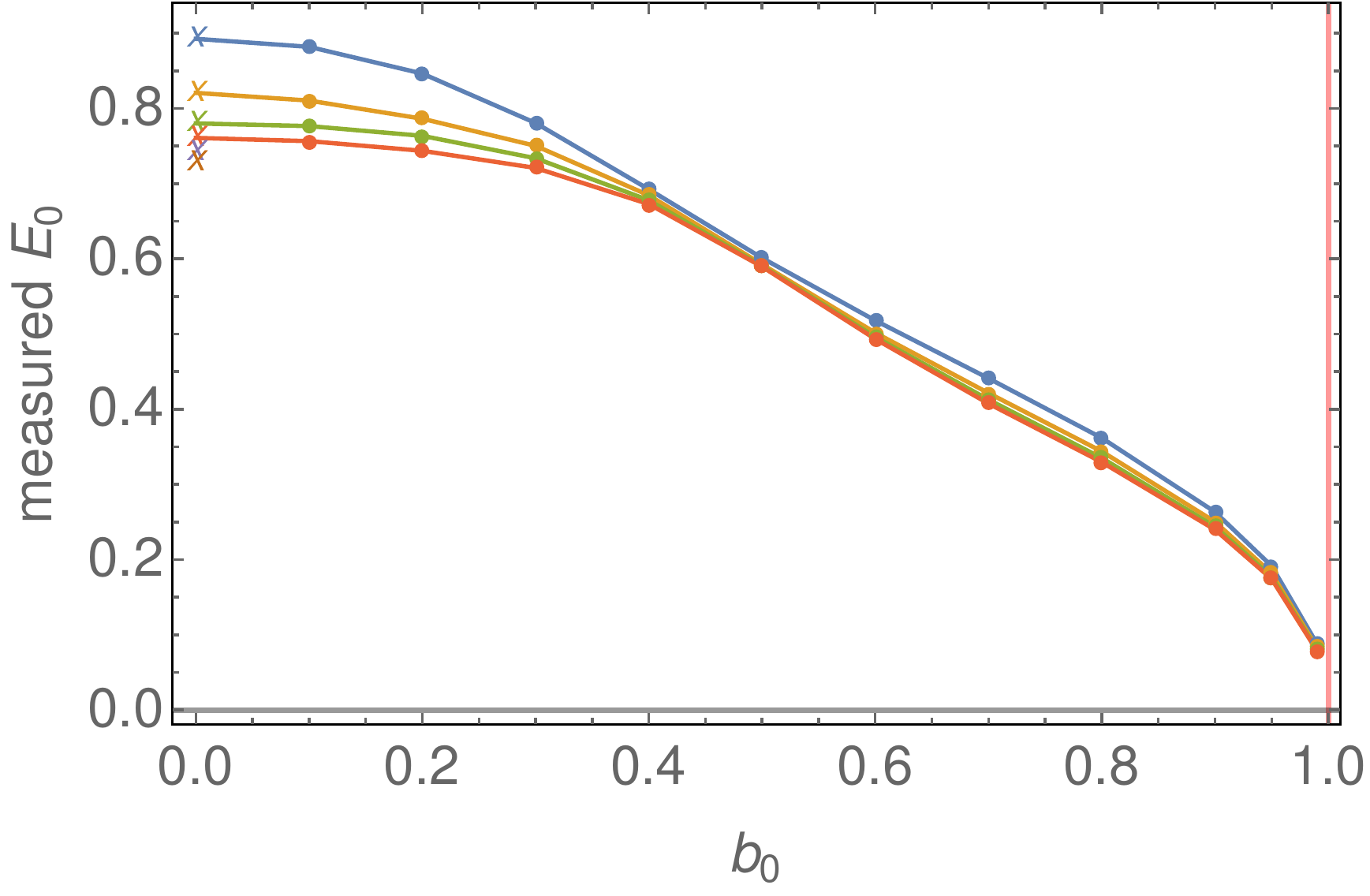}
\caption{$b_1=1$. Theoretical value is zero.\\ \ \\ \ }
\end{subfigure}
\caption{Cutoff dependence of ground state energy at coupling $b_1$, volume $R=6$, starting from different $b_0$ couplings.
Blue, orange, green, red, purple and brown symbols correspond to cutoffs $N_{\mathrm{max}}=2,4,6,8,10,12$,  respectively. Results from the free massive boson basis ($b_0=0$) are distinguished with x-marks.  $\mu_0=\frac{1}{2b_0^2}$, $\mu_1=\frac{1}{2b_1^2}$ }
\label{FigFFTCSA1}
\end{figure}

In Fig. \ref{FigFFTCSA1} we follow the precision of results as the TSM basis is changed from the free massive boson towards the self-dual point. Here we show explicitly the cutoff dependence of the ground state energy at $R=6$. The $b_0=0$ points result from using the free massive Hamiltonian \eqref{eq:Massivehamiltonian}. The largest cutoff considered for the form factor approach is $N_\mathrm{max}=8$, corresponding to an $1171$ dimensional (even sector) truncated Hamiltonian. In the free massive basis, we also show the point corresponding to $N_\mathrm{max}=10,12$, with basis sizes of $5830$ and $25488$, respectively. Not surprisingly, the result corresponding to any fixed cutoff improves as we start closer to $b_1$. However, it is striking that as we approximate $b_0$ to $b_1$, the power-law exponent of the cutoff dependence apparently worsens. As we have seen previously, when $b_1$ is infinitesimally close to $b_0$, the first order term of the form factor perturbation series dominates. However, as we move $b_0$ slightly away, corrections from all energy scales play a role.

\subsection{What Have We Learned: Massless Regime for $b>1$?}\label{sectionmasslessregime}

While one can see from the right panel of Fig.~\ref{FigFFTCSA1} that it is difficult to obtain accurate results (again because of slow convergence in the cutoff) for $b_1$ close to the self-dual point (even when $b_0$ itself is close to $b=1$), we can use the properties of this massive interacting basis to draw conclusions on the nature of the theory beyond $b>1$.
We begin, as before, with a starting point $b_0<1$ and aim to describe the theory for a fixed $b_1>1$.  We are completely free to choose $b_0<1$ and we so chose it such that it satisfies 
\be
\frac{Q(b_0)}{2} = \frac{1}{2} \left(b_0+\frac{1}{b_0}\right)\,=\, b_1.
\ee
However for this particular choice of $b_1$, the VEV of the vertex operator $e^{b\varphi}$ (relative to the $b_0$ theory) is exactly zero!  This in turn means that all matrix elements of $e^{b\varphi}$ vanish.  Hence according to 
eq.~\ref{newwww}, the perturbation consists in simply subtracting the original vertex operators from the original Hamiltonian, which (formally at least) leads to a massless Gaussian model.  This argument suggests then that for $b>1$, the ShG model, like its SG counterpart is trivial.  

Because $\frac{Q(b_0)}{2}\geq1,\:\forall b_0$, this argument only works when the target theory satisfies $b_1\geq1$.   We note that perturbing a theory with a vertex operator with $b>\frac{Q}{2}$ is meaningless as the vacuum expectation value becomes negative, thus (formally) making the spectrum unbounded.  A plausible explanation is that the VEV $\langle e^{a\varphi} \rangle$ in the Lagrangian formulation is only identical to the FFLZ formula in the domain $-\frac{Q}{2} \leq a \leq \frac{Q}{2} $ and vanishes identically outside this interval.   We note that this argument then requires us to use the FFLZ formula at the boundary of its validity.
These arguments are certainly heuristic but give nevertheless a hint on the phase of the ShG model in the strong coupling regime $b > 1$.

\section{Conclusions and Future Directions} \label{Conclusionssection}
In this paper we have studied in detail the ShG model in finite volume as a function of its coupling constant $b$.  This analysis has explored several important conceptual and numerical features of this model. At the conceptual level, we have seen that the model admits two different formulations, given in terms of (i) a Lagrangian and (ii) a $S$-matrix. 
The $S$-matrix formulation is inherently an infrared theory, giving rise in particular to a thermodynamic Bethe ansatz equation describing in finite volume the energy of the ground and excited states. Moreover, the $S$-matrix formulation is ecumenical as to the actual value $M_{ShG}$ of the mass of the physical particle.  The Lagrangian formulation, on the other hand, gives us the opportunity to study the theory from several directions.  Here the theory can be conceived as a perturbation of a massive free boson or as a perturbation of a conformal field theory, either Gaussian or Liouvillian.  While the $S$-matrix formulation is manifestly invariant under the weak/strong duality transformation, $b \rightarrow 1/b$, the Lagrangian formulation has no sign of such a symmetry.

Herein we have used truncated spectrum methods to study in finite volume the ShG model.  This proved to be particularly challenging. Indeed, while at small values of the coupling constant $b$, dynamical quantities of the model  -- such as physical mass, vacuum expectation values of vertex operators, finite volume energies of the ground state and excited states, the two-body $S$-matrix -- could be accurately determined and were found to coincide with their theoretical predictions, this ceased to be true upon approach to $b=1$.  There we observed that the values of these quantities started to deviate noticeably from their predicted exact results.  Indeed, in the vicinity of the self-dual point $b=1$, the TSM data showed a marked sensitivity to the cut-off $N_c$ related to the number of states employed by the TSM.  This sensitivity is of a different nature from other quantum field theories studied so far by means of TSM, a consequence of the unbounded exponential nature of its interaction.

Understanding and attempting to ameliorate these difficulties have had a series of positive by-products.  In the first, we were able to come to a detailed understanding of the small volume region of the theory through exploring the quantum mechanics of the zero mode of the field. This analysis led us, in particular, to derive an effective potential which took into account the effect of the oscillator modes.  This permitted a more precise measurement of the infrared parameters from the TSM through combining UV numerics with the small-volume expansion of the TBA.

In a second happy after effect, we were able to extend the usual RG scheme for improving TSM numerical data.  Normally these improvements are perturbative in the coupling of the theory, here  $\mu_{ShG}$.  This series proved however to be pathological in $\mu_{ShG}$ in the sense that its coefficients diverge absent a UV cutoff {\it and} the coefficients at order $n$ increase more rapidly than $n!$, making the series not even Borel resummable.  We were able to overcome these pathologies using generalized resummation techniques for asymptotic series -- here dubbed a supra-Borel resummation.  In particular we were able to show that summation of this series led to a finite result upon taking the UV cutoff to infinity and that we could partially explain the observed power law dependence in $N_c$ observed in our TSM data.  This is the first time that the leading corrections at all orders in this perturbative RG treatment have been explicitly summed.  This perhaps might be useful for studies of other theories where partial resummations have been considered \cite{PhysRevD.96.065024,Elias-Miro2017}.

It is important to stress that all results obtained to this point indicate that the Lagrangian of the theory provide a faithful definition of the theory for $b < 1$, even though the Lagrangian does not share the self-duality of the $S$-matrix.  A hint about the ShG model for $b > 1$ may however come from the third spinoff explored in this paper: treating the ShG at a given coupling $b_1$ as perturbation of a ShG model at another coupling, $b_0$. As explained in Section \ref{sectionmasslessregime}, for $b > 1$ there is {\em always} the possibility of identifying a point $b_0$ for which the perturbation is given in terms of a vertex operator $V_{a}$ with $a = (b+b^{-1})/2$. Since the VEV of this operator vanishes, it simultaneously ensures the vanishing of all of the matrix elements involving $\cosh(b\phi)$ and therefore leads to the remarkable conclusion that the ShG model in the strong coupling regime $b > 1$ is a free massless theory! 

We have already noted that in the strong coupling regime of the ShG model that all exact and analytic formulas for the physical mass and vacuum expectation values (see eqs.~\ref{ShGmassformula} and \ref{VEVexp}) have a singularity at $b=1$.  If one analytically continues these expressions beyond $b=1$, they give, in general, complex values which make their physical interpretation challenging.  One can take the point of view that for $b>1$ the Lagrangian should be ignored and one should rely only on the $S$-matrix (and its explicit duality) to define the theory.  In this way one uses explicitly results for $b<1$ to define the theory for $b>1$.  However this is a tautological way of defining the duality as it leads to no predictions with regards to the mass formula, the VEVs, and the energy levels.

It is worth stressing that the same conclusions may apply as well to all Toda field theories which share with the ShG model all of its basic features, namely an apparent duality of their $S$-matrix which is absent in their Lagrangian formulation.  Using the exact formulae reported in \cite{Ahn:1999dz} of the Toda field theories for the physical mass, vacuum expectation values, reflection amplitudes of the underlying (generalized) Liouville field theory, one can repeat indeed the analysis done in this paper, applying in particular the form factor basis to argue that the Toda field theories for $b > 1$ are massless models.   Indeed, establishing whether the ShG model and all Toda field theories are not in fact self-dual models as their $S$-matrix suggests, but on the contrary are massless, as conjectured in \cite{SKLYANIN1989719}, is one of the most interesting and important open problems for the future.

\subsection*{Acknowledgments}
We are grateful to Zolt\'an Bajnok, Sergei Lukyanov, G\'abor Tak\'acs, Davide Squizzato, J\"org Teschner, Eric Vernier, Igor Tupitsyn, Nikolai Prokof'ev, and Alexander Zamolodchikov for useful discussions. We also thank Zolt\'an Bajnok for comments on the manuscript. ML thanks SISSA for the kind hospitality during the early stage of the work. GM is grateful to the Simons Center for Geometry and Physics in Stony Brook and Brookhaven National Laboratories for the nice hospitality and partial support during part of this work. We also thank the International Institute of Physics in Natal for hospitality and partial support during the workshop ``Emergent Hydrodynamics in low dimensional quantum systems" where part of this work was carried out.
RMK is supported by the Office of Basic Energy Sciences, Material Sciences and Engineering Division, U.S. Department of Energy (DOE) under Contract No. de-sc0012704.  ML is supported by the National Research, Development and Innovation Office (NKFIH) Research Grant K116505.


\vspace{8mm}
\appendix

\section{Relating the Couplings Between the Gaussian and Massive Formulations of the ShG Model} \label{AppRelatingSchemes}

In this appendix we derive the relation shown in eq.~(\ref{lambdamm}),
\begin{equation}
\mu_{ShG}=\frac{gm^{2+2 b^{2}}}{2b^{2}}\left(\frac{e^{\gamma_{E}}}{2}\right)^{2 b^{2}},
\end{equation}
connecting the coupling $\mu_{ShG}$ in the Gaussian formulation of the ShG model with the mass $m$ that appears in using Feynman perturbation
theory in $b$.  We do so by recasting the finite volume Hamiltonian of the ShG theory where normal ordering is done with respect to a massless basis to a Hamiltonian 
where the normal ordering is done with respect to a massive oscillator basis.
We use the form of the massive basis form \eqref{eq:Massivehamiltonian} in Section \ref{subsecFFimpres} where we compare it to the limit of form factors computed in a massive interacting basis.

\subsection{Massless $c=1$ Boson} \label{subsecMasslessBoson}

We first recall how the Hamiltonian of the ShG model on the cylinder as a perturbed Gaussian theory is derived from the theory defined on the plane.  This will inform how we derive the ShG model on the cylinder using massive oscillators and ultimate how we arrive at eq.~(\ref{lambdamm}).

Consider the $c=1$ boson on the plane. The stress-energy tensor is
given by
\begin{equation}
T_{\mu\nu}\left(z,\bar{z}\right)=g\left(:\partial_{\mu}\varphi\partial_{\nu}\varphi-\frac{1}{2}\eta_{\mu\nu}\partial_{\rho}\varphi\partial^{\rho}\varphi:\right).
\end{equation}
In the main body of the paper we fixed $g=\frac{1}{8\pi}$.  In the following we keep this normalization parameter explicit. 
The generator of dilatations is defined
in terms of the stress-energy tensor as
\begin{equation}
D^{\left(0\right)}=\frac{1}{2\pi i}\left[\intop_{C}dz\:zT\left(z\right)+\intop_{C}d\bar{z}\:\bar{z}\bar{T}\left(\bar{z}\right)\right]=L_{0}+\bar{L}_{0},\label{eq:dilatop}
\end{equation}
where the contour $C$ may be chosen as the positively directed unit
circle around the origin.

If we now map the theory to a cylinder of circumference $R$,
\begin{equation}
z=e^{\frac{2\pi}{R}w},\quad w\equiv\tau+ix,\label{eq:planetocyl}
\end{equation}
where $\tau$ is (Euclidean) time and $x$ is the space coordinate, the dilaton operator
becomes the Hamiltonian on the cylinder:
\begin{equation}
H_{\mathrm{cyl}}^{\left(0\right)}=\frac{2\pi}{R}\left(L^{\rm cyl}_{0}+\bar{L}^{\rm cyl}_{0}-\frac{1}{12}\right).\label{eq:H0cyl}
\end{equation}
If we use the expansion of the field in terms of modes satisfying canonical commutation relations
\begin{equation}
\varphi\left(x,\tau\right)=\varphi_{0}+\frac{\tau\Pi_0}{gR} + \frac{1}{\sqrt{4\pi g}}\sum_{n\neq0}\frac{1}{\sqrt{\left|n\right|}}\left(a_{n}e^{ik_{n}x}+a_{n}^{\dagger}e^{-ik_{n}x}\right),\label{modeexp}
\end{equation}
where $\left(k_{n}=2\pi nR^{-1}\right)$, the Hamiltonian eq.~(\ref{eq:H0cyl}) can be written as
\begin{equation}
H_{\mathrm{cyl}}^{\left(0\right)}=\left(\frac{\Pi_{0}^{2}}{2gR}-\frac{\pi}{6R}\right)+\sum_{n\neq0}\frac{2\pi\left|n\right|}{R}a_{n}^{\dagger}a_{n}.\label{eq:H0cyl-1}
\end{equation}
While the $-\frac{\pi}{6R}$ term can be thought of as arising from the anomalous properties of the stress-energy tensor under conformal transformation, it can also be seen as a result of bringing the oscillator modes into normal order.
This will be important going forward.

We end this subsection by noting that under the
mapping eq.~(\ref{eq:planetocyl}), normal ordered vertex operators on the plane and on the cylinder are related by
\begin{equation}
\mathcal{V}_{b}^{\left(cyl\right)}\left(w,\bar{w}\right)=\left(\frac{R}{2\pi}\right)^{\frac{b^{2}}{4\pi g}}\mathcal{V}_{b}^{\left(pl\right)}\left(z\left(w\right),\bar{z}\left(\bar{w}\right)\right).\label{eq:VertexTransf}
\end{equation}

\subsection{Free Massive Boson as a Perturbed $c=1$ Boson}

We now consider how we can mimic the well understood procedure outlined in Appendix \ref{subsecMasslessBoson} but with a massive boson.  To this end we define the perturbed dilatation on the plane as 
\begin{equation}
D^{\left(m\right)}=L_{0}+\bar{L}_{0}+\lim_{b\rightarrow0}\frac{gm^{2+\frac{b^2}{4\pi g}}}{2b^{2}}\intop_{C}dz\left(\mathcal{V}_{b}^{\left(pl\right)}\left(z,z^{*}\right)+\mathcal{V}_{-b}^{\left(pl\right)}\left(z,z^{*}\right)-2\right).
\end{equation}
By mapping this operator to the cylinder, we obtain the following Hamiltonian
for the free massive boson:
\begin{equation}
H_{\mathrm{cyl}}^{\left(m\right)}=H_{\mathrm{cyl}}^{\left(0\right)}+\frac{gm^{2}}{2}\intop_{0}^{R}:\varphi^{2}\left(x,0\right):dx+\frac{m^{2}R}{4\pi}\ln\left(\frac{mR}{2\pi}\right).\label{eq:Hmcyl}
\end{equation}
Notice the additive constant, a result of applying
eq.~(\ref{eq:VertexTransf}) in the $b\rightarrow 0$ limit. This constant will be important for making contact with energy as determined by the thermodynamic Bethe ansatz (TBA). 

Using the mode expansion of eq.~(\ref{modeexp}), we can write our Hamiltonian as
\begin{eqnarray}
H_{\mathrm{cyl}}^{\left(m\right)} &=& H_{\mathrm{ZM}}^{\left(m\right)}+H_{\mathrm{osc}}^{\left(m\right)};\cr\cr
H_{\mathrm{ZM}}^{\left(m\right)} &=& \left(\frac{1}{2gR}\Pi_{0}^{2}+\frac{gm^{2}R}{2}\varphi_{0}^{2}-\frac{\pi}{6R}+\frac{m^{2}R}{4\pi}\ln\left(\frac{R}{2\pi}\right)\right);\cr\cr
H_{\mathrm{osc}}^{\left(m\right)} &=& \sum_{n\neq0}\left[\frac{2\pi\left|n\right|}{R}a_{n}^{\dagger}a_{n}+\frac{m^{2}R}{8\pi\left|n\right|}\left(2a_{n}^{\dagger}a_{n}+a_{n}^{\dagger}a_{-n}^{\dagger}+a_{n}a_{-n}\right)\right].\label{eq:HMosc}
\end{eqnarray}
We now rewrite this Hamiltonian in term of a massive oscillator basis.

We begin here with the zero mode Hamiltonian.  It is a harmonic oscillator of frequency $m$
and `mass' $gR$.  Introducing massive zero mode operators $a_0$ and $a^\dagger_0$, we can write:
\begin{equation}
\varphi_{0}=\frac{1}{\sqrt{2gmR}}\left(a_{0}+a_{0}^{\dagger}\right);\quad\Pi_{0}=i\sqrt{\frac{gmR}{2}}\left(a_{0}^{\dagger}-a\right).\label{eq:Zmodes}
\end{equation}
In terms of the creation operators the zero mode Hamiltonian takes
the form
\begin{equation}
H_{\mathrm{ZM}}^{\left(m\right)}=ma_{0}^{\dagger}a_{0}+\frac{m}{2}-\frac{\pi}{6R}+\frac{m^{2}R}{4\pi}\ln\left(\frac{Rm}{2\pi}\right).
\end{equation}

We now turn to the oscillator part, $H_{\mathrm{osc}}^{\left(m\right)}$, of the Hamiltonian.
This can be diagonalized by means of a Bogoliubov transformation implemented by the unitary operator
\begin{equation}
U=\exp\left\{ -\sum_{m>0}\chi_{m}\left(a_{m}a_{-m}-a_{m}^{\dagger}a_{-m}^{\dagger}\right)\right\}, \label{eq:Uop}
\end{equation}
where 
\begin{equation}
e^{\chi_{n}}=\left(\frac{\omega_{n}}{\left|k_{n}\right|}\right)^{\frac{1}{2}},\quad\omega_{n}=\sqrt{k_{n}^{2}+m^{2}}.
\end{equation}
This acts on the mode operators $a_{n}$ as
\begin{align}
U^{\dagger}a_{n}U & =\cosh\chi_{\left|n\right|}a_{n}-\sinh\chi_{\left|n\right|}a_{-n}^{\dagger};\nonumber \\
U^{\dagger}a_{-n}^{\dagger}U & =\cosh\chi_{\left|n\right|}a_{-n}^{\dagger}-\sinh\chi_{\left|n\right|}a_{n}.\label{eq:UaU}
\end{align}
Applying the Bogoliubov transformation to the Hamiltonian we obtain
\begin{eqnarray}
H_{\mathrm{cyl}}^{\left(m\right)} &=&\sum_{n=-\infty}^{\infty}\omega_{n}a_{n}^{\dagger}a_{n}+\frac{m}{2} - \frac{\pi}{6R} + \frac{m^2R}{4\pi}\ln \bigg(\frac{Rm}{2\pi}\bigg) + S_2(m,R);\cr\cr
S_{2}\left(m,R\right) &=&\sum_{n\neq0}\frac{\omega_{n}}{2}-\frac{\left|k_{n}\right|}{2}-\frac{m^{2}}{4\left|k_{n}\right|}.
\end{eqnarray}
The factor $S_2(m,R)$ arises from normal ordering the massive oscillator basis after the Bogoliubov transformation is performed.
In Appendix \ref{AppIntRep} we show that it admits an integral representation,
\begin{equation}
S_{2}\left(m,R\right)=\frac{\pi}{6R}-\frac{m}{2}+\frac{m^{2}R}{4\pi}\left(\frac{1}{2}+\ln\frac{2}{m}-\gamma_{E}\right)+\frac{m^{2}R}{4\pi}\ln\frac{2\pi}{R}+E_0^{(m)}(m,R),
\end{equation}
where $E_0^{(m)}(m,R)$ is defined as
\begin{equation}
E_0^{(m)}(m,R)=m\intop_{-\infty}^{\infty}\frac{du}{2\pi}\cosh u\ln\left(1-e^{-mR\cosh u}\right).
\end{equation}
Our final result for the Hamiltonian is
\begin{eqnarray}
H_{\mathrm{cyl}}^{\left(m\right)} &=&\sum_{n=-\infty}^{\infty}\omega_{n}a_{n}^{\dagger}a_{n}+\mathcal{E}_0R + E^{m}_0(m,R);\cr\cr
\mathcal{E}_{0}^{\left(m\right)} &= &\frac{m^{2}}{4\pi}\left(\frac{1}{2}+\ln\frac{2}{mR}-\gamma_{E}\right).\label{eq:freebulk}
\end{eqnarray}
The ground state energy of this Hamiltonian agrees with the energy $E_0^{(m)}(m,R)$ associated with a free massive boson computed in TBA \cite{Mussardo:2020rxh}.
Here we are able to pin down the bulk energy density, $\mathcal{E}_0$, that would be observed in a TSM computation.

\subsection{Expressing the Sinh-Gordon Model in Terms of a Massive Oscillator Basis}

Here we finally provide the derivation of eq.~(\ref{lambdamm}).  Our starting point again is
the perturbed dilatation operator that defines the sinh-Gordon theory on the plane to be:
\begin{equation}
D^{\left(\mathrm{shG}\right)}=L_{0}+\bar{L}_{0}+\mu_{ShG}\intop_{C}dz\left(\mathcal{V}_{b}^{\left(pl\right)}\left(z,z^{*}\right)+\mathcal{V}_{-b}^{\left(pl\right)}\left(z,z^{*}\right)\right).
\end{equation}
Mapped to the cylinder, this gives rise to the Hamiltonian
\begin{equation}
H^{\left(\mathrm{shG}\right)}=H_{\mathrm{cyl}}^{\left(0\right)}+\mu_{ShG}\left(\frac{R}{2\pi}\right)^{\frac{b^{2}}{4\pi g}}\intop_{0}^{R}dx:e^{b\varphi\left(x,0\right)}:+:e^{-b\varphi\left(x,0\right)}:.
\end{equation} The spatial integration can easily be carried out at the cost of imposing
momentum conservation explicitly, symbolized by the presence of $\delta_{P}$. Because we are interested in making contact with the analysis of the sinh-Gordon as a massive perturbed boson, we add
and subtract at the same time an auxiliary mass term
\begin{eqnarray} 
H^{\left(\mathrm{shG}\right)} &=& H_{\mathrm{aux}}^{\left(m\right)} +\mu_{ShG} R\left(\frac{R}{2\pi}\right)^{\frac{b^{2}}{4\pi g}}\delta_{P}\left(:e^{b\varphi\left(x,0\right)}:+:e^{-b\varphi\left(x,0\right)}:\right)-\frac{gm^{2}}{2}R\delta_{P}:\varphi^{2}\left(0,0\right):;\cr\cr
H_{\mathrm{aux}}^{\left(m\right)} &\equiv& H_{\mathrm{cyl}}^{\left(0\right)}+\frac{gm^{2}}{2}R\delta_{P}:\varphi^{2}\left(0,0\right):.\label{ShGmassivebasis}
\end{eqnarray}
$H_{\mathrm{aux}}^{\left(m\right)}$ is the same type of Hamiltonian as $H_{\mathrm{cyl}}^{\left(m\right)}$
of eq.~(\ref{eq:Hmcyl}).  After a Bogoliubov transformation, it can be rewritten as
\begin{equation}
H_{\mathrm{aux}}^{\left(m\right)}=\sum_{n=-\infty}^{\infty}\omega_{n}a_{n}^{\dagger}a_{n}+\mathcal{E}_{0}^{\left(m\right)}R+E_{0}^{\left(m\right)}\left(m,R\right)-\frac{m^{2}R}{4\pi}\ln\left(\frac{R}{2\pi}\right).
\end{equation}

We now perform the same Bogoliubov transformation on the remaining terms of eq.~(\ref{ShGmassivebasis}):
\begin{eqnarray}
U^{\dagger}\delta_{P}:\varphi^{2}\left(0,0\right):U &=& \delta_{P}:\varphi^{2}\left(0,0\right):_{m}+\frac{1}{2gmR}+\frac{1}{2Rg}\sum_{n\neq0}\left(\frac{1}{\omega_{n}}-\frac{1}{\left|k_{n}\right|}\right);\cr\cr
U^{\dagger}\delta_{P}:\cosh(b\varphi(0,0):U &= & e^{\frac{b^{2}}{4gmR}}e^{-\frac{b^{2}}{4gR}\sum_{q\neq0}\left(\frac{1}{\left|k_{q}\right|}-\frac{1}{\omega_{q}}\right)}\delta_{P}:\cosh(b\varphi(0,0)):_{m} . 
\end{eqnarray}
Here the normal ordering $::_{m}$ indicates the normal ordering is being done w.r.t. the massive oscillator modes.
The sum appearing in the above is evaluated in Appendix \ref{subsec:The-sumS1} to be
\begin{align}
S_1\equiv \sum_{n\neq0}\left(\frac{1}{\omega_{n}}-\frac{1}{\left|k_{n}\right|}\right) & =2R\rho\left(mR\right)-\frac{1}{m}+\frac{R}{\pi}\left(\ln\frac{4\pi}{mR}-\gamma_{E}\right),
\end{align}
where $\rho\left(x\right)$ is defined as
\begin{equation}
\rho\left(x\right)=\intop_{-\infty}^{\infty}\frac{du}{2\pi}\frac{1}{\left(e^{x\cosh u}-1\right)}.
\end{equation}
After transformation, our entire Hamiltonian appears as 
\begin{eqnarray}
H^{\left(\mathrm{shG}\right)} &=& \sum_{n=-\infty}^{\infty}\omega_{n}a_{n}^{\dagger}a_{n} + \frac{m^{2}R}{8\pi}+E_{0}^{\left(m\right)}\left(m,R\right)-\frac{m^{2}R}{2}\rho\left(mR\right)\cr\cr
&& \hskip -1.2in + \delta_P \bigg( -\frac{gm^{2}}{2}R\delta_{P}:\varphi^{2}\left(0,0\right):_{m} +\mu_{ShG} R\left(\frac{2}{m}\right)^{\frac{b^{2}}{4\pi g}}e^{\frac{b^{2}}{2g}\rho\left(mR\right)}e^{-\frac{b^{2}\gamma_{E}}{4\pi g}} :\cosh(b\varphi(0,0)):_{m}\bigg).
\label{eq:Massivehamiltonian}
\end{eqnarray}
We are now in a position to establish eq.~(\ref{lambdamm}).   Using the large $R$ limit of $\rho$, $\lim_{x\rightarrow\infty}\rho\left(x\right)=0$, we choose $m$
so that the quadratic term in the second line of eq.~(\ref{eq:Massivehamiltonian}) vanishes, resulting in
\begin{equation}
\mu_{ShG}=\frac{gm^{2+\frac{b^{2}}{4\pi g}}}{2b^{2}}\left(\frac{e^{\gamma_{E}}}{2}\right)^{\frac{b^{2}}{4\pi g}}.\label{eq:lambdamrel}
\end{equation}
We remark that substituting eq.~(\ref{eq:lambdamrel}) into the
exact sinh-Gordon bulk energy formula and expanding in $b$ results in the $O\left(b^{0}\right)$ term precisely coinciding with vacuum energy density of eq.~(\ref{eq:freebulk}).

\section{Calculating $S_1$ and $S_2$} \label{AppIntRep}

In this appendix we provide integral representations for the sums $S_1$ and $S_2$ introduced in Appendix \ref{AppRelatingSchemes}.

\subsection{$S_{1}\left(m,R\right)=\sum_{n\protect\neq0}\frac{1}{\omega_{n}}-\frac{1}{\left|k_{n}\right|}$\label{subsec:The-sumS1}}

To evaluate this sum, let us first introduce a cutoff $\Lambda=\frac{2\pi N}{R}$,
$N\gg1$, and then add and subtract an auxiliary integral term:
\begin{equation}
S_{1}=\lim_{\Lambda\rightarrow\infty}\left\{ \left(\sum_{n=-N}^{N}\frac{1}{\omega_{n}}-\frac{1}{m}-\frac{R}{2\pi}\intop_{-\Lambda}^{\Lambda}\frac{dk}{\sqrt{m^{2}+k^{2}}}\right)+\left(\frac{R}{2\pi}\intop_{-\Lambda}^{\Lambda}\frac{dk}{\sqrt{m^{2}+k^{2}}}-2\frac{R}{2\pi}\sum_{n=1}^{N}\frac{1}{n}\right)\right\} .\label{eq:SumS1}
\end{equation}
In the first term on the r.h.s. we have rewritten the sum so that it includes the $n=0$ term.   Using the definition of the
Euler-Mascheroni constant
\begin{equation}
\gamma_{E}=\lim_{N\rightarrow\infty}\left(\sum_{n=1}^{N}\frac{1}{n}-\ln N\right),
\end{equation}
and the explicit expression of the integral
\begin{equation}
\intop_{-\Lambda}^{\Lambda}\frac{dk}{\sqrt{m^{2}+k^{2}}}=2\ln\left(\frac{\Lambda}{m}+\sqrt{1+\frac{\Lambda^{2}}{m^{2}}}\right)=2\ln N+2\ln\frac{4\pi}{mR}+O\left(\frac{1}{\Lambda^{2}}\right),
\end{equation}
we can immediately evaluate the second term on the r.h.s. of eq.~(\ref{eq:SumS1}):
\begin{equation}
\frac{R}{2\pi}\lim_{\Lambda\rightarrow\infty}\left(\intop_{-\Lambda}^{\Lambda}\frac{dk}{\sqrt{m^{2}+k^{2}}}-2\sum_{n=1}^{N}\frac{1}{n}\right)=\frac{R}{\pi}\left(\ln\frac{4\pi}{mR}-\gamma_{E}\right).
\end{equation}
The first term of eq.~(\ref{eq:SumS1}) can be evaluated in the same way that Matsubara sums are:
\begin{equation}
\sum_{n=-N}^{N}\frac{1}{\omega_{n}}=\frac{R}{2\pi}\intop_{C}\frac{e^{ipR}}{e^{ipR}-1}\frac{1}{\sqrt{m^{2}+p^{2}}}.
\end{equation}
Here the contour $C$ consists of distinct, small circles around
all poles between $p=-\Lambda$ and $p=+\Lambda$ on the real axis.
We deform this disjoint contour to form two straight vertical line
sections running slightly above and below the real axis. The contour in the lower-half plane
is then combined with the first integral in eq.~(\ref{eq:SumS1}). At this point the limit $\Lambda\rightarrow\infty$
can be taken and the two contours tightened around the branch
cuts of the square roots. In this way we obtain the representation,
\begin{align}
\lim_{N\rightarrow\infty}\sum_{n=-N}^{N}\frac{1}{\omega_{n}}-\frac{R}{2\pi}\intop_{-\Lambda}^{\Lambda}\frac{dk}{\sqrt{m^{2}+k^{2}}} & =\frac{2R}{\pi}\intop_{1}^{\infty}\frac{d\tau}{\left(e^{mR\tau}-1\right)\sqrt{\tau^{2}-1}}\nonumber \\
 & =\frac{R}{\pi}\intop_{-\infty}^{\infty}\frac{du}{\left(e^{mR\cosh u}-1\right)}.\label{eq:1ovomegasum}
\end{align}
Collecting terms, we see that
\begin{equation}
S_{1}=\frac{R}{\pi}\intop_{-\infty}^{\infty}\frac{du}{\left(e^{mR\cosh u}-1\right)}-\frac{1}{m}+\frac{R}{\pi}\left(\ln\frac{4\pi}{mR}-\gamma_{E}\right).\label{eq:S1result}
\end{equation}

\subsection{$S_{2}\left(m,R\right)=\sum_{n\protect\neq0}\frac{\omega_{n}}{2}-\frac{\left|k_{n}\right|}{2}-\frac{m^{2}}{4\left|k_{n}\right|}$}

Let us begin with adding and subtracting $\sum_{n\neq0}\frac{m^{2}}{4\omega_{n}}$.
Then we separate the sum into two convergent parts:
\begin{eqnarray}
S_{2}\left(m,R\right) &=& \frac{m^{2}}{4}S_{1}\left(m,R\right)+\tilde{S}_{2}\left(m,R\right) \label{eq:S2tdef}\cr\cr
\tilde{S}_{2} &= & \sum_{n\neq0}s_{2}\left(k_{n}\right)=\sum_{n\neq0}\frac{\omega_{n}}{2}-\frac{\left|k_{n}\right|}{2}-\frac{m^{2}}{4\omega_{n}},
\end{eqnarray}
where $S_{1}$ is given by eq.~(\ref{eq:S1result}). The sum $\tilde{S}_{2}$
can again be rewritten as a Matsubara-type contour integral 
\begin{equation}
\tilde{S}_{2}=\frac{R}{2\pi}\intop_{C}\frac{e^{ipR}}{e^{ipR}-1}s_{2}\left(p\right),
\end{equation}
where now the contour $C$ encloses all poles (in the positive direction)
on the real axis except for the one at the origin. The function $s_{2}\left(p\right)$
has two overlapping pairs of branch cuts all along the imaginary axis
of the complex $p$ plane. The contour $C$ can be deformed in the
first step into two connected contours at each side of the branch
cut, starting and ending at infinity and enclosing all poles lying
on one half of the real axis. In the second step, this pair of contours
is unbent to form two straight vertical lines aligned in the immediate
vicinity of the branch cuts. The contour integrals can then be written
as a sum of integrals over the real line:
\begin{equation}
\tilde{S}_{2}=I_{1}+I_{2}
\end{equation}
\begin{align}
I_{1} & =-\frac{m^{2}R}{2\pi}\left[\intop_{-\infty}^{-1}\frac{\tau}{e^{-mR\tau}-1}+\intop_{-1}^{\infty}\frac{\tau}{1-e^{mR\tau}}\right]-\frac{m}{4}=\frac{\pi}{6R}+\frac{m^{2}R}{4\pi}-\frac{m}{4}\label{eq:I1int},\\
I_{2} & =\frac{m^{2}R}{4\pi}\left[\intop_{1}^{\infty}2\tau+\frac{1-2\tau^{2}}{\sqrt{\tau^{2}-1}}\coth\left(\frac{mR\tau}{2}\right)\right].
\end{align}
The explicit term $-m/4$ arises in eq.~(\ref{eq:I1int}) due to the pole
at the origin. After a change of variables $\tau=\cosh u$, the integral
$I_{2}$ becomes
\begin{equation}
I_{2}=-\frac{m^{2}R}{8\pi}-\frac{m^{2}R}{4\pi}\intop_{0}^{\infty}du\left(4\sinh u\frac{\sinh ue^{-mR\cosh u}}{1-e^{-mR\cosh u}}+\frac{2}{e^{mR\cosh u}-1}\right),
\end{equation}
where an explicit integration $\intop_{0}^{\infty}e^{-2u}du$ was
performed. After an integration by parts in the first term of the
integral and extending the integration range from $-\infty$ to $\infty$,
we get
\begin{equation}
I_{2}=-\frac{m^{2}R}{8\pi}-\frac{m^{2}R}{4\pi}\intop_{-\infty}^{\infty}du\left(-2\cosh u\ln\left(1-e^{-mR\cosh u}\right)+\frac{1}{e^{mR\cosh u}-1}\right).
\end{equation}
Combining all terms, we arrive at
\begin{align}
S_{2}\left(m,R\right) & =-\frac{m}{2}+\frac{m^{2}R}{4\pi}\left(\frac{1}{2}+\ln\frac{2}{m}-\gamma_{E}\right)+\frac{m^{2}R}{4\pi}\ln\frac{2\pi}{R}\nonumber \\
 & +\frac{\pi}{6R}+m\intop_{-\infty}^{\infty}\frac{du}{2\pi}\cosh u\ln\left(1-e^{-mR\cosh u}\right).
\end{align}

\section{Perturbative Expansion of Finite Volume Eigenvalues\label{sec:pertAppendix}}

In this appendix we will consider the perturbative expansion for the
ground state energy.   We begin generally and consider a Hamiltonian of the form
\begin{equation}
H\,=\,H_{0}+\mu V.
\end{equation}
A convenient representation of the expansion of the ground state energy, $E_0$, is given by the Brillouin-Wigner series
\begin{align}
E_{0} & =E_{0}^{0}+\mu V_{00}+\mu^{2}\sum_{m_{1}}\:^{\prime}\frac{V_{0m_{1}}V_{m_{1}0}}{E_{0}-E_{m_{1}}^{0}}+\dots\nonumber \\
 &
   +\mu^{n}\sum_{m_{1}}\:^{\prime}\sum_{m_{2}}\:^{\prime}\dots\sum_{m_{n-1}}\:^{\prime}\frac{V_{0m_{1}}V_{m_{1}m_{2}}\dots V_{m_{n-1}0}}{(E_{0}-E_{m_{1}}^{0})\cdots(E_{0}-E_{m_{n-1}}^{0})}+\dots ,
\end{align}
where the primes indicate that the ground state is to be omitted from the sum.
In the above series, the exact value of the energy, $E_0$, appears on both
sides. The disconnected terms of the usual Rayleigh-Schr\"odinger
series are generated by using the formula iteratively and expanding
in $\mu$.

Now we are interested in the divergences appearing in the sinh-Gordon
model.  It can easily seen that these disconnected terms are less
divergent.  We thus focus on the connected contributions of this series:
\begin{align}
E_{0} & =E_{0}^{0}+\mu_{ShG} V_{00}+\mu_{ShG}^{2}\sum_{m_{1}}\:^{\prime}\frac{V_{0m_{1}}V_{m_{1}0}}{E_{0}^{0}-E_{m_{1}}^{0}}+\dots+\nonumber \\
 & \mu_{ShG}^{n}\sum_{m_{1}}\:^{\prime}\sum_{m_{2}}\:^{\prime}\dots\sum_{m_{n-1}}\:^{\prime}\frac{V_{0m_{1}}V_{m_{1}m_{2}}\dots V_{m_{n-1}0}}{(E_{0}^{0}-E_{m_{1}}^{0})\cdots(E_{0}^{0}-E_{m_{n-1}}^{0})}+\dots,
\end{align}
where $V_{ij}$ denotes the matrix elements of the perturbation w.r.t.
the unperturbed theory.  The n-th order term in the above can be rewritten as
\begin{align}
\delta E_{0}^{(n)}= & \mu_{ShG}^{n}\sum_{m_{1}}\:^{\prime}\sum_{m_{2}}\:^{\prime}\dots\sum_{m_{n-1}}\:^{\prime}\frac{V_{0m_{1}}V_{m_{1}m_{2}}\dots V_{m_{n-1}0}}{(E_{0}^{0}-E_{m_{1}}^{0})\cdots(E_{0}^{0}-E_{m_{n-1}}^{0})}\nonumber \\
= & (-1)^{n-1}\mu_{ShG}^{n}\sum_{m_{1}}\:^{\prime}\sum_{m_{2}}\:^{\prime}\dots\sum_{m_{n-1}}\:^{\prime}\intop_{0}^{\infty}d\tau_{1}\dots d\tau_{n-1}e^{E_{0}^{0}(\tau_{1}+\dots+\tau_{n-1})}\nonumber \\
 & V_{0m_{1}}e^{-E_{m_{1}}^{(0)}\tau_{1}}V_{m_{1}m_{2}}e^{-E_{m_{2}}^{(0)}\tau_{2}}\dots e^{-E_{m_{n-1}}^{(0)}\tau_{n-1}}V_{m_{n-1}0},
\end{align}
where, if we introduce $\tau_{i}=t_{i}-t_{i+1}$, we can rewrite
the $n$th order term as an integral over an Euclidean correlator
of the unperturbed theory:
\begin{eqnarray}
\delta E_{0}^{(n)} &=& (-1)^{n-1}\mu_{ShG}^{n}\langle 0|V(t_{1})\dots V(t_{n}=0)|0\rangle; \cr\cr
V(t) &=& e^{H_{0}t}Ve^{-H_{0}t}.\label{eq:timedep}
\end{eqnarray}
From now on, the time dependence of any operator is understood in
the sense of eq.~(\ref{eq:timedep}).

We now turn to the Hamiltonian of direct concern:
\begin{equation}
H=H_{0}+\mu_{ShG}(\frac{R}{2\pi})^{2b^{2}}\intop_{0}^{R}dx(e^{b{\varphi_0}}:e^{b\tilde{\varphi}(x)}:+e^{-b{\varphi_0}}:e^{-b\tilde{\varphi}(x)}:)-\frac{R}{16\pi}\omega_{\varphi_0}^{2} {\varphi_0}^{2},
\end{equation}
with
\begin{equation}
H_{0}=\frac{4\pi}{R}\Pi_{0}^{2}+\frac{R}{16\pi}\omega_{\varphi_0}^{2} {\varphi_0}^{2}.\label{eq:ZMHO}
\end{equation}
Here we have added and subtracted a term quadratic in the zero mode allowing us to formulate the perturbation theory about a massive zero mode.
This choice allows for some additional explicit steps in the following calculations.

Let us begin by evaluating the second order term in this perturbative
expansion.  The vertex functions of the oscillator part of bosonic field, $\tilde\varphi$, can be written as
\begin{equation}
\langle 0|:e^{\alpha\tilde{\varphi}(\tau,\rho)}::e^{\beta\tilde{\varphi}(0,0)}:|0\rangle =\frac{1}{(1-e^{-\frac{2\pi}{R}(\tau+i\rho)})^{2\alpha\beta}(1-e^{-\frac{2\pi}{R}(\tau-i\rho)})^{2\alpha\beta}}.
\end{equation}
This correlator can be expanded into a binomial series
\begin{eqnarray}
\langle 0|:e^{b\tilde{\varphi}(t_{1},x_{1})}::e^{b\tilde{\varphi}(t_{2},x_{2})}:|0\rangle  
 &=& \sum_{p_{12},q_{12}=0}^{\infty}(-1)^{p_{12}+q_{12}}\binom{-2b^{2}}{p_{12}}\binom{-2b^{2}}{q_{12}}\cr\cr
&& \hskip -1.5in e^{-\frac{2\pi}{R}(p_{12}+q_{12})t_{1}}e^{\frac{2\pi}{R}(p_{12}+q_{12})t_{2}}e^{-\frac{2\pi i}{R}(p_{12}-q_{12})x_{1}}e^{-\frac{2\pi i}{R}(q_{12}-p_{12})x_{2}} ,\label{eq:expform}
\end{eqnarray}
The binomials admit the asymptotic behavior
\begin{equation}
\binom{z}{k}\sim\frac{(-1)^{k}}{\Gamma(-z)k^{z+1}},\quad k\gg1.\label{eq:Gamas}
\end{equation}
The leading singular part of the second-order perturbative correction
has the form
\begin{equation}
\delta E_{0}^{(2)} = -2\mu_{ShG}^{2}\left(\frac{R}{2\pi}\right)^{4b^{2}}\intop_{0}^{\infty}dt\intop_{0}^{R}dx_{1}dx_{2}\langle 0|e^{b{\varphi_0}(t)}e^{b{\varphi_0}(0)}|0\rangle \langle 0|:e^{b\tilde{\varphi}(t,x_{1})}::e^{b\tilde{\varphi}(0,x_{2})}:|0\rangle.
\end{equation}
Note that a factor $2$ arises due to an analogous term with all exponents
negative. The zero mode quantum mechanics \eqref{eq:ZMHO} is that
of a harmonic oscillator, so the correlator on the left is easily
evaluated. We introduce
\begin{equation}
{\varphi_0}={\varphi_0}_{+}+{\varphi_0}_{-},
\end{equation}
such that
\begin{eqnarray}
{\varphi_0}_{+}\,=\,\left(\frac{4\pi}{R\omega_{\varphi_0}}\right)^{\frac{1}{2}}a_{{\varphi_0}};~~ && {\varphi_0}_{-}\,=\,\left(\frac{4\pi}{R\omega_{\varphi_0}}\right)^{\frac{1}{2}}a_{{\varphi_0}}^{\dagger}\cr\cr
&&\hskip -.5in  [a_{{\varphi_0}},a_{{\varphi_0}}^{\dagger}]=1.
\end{eqnarray}
The Euclidean correlator takes the form
\begin{equation}
[{\varphi_0}_{+}(t),{\varphi_0}_{-}(0)]=\frac{4\pi}{R\omega_{\varphi_0}}e^{-\omega_{{\varphi_0}}t}.
\end{equation}
Using eq.~(\ref{eq:expform}), the $x$ integrals can be performed
immediately with the result
\begin{equation}
\delta E_{0}^{(2)}=-2\mu_{ShG}^{2}R^{2}\left(\frac{R}{2\pi}\right)^{4b^{2}}\intop_{0}^{\infty}dt\langle 0|e^{b{\varphi_0}(t)}e^{b{\varphi_0}(0)}|0\rangle \sum_{m=0}^{\infty}\binom{-2b^{2}}{m}^{2}e^{-\frac{4\pi}{R}mt}.\label{eq:2pert}
\end{equation}
For $b<\frac{1}{\sqrt{2}}$, eq.~(\ref{eq:2pert}) evaluates
to a finite value. For larger couplings, we need to introduce a chiral
cutoff $N_c$. In this parameter domain, we can approximate the
sum by an integral, and using the asymptotic eq.~(\ref{eq:Gamas}), we
obtain
\begin{equation}
\delta E_{0}^{(2)}\approx-2\mu_{ShG}^{2}\left(\frac{R}{2\pi}\right)^{4b^{2}}\frac{R^{2}e^{b^{2}\frac{4\pi}{R\omega_{\varphi_0}}}}{\Gamma(2b^{2})^{2}}\intop_{0}^{\infty}dte^{b^{2}\frac{4\pi}{R\omega_{\varphi_0}}e^{-\omega_{{\varphi_0}}t}}\intop_{0}^{N_c}dmm^{4b^{2}-2}e^{-\frac{4\pi}{R}mt}.
\end{equation}
The $\omega_{\varphi_0}^{-1/2}$ terms make an expansion singular around $\omega_{\varphi_0}=0$.
On the other hand, having in mind $b>\frac{1}{\sqrt{2}}$ and a fixed
$\omega_{\varphi_0}>0$, we argue that the corrections due to the double exponential
are subleading for large $N_c$. Let us expand the outer exponential
into a Taylor series. Then we need to perform simple exponential time
integrals, which yield $\frac{1}{m}$-like terms. Rescaling $m\rightarrow N_c\tilde{m}$
and factoring out $N_c$ leaves an explicit $k\omega_{{\varphi_0}} N_c^{-1}$
term in the denominator ($k$ is the order of the Taylor expansion
term). Even though $k$ eventually becomes comparable to $N_c$,
the corresponding term is multiplied by an overall $1/k!$ factor,
which renders it negligible. What remains is an effective $e^{b^{2}\frac{4\pi}{R\omega_{\varphi_0}}}$
constant multiplier
\begin{equation}
\delta E_{0}^{(2)}=-2\mu_{ShG}^{2}\left(\frac{R}{2\pi}\right)^{4b^{2}}\frac{R^{3}e^{b^{2}\frac{8\pi}{R\omega_{\varphi_0}}}}{4\pi\Gamma(2b^{2})^{2}}\frac{N_c^{4b^{2}-2}}{4b^{2}-2}(1+O(N_c^{-1}))+O(N_c^{0}).
\end{equation}
Note that the spurious singularity at $b=\frac{1}{\sqrt{2}}$ is due
to the asymptotic approximation and is unphysical.

Let us turn to the general case. The relevant correlation function
has the form
\begin{eqnarray}
\langle 0|:e^{b\tilde{\varphi}(x_{1},t_{1})}:\cdots:e^{b\tilde{\varphi}(x_{n},0)}:|0\rangle &=& \prod_{i=1}^{n-1}\prod_{j=i+1}^{n}e^{b^{2}[\tilde{\varphi}_{+}(t_{i},x_{i}),\tilde{\varphi}_{-}(t_{j},x_{j})]}\cr\cr
&& \hskip -2in = \sum_{\{ p_{ij}\} =0}^{\infty}\sum_{\{ q_{ij}\} =0}^{\infty}\bigg\{ \prod_{i=1}^{n-1}\prod_{j=i+1}^{n}(-1)^{p_{ij}}\binom{-2b^{2}}{p_{ij}}(-1)^{q_{ij}}\binom{-2b^{2}}{q_{ij}}\bigg\} e^{-\frac{2\pi}{R}\sum_{i}\sum_{j>i}(p_{ij}+q_{ij})t_{i}} \cr\cr
&& \hskip -1in \times  e^{\frac{2\pi}{R}\sum_{j}\sum_{i<j}(p_{ij}+q_{ij})t_{j}}e^{-\frac{2\pi i}{R}\sum_{i}[\sum_{j>i}(p_{ij}-q_{ij})-\sum_{j<i}(p_{ji}-q_{ji})]x_{i}}.
\end{eqnarray}
The spatial integrations yield the relations
\begin{equation}
\sum_{j>i}(p_{ij}-q_{ij})=\sum_{j<i}(p_{ji}-q_{ji}),\quad i=1\cdots n-1,
\end{equation}
which can be used to fix $q_{in},\:\forall i$:
\begin{equation}\label{qin}
\sum_{j>i}p_{ij}+\sum_{j<i}q_{ji}-\sum_{j<i}p_{ji}-\sum_{i<j<n}q_{ij}=q_{in}.
\end{equation}
The Euclidean time integrations are to be taken over the simplex $t_{1}>t_{2}>\dots>t_{n-1}>0$.
Using the simple formula
\begin{eqnarray}
\intop_{0}^{\infty}dt_{n-1}\dots\intop_{t_{3}}^{\infty}dt_{2}\intop_{t_{2}}^{\infty}dt_{1}e^{-(\sum_{i=1}^{n-1}m_{i}t_{i})} = \frac{1}{m_{1}(m_{1}+m_{2})\cdots(\sum_{i=1}^{n-1}m_{i})},
\end{eqnarray}
together with eq.~(\ref{qin}), allows us to write:
\begin{eqnarray}
\intop_{0}^{\infty}dt_{n-1}\dots\intop_{t_{3}}^{\infty}dt_{2}\intop_{t_{2}}^{\infty}dt_{1}\langle 0|V(t_{1})\dots V(t_{n}=0)|0\rangle &= & 2(\frac{R}{2\pi})^{2nb^{2}}\frac{e^{\frac{2\pi b^{2}}{R\omega_{\varphi_0}}n^{2}}R^{2n-1}}{(4\pi)^{n-1}\Gamma(2b^{2})^{n(n-1)}}\cr\cr
&& \hskip -2.25in \times \intop_{0}^{N_c}\frac{\prod_{i=1}^{n-1}\prod_{j=i+1}^{n}dp_{ij}p_{ij}^{\xi}}{\prod_{k=1}^{n-1}(\sum_{j=k+1}^{n}\sum_{i=1}^{k}p_{ij})}\intop_{0}^{N_c}\prod_{i=1}^{n-2}\prod_{j=i+1}^{n-1}dq_{ij}q_{ij}^{\xi}\cr\cr
&& \hskip -2.25in \times \prod_{l=1}^{n-1}(\sum_{m>l}p_{lm}+\sum_{m<l}q_{ml}-\sum_{m<l}p_{ml}-\sum_{l<m<n}q_{lm})^{\xi}\cr\cr
&& \hskip -2.25in \times\Theta_{H}(\sum_{m>l}p_{lm}+\sum_{m<l}q_{ml}-\sum_{m<l}p_{ml}-\sum_{l<m<n}q_{lm})\cr\cr
&& \hskip -2.25in \times \Theta_{H}(1-(\sum_{m>l}p_{lm}+\sum_{m<l}q_{ml}-\sum_{m<l}p_{ml}-\sum_{l<m<n}q_{lm})),\label{eq:genform}
\end{eqnarray}
where
\[
\xi=2b^{2}-1.
\]
In eq.~(\ref{eq:genform}) the time dependence of the zero mode correlators
were neglected by an analogous argument to that of the $\delta E_{0}^{(2)}$
case. The cutoff $N_c$ can be scaled out by transforming to new
variables $p_{ij}=N_c\tilde{p}_{ij}$, $q_{ij}=N_c\tilde{q}_{ij}$,
resulting in the final form of the leading order asymptotic
\begin{align}
\delta E_{0}^{n)}\approx & (-1)^{n-1}2\mu_{ShG}^{n}\left(\frac{R}{2\pi}\right)^{2nb^{2}}e^{\frac{2\pi b^{2}}{R\omega_{\varphi_0}}n^{2}}\frac{R^{2n-1}N_c^{2(1-n)+2b^{2}(n^{2}-n)}}{(4\pi)^{n-1}\Gamma(2b^{2})^{n(n-1)}}I_{n},
\end{align}
with the integral $I_n$ defined as
\begin{eqnarray}
I_{n} &=& \intop_{0}^{1}\frac{\prod_{i=1}^{n-1}\prod_{j=i+1}^{n}dp_{ij}p_{ij}^{\xi}}{\prod_{k=1}^{n-1}(\sum_{j=k+1}^{n}\sum_{i=1}^{k}p_{ij})}\intop_{0}^{1}\prod_{i=1}^{n-2}\prod_{j=i+1}^{n-1}dq_{ij}q_{ij}^{\xi}\cr\cr
 && \hskip -0.25in \times \prod_{l=1}^{n-1}(\sum_{m>l}p_{lm}+\sum_{m<l}q_{ml}-\sum_{m<l}p_{ml}-\sum_{l<m<n}q_{lm})^{\xi}\cr\cr
&& \hskip -0.25in \times \Theta_{H}(\sum_{m>l}p_{lm}+\sum_{m<l}q_{ml}-\sum_{m<l}p_{ml}-\sum_{l<m<n}q_{lm})\cr\cr
 && \hskip -0.25in \times \Theta_{H}(1-(\sum_{m>l}p_{lm}+\sum_{m<l}q_{ml}-\sum_{m<l}p_{ml}-\sum_{l<m<n}q_{lm})).
\end{eqnarray}
This is the form that we use as a starting point in Section \ref{min_resum}.

\section{Form Factors of the Sinh-Gordon Model} \label{sec:FFappendix}
In this appendix we provide the explicit form of infinite volume form factors needed for the implementation of the form factor TSM in Section \ref{FFTCSASECTION}.

Using the $S$-matrix (eq.~\ref{SmatrixShG}) and the integrability of the model, one can compute the exact form factors of the local operators ${\mathcal O}$ on the multi-particle states of the theory \cite{Fring:1992pt,Koubek:1993ke}:
\be
\langle 0 | {\mathcal O}(0) |\theta_1,\ldots,\theta_n \rangle \equiv F_k^{\mathcal O}(\theta_1,\ldots,\theta_n ) .
\label{FFdef}
\ee
Introducing the notation
\be
[k] \,=\,\frac{\sin k\,\pi B}{\sin \pi B},
\ee
where the function $B(b)$ is given in eq.~\ref{importantB},
the form factors of the exponential fields in the ShG model can be written as
\be
\langle 0 | e^{k b \phi(0)} | \theta_1,\ldots,\theta_n \rangle \,=\, \langle e^{k b \phi} \rangle \, F_n(\theta_1,\ldots,\theta_n),
\ee
where $\langle e^{k b \phi} \rangle$ is the VEV given in
eq.~\ref{VEVexp} and $F_n(\theta_1,\dots,\theta_n)$ is given by
\begin{equation}
F_n(\theta_1,\dots,\theta_n)\,=\,H_n\,Q_n(x_1,\dots,x_n)\,\prod^n_{i<j}\frac{F_{min}(\theta_{ij})}{x_i+x_j}\,,
\label{F_n}
\ee
where
\begin{eqnarray}
F_{min}(\theta) &=& \exp\bigg[-2\int_0^\infty \frac{d t}{t}\,\frac{\sinh\left(\frac{t B}{2}\right) \sinh\left(\frac{t(1- B)}{2}\right)}{\sinh(t)\cosh\left(\frac{t}{2}\right)} \cos\left(\frac{t(i \pi - \theta)}{\pi}\right)\bigg]\,,\cr\cr
H_n &=& \left(\frac{4 \sin \pi B}{F_{min}(i \pi)} \right)^\frac{n}{2} \, [k], ~~~ x_i=e^{\theta_i}, \label{FMINSHG}
\end{eqnarray}
and $Q_n(x_1,\dots,x_n)$ are symmetric polynomials in $x_i$ given by
\begin{equation}
Q_n(x_1,\ldots,x_n) \,=\, {\rm det} \, M, ~~~ M_{i,j} \,=\, [i - j +k]\, \sigma_{2 i - j}^{(n)},
\end{equation}
with $\sigma^{(n)}_s$ the elementary symmetric polynomials in $n$ variables of total degree $s$. Form factors of the ShG model were also studied in \cite{Lukyanov:1997bp,Lashkevich:2013yja}.


\section{Reverse Communication Protocols and Chiral Factorization\label{sec:Improved-exact-TCSA}}
There are several iterative numerical methods (L\'anczos, Arnoldi, Jacobi-Davidson,
etc.) to obtain the smallest eigenvalues (and the corresponding eigenvectors) of a Hamiltonian. A common
feature of these algorithms is that they do not need all of the Hamiltonian's matrix elements
to be stored in memory. What they do require is a routine by which 
arbitrary vectors are acted upon by the Hamiltonian matrix.  We will
describe in this section how the tensor product nature of the Hilbert space for the sinh-Gordon model makes this matrix-vector multiplication
remarkably efficient. 


\subsection{Separating Zero and Oscillator Modes}
As a demonstration of the numerical efficiency that can be obtained, we begin by factorizing the Hilbert space into (merely) two pieces,
\begin{equation}
\mathcal{H} = \mathcal{H}_{ZM}\otimes\mathcal{H}_{\rm osc},
\end{equation}
one encompassing the zero mode $\mathcal{H}_{ZM}$ and one the oscillator modes, $\mathcal{H}_{\rm osc}=\mathcal{H}_L\otimes\mathcal{H}_R$.
Let $N={\rm dim}\mathcal{H}$, $N_{ZM}={\rm dim}\mathcal{H}_{ZM}$, $N_{osc}={\rm dim}\mathcal{H}_{osc}$ be the sizes of these various Hilbert spaces.

We want to compute the action of our Hamiltonian upon a vector $|v\rangle$.  Normally this would require $N^2$ multiplications.  We show that this can be reduced by a factor of $N_{ZM}$.  To do so, we write $|v\rangle$ in terms of our tensored Hilbert space:
\begin{equation}
|v\rangle = \sum_Jv_J|j\rangle \equiv \sum_{j_0,\tilde j} v_{j_0\tilde j} |j_0\rangle\otimes |\tilde j\rangle.
\end{equation}
with $|j_0\rangle$ a basis vector for $\mathcal{H}_{ZM}$ and $|\tilde j\rangle$ a basis vector of $\mathcal{H}_{osc}$.
The Hamiltonian consists of a sum of matrices $\Phi$ whose matrix elements respect the tensor product structure:
\begin{equation}
\Phi_{IJ} = (\Phi^0)_{i_0j_0}(\tilde{\Phi})_{\tilde k\tilde l},
\end{equation}
where $\Phi^0/\tilde\Phi$ act in $\mathcal{H}_0/\mathcal{H}_{osc}$.
Applying $\Phi$ to $|v\rangle$ then leads to the need to evaluate:
\begin{equation}
\sum_{J=1}^N\Phi_{IJ}v_{J} = \sum_{j_0=1}^{N_{ZM}}  \sum_{\tilde j=1}^{N_{osc}}  (\Phi^0)_{i_0j_0}(\tilde{\Phi})_{\tilde i \tilde j}v_{j_0\tilde j} .
\end{equation}
To do so we first perform the matrix-vector multiplication involving the oscillator modes:
\begin{equation}
W_{j_0\tilde i}=\sum_{\tilde j=1}^{N_{osc}}(\tilde{\Phi})_{\tilde i \tilde j} v_{j_0\tilde j}.
\end{equation}
This involves $N_{ZM}N_{osc}^2$ multiplications.  We now follow this by the matrix-vector product over the zero mode space:
\begin{equation}
U_{i_0\tilde i}=\sum_{j_0=1}^{N_{ZM}}(\Phi_{0})_{i_0j_0}W_{j_0\tilde i}.
\end{equation}
This involves another $N_{ZM}^2N_{osc}$ multiplications.  If we write $|u\rangle= \Phi |v\rangle$, its components, $u_I$ are defined as
\begin{equation}
u_{I}\equiv u_{i_0\tilde i}=U_{i_0\tilde i}.
\end{equation}
As is typical in our computations, $N_{ZM} \ll N_{osc}$, we see that we have reduced the total number of multiplications by an approximate factor of $N_{ZM}$.  The above algorithm is a variant of the \emph{Shuffle Algorithm} where `shuffle' refers to the reshuffling
of elements of the vector $|v\rangle$ into a multi-index tensor $v_{j_0\tilde j}$.

\subsection{Exploiting the Structure of $\mathcal{H}_{osc}$}

The above multiplication algorithm can be further optimized by taking into
account the chiral structure of the oscillator Hilbert space. Here
we will account for momentum conservation. Up to a chiral cutoff $N_{c}$, the oscillator Hilbert space in the momentum zero 
sector takes the form:
\begin{equation}
(\mathcal{H}^{(0)}_L\otimes{\mathcal{H}}^{(0)}_R)\oplus(\mathcal{H}^{(1)}_L\otimes{\mathcal{H}}^{(1)}_R)\oplus\dots\oplus(\mathcal{H}^{(N_{c})}_L\otimes{\mathcal{H}}^{(N_{c})}_R).
\end{equation}
Here $\mathcal{H}^{(m)}_{L,R}$ is the Hilbert space of all states in the oscillator space of level $m$ (i.e. the state $a^\dagger_{n_1}\cdots a^\dagger_{n_k}|0\rangle \in \mathcal{H}_R$ belongs to $\mathcal{H}^{(m)}_R$ if $\sum n_l = m$).
The dimensionality of $\mathcal{H}^{(i)}_{L,R}$ is
\begin{equation}
{\rm dim} \mathcal{H}^{(i)}_{L/R}=\mathcal{P}(i)
\end{equation}
where $\mathcal{P}(i)$ denotes the number of integer partitions
of $i$ with $\mathcal{P}(0)=1$.  The dimension of the chiral/anti-chiral Hilbert space is
\begin{equation}
N_{L/R} = \sum_{j=0}^{N_c}\mathcal{P}(j),
\end{equation}
while the dimension of the oscillator Hilbert space (of zero momentum states) is 
\begin{equation}
N_{osc} = \sum_{j=0}^{N_c}\mathcal{P}(j)^2.
\end{equation}

We can proceed just as in the previous section.  We write the vector $|v\rangle$ that we want to multiply as
\begin{equation}
|v\rangle = \sum_{j_0=1}^{N_{ZM}}\sum_{k=0}^{N_c}\sum_{n_k,\bar{n}_k=1}^{\mathcal{P}(k)} v_{j_0n_k\bar{n}_k}^{(k)}|j_0\rangle_0\otimes|k,n_k\rangle_L \otimes |k,\bar{n}_k\rangle_R 
\end{equation}
where the same level index $k$ appearing in the left/right chiral vectors encodes momentum conservation. The matrix $\Phi$ representing one of the terms in the Hamiltonian is
\begin{align}
\Phi = &\sum_{i_0,j_0=1}^{N_{ZM}}\sum_{i,j=0}^{N_c}\sum_{n_i,\bar{n}_i=1}^{\mathcal{P}(i)}\sum_{n_j,\bar{n}_j=1}^{\mathcal{P}(j)}(\Phi_{0})_{i_0j_0}\left(\Phi_L^{(i,j)}\right)_{n_i,n_j} \left(\Phi_R^{(i,j)}\right)_{\bar{n}_i,\bar{n}_j} \nonumber \\
& \hskip .2in \times |i_0\rangle_0 \langle j_0|_0\otimes|i,n_i\rangle_L \langle j,n_j |_L\otimes|i,\bar{n}_i\rangle_R \langle j,\bar{n}_j |_R .
\end{align}
Applying $\Phi$ to $|v\rangle$ yields 
\begin{eqnarray}
\Phi_{IJ}v_{J} =\sum_{j_0=1}^{N_{ZM}}\sum_{j=0}^{N_c}\sum_{n_j,\bar{n}_j=1}^{\mathcal{P}(j)}\left(\Phi_{0}\right)_{i_0j_0}\left(\Phi_L^{(i,j)}\right)_{n_i,n_j} \left(\Phi_R^{(i,j)}\right)_{\bar{n}_i,\bar{n}_j}v_{j_0n_j\bar{n}_j}^{(j)} .
\end{eqnarray}
We now determine the number of multiplications necessary to evaluate this expression.  The first step is to evaluate the action of $\Phi_{R}$.  This costs $N_{osc}N_LN_{ZM}$ multiplications.
Similarly the application of the $\Phi_{L}$ matrix involves $N_{osc} N_RN_{ZM}$ multiplications.  Finally the action of $\Phi_0$ costs $N_LN_RN_{ZM}^2$ multiplications.  The total number of required multiplications is thus
\begin{equation}
N_{osc}(N_L+N_R)N_{ZM} + N_LN_RN_{ZM}^2.
\end{equation}
For larger values of $N_c$, the number of needed multiplications can be several hundred times smaller than
using the naive matrix-vector multiplication involving $(N_{osc}N_{ZM})^2$ multiplications.

\newpage 

\bibliographystyle{elsarticle-num}
\bibliography{sinh_gordon_paper}
\end{document}